\documentclass[11pt]{article}

\usepackage{threeparttable}
\usepackage{amsfonts}
\usepackage{amsmath}
\usepackage{fancyhdr}
\usepackage{epsfig,morefloats}
\usepackage{booktabs,multirow,multicol,longtable}
\usepackage{mathrsfs} 
\usepackage{amsbsy}
\usepackage{amsthm}
\usepackage{amssymb}
\usepackage{color}
\usepackage{lscape}
\usepackage{rotating}
\usepackage{caption}
\usepackage{subfigure}
\usepackage{graphicx}
\usepackage{subcaption}
\usepackage[compact]{titlesec}
\usepackage[breaklinks,colorlinks,linkcolor=black,citecolor=blue,urlcolor=black]{hyperref}
\usepackage{lineno}
\usepackage{cancel}
\usepackage{multirow}

\usepackage{makecell} 
\usepackage{array}    
\usepackage{appendix} 

\usepackage{setspace}
\newcommand{\p}{{\rm p}}

\definecolor{mygray}{gray}{0.6}

\def\be{\begin{equation}}
	\def\ee{\end{equation}}
\def\bea{\begin{eqnarray}}
	\def\eea{\end{eqnarray}}
\def\bd{\begin{displaymath}}
	\def\ed{\end{displaymath}}
\def\bda{\begin{eqnarray*}}
	\def\eda{\end{eqnarray*}}
\def\bsm{\begin{small}}
	\def\esm{\end{small}}

\def\t0{\theta_0}

\def\ha1{\hat \beta_1}

\def\bnt{\begin{enumerate}}
	\def\ent{\end{enumerate}}
\def\p{{ \mathrm{p} }}

\def\calL{{ \mathcal{\scriptscriptstyle L} }}
\def\calF{{ \mathcal{\scriptscriptstyle F} }}

\def\calR{{ \mathcal{\scriptscriptstyle R} }}

\def\calM{{ \mathcal{\scriptscriptstyle M} }}

\def\calS{{ \mathcal{\scriptscriptstyle S} }}
\def\calG{{ \mathcal{\scriptscriptstyle G} }}
\def\calF{{ \mathcal{\scriptscriptstyle F} }}

\def\bsc{\begin{scriptsize}}
	\def\esc{\end{scriptsize}}

\newtheorem{theorem}{Theorem}

\newtheorem{lemma}{Lemma}

\newtheorem{proposition}{Proposition}

\theoremstyle{definition}

\newtheorem{remark}{Remark}

\makeatletter
\newcommand{\figcaption}{\def\@captype{figure}\caption}
\newcommand{\tabcaption}{\def\@captype{table}\caption}
\makeatother

\newcommand{\sgn}{\mbox{\rm sgn}}

\newcommand{\supp}{\mathrm{supp}}

\newcommand{\bA}{{\mathbf A}}
\newcommand{\bB}{{\mathbf B}}

\newcommand{\bJ}{{\mathbf J}}
\newcommand{\bG}{{\mathbf G}}
\newcommand{\bH}{{\mathbf H}}
\newcommand{\bI}{{\mathbf I}}

\newcommand{\bM}{{\mathbf M}}
\newcommand{\bQ}{{\mathbf Q}}

\newcommand{\bU}{{\mathbf U}}
\newcommand{\bV}{{\mathbf V}}

\newcommand{\ba}{{\mathbf a}}
\newcommand{\bb}{{\mathbf b}}

\newcommand{\bfe}{{\mathbf e}}
\newcommand{\bff}{{\mathbf f}}
\newcommand{\bh}{{\mathbf h}}
\newcommand{\bg}{{\mathbf g}}

\newcommand{\bw}{{\mathbf w}}
\newcommand{\bx}{{\mathbf x}}
\newcommand{\by}{{\mathbf y}}
\newcommand{\bz}{{\mathbf z}}

\newcommand{\bbeta}  {\boldsymbol{\beta}}
\newcommand{\bfeta}  {\boldsymbol{\eta}}

\newcommand{\bdelta} {\boldsymbol{\delta}}
\newcommand{\blambda}{\boldsymbol{\lambda}}

\newcommand{\bSigma}{\boldsymbol{\Sigma}}

\newcommand{\bTheta} {\boldsymbol{\Theta}}

\newcommand{\bpsi}{\boldsymbol{\psi}}

\newcommand{\btheta} {\boldsymbol{\theta}}
\newcommand{\bxi} {\boldsymbol{\xi}}
\newcommand{\bXi} {\boldsymbol{\Xi}}

\newcommand{\bGamma} {\boldsymbol{\Gamma}}

\newcommand{\bC}{{\mathbf C}}
\newcommand{\bD}{{\mathbf D}}
\newcommand{\bzero}{{\mathbf 0}}

\newcommand{\bthetazero} {\boldsymbol{\theta}_{0}}
\newcommand{\hbthetan} {\hat{\boldsymbol{\theta}}_{n}}

\newcommand{\bseta} {\boldsymbol{\eta}}
\newcommand{\mm} {\mathcal{M}}

\newcommand{\bfP}{{\mathbf P}}

\newcommand{\bfA}{{\mathbf A}}
\newcommand{\bfB}{{\mathbf B}}
\newcommand{\bfC}{{\mathbf C}}
\newcommand{\bfD}{{\mathbf D}}

\newcommand{\bfH}{{\mathbf H}}
\newcommand{\bfJ}{{\mathbf J}}
\newcommand{\bfR}{{\mathbf R}}

\newcommand{\bfV}{{\mathbf V}}

\newcommand{\bfx}{{\mathbf x}}

\newcommand{\bfy}{{\mathbf y}}

\newcommand{\bfg}{{\mathbf g}}

\newcommand{\bfw}{{\mathbf w}}
\newcommand{\bfz}{{\mathbf z}}

\newtheorem{condition}{Condition}
\theoremstyle{definition}

\newcommand{\blind}{1}

\renewcommand{\theequation}{\arabic{equation}}

\allowdisplaybreaks

\usepackage{amsmath, amssymb, amsthm, latexsym,color}
\usepackage[square,sort,comma,numbers,sectionbib,round]{natbib}
\usepackage{graphicx}
\usepackage{fancyhdr}
\usepackage{algpseudocode}
\usepackage{algorithm}
\usepackage{algpseudocode}
\usepackage{enumitem}
\usepackage{geometry}
\theoremstyle{definition}

\def\T{{ \mathrm{\scriptscriptstyle \top} }}

\newcommand{\bq}{{\mathbf q}}
\newcommand{\bepsilon}{\boldsymbol{\varepsilon}}

\def\6bullets{\bullet\bullet\bullet\bullet\bullet\bullet}

\DeclareMathAlphabet\EuScriptBF{U}{eus}{b}{n}

\usepackage{authblk}

\def\today{\number\day~\ifcase\month\or
	January\or February\or March\or April\or May\or June\or
	July\or August\or September\or October\or November\or December\fi~\number\year}

\def \0{\scriptscriptstyle 0}
\def \mT{\mathcal{\scriptscriptstyle T}}

\theoremstyle{definition}
\allowdisplaybreaks
\begin{document}

\bibliographystyle{agsm}
\bibpunct{(}{)}{,}{a}{}{;}
	

	
	
	\if1\blind
	{
		\title{\bf  Empirical likelihood approach for high-dimensional moment restrictions with dependent data}
		

\author[1,2]{Jinyuan Chang}
\author[1]{Qiao Hu}
\author[3]{Zhentao Shi}
\author[1]{Jia Zhang}

\affil[1]{\it \small Joint Laboratory of Data Science and Business Intelligence, Southwestern University of Finance and Economics, Chengdu, Sichuan, China}
\affil[2]{\it \small Academy of Mathematics and Systems Science, Chinese Academy of Sciences, Beijing, China}
\affil[3]{\it \small Department of Economics, Chinese University of
Hong Kong, Hong Kong SAR, China}

		\date{}
		
		\maketitle
	} \fi
	
	\if0\blind
	{
		\bigskip
		\bigskip
		\bigskip
		\begin{center}
			{\LARGE\bf  Empirical likelihood approach for high-dimensional moment restrictions with dependent data }
		\end{center}
	} \fi
	
	

	
	
	

	\onehalfspacing

	
	
\begin{abstract}
  
Economic and financial models---such as vector autoregressions, local projections, and multivariate volatility models---feature complex dynamic interactions and spillovers across many time series.
These models can be integrated into a unified framework, with high-dimensional parameters identified by moment conditions. As the number of parameters and moment conditions may surpass the sample size, we propose adding a double penalty to the empirical likelihood criterion to induce sparsity and facilitate dimension reduction. Notably, we utilize a marginal empirical likelihood approach despite temporal dependence in the data. Under regularity conditions, we provide asymptotic guarantees for our method, making it an attractive option for estimating large-scale multivariate time series models. We demonstrate the versatility of our procedure through extensive Monte Carlo simulations and three empirical applications, including analyses of US sectoral inflation rates, fiscal multipliers, and volatility spillover in China's banking sector.

	\end{abstract}
	\bigskip
	\noindent
	{\sl keywords}: $\alpha$-mixing, asymptotic analysis, confidence region,	high dimensionality, penalized likelihood  
	
\bigskip \bigskip


\section{Introduction}


Econometrics began almost a hundred years ago with a focus on deciphering business cycles and economic fluctuations, as demonstrated by early researchers \cite{fisher1925our}, \cite{frisch1933propagation}, and \cite{tinbergen1939}. These pioneers carried out their quantitative analyses with limited macro-level, low-frequency datasets. However, with the progress in information technology, economists now have access to extensive long-term data series that reflect the state of real economies and financial markets from diverse angles. Despite this, the complexity of the economic landscape continues to escalate, characterized over the last century by exceptional growth interspersed with recessions, wars, and crises. The age-old inquiry into the interplay between economic variables and their evolving paths still holds a prominent place in the current big data era.

An important class of economic models is characterized by moment conditions.
To fix ideas, let $\{\bfx_t\}_{t=1}^n$ be  $d$-dimensional random vectors, and $\btheta=(\theta_1,\ldots,\theta_p)^{\T}$ be a $p$-dimensional parameter in a parameter space $\bTheta$. For an $r$-dimensional estimating function $\bfg(\bfx;\btheta)=\{g_1(\bfx;\btheta),\dots,g_r(\bfx;\btheta)\}^\T$, the information for the true parameter $\btheta_0$ is identified by the  moment condition 
	\begin{align}\label{eq:identification}
	 \mathbb{E}\{{\bfg(\bfx_t;\btheta_0)}\}={\bf0}
	\end{align}
    for each $t=1,\ldots,n$.
In the traditional asymptotic analysis, it is assumed that the sample size $n$ approaches infinity while the model remains unchanged. Economic theory, however, often relies on the orthogonality of variables as a justification for moment conditions, but it generally lacks clarity regarding which variables should be included or which moment conditions should be utilized. 
For example, \cite{angrist1991does} and \cite{eaton2011} involve a large number of moments, and \cite{blundell1993we} and \cite{fan2014endogeneity} further contain many parameters. 
Beyond these applications in labor economics, international trade, and household consumption, multivariate time series models provide fertile ground for the development of complex economic models.

This study is driven by three commonly utilized multivariate time series models, as detailed in \cite{lutkepohl2005new}'s monograph. First, 
the vector autoregressive (VAR) model \citep{sims1980macroeconomics} is a multivariate extension of the univariate autoregressive (AR) model. In a VAR model composed of $d$ variables, the total number of slope coefficients is $p= O(d^2)$. This is referenced as our Example 1 in Section \ref{sec:simulations}. Conventionally, the VAR model only includes a limited number of variables, which is not designed to handle a modern large system of variables.

The second example is concerning the impulse response function (IRF), which describes the effect of an external shock to a target variable. IRF is a key object of macroeconomic interest; see the survey by \cite{nakamura2018identification}. 
Traditionally, IRF is implied by a fully and ``correctly specified'' VAR model. Recently, \citet{Jorda2005} introduced \emph{local projection} (LP), 
a much simpler method which involves a series of single-equation regressions over the forecast horizons $h = 0, 1, \ldots, H$. If each individual regression includes $d$ regressors, the system yields $p = d(H+1)$ slope coefficients to be estimated.

In addition to the above two examples of conditional mean models, volatility is fundamental for financial risk management. The most recognized 
univariate time series volatility models are the autoregressive conditional heteroskedasticity (ARCH) model \citep{engle1982autoregressive} and its generalized version (GARCH) \citep{bollerslev1986generalized}. In typical financial markets, numerous assets are traded daily. To handle this complexity, multivariate volatility models, such as the multivariate ARCH (MARCH) and multivariate GARCH (MGARCH) \citep{EngleandKroner1995}, have been developed. As will be explained in Example 3 in Section \ref{sec:simulations}, these models require high-dimensional coefficients to capture the interdependencies among various assets.

Moment restrictions serve as a comprehensive framework for identifying the parameters of interest across all the above-mentioned three multivariate time series models. Within the context of these moment constraints, the generalized method of moments (GMM) is widely regarded as the default econometric estimation approach \citep{hansen1982large, hansen1982generalized}. Although Hansen's two-step GMM achieves asymptotic efficiency under the traditional asymptotic scenario with fixed $r$ and $p$, its performance in finite samples can be unsatisfactory, as is highlighted by \citet{altonji1996small}. This issue primarily arises from the inversion of the estimated covariance matrix of the estimating functions. Recently, \cite{cheng2023weight} investigate the bias of GMM in cases where $p$ is proportional to, but smaller than $n$. When $p$ exceeds $n$, \cite{belloni2018high} propose using a sup-norm objective criterion function to manage many moments, departing from GMM’s quadratic form.

An alternative to GMM is empirical likelihood (EL) \citep{Owen(1988), QinandLawless(1994)}, which can be viewed through the lens of information theory \citep{kitamura1997information}. Unlike methods that require explicit computation of the covariance matrix of the estimating functions, EL benefits from reduced variance due to higher-order enhancements \citep{newey2003higher}. When dealing with models incorporating many moments, the EL criterion function can be augmented with a penalized approach to regularize both the multitude of moments and high-dimensional coefficients. This branch of theoretical properties, specifically for independently and identically distributed (i.i.d.)~observations, has been extended by \cite{otsu2007penalized}, \cite{leng2012penalized}, \cite{shi2016econometric}, and \citet{CTW_2018}. Penalized method theory is relatively straightforward with i.i.d.~data. To our knowledge, this paper is the first to explore the EL methodology in the context of high-dimensional temporally dependent data.

In classical low-dimensional settings, \cite{kitamura1997empirical} 
advocates using blocks of time series observations in EL to preserve the temporal dependence and recover the Wilks' phenomenon.
This blocking technique is employed by \cite{CCC(2015)} under a moderately high-dimensional environment where $r/n\to 0$.
However, the blocking technique is inconvenient when 
dealing with high-dimensionality in both the parameters and moments. 
This paper maintains the simplest approach: treating the EL as if the data is i.i.d., 
which can be interpreted as the \emph{marginal} EL for time series. 
Though marginal likelihood estimation has been used in low-dimensional models \citep{levenbach1972estimation} and has been applied in financial applications (\citealp{stambaugh1997analyzing}, \citealp{patton2006estimation})
for integration of varying lengths, this paper is the first to investigate the scheme of marginalization in the framework of EL for time series, distinct from the approaches by \cite{ChangTangWu2013,chang2016local}.

We establish rigorous asymptotic theory for the penalized EL (PEL) with many parameters and many moments under time dependence of $\alpha$-mixing.
To accommodate stronger temporal dependence, we introduce (in Condition \ref{con:mixingdecay}) a diverging quantity $L_n$ within the $\alpha$-mixing coefficient.
To address the high-dimensional challenges in this framework, 
we bound the tail probabilities of certain key statistics by novel inequalities,
which are constructed via self-normalized sum inequalities \citep{JingShaoWang2003}.
We show that under sparsity conditions, the PEL approach delivers consistent estimation, and the resulting PEL estimator is asymptotically normally distributed. 
For investigations focused on a low-dimensional parameter, we further employ a projected PEL (PPEL) method \citep{Chang2020} to eliminate the bias induced by the high-dimensional nuisance parameters, thereby reinstating the usual inference method using the $t$-statistic. 

This procedure enables the application of the method to a wide range of multivariate time series models with many parameters, including VAR, LP, and MARCH/MGARCH models as discussed. Extensive Monte Carlo simulations show that our method performs well in finite sample. We employ it to further study the persistence and spillover of the USA's sectoral inflation, the magnitude of the fiscal multiplier, and the volatility network of China's banking industry.





This paper stands on the large literature of time series, empirical likelihood, and high-dimensional estimation. 
It leverages the penalized estimation in \cite{CTW_2018} and \cite{Chang2020}, developed under the i.i.d.~setting.
These steps are carried over and adapted to the environment with temporal dependence. 
Compared to the GMM-type alternative, \cite{belloni2018high} is developed in the i.i.d.~environment, which is critical for their sample splitting;
sample splitting in time series is much more challenging and may adversely affect finite sample performance when the time length is moderate in practice.
On the other hand, the literature of high-dimensional time series has witnessed specific proposals for standalone models. For example, 
\cite{shi2022} and \cite{mei2024lasso} focus on high-dimensional time series dense regressions and sparse regressions, respectively.
Under the assumption of Gaussian errors, \cite{kock2015oracle} develop Lasso-type penalized estimation for VAR models. 
\cite{caner2018high} attack high-dimensional GMM-type models, where their linear setting facilitates the estimation of the large weighting matrix.
\cite{adamek2024local} provide theory for a single-equation regression with many covariates for local projection, and deal with ridge-type regularization in fixed dimension.  
Our procedure provides a unified framework to handle models defined by moment conditions.





The rest of the paper is organized as follows. 
Section \ref{sec:preliminaries} sets up the model and the technical conditions. 
The consistency and asymptotic distribution of the PEL estimator are established in Section \ref{sec:consistency}, and the asymptotic normality of PPEL is presented in Section \ref{sec:Inference}.
We carry out Monte Carlo simulations in Section \ref{sec:simulations}, and showcase our method in three empirical applications in Section \ref{sec:empirical-app}. 
The code and data are available at the  GitHub repository: \url{https://github.com/JinyuanChang-Lab/PenalizedELwithDependentData}.
All proofs are relegated to the Appendices.

{\bf Notation}. We use the abbreviations ``w.p.a.1'' and ``w.r.t'' to denote, respectively, \emph{with probability approaching one} and \emph{with respect to}. For any real number $x$, define
$
\lfloor x \rfloor = \max \{ q \in  \mathbb{Z} : q \leq  x \}
$,
where $\mathbb{Z}$ denotes the set of all integers.
For two sequences of positive numbers $\{a_n\}$ and $\{b_n\}$, we write $a_n\lesssim b_n$ or $b_n\gtrsim a_n$ if there exists a positive constant $c$ such that $\lim\sup_{n\to \infty} a_n/b_n\leq c$, and $a_n\asymp b_n$ if and only if $a_n\lesssim b_n$ and $b_n\lesssim a_n$  hold simultaneously.
We write $a_n\ll b_n$ or $b_n\gg a_n$ if $\lim\sup_{n\to \infty} a_n/b_n=0$. 
Let ``vec'' and ``vech'' be the vector operators that stack the columns of a matrix and the upper triangular part of a matrix, respectively, into a vector.
For a positive integer $q$, we write $[q]=\{1,\ldots,q\}$, and let $\bI_{q}$ be the  $q\times q$ identity matrix. 
For a $q\times q$ symmetric matrix $\bQ$, denote by $\lambda_{\min}(\bQ)$ and $\lambda_{\max}(\bQ)$ the smallest and largest eigenvalues of $\bQ$, respectively. For a $q_1\times q_2$ matrix $\bB=(b_{i,j})_{q_1\times q_2}$, let $\bB^{\T}$ be its transpose, 
 $|\bB|_{\infty}=\max_{i\in[q_1],j\in[q_2]}|b_{i,j}|$ be the sup-norm, and $\|\bB\|_2=\lambda_{\max}^{1/2}(\bB^{\otimes2})$ be the spectral norm with $\bB^{\otimes2}=\bB\bB^{\T}$.  
Specifically, when $q=1$,
we use $|\bB|_{1}=\sum_{i=1}^{q_1}|b_{i,1}|$ and $|\bB|_2=(\sum_{i=1}^{q_1}b_{i,1}^2)^{1/2}$ to denote the $L_1$-norm and $L_2$-norm of the vector $\bB$.  For two square matrices $\bQ_1$ and $\bQ_2$, we say $\bQ_1\leq \bQ_2$ if $(\bQ_2-\bQ_1)$ is a positive semi-definite matrix.
The population mean is denoted by $\mathbb{E}(\cdot)$, and the sample mean is denoted  by $\mathbb{E}_n(\cdot)=n^{-1}\sum_{t=1}^{n}(\cdot)$. For a given index set $\mathcal{L}$, let $|\mathcal{L}|$ be its cardinality. For a generic multivariate function $\bh(\cdot;\cdot)$, we denote by $\bh_{\calL}(\cdot;\cdot)$ the subvector of $\bh(\cdot;\cdot)$ collecting the components indexed by $\mathcal{L}$. Analogously, we write $\ba_{\mathcal{L}}$ as the corresponding subvector of a vector $\ba$. For simplicity and when no confusion arises, we use the generic notation $\bh_t(\btheta)$ as equivalent to $\bh(\bfx_t;\btheta)$, and $\nabla_{\btheta}\bh_t(\btheta)$ for the first-order partial derivative of $\bh_t(\btheta)$ w.r.t $\btheta$. Denote by $h_{t,k}(\btheta)$ the $k$-th component of $\bh_t(\btheta)$. Let $\bar{\bh}(\btheta)=\mathbb{E}_n\{\bh_t(\btheta)\}$, and  write its $k$-th component as $\bar{h}_k(\btheta)=\mathbb{E}_n\{h_{t,k}(\btheta)\}$. Analogously, let $\bh_{t,\calL}(\btheta)=\bh_{\calL}(\bfx_t;\btheta)$ and $\bar{\bh}_{\calL}(\btheta)=\mathbb{E}_n\{\bh_{t,\calL}(\btheta)\}$.

\section{Preliminaries}\label{sec:preliminaries}

In this paper, we build up our theory with $\alpha$-mixing time dependence. 
Let  $\mathcal{F}_{-\infty}^u$ and $\mathcal{F}_{u}^{\infty}$ be the $\sigma$-fields generated by $\{\bfx_{t}\}_{t\leq u}$ and $\{\bfx_{t}\}_{t\geq u}$, respectively. The $\alpha$-mixing coefficient of the sequence $\{\bfx_t\}$  at lag $k$ is defined as 
	\begin{align}\label{eq:alpha-mixing}
	\alpha_{n}(k):=\sup_{t}\sup_{A\in\mathcal{F}_{-\infty}^t,\,B\in \mathcal{F}_{t+k}^\infty}|\mathbb{P}(A\cap B)-\mathbb{P}(A)\mathbb{P}(B)|
	\end{align}
    for each $k\geq 1$.
The notion of $\alpha$-mixing in $\eqref{eq:alpha-mixing}$  broadly characterizes serial dependence. 
Specifically, we impose the following assumption as in \cite{chang2024optimal} in our study.

\begin{condition} \label{con:mixingdecay}
	$\{\bfx_t \}_{t=1}^n$ is an $\alpha$-mixing sequence, and there exist some universal constants $K_1>1$, $K_2>0$ and $\varphi\geq 1$ such that $\alpha_{n}(k)\leq K_1\exp\{-K_2(L_n^{-1}k)^{\varphi}\}$ for any $k\geq 1$,  
    where $L_n>0$ can stay finite or diverge to infinity with $n$.
\end{condition}

Condition \ref{con:mixingdecay} does not require $\{\bfx_t\}_{t=1}^{n}$ to be strictly stationary.
For an independent sequence $\{\bfx_t\}_{t=1}^{n}$, we can select $L_n=1/2$ and $\varphi=\infty$.
For an $L_n$-dependent sequence $\{\bfx_t\}_{t=1}^{n}$, we can select $\varphi=\infty$.
A variety of time series models that are routinely used in economics and finance
are covered by  Condition \ref{con:mixingdecay}. 
For example, under some regularity conditions, the autoregressive-moving-average (ARMA) processes, the stationary Markov chains 
\citep{FanYao_2003}, and the stationary GARCH models \citep{CarrascoChen_2002}
satisfy $\alpha$-mixing with the exponentially decaying coefficient ($L_n=1$ and $\varphi=1$).
This condition further covers their multivariate generalizations of VAR and MGARCH (MGARCH includes MARCH as a special case); 
see \cite{HP09}, \cite{BFS11} and \cite{Wong2020}.

The quantity $L_n$ involved in Condition \ref{con:mixingdecay} accommodates practical scenarios of big data collected over time. Write $\bfx_t=(x_{t,1},\ldots,x_{t,d})^{\T}$. Consider the simple case when each univariate time series $\{x_{t,i}\}_{t=1}^n$ for $i \in [d]$ is $\alpha$-mixing with exponentially decaying $\alpha$-mixing coefficients, while those $d$ sequences are mutually independent. Theorem 5.1 of \cite{bradley2005basic} indicates that $\alpha_n(k)$ defined in \eqref{eq:alpha-mixing} satisfies $\alpha_n(k)\leq d\exp(-ck)$ for some universal constant $c>0$, which implies Condition \ref{con:mixingdecay} holds for $\varphi=1$ and $L_n\asymp \log d$. 
Our novel framework also covers high-frequency time series models.
Suppose the observed vector process is generated from $\bfx_{t}=\bfP\bz_{t\delta}$,
for $t\in[n]$, 
where $\bfP\in\mathbb{R}^{ d\times q}$ is a loading matrix and $\delta>0$ is the sampling interval.
Let the latent vector process $\bz_s=(z_{s,1},\ldots,z_{s,q})^{\T}$ consist of $q$ independent processes,
where each $z_{s,i}$ for $i\in[q]$ follows the diffusion model 
${\rm d} z_{s,i}
= \tilde{\mu}_i(z_{s,i})\,{\rm d}s+\tilde{\sigma}_i(z_{s,i})\,{\rm d}{W}_{s,i}$, 
with a univariate standard Brownian motion ${W}_{s,i}$ and two parametric functions $\tilde{\mu}_i(\cdot)$ and $\tilde{\sigma}_i(\cdot)$.
When $\tilde{\mu}_i(\cdot)$ and $\tilde{\sigma}_i(\cdot)$ satisfy certain conditions as those in Lemma 4 of \cite{ait2004estimators},
the observed $\{\bfx_t\}_{t=1}^{n}$ satisfies Condition \ref{con:mixingdecay} with $\varphi=1$ and $L_n=\delta^{-1}$, where $L_n$ diverges if $\delta\rightarrow 0$ as $n\rightarrow\infty$.

\subsection{Penalized empirical likelihood}
 
We first set up the estimation procedure.
We are interested in the $p$-dimensional parameter $\btheta_{0}\in \bTheta$ defined as the solution of the $r$ moment conditions \eqref{eq:identification}.
Based on \citet{Owen(1988), Owen(1990)}'s seminal idea, 
given the estimating equations $\{\bfg_t(\cdot)\}_{t=1}^n$, 
\cite{QinandLawless(1994)} define the EL as
	\begin{equation*}
	L(\btheta) = \sup\bigg\{ \prod_{t=1}^{n}\pi_t:\pi_t > 0,  \, \sum_{t=1}^{n}\pi_t = 1, \,  \sum_{t=1}^{n}\pi_t\bfg_t(\btheta)  =\mathbf{0}\bigg \}\,. 
	\end{equation*}
Maximizing $L(\btheta)$ can be equivalently carried out via the corresponding dual problem, and its optimizer is the EL estimator:
	\begin{equation}\label{eq:moment condition}
    	\bar{\btheta}_n = \arg\min_{\btheta\in\bTheta}\max_{\blambda\in \hat{\Lambda}_n(\btheta)}\sum_{t=1}^{n}\log\{1+\blambda^{\T}\bfg_t(\btheta)\}\,,
	\end{equation}
where  $\blambda=(\lambda_1,\ldots,\lambda_r)^{\T}$ and  $\hat{\Lambda}_n(\btheta)=\{\blambda\in\mathbb{R}^r:\blambda^{\T}\bfg_t(\btheta)\in \mathcal{V} \text{ for any }t\in[n]\}$ for an open interval $\mathcal{V}$   containing zero.

Most economic and financial time series typically consist of a few hundred observations or more. Meanwhile, models that characterize economic interactions are inherently complex and high-dimensional. Regularization is crucial for accurately estimating many parameters with limited time datasets. 
Shrinkage serves as a useful statistical technique for dimension reduction, with its effectiveness depending on the nature of the data and models. 
Arguably, sparsity is an extensively used assumption in high-dimensional models, exemplified by the success of Lasso \citep{tibshirani1996regression}, SCAD \citep{FanLi2001} and MCP \citep{Zhang2010}, which have been applied across various scientific fields.


Consider a model with the number of estimating equations $r$ and the number of parameters $p$ both potentially larger than the sample size $n$. 
Such a high-dimensional model can be estimated by \cite{CTW_2018}'s PEL method:
\begin{align}\label{eq:double-pen}
	\hat{\btheta}_n = \arg\min_{\btheta\in\bTheta}\max_{\blambda\in \hat{\Lambda}_n(\btheta)}\bigg[\frac{1}{n}\sum_{t=1}^{n}\log\{1+\blambda^{\T}\bfg_t(\btheta)\}-\sum_{j=1}^{r}P_{2,\nu}(|\lambda_j|)+\sum_{k=1}^{p}P_{1,\pi}(|\theta_k|)\bigg]\,,
\end{align}
where two penalty functions $P_{1,\pi}(\cdot)$ and $P_{2,\nu}(\cdot)$ with tuning parameters $\pi$ and $\nu$ are appended to the dual problem \eqref{eq:moment condition}. 
For any  penalty function $P_\tau(\cdot)$ with a tuning parameter $\tau$, let $\rho(t;\tau)=\tau^{-1}P_{\tau}(t)$ for any $t \in [0,\infty)$ and $\tau \in (0,\infty)$. 
Assume the penalty functions $P_{1,\pi}(\cdot)$ and $P_{2,\nu}(\cdot)$ belong to the following class as in \cite{LvandFan(2009)}:
	\begin{equation}\label{penalty}
	\begin{split}
	\mathcal{P}=\{P_{\tau}(\cdot): \rho(t;\tau) \text{ is increasing in } t \in [0,\infty) \text{ and has continuous derivative }\rho'(t;\tau) \\ \text{ for }t \in (0,\infty) \text{ with }  \rho'(0^+;\tau)>0, \text{where } \rho'(0^+;\tau)\text{ is independent of } \tau \}\,.
	\end{split}
	\end{equation}
	

The PEL estimator \eqref{eq:double-pen} is formulated as if $\bfx_t$ is i.i.d.,
which contrasts with \cite{kitamura1997empirical}:
``\emph{... studies the method of empirical likelihood in models with
weakly dependent processes. In such cases, if the likelihood function is
formulated as if the data process were independent, obviously empirical
likelihood fails.}''
To restore the Wilks' phenomenon, \cite{kitamura1997empirical} proposes the blocking technique for low-dimensional EL estimation under a fixed $r$, 
and this method is adopted by \cite{CCC(2015)} under $r\to \infty $ with $r/n\to 0$.
However, blocking with a length $b$ reduces the effective sample size from $n$ to $\lfloor n/b \rfloor$.
Asymptotic theory for a long vector of $\bfx_t$ would request a large block size to cope with the variable in $\bfx_t$ of the maximum temporal dependence. 
In the finite sample, a big $b$ for all time series would substantially reduce the effective sample size; 
on the other hand, choosing a block size for each time series 
would involve many more additional tuning parameters.
Block preserves nice statistical properties in low-dimensional time series models, but it is inconvenient in high-dimensional contexts.  
Our marginal EL in \eqref{eq:double-pen} circumvents the choice of block sizes.
We will maintain the PEL formulation as for the i.i.d.~data and develop the theory accordingly.

\subsection{Technical conditions}

When economic theory provides no clear guidance about which variables are the most relevant or which  moments are the most informative,  shrinkage methods are helpful as a data-driven device for variable and moment selection.
Given the $p$-dimensional true parameter $\btheta_0
=(\theta_{0,1},\dots,\theta_{0,p})^{\T}
$, let $\mathcal{S} = \{k\in[p]:\theta_{0,k}\neq 0 \}$ be the active set of cardinality $s=|\mathcal{S}|$,
where $\mathcal{S}$ marks the location of the non-zero parameters with  $s\ll p$. 
Before any attempt at estimation, the parameter of interest must be identifiable from the population model. 

	\begin{condition}\label{con:identity}
For any $\epsilon>0$,	 there is a universal constant $K_3>0$ such that 
	\begin{align*}
	\inf_{\btheta \in \bTheta: \,|\btheta_{\calS}-\btheta_{0,\calS}|_{\infty} > \epsilon,\,\btheta_{\calS^{\rm c}}={\bf0}} |\mathbb{E}\{\bar{\bfg}( \btheta)\}|_{\infty} \geq K_3\epsilon\,.
	\end{align*}
	\end{condition}
    

This assumption means that the expected values of the estimating functions at the true parameter $\btheta_0$ are significantly different from those outside a narrow vicinity of the active coefficient $\btheta_{0,\calS}$. The sup-norm on moment conditions is a basic necessity that allows for the inclusion of many weak or entirely irrelevant moments. Under this condition, the parameter is identified locally as described by \cite{chen2014local}.

We move on to the proceeding conditions which are standard regularity assumptions in the literature. 
For any index set $\mathcal{F}\subset[r]$ and $\btheta\in\bTheta$, define $\widehat{\bV}_{\calF}(\btheta)=\mathbb{E}_n\{\bfg_{t,\calF}(\btheta)^{\otimes2}\}$. When $\mathcal{F}=[r]$, we write $\widehat{\bV}_{\calF}(\btheta)=\widehat{\bV}(\btheta)$ for conciseness.

\begin{condition}\label{con:moments1}

({\rm a}) There exist some universal constants $K_4>0$ and $\gamma>4+4/(3\varphi)$ with $\varphi$ specified in Condition \ref{con:mixingdecay} such that 
\begin{align*}
\max_{t\in[n]}\max_{j\in[r]}\mathbb{E}\bigg\{\sup_{\btheta\in \bTheta}|g_{t,j}(\btheta)|^{{2\gamma}}    \bigg\} \leq  K_4~~\textrm{and}~~
\max_{j\in[r]}\sup_{\btheta\in\bTheta}\mathbb{E}_n\{|g_{t,j}(\btheta)|^{\gamma}\}=O_{\p}(1)\,.
\end{align*}
({\rm b}) There exists some universal constant $K_5>0$ such that 
\begin{align*}
\min_{j\in[r]} {\rm{Var}}\bigg\{\frac{1}{\sqrt{k}}\sum_{t=\ell+1}^{\ell+k}g_{t,j}(\btheta_0)\bigg\}\geq K_5~~\textrm{and}~~\min_{j_1,j_2\in[r]}{\rm{Var}}\bigg\{\frac{1}{\sqrt{k}}\sum_{t=\ell+1}^{\ell+k}g_{t,j_1}(\btheta_0)g_{t,j_2}(\btheta_0)\bigg\}
\geq K_{5}
\end{align*}
for all integers  $ k\geq 1$ and $\ell\geq 0$. ({\rm c}) There exists some universal constant $K_6>1$ such that 
\begin{align*}
K_6^{-1}<\lambda_{\min}[\mathbb{E}\{\widehat{\bV}(\bthetazero)\}] \leq \lambda_{\max}[\mathbb{E}\{\widehat{\bV}(\bthetazero)\}] <K_6\,.
\end{align*}
\end{condition}

Condition \ref{con:moments1}(a) restricts to be finite the $(2\gamma)$-th population moments 
of the estimating functions, and the $\gamma$-th sample moments to be $O_{\rm p}(1)$ in a uniform manner. At the true parameter $\btheta_0$, Condition \ref{con:moments1}(b) requires non-degenerate long-run variance, and 
Condition \ref{con:moments1}(c) ensures a well-behaved covariance matrix $\mathbb{E}\{\widehat{\bV}(\bthetazero)\}$ for the estimating functions, which is satisfied if the eigenvalues of ${\rm Var}\{\bg_1(\btheta_0)\},\ldots,{\rm Var}\{\bg_n(\btheta_0)\}$
are uniformly bounded away from zero and infinity. Next, Condition \ref{con:moments2} mimics those in Condition \ref{con:moments1} by 
regularizing the derivatives of the estimating functions.

\begin{condition}\label{con:moments2}
Each $g_j(\bfx;\btheta)$ for  $j\in[r]$ is twice continuously differentiable w.r.t $\btheta\in\bTheta$ for any $\bfx$ on the support. ({\rm a}) There exists some universal constant $K_7>0$ such that 
\begin{align*}
\max_{t\in[n]}\max_{j\in[r]}\max_{l\in[p]}\mathbb{E}\bigg\{\sup_{\btheta\in\bTheta}\bigg|\frac{\partial g_{t,j}(\btheta)}{\partial \theta_l}\bigg|\bigg\}\leq K_7~~ \text{and}~~\max_{j\in[r]}\max_{l\in[p]}\sup_{\btheta\in\bTheta}\mathbb{E}_n\bigg\{\bigg|\frac{\partial g_{t,j}(\btheta)}{\partial \theta_l}\bigg|^2\bigg\}=O_{\p}(1)\,.
\end{align*}  
({\rm b}) There exist some universal constants $K_8>0$ and $K_9>0$ such that 
    \begin{align*}
&\max_{t\in[n]}\max_{j\in[r]}\max_{l\in[p]}\mathbb{E}\bigg\{\bigg|\frac{\partial g_{t,j}(\btheta_0)}{\partial \theta_l}\bigg|^{\gamma}
    \bigg\}\leq K_8~~\text{and}~~\min_{j\in[r]}\min_{l\in\mathcal{S}}  {\rm Var}\bigg\{\frac{1}{\sqrt{k}}\sum_{t=\ell+1}^{\ell+k}\frac{\partial g_{t,j}(\btheta_0)}{\partial\theta_{l}}\bigg\}\geq K_9\,,
\end{align*}
for all integers  $k\geq1$ and $\ell\geq 0$. ({\rm c}) It holds that 
\begin{align*}
\max_{j\in[r]}\max_{l_1,l_2\in[p]}\sup_{\btheta\in\bTheta}\mathbb{E}_n\bigg\{\bigg|\frac{\partial ^2g_{t,j}(\btheta)}{\partial \theta_{l_1}\partial \theta_{l_2}}\bigg|^2\bigg\}=O_{\p}(1)\,.
\end{align*}
\end{condition}

\begin{remark}
We provide low-level conditions that imply some inequalities in Conditions \ref{con:moments1} and \ref{con:moments2}. 
Recall $g_{t,j}(\btheta)=g_j(\bfx_t;\btheta)$.
If there exist functions $B_k(\cdot)$ with $\max_{t\in[n]}\mathbb{E}\{B_k(\bfx_t)\}<\infty$, $k=1,2,3$,
such that  $|g_{t,j}(\btheta)|^{\gamma}\leq B_1(\bx_t)$,  $|\partial g_{t,j}(\btheta)/\partial \theta_l|^2\leq B_2(\bfx_t)$ and  $|\partial^2 g_{t,j}(\btheta)/\partial \theta_{l_1}\partial \theta_{l_2}|^2\leq B_3(\bfx_t)$ for all $j\in[r]$ and $\btheta\in\bTheta$, then the second requirement in Condition \ref{con:moments1}(a), the second requirement in Condition \ref{con:moments2}(a),  and the  requirement in Condition \ref{con:moments2}(c) are satisfied. All these three $O_{\p}(1)$ quantities here are for the ease of presentation. 
They can be relaxed by $O_{\p}(\varpi_n)$ with some diverging sequence $\varpi_n$ in view of Lemma 2 of \cite{chang2024optimal}.
Condition \ref{con:moments1}(b) and the second requirement  in Condition \ref{con:moments2}(b) are important assumptions to derive self-normalized sums inequality under dependent sequences; see Lemma \ref{self-normalized alpha-mixing} in the supplementary material. A similar assumption can be found in (4.2) of \cite{chen2016}. Specifically,
Condition \ref{con:moments1}(b) is used to derive the convergence rates of $|\bar{\bg}(\btheta_0)|_{\infty}$ and $|\widehat{\bV}(\btheta_0)-\mathbb{E}\{\widehat{\bV}(\btheta_0)\}|_{\infty}$; see Lemma \ref{tailprabog0} in Appendix \ref{sec:prothe1}. The second requirement in Condition \ref{con:moments2}(b) is used to derive the convergence rate of 
$\max_{j\in[r]}\max_{k\in\mathcal{S}}|\mathbb{E}_n[ 
 \partial g_{t,j}(\btheta_{0}) / \partial\theta_k-\mathbb{E}\{
 \partial g_{t,j}(\btheta_{0})/\partial\theta_k\}]| $; 
see the proof of Lemma \ref{gammahatL2norm} in the supplementary material.     
\end{remark}

Finally, we deal with the choice of the penalty functions. 
Consider a generic penalty $P_{1,\pi}(\cdot)$ and let $a_n = \sum_{k=1}^{p}P_{1,\pi}(|\theta_{0,k}|)$.
Define $b_n$ = $\max\{a_n,\nu^2\}$, 
which is the larger value between $a_n$ and the square of the tuning parameter $\nu$ attached to the second penalty function $P_{2,\nu}(\cdot)$. 
In order to control the shrinkage bias induced by $P_{1,\pi}(\cdot)$ on $\hat{\btheta}_n$, suppose there exist $\chi_n\to 0$ and $c_n\to 0$ with $c_n  \gg b_n^{1/2} $ such that
\begin{align}\label{eq:chi0}
	\max_{k\in \mathcal{S}}\sup_{0<t<|\theta_{0,k}|+c_n}P_{1,\pi}'(t)=O(\chi_n)\,.
\end{align}
Under the assumption $b_n\ll \min_{k\in\mathcal{S}}|\theta_{0,k}|^2$, 
we can replace \eqref{eq:chi0} by 
\begin{align}\label{eq:chi1}
\max_{k\in \mathcal{S}}\sup_{c|\theta_{0,k}|<t<c^{-1}|\theta_{0,k}|}P_{1,\pi}'(t)=O(\chi_n)\
\end{align}
for some constant $c\in(0,1)$. If we select $P_{1,\pi}(\cdot)$
as an asymptotically unbiased penalty such as SCAD or MCP, we have $\chi_n=0$ in  \eqref{eq:chi1} when 
\begin{align}\label{eq:chi2}
\min_{k\in\mathcal{S}}|\theta_{0,k}|\gg \max\{b_n^{1/2},\pi\}\,.
\end{align}
To simplify the presentation, we assume that \eqref{eq:chi2} holds and $\chi_n=0$ in \eqref{eq:chi1}. 
It provides the minimum signal level on the nonzero components in $\btheta_0$. 

Regarding the second penalty term, we write  $\rho_2 (t;\nu)=\nu^{-1}P_{2,\nu}(t)$ for $P_{2,\nu}(t) \in \mathcal{P}$. Since  $\rho_2 '(0^{+};\nu)$ is independent of $\nu$, we denote $\rho_2 '(0^{+};\nu)$ by $\rho_2 '(0^{+})$ for simplicity.
For any $\btheta \in \bTheta$, define
\begin{align}\label{C*}
\mathcal{M}_{\btheta}^*=\{j\in[r] : |\bar {g}_{j}(\btheta)|\geq   C_*\nu \rho_2'(0^{+})\}
\end{align}
for some constant $C_* \in (0,1)$.
Given $\nu$, the complexity of the moment conditions can be controlled by a ``sparsity of moments'' index $\ell_n$ such that
\begin{align}\label{eq:aleph}
\mathbb{P}\bigg(\max _{\btheta \in \bTheta: \,|\btheta_{\mathcal{S}}-\btheta_{0,\mathcal{S}}|_{\infty} \leq c_{n},\,|\btheta_{\mathcal{S}^{\rm c}}|_1\leq \aleph_n}|\mathcal{M}_{\btheta}^*|\leq \ell_n\bigg)\to 1
\end{align}
with $c_{n}\rightarrow 0$ satisfying $ c_{n} \gg b_n^{1/2}  $, where $\aleph_n=n^{-3\varphi/(6\varphi+2)}(\log r)^{1/2}$.

\begin{remark}
To understand the above expression, for any $j\in[r]$, let $\bar{\mu}_j(\btheta)=\mathbb{E}\{\bar {g}_{j}(\btheta)\}$, so that  $\bar{\mu}_j(\btheta_0)=0$ and 
$|\bar {g}_{j}(\btheta)-\bar{\mu}_j(\btheta)|=o_{\p}(1)$. 
If $\btheta$ is in the neighborhood of $\btheta_0$, then a Taylor expansion yields $\bar{\mu}_j(\btheta)-\bar{\mu}_j(\btheta_0)=\{\nabla_{\btheta}\bar{\mu}_j(\btheta^*)\}^{\T}(\btheta-\btheta_0)$ for some  $\btheta^*$ between $\btheta$ and $\btheta_0$. Hence, those estimating functions with large expected derivatives will be selected. 
In addition, since $\max_{j\in[r]}|\bar {g}_{j}(\btheta_0)|=O_{\p}(\aleph_n)$ (see Lemma \ref{tailprabog0} in Appendix \ref{sec:prothe1}), 
if the tuning parameter $\nu\gg\aleph_n$, then  $\mathcal{M}_{\btheta_0}^*=\emptyset$ w.p.a.1. Therefore, \eqref{eq:aleph} requires that given the level of the tuning parameter $\nu$, the model should not be excessively complex so that the cardinality of moments with non-trivial derivatives can be controlled by some sequence $\ell_n$ in a neighborhood of  $\btheta_0$.
\end{remark}

\section{Consistency and asymptotic normality}\label{sec:consistency}

The technical conditions in the previous section allow us to proceed with the asymptotic properties of PEL. 
To facilitate  analysis under the listed assumptions,  we refine the problem \eqref{eq:double-pen} as
  \begin{align}\label{eq:est1}
	\hat{\btheta}_n = \arg\min_{\btheta\in\bTheta_*}\max_{\blambda\in \hat{\Lambda}_n(\btheta)}\bigg[\frac{1}{n}\sum_{t=1}^{n}\log\{1+\blambda^{\T}\bfg_t(\btheta)\}-\sum_{j=1}^{r}P_{2,\nu}(|\lambda_j|)+\sum_{k=1}^{p}P_{1,\pi}(|\theta_k|)\bigg]\,,
	\end{align}
where  $\bTheta_*=\{\btheta\in \bTheta: |\btheta_{\calS}-\btheta_{0,\calS}|_{\infty} \leq c_*,|\btheta_{\calS^{\rm c}}|_1\leq \aleph_n\}$ for a fixed $c_*>0$, where $\aleph_n=n^{-3\varphi/(6\varphi+2)}(\log r)^{1/2}$ has been defined in \eqref{eq:aleph}.
The refinement comes from $c_*$, which restricts  the true value to be a point in the closed parameter space, and the sparsity is controlled by $\aleph_n$.
Next, let 
 $\hat{\blambda} = (\hat{\lambda}_{1},\ldots,\hat{\lambda}_{r})^\T$ be the $r$-dimensional vector of Lagrange multiplier defined at $\hat{\btheta}_n${\rm:}
\begin{align*}
\hat{\blambda} := 
\hat{\blambda}(\hat{\btheta}_n)=\arg\max_{\blambda\in\hat{\Lambda}(\hat{\btheta}_n)}\bigg[\frac{1}{n}\sum_{t=1}^{n}\log \{1+\blambda^{\T}\bg_{t}(\hat{\btheta}_n)\}-\sum_{j=1}^{r}P_{2,\nu}(|\lambda_j|)\bigg]\,,
\end{align*}
and $\mathcal{R}_{n}={\rm supp}(\hat{\blambda})$ be its support.
Define $\hat{\bfeta}=(\hat{\eta}_1,\ldots,\hat{\eta}_{r})^{\T}$ with 
\begin{align}\label{etaj}
\hat{\eta}_j:=\frac{1}{n}\sum_{t=1}^{n}\frac{g_{t,j}(\hat{\btheta}_n)}{1+\hat{\blambda}_{\calR_n}^{\T}\bfg_{t,\calR_{n}}(\hat{\btheta}_n)}\,.
\end{align}
It holds by the Karush-Kuhn-Tucker condition that 
$\eta_j=\nu\rho_2'(|\hat{\lambda}_j|;\nu)\sgn(\hat{\lambda}_j)$
for $\hat{\lambda}_{j}\neq 0$, and 
the subdifferential at $\hat{\lambda}_j$ must include the zero element \citep{Bertsekas(1997)} for $\hat{\lambda}_j=0$. 
We impose the next condition.
\begin{condition}\label{eq:eta}
As $n\to \infty$, it holds that
\begin{enumerate}
    \item[(a)] 
   $ 
     \mathbb{P}\big[\bigcup_{j=1}^r\{\tilde{c}\nu\rho_2'(0^+)\leq |\bar{g}_{j}(\hat{\btheta}_n)|< \nu\rho_2'(0^+)\}\big]\to 0  
    $ for some constant $\tilde{c}\in(C_*,1)$\,,
\item[(b)] 
$
    \mathbb{P} \big[\bigcup_{j\in\mathcal{R}_n^{\rm c}}\{|\hat{\eta}_j|=\nu\rho_2'(0^+)\}  \big]\to 0
$\,.
\end{enumerate}

\end{condition}


Condition \ref{eq:eta}(a) is a technical assumption used to derive the convergence rate of the Lagrange multiplier $\hat{\blambda}(\hat{\btheta}_n)$ associated with $\hat{\btheta}_n$; see the proof of Theorem \ref{The.1} in Appendix \ref{sec:the1ii}.
Condition \ref{eq:eta}(b) requires that w.p.a.1~the nonzero $\hat{\eta}_j$ 
does not lie on the boundary, which is satisfied by continuous random variables.
Condition \ref{eq:eta} makes sure that $\hat{\blambda}(\btheta)$ is continuously differentiable at $\hat{\btheta}_n$ w.p.a.1; see Lemma \ref{eq:etaj0} in Appendix \ref{sec:profthe2}.

The conditions up to this point are sufficient to guarantee the consistency of the PEL estimator.

  \begin{theorem}\label{The.1}
		Let $P_{1,\pi}(\cdot)$, $P_{2,\nu}(\cdot)\in \mathcal{P}$ for $\mathcal{P}$ defined in \eqref{penalty}, and $P_{2,\nu}(\cdot)$ be a convex function with bounded second-order derivatives around $0$.	
		Under Conditions {\rm \ref{con:mixingdecay}--\ref{con:moments1}, \ref{con:moments2}(a)} and  {\rm  \ref{eq:eta}(a)}, if  $L_n^{\varphi}\log (rn)\ll n^{\varphi/(3\varphi+1)}$, $ b_n\ll\min\{n^{-2/\gamma},s^{-2}\ell_n^{-1}\}$ and $\ell_n\aleph_n\ll \min\{\nu,\pi\}$, then the local minimizer $\hat{\btheta}_n$ defined in  \eqref{eq:est1} satisfies $|\hat{\btheta}_{n,\calS}-\btheta_{0,\calS}|_{\infty}=O_{\p}(b_n^{1/2})$ and $\mathbb{P}(\hat{\btheta}_{n,\calS^{\rm c}}={\bf0})\to  1$ as $n\to \infty$.
	\end{theorem}

Theorem \ref{The.1} provides the consistency of the PEL estimator: $\hat{\btheta}_{n}$ in the true active set $\mathcal{S}$ converges in probability to its true value at rate $b_n^{1/2}$, whereas the coefficients in the inactive set $\mathcal{S}^{\rm c}$ are shrunken to exactly zero w.p.a.1. 
The orders involved in the statement give the admissible range of the tuning parameters.
Recall $\aleph_n=n^{-3\varphi/(6\varphi+2)}(\log r)^{1/2}$ and $b_n$ = $\max\{a_n,\nu^2\}$ with $a_n = \sum_{k=1}^{p}P_{1,\pi}(|\theta_{0,k}|)$. Due to $a_n \lesssim s\pi$,
Theorem \ref{The.1} requires that  the tuning parameters $(\nu,\pi)$ satisfy
\begin{equation}\label{eq:tuning_range_thm1}
  \ell_n \aleph_n \ll \nu\ll \min\{n^{-1/\gamma}, s^{-1} \ell_n^{-1/2} \}, \quad
\ell_n \aleph_n \ll \pi\ll \min\{s^{-1}n^{-2/\gamma}, s^{-3} \ell_n^{-1} \}  
\end{equation}
with  
$\ell_n\ll\min\{s^{-1}n^{-2/\gamma} \aleph_n^{-1}, s^{-3/2} \aleph_n^{-1/2} \}$ and   $s\ll n^{\delta}$ for some $\delta=
\min\{3\varphi/(6\varphi+2)-2/\gamma, \varphi/(6\varphi+2)\}$. 
There is a tradeoff between $s$ and $r$. For example, if  $s\asymp n^{\kappa}$ with $\kappa\in[0,\delta)$ is of polynomial order of $n$, 
Theorem \ref{The.1} ensures the consistency of $\hat{\btheta}_n$ even if the number of moments $r$ diverges exponentially fast under the rate
$
\log r\ll \min \{ n^{3\varphi/(3\varphi+1)-4/\gamma-2\kappa}, n^{3\varphi/(3\varphi+1)-6\kappa}, n^{\varphi/(3\varphi+1)}L_n^{-\varphi} \}
$
with $L_n\ll n^{1/(3\varphi+1)}(\log n)^{-1/\varphi}$.


While Theorem \ref{The.1} gives the rate of convergence, we further characterize the asymptotic distribution of $\hat{\btheta}_{n,\calS}$. 
For any index set $\mathcal{F}\subset[r]$ and $\btheta\in\bTheta$,  denote the long-run covariance matrix of $\{\bg_{t,\calF}(\btheta)\}_{t=1}^{n}$  by  
$\bXi_{\calF}(\btheta)={\rm Var}\{n^{-1/2} \sum_{t=1}^{n}\bg_{t,\calF}(\btheta)\},$
and denote $\bXi(\btheta):=\bXi_{\calF}(\btheta)$ when $\mathcal{F}=[r]$.

\begin{condition}\label{con:Gamma} 
(a) There exists some universal  constant $K_{10}>1$ such that
\begin{align*}
K_{10}^{-1}<\lambda_{\min}([\mathbb{E}  \{\nabla_{\btheta_{\calS}} \bar{\bfg}_{\calF}( \btheta_0) \}]^{\T,\otimes2})\le\lambda_{\max}([\mathbb{E}  \{\nabla_{\btheta_{\calS}} \bar{\bfg}_{\calF}( \btheta_0) \}]^{\T,\otimes 2})<K_{10}
\end{align*}
for any subset $\mathcal{F}\subset[r]$ with 
$s\leq |\mathcal{F}|\leq \ell_n$. (b) There exists some universal constant $K_{11}>1$ such that  
\begin{align*}
K_{11}^{-1}<\lambda_{\min}\{\bXi(\btheta_{0})\}\leq \lambda_{\max}\{\bXi(\btheta_{0})\}< K_{11}\,.
\end{align*}

\end{condition}

Condition \ref{con:Gamma}(a) corresponds to the sparse Riesz condition (\citealp{chen2008extended}, \citealp{zhang2008sparsity}, \citealp{chang2022culling}) and it is adapted to our setting for handling high-dimensional dependent data. 
Condition \ref{con:Gamma}(b) ensures that the eigenvalues of the long-run covariance matrix of the sequence $ \{\bg_{t}(\btheta_0)\}_{t=1}^{n}$
are uniformly bounded away from zero and infinity. 

Theorem \ref{The.2} below shows that the  PEL estimator for the nonzero components of $\btheta_0$ is  asymptotically normal,
with an asymptotic bias that deviates from zero due to the high dimensionality. 
In addition to \eqref{eq:tuning_range_thm1}, the admissible range of the tuning parameters $(\nu,\pi)$ in Theorem \ref{The.2} is slightly narrowed. 
Define 
\begin{equation}\label{eq:JM}
\begin{split}
     &~~{\bJ}_{\calR_n}=\big\{[\mathbb{E}  \{\nabla_{\btheta_{\calS}} \bar{\bfg}_{ \calR_n}( \btheta_0) \}]^{\T}[\mathbb{E}\{\widehat{\bfV}_{\calR_n}(\btheta_0)\}]^{-1}\bXi_{\calR_n}^{1/2}(\btheta_{0})\big\}^{\otimes2}\,,
\\ &~~~~~\bM_{\calR_n}=\big\{[\mathbb{E}  \{\nabla_{\btheta_{\calS}} \bar{\bfg}_{ \calR_n}( \btheta_0) \}]^{\T}[\mathbb{E}\{\widehat{\bfV}_{\calR_n}(\btheta_0)\}]^{-1/2}\big\}^{\otimes 2}\,.
\end{split}
\end{equation}
Here is the statement.

 \begin{theorem}\label{The.2}
		Let $P_{1,\pi}(\cdot)$, $P_{2,\nu}(\cdot)\in \mathcal{P}$ for $\mathcal{P}$ defined in \eqref{penalty}, and $P_{2,\nu}(\cdot)$ be a convex function with bounded second derivatives around $0$. 
	Under Conditions {\rm \ref{con:mixingdecay}--\ref{con:Gamma}}, if $L_n^{\varphi}\log r\ll n^{\varphi/(3\varphi+1)}$, $\ell_n^{3}L_{n}^{4\varphi}(\log n)^{4}\ll n$,
$\ell_n^3(\log r)^2\ll n^{(3\varphi-1)/(3\varphi+1)-2/\gamma}$, $b_n\ll \min\{n^{-2/\gamma}, n^{-1}\ell_n^{-2}s^{-3}\nu^{-2}\}$ and 
$\ell_n\aleph_n\ll \min\{\nu,\pi,s^{-1/2}\}$, then for any $\bz\in\mathbb{R}^s$ with $|\bz|_2=1$,   the local minimizer $\hat{\btheta}_n$ defined in  \eqref{eq:est1} satisfies 
	\begin{align*}
		n^{1/2}\bfz^\T{\bJ}_{\calR_n}^{-1/2}\bM_{\calR_n}(\hat{\btheta}_{n,\calS}-\btheta_{0,\calS}-\hat{\bpsi}_{\calR_n})\rightarrow N(0,1)
	\end{align*}
 in distribution as $n\rightarrow\infty$, where ${\bJ}_{\calR_n}$ and $\bM_{\calR_n}$ are defined in \eqref{eq:JM}
and the bias is \begin{align*}
    \hat{\bpsi}_{\calR_n}={\bM}_{\calR_n}^{-1}[\mathbb{E}  \{\nabla_{\btheta_{\calS}} \bar{\bfg}_{ \calR_n}( \btheta_0) \}]^{\T}[\mathbb{E}\{\widehat{\bfV}_{\calR_n}(\btheta_0)\}]^{-1}\hat{\bfeta}_{\calR_n}
\end{align*} with $\hat{\bseta}$ defined in {\rm \eqref{etaj}}.
 \end{theorem}

Though in principle the bias term $\hat{\bpsi}_{\calR_n}$ is estimable, it involves approximation to multiple components that are difficult to formulate and compute.  Theorem \ref{The.2} is a property of the PEL estimator for $\btheta_{0,\calS}$, and it is uninformative about the model parameters indexed in $\mathcal{S}^{\rm c}$. 
Consequently, we do not advocate using Theorem \ref{The.2} for statistical inference. 
Instead, if we are interested in testing  low-dimensional components of the model parameter, which is commonly the use case in applied econometrics, we recommend  PPEL in the following section.


\section{Inference for low-dimensional components of model parameters}
\label{sec:Inference}
	
We focus on the inference about $\btheta_{\calM}\in\mathbb{R}^m$ for some small subset $\mathcal{M}\subset[p]$ with $|\mathcal{M}|=m$.	
In economic and financial applications, most often the inference falls in a single parameter under which $m=1$.
Hence, we assume $m$ is a fixed integer for simplicity; there is no technical difficulty in allowing it to diverge with the sample size $n$. 
We write $\btheta=(\btheta_{\calM}^\T,\btheta_{\calM^{\rm c}}^\T)^\T$, where $\btheta_{\calM}\in\mathbb{R}^m$ contains the low-dimensional components of interest, and $\btheta_{\calM^{\rm c}}\in\mathbb{R}^{p-m}$ contains the nuisance parameters. 
We consider an inferential procedure similar to that in 
\cite{chang2022culling}.
Let $\bfA_n=(\ba_1^n,\dots,\ba_m^n)^\T\in\mathbb{R}^{m\times r}$, which is found row by row via the optimization problem
	\begin{equation}\label{eq:ifind}
	\ba_k^n=\arg\min_{\mathbf{u}\in\mathbb{R}^r} |\mathbf{u}|_1 \  \ \text{ s.t. } 
	|\{\nabla_{\btheta}\bar{\bfg}(\hat{\btheta}_n)\}^\T \mathbf{u}-\bxi_{k}|_\infty \leq \varsigma\,,
	\end{equation}
where $\varsigma$ is a tuning parameter, $\hat{\btheta}_n
$ is the PEL estimator from (\ref{eq:est1}), 
and $\{\bxi_{k}\}_{k=1}^m$ consists of the canonical basis of the linear space $\mm_{\bxi}=\{\mathbf{b}=(b_1,\dots,b_p)^\T:b_j=0 \text{ for any } j=m+1,\dots,p\}$, i.e., $\bxi_k$ is chosen such that its $k$-th component is $1$ and all other components are $0$. Define $\bff^{\bfA_n}(\cdot;\cdot)=\bfA_n\bfg(\cdot;\cdot)$, and then $ \bff^{\bfA_n}(\cdot;\cdot)$ are the new $m$-dimensional estimating functions. 
By construction, 
the influence of the nuisance parameters is projected out by  $\ba^n_k$.

The low-dimensional moment functions
$\bff^{\bfA_n}(\cdot;\cdot)$ substantially reduce the dimension $r$ all the way to a much smaller $m$. 
The EL constructed with $\bff^{\bfA_{n}}(\cdot;\cdot)$, instead of $\bfg(\cdot;\cdot)$, is used for the inference about $\btheta_{\calM}$. Specifically, let 
	\begin{align}\label{eq:model2}
	{L}_{\bfA_n}(\btheta_{\calM};\hat{\btheta}_{n,{\calM}^{\rm c}})=\sup\biggl\{\prod_{t=1}^{n}\pi_t:\pi_t>0, \, \sum_{t=1}^{n}\pi_t=1,
    \, \sum_{t=1}^{n}\pi_t\bff_t^{\bfA_n}(\btheta_{\calM},\hat{\btheta}_{n,{\calM}^{\rm c}})={\bf0}\biggr\}\,,
	\end{align}
where $\hat{\btheta}_{n,{\calM}^{\rm c}}$ is the nuisance component of 
$\hat{\btheta}_n$ from \eqref{eq:est1}.
Maximizing (\ref{eq:model2}) yields the PPEL estimator $\tilde{\btheta}_{\calM}$, which can also be obtained by solving the corresponding dual problem
\begin{align}
\tilde{\btheta}_{\calM}=\arg\min_{\btheta_{\calM}\in\widehat{\bTheta}_{\calM}}\max_{\blambda\in{\tilde{\Lambda}}_n(\btheta_{\calM})}\sum_{t=1}^{n}\log\{1+\blambda^\T\bff_t^{\bfA_n}(\btheta_{\calM},
	\hat{\btheta}_{n,\calM^{\rm c}})\}\,,
	\end{align}
	where $\widehat{\bTheta}_{\calM}=\{\btheta_{\calM}\in\mathbb{R}^m:|\btheta_{\calM}-\hat{\btheta}_{n,\calM}|_{\infty}\leq O_{\p}(\nu)\}$, and $\tilde{\Lambda}_n(\btheta_{\calM})=\{\blambda\in \mathbb{R}^m:\blambda^\T\bff_t^{\bfA_n}(\btheta_{\calM},\hat{\btheta}_{n,\calM^{\rm c}})\in \mathcal{V} \text{ for any }t\in[n]\}$. 
To justify this PPEL estimator, we impose one more assumption.
\begin{condition}\label{con:alpha}
 (a) For each $k\in[m]$, there exists a nonrandom $\ba_k^0\in\mathbb{R}^r$ satisfying $[\mathbb{E}\{\nabla_{\btheta}\bar{\bfg}(\btheta_{0})\}]^{\T}\ba_k^0=\bxi_{k}$, $|\ba_k^0|_1\leq K_{12}$ for some universal constant $K_{12}>0$, and $\max_{k\in[m]}|\ba_k^n-\ba_k^0|_1=O_{\p}(\omega_n)$ for some $\omega_n\to 0$. 
 (b)  All singular values of $\bfA=(\ba_1^0,\dots, \ba_m^0)^{\T}$ are uniformly bounded away from zero and infinity.
\end{condition}

By definition, the population quantity $\ba_k^0$ is the counterpart of the sample's $\ba^n_k$. Condition \ref{con:alpha}(a) guarantees that these $\ba_k^0$ are $L_1$-sparse and they are well estimated by $\ba^n_k$ from (\ref{eq:ifind}). Condition \ref{con:alpha}(b) ensures that $\bfA$ spans an $m$-dimensional linear space so the information provided by the moments concerning the $m$-dimensional parameters of interest does not degenerate. 

\begin{proposition}\label{p.1}
	Suppose the conditions of Theorem $\ref{The.1}$ hold. Under Conditions {\rm \ref{con:moments2}(b)}, {\rm \ref{con:moments2}(c)}, {\rm \ref{con:Gamma}(a)} and  {\rm \ref{con:alpha}},  if $n\aleph_n^2\omega_n^2=O(1)$, $b_n\ll s^{-3}\ell_n^{-1}$
 and 
$ns\ell_n\nu^2\max\{s\ell_n\nu^2, \varsigma^2\}=O(1)$, then 
$|\tilde{\btheta}_{\calM}-\btheta_{0,\calM}|_2=O_{\p}(L_n^{1/2}n^{-1/2})$. 
\end{proposition}

Proposition \ref{p.1} provides the convergence rate of the PPEL estimator 
$\tilde{\btheta}_{\calM}$. 
If $L_n$ is a constant, $\tilde{\btheta}_{\calM}$ reaches the usual optimal rate of convergence $n^{-1/2}$.
To find its asymptotic distribution, we must cope with the potential serial correlation in the projected moment functions 
$\bff^{\bfA_n}(\cdot;\cdot)$. Define  \begin{align*}
    \widehat{\bXi}_{\bff^{\bA_n}}(\tilde{\btheta}_{\calM},\hat{\btheta}_{n,\calM^{\rm c}})=\sum_{j=-(n-1)}^{n-1}\mathcal{K} \bigg(\frac{j}{h_n}\bigg) \widehat{\bH}_j(\tilde{\btheta}_{\calM},\hat{\btheta}_{n,\calM^{\rm c}})
\end{align*}
as an estimator of the long-run covariance matrix, where
\begin{equation*}
\widehat{\bH}_j(\tilde{\btheta}_{\calM},\hat{\btheta}_{n,\calM^{\rm c}})=\left\{
	\begin{aligned}
		\frac{1}{n}\sum_{t=j+1}^{n}\bff_t^{\bfA_n}(\tilde{\btheta}_{\calM},\hat{\btheta}_{n,\calM^{\rm c}})\bff_{t-j}^{\bfA_n}(\tilde{\btheta}_{\calM},\hat{\btheta}_{n,\calM^{\rm c}})^{\T}\,, ~~&\textrm{if}~j\geq 0\,, 
		\\
		\frac{1}{n}\sum_{t=1-j}^{n}\bff_{t+j}^{\bfA_n}(\tilde{\btheta}_{\calM},\hat{\btheta}_{n,\calM^{\rm c}})\bff_t^{\bfA_n}(\tilde{\btheta}_{\calM},\hat{\btheta}_{n,\calM^{\rm c}})^{\T}\,,  ~~& \textrm{if}~j<0\,.
	\end{aligned}
	\right.
\end{equation*}
Here $\mathcal{K}(\cdot)$ is a symmetric kernel 
for the estimation of the long-run covariance matrix (\citealp{newey1987simple}, \citealp{Andrews(1991)}), 
inside of which lies the diverging bandwidth $h_n$. Condition \ref{con:kernel} consists of standard conditions on these kernels,
which are satisfied by the widely used choices such as the Parzen kernel, the Tukey-Hanning kernel, and the QS kernel.

\begin{condition}\label{con:kernel}
The kernel function $\mathcal{K}(\cdot):\mathbb{R}\to [-1,1]$ is  continuously differentiable with bounded derivatives on $\mathbb{R}$ and satisfies {\rm (a)} $\mathcal{K}(0)=1$, {\rm (b)} $\mathcal{K}(x)=\mathcal{K}(-x)$ for any $x\in\mathbb{R}$, and {\rm (c)} 
$\int_{-\infty}^{\infty}|\mathcal{K}(x)|\,{\rm d}x\leq  K_{13}$ for some  universal constant $K_{13}>0$.
\end{condition}

As the counterpart of \eqref{eq:JM}, we define
\begin{equation}
    \label{eq:JMfa}
\begin{split}&\widehat{\bJ}_{\bff^{\bfA_{n}}}  =\big[\{\nabla_{\btheta_{\calM}}\bar{\bff}^{\bfA_n}(\tilde{\btheta}_{\calM},\hat{\btheta}_{n,\calM^{\rm c}})\}^{\T}\widehat{\bfV}^{-1}_{\bff^{\bfA_n}}(\tilde{\btheta}_{\calM},\hat{\btheta}_{n,\calM^{\rm c}})\widehat{\bXi}_{\bff^{\bA_n}}^{1/2}(\tilde{\btheta}_{\calM},\hat{\btheta}_{n,\calM^{\rm c}})\big]^{\otimes2}\,,\\ &~~~~~~~~~~
 \widehat{\bM}_{\bff^{\bA_n}} =\big[\{\nabla_{\btheta_{\calM}}\bar{\bff}^{\bfA_n}(\tilde{\btheta}_{\calM},\hat{\btheta}_{n,\calM^{\rm c}})\}^{\T}\widehat{\bfV}^{-1/2}_{\bff^{\bfA_n}}(\tilde{\btheta}_{\calM},\hat{\btheta}_{n,\calM^{\rm c}})\big]^{\otimes2}
 \end{split}
\end{equation}
with $\widehat{\bfV}_{\bff^{\bfA_n}}(\tilde{\btheta}_{\calM},\hat{\btheta}_{n,\calM^{\rm c}})=\mathbb{E}_n[\{\bff^{\bfA_{n}}_t(\tilde{\btheta}_{\calM},\hat{\btheta}_{n,\calM^{\rm c}})\}^{\otimes2}]$.  
Now we are ready to state the asymptotic normality of the PPEL estimator.

\begin{theorem}\label{The.3}
		Assume the conditions of Proposition $\ref{p.1}$ hold. Under Conditions {\rm \ref{con:Gamma}(b)} and  {\rm \ref{con:kernel}}, if  $L_{n}^{4\varphi}(\log n)^{4}\ll n$, $n\aleph_n^2\omega_n^2=o(1)$,
$L_nh_n^2\max\{s\ell_n\nu^2,\omega_n^2, L_n n^{-1}\}=o(1)$,  $L_n^5\ll h_n^2$ and $ns\ell_n\nu^2\max\{s\ell_n\nu^2, \varsigma^2\}=o(1)$, 	for any $\bfz\in\mathbb{R}^{m}$ with $|\bfz|_2=1$,  it holds that 
\begin{align*}
n^{1/2}\bz^{\T}\widehat{\bJ}_{\bff^{\bfA_{n}}}^{-1/2}\widehat{\bM}_{\bff^{\bA_n}}(\tilde{\btheta}_{\calM}-\btheta_{0,\calM})\rightarrow N(0,1)
\end{align*}
in distribution as $n\to\infty$,
where $\widehat{\bJ}_{\bff^{\bfA_{n}}}$ and $\widehat{\bM}_{\bff^{\bA_n}}$ are defined in \eqref{eq:JMfa}.
\end{theorem}

Theorem \ref{The.3} establishes the asymptotic normality of $\tilde{\btheta}_{\calM}$.
Unlike Theorem \ref{The.2}, here the asymptotic distribution is well centered around zero.
Standardized by a consistent estimate of the asymptotic variance, the limiting  distribution is the desirable $N(0,1)$.
In particular, when researchers are interested in the coefficient of one variable at a time, we can use $\bz$ to pick out the coordinate and the left-hand side of the statement becomes a $t$-statistic;
we then refer to a corresponding quantile of $N(0,1)$ to decide whether we shall reject the null hypothesis at a pre-specified test size.

Compared with Theorem \ref{The.1},
Theorem \ref{The.3}  further requires the tuning parameters $(\nu,\pi,\varsigma)$ and the bandwidth $h_n$ in the kernel to satisfy
\begin{equation*}
\begin{split}
& \nu\ll \min\{ s^{-3/2}\ell_n^{-1/2},  n^{-1/4}s^{-1/2}\ell_n^{-1/2}\}\,, \pi\ll \min\{s^{-1}n^{-2/\gamma}, s^{-4}\ell_n^{-1}\}\,,
\\& ~~~L_n^{5/2}\ll h_n\ll \min\{n^{1/2}L_n^{-1}, L_n^{-1/2}\omega_n^{-1}\}\,,~h_n\nu\ll L_n^{-1/2}s^{-1/2}\ell_n^{-1/2}
\end{split}
\end{equation*} 
and $\nu\varsigma\ll n^{-1/2}s^{-1/2}\ell_n^{-1/2}$
with  $\ell_n \ll \min\{s^{-1}n^{-2/\gamma}\aleph_n^{-1},s^{-2}\aleph_n^{-1/2}, s^{-1/3}n^{-1/6}\aleph_n^{-2/3}, L_n^{-2} s^{-1/3} \aleph_n^{-2/3}\}$
and  {$s\ll \min\{n^{\delta}, n^{3\varphi/(3\varphi+1)}L_n^{-6}\}$ for {$\delta=
\min\{3\varphi/(6\varphi+2)-2/\gamma,3\varphi/(24\varphi+8)\}$}}.
If we allow $s$ to grow in a  polynomial order of $n$ in that 
$s\asymp n^{\kappa}$ satisfying 
 $n^{\kappa}\ll n^{\psi}L_n^{-6} $, where $\kappa\in[0,{\delta})$, and $\psi=3\varphi/(3\varphi+1)$, then our proposed  estimator $\tilde{\btheta}_{\calM}$ is asymptotically normal even if
the number of moment conditions $r$ diverges at an exponential order  in that
$
\log r\ll \min \{ n^{\psi-4/\gamma-2\kappa}, n^{\psi-8\kappa},n^{(3\varphi-1)/(6\varphi+2)-\kappa}, n^{\psi-\kappa}L_n^{-6},
n^{\psi/3}L_n^{-\varphi}, n^{-\psi/(3\varphi)}\omega_n^{-2} \},
$
where $L_n\ll \min\{n^{1/(4\varphi)}(\log n)^{-1/\varphi},n^{\psi/6}, n^{1/7}\}$ and  $\omega_n\ll \min\{L_n^{-3},n^{-1/(6\varphi+2)}\}$.
The ranges of admissible $r$ and $p$ highlight the adaptivity to high-dimensional models.


To summarize the theoretical results, Theorem \ref{The.1} shows that the PEL estimator $\hat{\btheta}_n$ is a consistent estimator for the true parameter $\btheta_0$, and 
Theorem \ref{The.2} verifies its asymptotic normality of the nonzero components. However, the limiting normal distribution is not centered at zero due to the influence of the high dimensional moments, making it difficult to use for statistical inference. 
While most applied econometric use cases of inference focus on a low-dimensional parameter of interest, say $\btheta_{0,\calM}$,
the PPEL estimator $\tilde{\btheta}_{\calM}$ projects out the influence of the nuisance parameters in $\mathcal{M}^{\mathrm c}$ from PEL, and it restores in Theorem \ref{The.3} the standard inferential procedure based on a zero-mean limiting normal distribution.

\section{Simulations}\label{sec:simulations}

In this section we demonstrate our estimation and inference procedures in three important econometric models which fit seamlessly into our framework. 
The VAR model conventionally features a small cross section and a relatively long time dimension, thereby taking $n\to \infty$ only in the asymptotic framework.
Such asymptotics fails to provide satisfactory approximation to the finite sample behavior when the cross section is non-trivial.
The second example is the local projection method that is closely related to VAR,
where the number of unknown parameters accumulates when the prediction horizon moves forward.
The third example is the MGARCH that mimics VAR in modeling the dynamics and interaction of the volatility over a cross section.

Estimation for high-dimensional models involves meticulous tuning and optimization. 
PEL's $P_{1,\pi}(\cdot)$ and $P_{2,\nu}(\cdot)$ are chosen as the SCAD penalty and the $L_1$-norm (Lasso) penalty, respectively. We employ the
interior-point method \citep{koh2007interior,koh2007efficient} to efficiently solve the inner layer optimization for $\blambda$ in \eqref{eq:double-pen}. For the outer layer, to handle the non-differentiability of the SCAD penalty at 0, we utilize a strategy that synthesizes adaptive moment estimation (ADAM) 
algorithm \citep{Kingma2014AdamAM} with proximal gradient descent.
The tuning parameters $\pi$ and $\nu$ are chosen by minimizing the BIC-type function
\begin{align*}
    \text{BIC}(\nu, \pi) = \log\{|\bar{\bg}(\hat{\btheta}_n^{(\nu,\pi)})|_2^2\}+ \frac{\log n}{n}\cdot \{\text{df}(\hat{\btheta}_n^{(\nu,\pi)})+\text{df}(\hat\blambda(\hat{\btheta}_n^{(\nu,\pi)}))\} \,,
\end{align*}
where $\hat{\btheta}_n^{(\nu,\pi)}$ is a local minimizer of \eqref{eq:double-pen} for 
 a given pair $(\nu,\pi)$, and $\text{df}(\hat{\btheta}_n^{(\nu,\pi)})$ and $\text{df}(\hat\blambda(\hat{\btheta}_n^{(\nu,\pi)}))$ denote the number of nonzero elements in $\hat{\btheta}_n^{(\nu,\pi)}$ and $\hat\blambda(\hat{\btheta}_n^{(\nu,\pi)})$, respectively.
For the PPEL method, we solve \eqref{eq:ifind} for a properly chosen tuning parameter $\varsigma = 0.2n^{-1/3}$ which satisfies the theoretical assumptions. 
When we experiment with the Parzen kernel, the Tukey-Hanning kernel, and the QS kernel to compute the asymptotic variance of the PPEL estimator under the bandwidth
$h_n=n^{1/5}$, we find the numerical results are robust to all kernels; we therefore report those under the Parzen kernel only.
For each data generation process, we repeat $N=500$ times for computing $\text{MSE}=p^{-1}N^{-1}\sum_{i=1}^{N} |\hat\btheta^{(i)}-\btheta_0|_2^2$, $\text{Bias}^2 =p^{-1} |N^{-1}\sum_{i=1}^{N}\hat\btheta^{(i)}-\btheta_0|_2^2$ and $\text{Var} = \text{MSE} - \text{Bias}^2$, where $\hat\btheta^{(i)}$ is the estimate of $\btheta_0$ in the $i$-th repetition. 

\subsection{VAR}\label{subsec:sim-var}

VAR is an off-the-shelf multivariate time series model. We first present a VAR($l$) in the following example to fix the notations. 


\noindent{\bf Example 1: VAR($l$)}.  A VAR of lag $l$ for a vector $\bfz_t =(z_{t,1},\ldots,z_{t,d_z})^{\T}\in\mathbb{R}^{d_z}$ follows
\begin{align*}
	\bfz_t=\bG_1\bfz_{t-1}+\cdots+\bG_l\bfz_{t-l}+\boldsymbol{\varepsilon}_t\,,
\end{align*}
where $\bG_1,\ldots,\bG_l$ are $l$ coefficient matrices and $\bepsilon_t$ is a white noise series.
We collect $\bfx_t=(\bfz_t^{\T},\ldots, \bfz_{t-l}^{\T})^{\T}$, 
 $\btheta=\{{\rm vec}(\bG_1)^{\T},\ldots,{\rm vec}(\bG_l)^{\T}\}^{\T}$,
 and 
 \begin{align*}
    \bg(\bfx_t;\btheta_0) 
    =(\bfz_t-\bG_1\bfz_{t-1}-\cdots-\bG_l\bfz_{t-l})\otimes (1,\bfz_{t-1}^{\T},\ldots,\bfz_{t-l}^{\T})^{\T}\,.
\end{align*}
Given $\bepsilon_t$ is of zero mean and uncorrelated with $\bfz_{t-1},\ldots,\bfz_{t-l}$, we have
$
\mathbb{E}\{\bg(\bfx_t;\btheta_0)\}={\bf0}$, which is over-identified for  $\btheta_0$.

\bigskip

While AR(1) is the prototype of the univariate AR model, VAR(1) is the most used VAR specification in practice.
We generate  $\bfz_t$ from
$\bfz_t = \bG_1\bfz_{t-1}+\bepsilon_t$ with $\bepsilon_t\overset{{\rm i.i.d.}}{\sim}N(\bzero, \bSigma_{\bepsilon})$,
where $\bG_1$ is a $d_z\times d_z$ parameter matrix and $\bSigma_{\bepsilon}$ is specified as 
\begin{itemize}
\item {\bf Case I} (Isotropic): $\bSigma_{\bepsilon}=\bI_{d_z}$.

\item {\bf Case II} (Toeplitz): $\bSigma_{\bepsilon}=(\sigma_{\bepsilon,i,j})_{d_z\times d_z}$ with $\sigma_{\bepsilon,i,j} = 0.2^{|i-j|}$. 
\end{itemize}
The VAR(1) implies the moments
$
\mathbb{E}(\bepsilon_t)=\mathbb{E}(\bfz_t - \bG_1\bfz_{t-1})=\bzero\
$
and 
$\mathbb{E}(\bepsilon_t\otimes\bfz_{t-1})=\mathbb{E}\{(\bfz_t - \bG_1\bfz_{t-1})\otimes\bfz_{t-1}\}=\bzero$.

\begin{table}[htbp]
\centering
\caption{The performance of the PEL estimator in the VAR model}
	\label{tab1}
	\smallskip\small
	\begin{threeparttable}
\begin{tabular}{clcccccc}
\toprule
  &  & \multicolumn{3}{c}{Case I}                                                    & \multicolumn{3}{c}{Case II}                                                   \\ \cmidrule(lr){3-5} \cmidrule(lr){6-8}
    $(n,d_z)$                     &  Method                     & \multicolumn{1}{c}{MSE} & \multicolumn{1}{c}{Bias$^2$} & \multicolumn{1}{c}{Var} & \multicolumn{1}{c}{MSE} & \multicolumn{1}{c}{Bias$^2$} & \multicolumn{1}{c}{Var} \\ \midrule
(50,10) & PEL                     & 0.004                   & 0.000                     & 0.004                   & 0.005                   & 0.000                     & 0.005                   \\
                         & OLS                     & 0.024                   & 0.000                     & 0.024                   & 0.026                   & 0.000                     & 0.026                   \\
                         & $\ell_1$-LS                   & 0.065                   & 0.063                     & 0.002                   & 0.061                   & 0.059                     & 0.002                   \\
                         \hline
(80,10) & PEL                     & 0.002                   & 0.000                     & 0.002                   & 0.001                   & 0.000                     & 0.001                   \\
                         & OLS                     & 0.013                   & 0.000                     & 0.013                   & 0.014                   & 0.000                     & 0.014                   \\
                         & $\ell_1$-LS                   & 0.070                   & 0.069                     & 0.001                   & 0.065                   & 0.064                     & 0.001                   \\
\hline
(50,30) & PEL                     & 0.004                   & 0.003                     & 0.002                   & 0.004                   & 0.003                     & 0.001                   \\
                         & OLS                     & 0.054                   & 0.000                     & 0.054                   & 0.058                   & 0.000                     & 0.058                   \\
                         & $\ell_1$-LS                   & 0.003                   & 0.003                     & 0.000                   & 0.003                   & 0.003                     & 0.000                   \\
\hline
(80,30) & PEL                     & 0.004                   & 0.002                     & 0.001                   & 0.004                   & 0.002                     & 0.001                   \\
                         & OLS                     & 0.020                   & 0.000                     & 0.020                   & 0.022                   & 0.000                     & 0.022                   \\
                         & $\ell_1$-LS                   & 0.003                   & 0.003                     & 0.000                   & 0.003                   & 0.003                     & 0.000             \\ \bottomrule      
\end{tabular}
\end{threeparttable}
\end{table}

We experiment with four combinations $(n,d_z) = (50,10), (80,10), (50,30)$ and $(80,30)$.
The model complexity $d_z$ is comparable with the dimension in the empirical application, whereas $n$ here is sufficient to illustrate the performance. 
Under each $(n,d_z)$, we generate the coefficient matrix $\bG_1$ with $10\%$ nonzero elements at random and rescale it to ensure that the process is stable with the signal-to-noise ratio $2:1$.

We compare the PEL estimator of $\btheta_0$ with the OLS estimator and \cite{basu2015regularized}'s $\ell_1$-LS estimator under their tuning procedure. 
We choose the OLS estimate as the initial value of PEL.
The results are reported in Table \ref{tab1}. 
Utilizing the sparsity of the parameters, the penalized methods significantly outperform OLS in various settings.
When $d_z = 10$, PEL attains an edge over the competitive methods. 
When $d_z = 30$, the performances of the PEL estimator and the  $\ell_1$-LS estimator are comparable, but PEL has the advantage of exhibiting lower bias.

As \cite{basu2015regularized} does not provide an asymptotic distribution for inference, we can only focus on our PPEL method. The coverage frequency and median length
of the PPEL-based CI for $\theta_{0,k}$ are tabulated in Table \ref{tab2}. Without loss of generality, we simply choose $k$ as the index of the first nonzero element of $\btheta_0$. Three confidence levels, $90\%$, $95\%$ and $99\%$, are considered.
Table \ref{tab2} shows that the coverage frequency approaches its nominal levels, and the median length of the CIs decreases as $n$ gets bigger with $d_z$ fixed. In addition, the median length of the CIs increases as $d_z$ gets bigger with $n$ fixed. These observations reflect the interaction between the complexity of the models and the sample sizes.

\begin{table}[htbp]
\centering
\caption{Coverage frequency and median length of PPEL-based CIs in VAR}
 \label{tab2}
	\smallskip\small
	\begin{threeparttable}
\begin{tabular}{ccccccccccccc} \toprule
& \multicolumn{6}{c}{Coverage frequency}                   & \multicolumn{6}{c}{Median CI length}                     \\ \cmidrule(lr){2-7} \cmidrule(lr){8-13}
 & \multicolumn{3}{c}{Case I} & \multicolumn{3}{c}{Case II} & \multicolumn{3}{c}{Case I} & \multicolumn{3}{c}{Case II} \\ \cmidrule(lr){2-4} \cmidrule(lr){5-7}\cmidrule(lr){8-10} \cmidrule(lr){11-13}
                   $(n, d_z)$      & 90\%    & 95\%    & 99\%   & 90\%    & 95\%    & 99\%    & 90\%    & 95\%    & 99\%   & 90\%    & 95\%    & 99\%    \\ \midrule
                                             
(50, 10)                                     & 0.892   & 0.954   & 0.994  & 0.886   & 0.954   & 0.994   & 0.384   & 0.457   & 0.601  & 0.418   & 0.498   & 0.655   \\
(80, 10)                                     & 0.902   & 0.948   & 0.992  & 0.906   & 0.960   & 0.994   & 0.298   & 0.355   & 0.467  & 0.321   & 0.382   & 0.502   \\
(50, 30)                                     & 0.912   & 0.958   & 0.984  & 0.916   & 0.954   & 0.990   & 0.654   & 0.779   & 1.024  & 0.693   & 0.826   & 1.085   \\
(80, 30)                                     & 0.908   & 0.942   & 0.988  & 0.922   & 0.950   & 0.978   & 0.464   & 0.553   & 0.727  & 0.481   & 0.573   & 0.753  
\\ \bottomrule
\end{tabular}
\end{threeparttable}
\end{table}

\subsection{Local projection}\label{subsec:lp}

IRFs, interpreted as causal effects in dynamic modeling, are of central economic interest in macroeconomics.
IRF can be elicited from the multi-equation VAR, or from a sequence of single-equation regressions. Let us continue with Example 1.


\noindent{\bf Example 2: Local Projection}. 
Without loss of generality, suppose that we are interested in the first variable $z_{t,1}$ of the vector $\bfz_t$. The VAR system's first equation is a one-period-ahead predictive regression 
$
z_{t,1} = \bG_{1,\cdot} \bfz_{t-1} + \cdots + \bG_{l,\cdot} \bfz_{t-l} + \varepsilon_{t,1},
$
where $\bG_{k,\cdot}$ is the $k$-th row of $\bG_k$. 
The key idea of LP is \emph{projecting} the $h$-period-ahead $z_{t+h,1}$ to the past variables of time $(t-1, \ldots, t-l)$:
$$
z_{t+h,1} =  \bfz_{t-1}^{\T} \bbeta_1^{(h)} + \cdots + \bfz_{t-l}^{\T} \bbeta_l^{(h)} + \varepsilon^{(h)}_{t+h,1}
$$
for each $h\in \{0\}\cup [H]$,
where $H$ is the longest prediction horizon of interest, and  $\{\bbeta_1^{(h)}, \ldots, \bbeta_l^{(h)} \}_{h=0}^H$ are functions of the VAR parameters by simple iterative substitution.
It turns out that the sequence of $\{\bbeta_1^{(h)}\}_{h=0}^H$ is the IRF \citep{Jorda2005}. 
These regressions imply the $(H+1)\times l$ moment conditions
\begin{align}\label{eq:lp_moments}
    \mathbb{E}\{ 
     (z_{t+h,1}   - \bfz_{t-1}^{\T} \bbeta_1^{(h)} - \cdots  -\bfz_{t-l}^{\T} \bbeta_l^{(h)})
         \bfz_{t-q}
      \} = \bzero\,\quad 
\end{align}
for $h\in \{0\}\cup [H]$ and $q\in [l]$,  where $\btheta = 
\textrm{vec} ( \{\bbeta_1^{(h)}, \ldots, \bbeta_l^{(h)} \}_{h=0}^H )$. 

\bigskip

Our simulation design mimics the empirical application based on \cite{RameyZubairy2018}.
They model two variables, $z_{t,1}$ for the GDP growth and $z_{t,2}$ for the growth of government spending, via a VAR(4) for ``quarterly data''
\begin{align*}
    \bigg(\begin{array}{c}
        z_{t,1}  \\ z_{t,2}
    \end{array}\bigg) = \sum_{l=1}^4
\bG_l  \bigg(\begin{array}{c}
        z_{t-l, 1}  \\ z_{t-l,2}
    \end{array}\bigg) + \bb_0 \cdot \mathrm{shock}_t + \sum_{l=1}^4\bb_l \cdot \mathrm{shock}_{t-l} + \bigg(\begin{array}{c}
        \epsilon_{t, 1} \\ \epsilon_{t,2}
    \end{array}\bigg)\,.
\end{align*}
We set the true parameters
$    \bG_1 = \bigg(\begin{array}{cc}
      0.5   &  0.2\\
       0  &  0.5
    \end{array}\bigg)$ and 
    $\bb_0 = (0.5, 0.5)^{\T}$,
along with the sparse coefficients $\bb_1=\bzero$,
$\bG_l=\bzero$ and $\bb_l=\bzero$ for $l=2,3,4$, and the error term is generated from
$
(\epsilon_{t,1}, \epsilon_{t,2})^{\T} \sim N(\bzero, \bSigma_\epsilon)$
with
$\bSigma_\epsilon = \bigg(\begin{array}{cc}
        1 & 0.5 \\
        0.5 & 1
\end{array}\bigg)$. 
We use the real data of \cite{RameyZubairy2018}'s exogenous news shock 
for the variable ``$\mathrm{shock}_t$''.
Without loss of generality, we look at the results for $z_{t,1}$.
Given the simulated data, we estimate the  linear regression
\begin{align}\label{eq:lp1}
z_{t+h,1} = \alpha_1^{(h)} + \beta_1^{(h)} \cdot \text{shock}_t + \Psi_1^{(h)} (L)\bw_{t-1} + \varepsilon^{(h)}_{t+h,1}\,,
\end{align}
where $\bw_{t-1}$ is a vector of control variables, 
and $\Psi_1^{(h)}(L)$ is a polynomial in the lag operator.

The LP model features the moment conditions as in \eqref{eq:lp_moments}.
The default implementation of LP is via OLS. 
We will compare the numerical performances of OLS and PEL/PPEL. Following the specification in \citet{RameyZubairy2018}, we set $h=0,1,\ldots,20$ in \eqref{eq:lp1},
the control variable vector $\bw_{t-1}$ includes lags of $(z_{t,1}, z_{t,2}, \mathrm{shock}_t)$, and the number of the lags is $4$. Notice that here we have  $p = 294$ parameters to handle, as each of the $h$-period ahead regressions involves an $\alpha^{(h)}_1$, a $\beta^{(h)}_1$, and $3 \times4 $ coefficients for lag terms,
totaling $294 =21\times (1+1+12)$.

\begin{table}[htbp]
\centering
\caption{The performance of PEL/PPEL in linear models}
	\label{tab:lp1}
	\smallskip\small
	\begin{threeparttable}
\begin{tabular}{lllllllllllll} \toprule
                      
                      & \multicolumn{12}{c}{Estimation}                                                                                                                                                                                                                                                                                               \\ \cmidrule(lr){2-13}
                      & \multicolumn{6}{c}{$n=300$}                                                                                                                                     & \multicolumn{6}{c}{$n=500$}                                                                                                                                     \\ \cmidrule(lr){2-7}\cmidrule(lr){8-13}
                      & \multicolumn{3}{c}{LP}                                                        & \multicolumn{3}{c}{PEL}                                                       & \multicolumn{3}{c}{LP}                                                        & \multicolumn{3}{c}{PEL}                                                       \\ \cmidrule(lr){2-4}\cmidrule(lr){5-7}\cmidrule(lr){8-10}\cmidrule(lr){11-13}
                      & \multicolumn{1}{c}{MSE} & \multicolumn{1}{c}{Bias$^2$} & \multicolumn{1}{c}{Var} & \multicolumn{1}{c}{MSE} & \multicolumn{1}{c}{Bias$^2$} & \multicolumn{1}{c}{Var} & \multicolumn{1}{c}{MSE} & \multicolumn{1}{c}{Bias$^2$} & \multicolumn{1}{c}{Var} & \multicolumn{1}{c}{MSE} & \multicolumn{1}{c}{Bias$^2$} & \multicolumn{1}{c}{Var} \\ \cmidrule(lr){2-13}
                      & 0.282                   & 0.001                     & 0.281                   & 0.064                   & 0.002                     & 0.062                   & 0.273                   & 0.001                     & 0.272                   & 0.050                   & 0.002                     & 0.048                   \\ \toprule
                      & \multicolumn{12}{c}{Coverage frequency}                                                                                                                                                                                                                                                                                       \\ \cmidrule(lr){2-13} & \multicolumn{6}{c}{$n=300$}                                                                                                                                     & \multicolumn{6}{c}{$n=500$}                                                                                                                                     \\ \cmidrule(lr){2-7}\cmidrule(lr){8-13}
                      & \multicolumn{3}{c}{LP}                                                        & \multicolumn{3}{c}{PPEL}                                                      & \multicolumn{3}{c}{LP}                                                        & \multicolumn{3}{c}{PPEL}                                                      \\ \cmidrule(lr){2-4}\cmidrule(lr){5-7}\cmidrule(lr){8-10}\cmidrule(lr){11-13}
$h$ & 90\%                    & 95\%                      & 99\%                    & 90\%                    & 95\%                      & 99\%                    & 90\%                    & 95\%                      & 99\%                    & 90\%                    & 95\%                      & 99\%                    \\ \midrule
0                     & 0.782                   & 0.850                     & 0.930                   & 0.900                   & 0.940                     & 0.980                   & 0.770                   & 0.838                     & 0.924                   & 0.904                   & 0.954                     & 0.984                   \\
2                     & 0.792                   & 0.852                     & 0.930                   & 0.900                   & 0.930                     & 0.970                   & 0.792                   & 0.844                     & 0.914                   & 0.906                   & 0.948                     & 0.978                   \\
4                     & 0.788                   & 0.844                     & 0.922                   & 0.908                   & 0.940                     & 0.970                   & 0.764                   & 0.834                     & 0.924                   & 0.910                   & 0.952                     & 0.970                   \\
6                     & 0.772                   & 0.844                     & 0.928                   & 0.912                   & 0.952                     & 0.982                   & 0.784                   & 0.846                     & 0.918                   & 0.914                   & 0.956                     & 0.982                   \\
8                     & 0.798                   & 0.852                     & 0.926                   & 0.916                   & 0.950                     & 0.974                   & 0.772                   & 0.844                     & 0.924                   & 0.902                   & 0.944                     & 0.968                   \\
10                    & 0.772                   & 0.832                     & 0.924                   & 0.910                   & 0.946                     & 0.964                   & 0.770                   & 0.830                     & 0.918                   & 0.906                   & 0.952                     & 0.974                   \\
12                    & 0.788                   & 0.860                     & 0.946                   & 0.920                   & 0.938                     & 0.978                   & 0.796                   & 0.858                     & 0.936                   & 0.924                   & 0.956                     & 0.974                   \\
14                    & 0.796                   & 0.864                     & 0.934                   & 0.906                   & 0.952                     & 0.968                   & 0.794                   & 0.844                     & 0.932                   & 0.914                   & 0.954                     & 0.982                   \\
16                    & 0.824                   & 0.884                     & 0.946                   & 0.936                   & 0.962                     & 0.978                   & 0.776                   & 0.836                     & 0.920                   & 0.910                   & 0.954                     & 0.986                   \\
18                    & 0.788                   & 0.852                     & 0.924                   & 0.924                   & 0.956                     & 0.974                   & 0.786                   & 0.842                     & 0.912                   & 0.894                   & 0.944                     & 0.970                   \\
20                    & 0.790                   & 0.860                     & 0.934                   & 0.900                   & 0.952                     & 0.964                   & 0.762                   & 0.838                     & 0.930                   & 0.906                   & 0.954                     & 0.978                   \\ \toprule
                      &  \multicolumn{12}{c}{Median CI length}                                              \\ \cmidrule(lr){2-13}
                      & \multicolumn{6}{c}{$n=300$}                                                                                                                                     & \multicolumn{6}{c}{$n=500$}                                                                                                                                     \\ \cmidrule(lr){2-7}\cmidrule(lr){8-13}
                      & \multicolumn{3}{c}{LP}                                                        & \multicolumn{3}{c}{PPEL}                                                      & \multicolumn{3}{c}{LP}                                                        & \multicolumn{3}{c}{PPEL}                                                      \\ \cmidrule(lr){2-4}\cmidrule(lr){5-7}\cmidrule(lr){8-10}\cmidrule(lr){11-13}
                      $h$ & 90\%                    & 95\%                      & 99\%                    & 90\%                    & 95\%                      & 99\%                    & 90\%                    & 95\%                      & 99\%                    & 90\%                    & 95\%                      & 99\%                    \\ \midrule
0                     & 2.049                   & 2.444                     & 3.221                   & 2.624                   & 3.127                     & 4.110                   & 2.006                   & 2.392                     & 3.148                   & 2.583                   & 3.078                     & 4.046                   \\
2                     & 2.311                   & 2.757                     & 3.633                   & 2.608                   & 3.108                     & 4.084                   & 2.272                   & 2.709                     & 3.565                   & 2.595                   & 3.092                     & 4.064                   \\
4                     & 2.295                   & 2.737                     & 3.607                   & 2.521                   & 3.004                     & 3.948                   & 2.247                   & 2.679                     & 3.526                   & 2.532                   & 3.017                     & 3.965                   \\
6                     & 2.312                   & 2.758                     & 3.634                   & 2.580                   & 3.074                     & 4.040                   & 2.263                   & 2.698                     & 3.551                   & 2.466                   & 2.939                     & 3.862                   \\
8                     & 2.351                   & 2.804                     & 3.695                   & 2.600                   & 3.098                     & 4.071                   & 2.259                   & 2.694                     & 3.546                   & 2.525                   & 3.009                     & 3.954                   \\
10                    & 2.325                   & 2.774                     & 3.655                   & 2.588                   & 3.084                     & 4.053                   & 2.226                   & 2.654                     & 3.493                   & 2.498                   & 2.977                     & 3.912                   \\
12                    & 2.382                   & 2.842                     & 3.744                   & 2.617                   & 3.118                     & 4.098                   & 2.259                   & 2.694                     & 3.546                   & 2.517                   & 2.999                     & 3.942                   \\
14                    & 2.330                   & 2.779                     & 3.662                   & 2.527                   & 3.011                     & 3.957                   & 2.284                   & 2.723                     & 3.585                   & 2.555                   & 3.044                     & 4.001                   \\
16                    & 2.326                   & 2.774                     & 3.656                   & 2.555                   & 3.044                     & 4.001                   & 2.263                   & 2.699                     & 3.552                   & 2.493                   & 2.971                     & 3.904                   \\
18                    & 2.321                   & 2.769                     & 3.649                   & 2.556                   & 3.045                     & 4.002                   & 2.318                   & 2.764                     & 3.638                   & 2.545                   & 3.033                     & 3.985                   \\
20                    & 2.358                   & 2.812                     & 3.706                   & 2.578                   & 3.072                     & 4.037                   & 2.240                   & 2.671                     & 3.516                   & 2.480                   & 2.955                     & 3.884    
\\ \bottomrule
\end{tabular}
\end{threeparttable}
\end{table}

\begin{figure}[!htbp]
\centering
\includegraphics[width=0.75\textwidth]{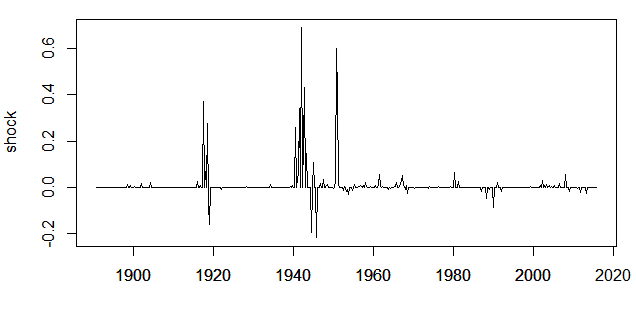}
\caption{Military news shock (proportion of GDP) from \cite{RameyZubairy2018}}\label{fig:lp2}
\end{figure}

The simulation results are summarized in Table \ref{tab:lp1} for $n=300$ and $500$, where we use ``LP'' to denote the standard LP's OLS estimator.
Focusing on the IRF, here we report the coverage frequency and median CI length for each $\beta_1^{(h)}$. 
Under the panel ``Estimation'', PEL enjoys much smaller variance and overall MSE than LP, showing the benefit of taking advantage of the sparsity. 
In terms of the coverage frequency for each coefficient $\beta_1^{(h)}$, 
here we report those of the even $h$ as the results under the odd $h$ have virtually no difference in terms of the patterns. The empirical coverage of LP  is lower than the nominal counterpart, while PPEL (with long-run variance estimated under the Parzen kernel) performs much better. 
The observation that  LP's coverage frequency does not improve as the sample size expands from $n=300$ to $n=500$ stems from actual data on military news shocks, as depicted in Figure \ref{fig:lp2}. Historically, a significant portion of America's major shocks --- 88\% of the total 500 shocks by absolute magnitude --- happened within the initial 300 observations (prior to 1960). The standard deviation of these first 300 shocks is 0.077, whereas it notably drops to 0.012 for the following 200 shocks.

\subsection{MGARCH model}\label{subsec:sim_MARCH}

While VAR uses historical data to predict the means,  MGARCH forecasts future volatility. 

 
\noindent{\bf Example 3: MGARCH model}.
Suppose a stochastic vector process $\bfy_t\in\mathbb{R}^{d_y}$ with $\mathbb{E}(\bfy_t)={\bf0}$ follows \citet{EngleandKroner1995}'s MGARCH-BEKK(1,1) model
\begin{align}\label{eq:Ht}
\bfy_t=\bfH_t^{1/2}\bepsilon_t\,, 
~~\text{where}~~\bepsilon_t\overset{{\rm i.i.d.}}{\sim}N(\bzero,\bI_{d_y})\,,~~\text{and}~~
\bfH_t=\bC^{\T}\bC+\bD\bfy_{t-1}\bfy_{t-1}^{\T}\bD^{\T}+\bB\bfH_{t-1}\bB^{\T}\,,
\end{align}
where $\bD$ and $\bB$ are $d_{y}\times d_{y}$ parameter matrices,  $\bC$ is a $d_{y}\times d_{y}$ triangular matrix.
Denote $\mathcal{J}_{-\infty}^{t}$ as the $\sigma$-filed generated by $\{\bfy_s\}$ up to and including time $t$. Then by \eqref{eq:Ht}, we have
$
\mathbb{E}(\bfy_{t}\bfy_{t}^{\T}|\,\mathcal{J}_{-\infty}^{t-2})=\bC^{\T}\bC+\bD\mathbb{E}(\bfy_{t-1}\bfy_{t-1}^{\T}|\,\mathcal{J}_{-\infty}^{t-2})\bD^{\T}+\bB\mathbb{E}(\bfy_{t-1}\bfy_{t-1}^{\T}|\,\mathcal{J}_{-\infty}^{t-2})\bB^{\T}
$,
which implies 
\begin{align}\label{eq:r1}
\mathbb{E}(\bfy_{t}\bfy_{t}^{\T}-\bC^{\T}\bC-\bD\bfy_{t-1}\bfy_{t-1}^{\T}\bD^{\T}-\bB\bfy_{t-1}\bfy_{t-1}^{\T}\bB^{\T}|\,\mathcal{J}_{-\infty}^{t-2})={\bf 0}\,.
\end{align}	
Let $\bq^K(\by_{t-2})=\{q^{K}_{1}(\by_{t-2}),\ldots,q^{K}_{K}(\by_{t-2})\}^{\T}$ denote a $K\times 1$ vector of known basis functions which, as $K\to \infty$, well approximate square integrable functions of $\by_{t-2}$, such as polynomial splines, B-splines, and power series. Then the conditional moment restrictions in \eqref{eq:r1} lead to  a large number of unconditional moments
\begin{align}\label{eq:r1_1}
\mathbb{E}\{\bg_1(\bfx_t;\btheta_0)\}:=\mathbb{E}\{{\rm vech}(\bfy_{t}\bfy_{t}^{\T}-\bC^{\T}\bC-\bD\bfy_{t-1}\bfy_{t-1}^{\T}\bD^{\T}-\bB\bfy_{t-1}\bfy_{t-1}^{\T}\bB^{\T})\otimes \bq^K(\by_{t-2})\}={\bf 0}\,, 
\end{align}
where $\bfx_t = (\bfy_t^{\T},\bfy_{t-1}^{\T},\bfy_{t-2}^{\T})^{\T}$ and $\btheta_{0}=\{{\rm vech}(\bC)^{\T},{\rm vec}(\bD)^{\T}, {\rm vec}(\bB)^{\T}\}^{\T}$.  
Moreover, \eqref{eq:Ht} also implies 
\begin{align}
\mathbb{E}\{\bg_2(\bfx_t;\btheta_0)\} := \mathbb{E}\{(\bfy_{t}\bfy_{t}^{\T}-\bC^{\T}\bC-\bD\bfy_{t-1}\bfy_{t-1}^{\T}\bD^{\T})\bfy_{t-1}\} ={\bf 0}\,. \notag
\end{align}
Together, they produce the moment constraints $\bg(\bfx_t;\btheta_0)=\{\bg_1(\bfx_t;\btheta_0)^{\T}, \bg_2(\bfx_t;\btheta_0)^{\T}\}^{\T}$.



\begin{figure}[!htbp]
\centering
\includegraphics[width=0.8\textwidth]{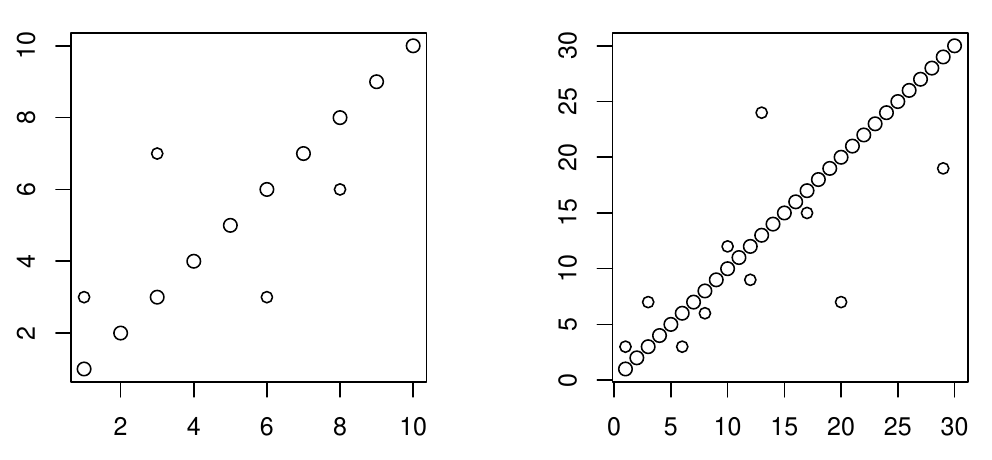}
\caption{The parameter matrix $\bD$ in the MGARCH model. Left: $d_y=10$, right: $d_y=30$}\label{fig1}
\end{figure}

\begin{table}[htbp]
\centering
\caption{The performance of the PEL estimator in the MGARCH model}
	\label{tab3}
	\smallskip\small
	\begin{threeparttable}
\begin{tabular}{clcccccc} \toprule
  &  & \multicolumn{3}{c}{Case I}                                                    & \multicolumn{3}{c}{Case II}                                                   \\ \cmidrule(lr){3-5} \cmidrule(lr){6-8}
                    $(n,d_y)$     &    Method                     & \multicolumn{1}{c}{MSE} & \multicolumn{1}{c}{Bias$^2$} & \multicolumn{1}{c}{Var} & \multicolumn{1}{c}{MSE} & \multicolumn{1}{c}{Bias$^2$} & \multicolumn{1}{c}{Var} \\ \midrule
(50,10) & PEL                     & 0.139                   & 0.003                     & 0.136                   & 0.141                   & 0.004                     & 0.137                   \\
                         & MLE                     & ×                       & ×                         & ×                       & ×                       & ×                         & ×                       \\
                         & $\ell_1$-MLE                  & ×                       & ×                         & ×                       & ×                       & ×                         & ×                       \\
\hline
(80,10) & PEL                     & 0.060                   & 0.025                     & 0.035                   & 0.060                   & 0.025                     & 0.035                   \\
                         & MLE                     & ×                       & ×                         & ×                       & ×                       & ×                         & ×                       \\
                         & $\ell_1$-MLE                  & ×                       & ×                         & ×                       & ×                       & ×                         & ×                       \\
\hline
(50,30) & PEL                     & 0.125                   & 0.005                     & 0.119                   & 0.130                   & 0.005                     & 0.125                   \\
                         & MLE                     & ×                       & ×                         & ×                       & ×                       & ×                         & ×                       \\
                         & $\ell_1$-MLE                  & ×                       & ×                         & ×                       & ×                       & ×                         & ×                       \\
\hline
(80,30) & PEL                     & 0.050                   & 0.008                     & 0.043                   & 0.050                   & 0.008                     & 0.043                   \\
                         & MLE                     & ×                       & ×                         & ×                       & ×                       & ×                         & ×                       \\
                         & $\ell_1$-MLE                  & ×                       & ×                         & ×                       & ×                       & ×                         & ×                      
\\ \bottomrule
\end{tabular}
{\footnotesize Note: Marked by ``×'', MLE and $\ell_1$-MLE fail to converge numerically.}
\end{threeparttable}
\end{table}

\begin{table}[htbp]
\centering
\caption{Coverage frequency and median length of PPEL-based CIs in the MGARCH model}
	\label{tab4}
	\smallskip\small
	\begin{threeparttable}
\begin{tabular}{ccccccccccccc} \toprule
  & \multicolumn{6}{c}{Coverage frequency}                   & \multicolumn{6}{c}{Median CI length}                     \\ \cmidrule(lr){2-7} \cmidrule(lr){8-13}
 & \multicolumn{3}{c}{Case I} & \multicolumn{3}{c}{Case II} & \multicolumn{3}{c}{Case I} & \multicolumn{3}{c}{Case II} \\ \cmidrule(lr){2-4} \cmidrule(lr){5-7} \cmidrule(lr){8-10} \cmidrule(lr){11-13}
$(n, d_y)$                        & 90\%    & 95\%    & 99\%   & 90\%    & 95\%    & 99\%    & 90\%    & 95\%    & 99\%   & 90\%    & 95\%    & 99\%    \\ \midrule
                                           
(50, 10)                                     & 0.830   & 0.902   & 0.978  & 0.882   & 0.922   & 0.960   & 2.634   & 3.138   & 4.124  & 4.692   & 5.591   & 7.347   \\
(80, 10)                                     & 0.880   & 0.932   & 0.988  & 0.896   & 0.938   & 0.982   & 2.730   & 3.253   & 4.275  & 2.827   & 3.369   & 4.428   \\
(50, 30)                                     & 0.874   & 0.934   & 0.985  & 0.878   & 0.929   & 0.976   & 2.520   & 3.003   & 3.947  & 2.457   & 2.928   & 3.848   \\
(80, 30)                                     & 0.917   & 0.937   & 0.973  & 0.914   & 0.943   & 0.971   & 2.133   & 2.541   & 3.340  & 2.215   & 2.639   & 3.468  
\\ \bottomrule
\end{tabular}
\end{threeparttable}
\end{table}




We experiment with MGARCH in \eqref{eq:Ht} under the following parameter matrices $\bC,\bD,\bB$: 
\begin{itemize}
\item {\bf Case I} (Diagonal): $\bC = \bI_{d_y}$, $\bB = 0.6 \bI_{d_y}$, and $\bD = 0.6 \bI_{d_y}$. 

\item {\bf Case II} (Sparse): $\bC = \bI_{d_y}$, $\bB = 0.6 \bI_{d_y}$, and $\bD$ is a general sparse matrix with nonzero elements of the value $0.6$ on the diagonal and $0.1$ off the diagonal, as shown in Figure \ref{fig1}.
\end{itemize}

In our estimation, we specifically choose $\bq^K(\by_{t-2}) = (y_{t-2,1},\ldots, y_{t-2,5})^{\T}$ in \eqref{eq:r1_1}.
For  convenience in the simulation, we randomly choose a point near $\btheta_0$  as the initial value, $\btheta^0 = \btheta_0 + \bepsilon$ where $\bepsilon \sim N(\bzero, 0.5^2\bI_p)$, for all three estimators: PEL, MLE and $\ell_1$-penalized MLE ($\ell_1$-MLE).
Table \ref{tab3} summarizes the performances of the estimators. 
MLE and $\ell_1$-MLE fail to numerically converge in all settings --- 
in our experiments, MLE and $\ell_1$-MLE numerically break down even if the initial value is set as the true $\btheta_0$, for
the dimension of the parameter here surpasses the capacity of these two methods. 
In sharp contrast, PEL maintains robustness. The MSE of PEL decreases as $d_y$ gets bigger under the same $n$, because the significantly higher parameter dimension $p = 2d_y^2 + d_y(d_y+1)/2$ of MGARCH enhances relative sparsity and thereby the effectiveness of the penalization.
Moreover, Table \ref{tab4} shows that PPEL's coverage frequency aligns well with the nominal ones. 

\section{Empirical applications}
\label{sec:empirical-app}

Given the reasonable performance of our PEL/PPEL procedure in the simulations, we apply this method to three real-data economic and financial applications. 

\subsection{Sectoral inflation}

Inflation is a key macroeconomic indicator which serves as a signal of broad economic conditions, helping policymakers, businesses, and households make informed decisions. Controlling inflation, as one of the Federal Reserve's dual mandates, is vital for maintaining economic stability and overall well-being.
Inflation is measured by the growth rate of a price index. 
Besides the \emph{consumer price index (CPI)}, the \emph{personal consumption expenditures (PCE) price index} is the Fed's primary measure for monetary policies. 
PCE has a wide coverage, with its sectoral indices breaking down overall price changes into sectors such as food, housing, energy, healthcare, and so on. Examining how prices evolve and co-move across sectors helps pinpoint the source of inflation.

Following the onset of the COVID-19 pandemic, the United States has seen considerable variations in its inflation rate. Notably, in December 2021, inflation peaked at 7.0\%, marking the highest level in several decades, driven by surging demand and disruptions in the supply chain. This prompted the Federal Reserve to implement stricter monetary policies. By late 2023, the inflation rate had decreased to approximately 3.0\%. The data utilized for this empirical analysis is sourced directly from the Bureau of Economic Analysis, which publishes the PCE. It includes quarter-to-quarter changes in 16 sectoral PCE indices that are seasonally adjusted, covering the period from the first quarter of 1959 to the third quarter of 2023 (259 quarters).

\begin{figure}[htbp]
\begin{center}

\includegraphics[width=0.8\textwidth]{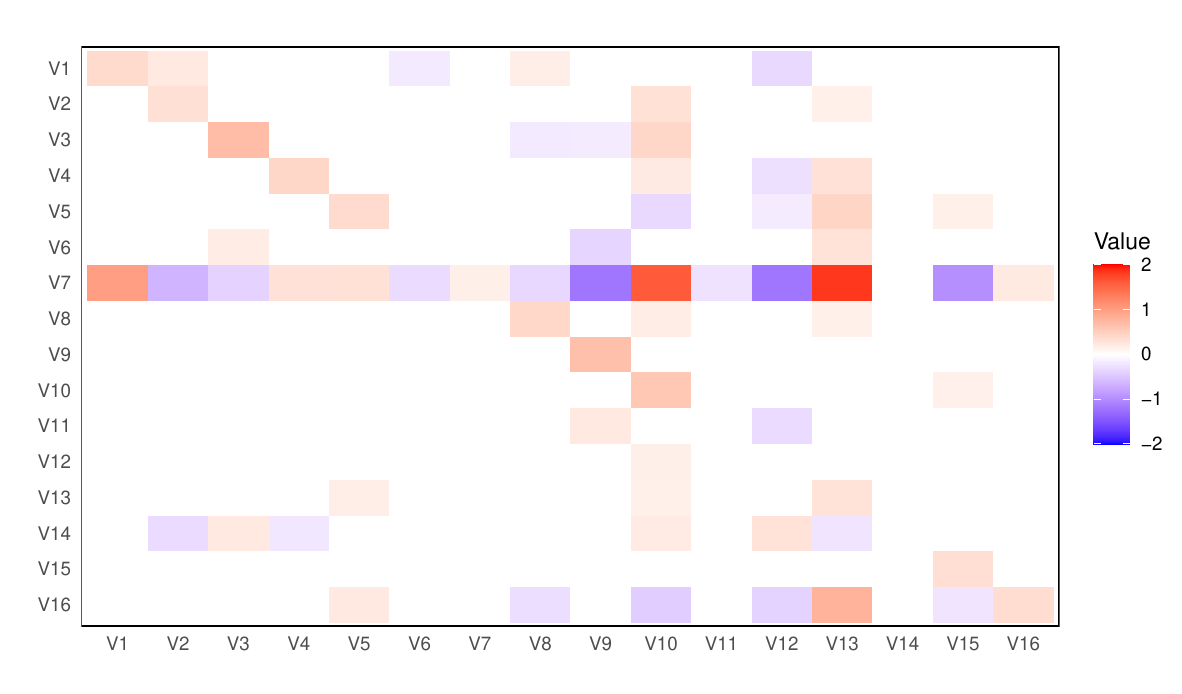} 
\caption{The AR(1) coefficient matrix of the 16-sector VAR(1) model}
\label{fig:pce_coef}
\end{center}

\begin{flushleft}
Notes: List of the code and name of each sector.\\
\texttt{V1}: \texttt{Motor vehicles and parts}, 
\texttt{V2}: \texttt{Furnishings and durable household equipment}, \\
\texttt{V3}: \texttt{Recreational goods and vehicles}, 
\texttt{V4}: \texttt{Other durable goods}, 
\texttt{V5}: \texttt{Food and beverages purchased for off-premises consumption}, 
\texttt{V6}: \texttt{Clothing and footwear}, 
\texttt{V7}: \texttt{Gasoline and other energy goods}, 
\texttt{V8}: \texttt{Other nondurable goods}, 
\texttt{V9}: \texttt{Housing and utilities}, 
\texttt{V10}: \texttt{Health care}, \\
\texttt{V11}: \texttt{Transportation services}, 
\texttt{V12}: \texttt{Recreation services}, 
\texttt{V13}: \texttt{Food services and accommodations}, 
\texttt{V14}: \texttt{Financial services and insurance}, 
\texttt{V15}: \texttt{Other services}, 
\texttt{V16}: \texttt{Final consumption expenditures of nonprofit institutions serving households (NPISHs)}.

\end{flushleft}

\end{figure}

To understand the dynamics and sectoral spillover effects, we follow the implementation of our simulation in Section \ref{subsec:sim-var} to 
fit a 16-sector VAR(1). The estimate of  coefficient matrix is shown in Figure \ref{fig:pce_coef}.
While some sectors have upstream-downstream relationships, other sectors are less connected. 
One salient feature is that most autoregressive coefficients (those on the diagonal) are non-zeros, which is a key driving force of the persistence of inflation. Secondly, the estimated matrix is overall quite sparse.
Notice that \texttt{V7}: \texttt{Gasoline and other energy goods} is much more volatile than other sectors, due to weather conditions, geopolitical uncertainty, and occasional energy crises, and therefore the fitting mechanism delivers many more active coefficients along its row to reduce the magnitude of the corresponding residual. 

\begin{figure}[htb]
	\centering
    \includegraphics[width=0.8\textwidth]{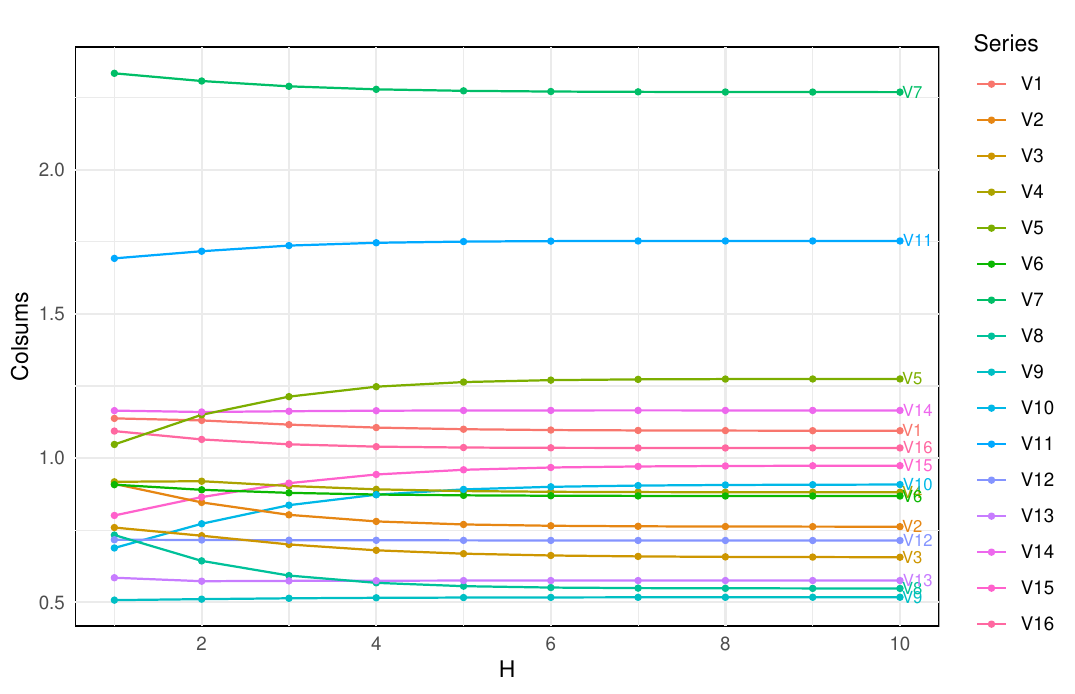}
 
	\caption{Directional connectedness:
    The colsums of $\widetilde\bD^{g,h}$ for $h=1,\ldots,10$}\label{fig:dc2} 
\end{figure}

We further follow \citet{DIEBOLD2014119}'s decomposition 
to measure the relative weight of the shocks. 
The $h$-step generalized variance decomposition matrix $\bD^{g,h} = (d_{i,j}^{{g,h}})$ is
$$
    d_{i,j}^{g,h} = \frac{\sigma_{\varepsilon,j,j}^{-1}\sum_{\ell=0}^{h-1}(\boldsymbol{\iota}_i^{\T} \bG_1^{\ell} \bSigma_{\varepsilon}\boldsymbol{\iota}_j)^2 }{\sum_{\ell=0}^{h-1}\boldsymbol{\iota}_i^{\T}\bG_1^{\ell}\bSigma_{\varepsilon}\bG_1^{\ell,\T}\boldsymbol{\iota}_i  }\,,
$$
where $\bSigma_{\varepsilon}$ is the covariance matrix of the disturbance vector, $\sigma_{\varepsilon,j,j}$ is the $j$-th diagonal element of $\bSigma_{\varepsilon}$, and $\boldsymbol{\iota}_i$ is the selection vector with one as the $i$-th element and zeros otherwise. We normalize each entry of the generalized variance decomposition matrix $\bD^{g, h}$ by the row sum, denoted by $\widetilde\bD^{g, h}$, and then calculate the column sums to measure the out-variance of each sector for different horizons $h$. The results are shown in Figure \ref{fig:dc2}. Each dot represents the relative fraction source of the variation that is originated from a particular sector over $h=1,\ldots, 10$. 
The two major sources of variation come
from \texttt{V7:Gasoline and other energy goods}  and \texttt{V11:Transportation services}.
The former exhibits the highest variance among all sectors, as mentioned above,  
and the latter is tightly linked to the energy sector. They spread out the variations to other sectors via the interconnection of the network.
This example illustrates that our PEL method allows us to work with the granular sectoral level indices to inspect the relative importance and dynamics, which is much richer than simply looking at an overall univariate inflation measure at the national level.

\subsection{Government spending multipliers}

\begin{figure}[htbp]
\centering
\includegraphics[width=1.0\textwidth]{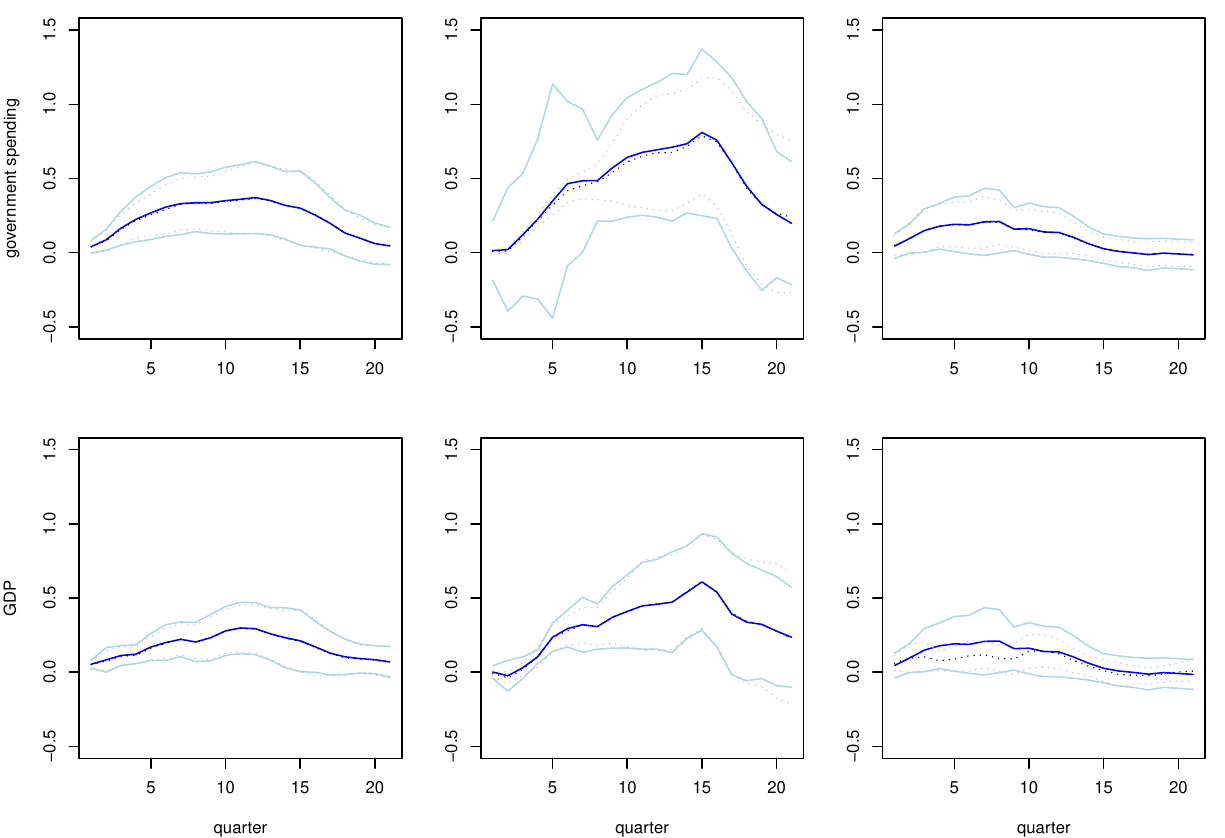}
\caption{Point estimates of IRF and 95\% confidence intervals computed from LP (gray dotted line) and PPEL (blue solid line). The first column: the full sample; the second column: subsample of high-unemployment state; the third column: subsample of low-unemployment state}\label{fig:lp1}

\end{figure}

John Maynard Keynes introduced the government spending multiplier in his \emph{General Theory},
which is foundational to his broader theories on fiscal policy and its role in managing economic activities, especially during recessions. 
Despite the importance of the concept, the measurement of the multiplier is by no means straightforward, because it is difficult to isolate it from confounding factors. 
An influential recent empirical study of the fiscal multiplier by \cite{RameyZubairy2018} uses quarterly data from 1889 to 2015. 
We follow the same specification to replicate their Figure 5,
and then use our method to re-evaluate the IRFs.
While our simulation in Section \ref{subsec:lp} utilizes only the real data of news shock shown in Figure \ref{fig:lp2}, here in the empirical application we  use the real time series of government spending and GDP. The LP specification in \eqref{eq:lp1} sets  either government spending or GDP as the target variable, and $\text{shock}_t$ is again the military spending news. The control variable $\bfw_{t-1}$ includes four lags of the news shock, government spending, and GDP. A state-dependent alternative version of the model \eqref{eq:lp1} is also considered for the subsamples of high/low-unemployment states, respectively. 



The empirical results are reported in Figure \ref{fig:lp1}.
It consists of six subplots, each being the IRF of the fiscal multiplier of the news shock to government spending (the first row) and GDP (the second row). The three columns represent the full sample (the first column), the high unemployment periods (the second column, 181 observations), and the low-unemployment periods (the third column, 320 observations), following the original empirical study. The split of ``bad times'' and ``good times'' is to check the structural stability of the IRFs under distinctive states of the economy. 

The two methods, \cite{RameyZubairy2018}'s LP estimated by OLS (which is the same as the original paper) and this paper's PPEL, return very close point estimates. 
The main message is that the fiscal multipliers are mostly much smaller than unity in American history through booms and recessions, so that counter-cyclical policies likely have dampened effects.
However, there are visually salient gaps in terms of the interval estimates.
Recall that our simulation in Section \ref{subsec:lp} has shown that, despite the narrower length of CI, the coverage of  LP may deviate from the nominal coverage frequency. 
It is echoed by Figure \ref{fig:lp1} where PPEL has slightly wider CIs than  LP  for obtaining nominal CIs.
In particular, in the second column of the subplots,
the LP with a small $h =0, 1,\ldots,5$ 
has extremely narrow confidence intervals; given a sample size of 181 observations, it is surprising to see that the confidence intervals are even much narrower than those from the full sample of 501 as in the first column. The counterintuitive observation is likely the consequence of the low empirical coverage probability that deviates from the designated nominal coverage rate. On the other hand, the length of the confidence intervals from PPEL suitably reflects the statistical randomness that corresponds to the sample sizes. 
The evidence suggests that the latter provides more reasonable quantification of the underlying uncertainty. 


\subsection{Volatility spillover across banks}

In the last empirical application, we explore the volatility spillover effect of 16 stocks from China's banking industry,
with data from the CSMAR database (\url{https://data.csmar.com/}). We investigate the daily log-returns of these stocks from January 2, 2024, to April 19, 2024. 
Table \ref{tab8} summarizes the descriptive statistics of the time series and their pairwise sample correlation. The sample means of all stocks are very close to 0, and the sample kurtosises of several stocks are much larger than 3.
Moreover, all pairwise sample correlations are positive. These findings motivate us to employ the MGARCH-BEKK(1,1) model specified in \eqref{eq:Ht} to study the interconnection of volatility.

\begin{table}[]
\setlength\tabcolsep{4pt}
\centering
\caption{Summary of 16 stocks}
	\label{tab8}
	\smallskip\small
	\begin{threeparttable}
\begin{tabular}{lcccccccccccccccc}\toprule
         & PA       & NB       & PF       & HX       & MS       & ZS       & NJ       & XY        & BJ       & NY       & JT       & GS       & GD       & JS       & ZG       & ZX       \\\midrule
mean     & 0.00  & 0.00 & 0.00 & 0.00  & 0.00 & 0.00 & 0.00  & 0.00  & 0.00  & 0.00  & 0.00  & 0.00  & 0.00  & 0.00  & 0.00 & 0.00 \\

sd       & 0.02     & 0.02     & 0.01     & 0.01     & 0.01     & 0.01     & 0.01     & 0.01      & 0.01     & 0.01     & 0.01     & 0.01     & 0.01     & 0.01     & 0.01     & 0.02     \\
skew     & 2.32     & 1.01     & 0.09     & -0.06    & 0.45     & 1.06     & -0.06    & -3.20     & -0.34    & -0.52    & -0.18    & -0.76    & -2.30    & -0.41    & 0.21     & 0.58     \\
kurtosis & 10.32    & 1.75     & 0.32     & 1.04     & 0.49     & 1.96     & 0.63     & 17.43     & 0.45     & 0.07     & 1.31     & 1.14     & 11.79    & 1.36     & 1.14     & 4.40     \\ \hline
PA       &          & 0.64     & 0.69     & 0.47     & 0.50     & 0.77     & 0.46     & 0.54      & 0.43     & 0.27     & 0.37     & 0.34     & 0.50     & 0.31     & 0.28     & 0.50     \\
NB       &          &          & 0.56     & 0.46     & 0.45     & 0.65     & 0.42     & 0.41      & 0.34     & 0.24     & 0.39     & 0.29     & 0.37     & 0.33     & 0.31     & 0.50     \\
PF       &          &          &          & 0.71     & 0.68     & 0.71     & 0.65     & 0.50      & 0.57     & 0.55     & 0.65     & 0.59     & 0.52     & 0.62     & 0.62     & 0.47     \\
HX       &          &          &          &          & 0.70     & 0.43     & 0.61     & 0.22      & 0.73     & 0.69     & 0.77     & 0.70     & 0.55     & 0.73     & 0.69     & 0.54     \\
MS       &          &          &          &          &          & 0.41     & 0.59     & 0.31      & 0.58     & 0.56     & 0.57     & 0.57     & 0.57     & 0.60     & 0.58     & 0.52     \\
ZS       &          &          &          &          &          &          & 0.34     & 0.54      & 0.33     & 0.30     & 0.44     & 0.37     & 0.44     & 0.34     & 0.36     & 0.47     \\
NJ       &          &          &          &          &          &          &          & 0.30      & 0.78     & 0.56     & 0.63     & 0.66     & 0.55     & 0.67     & 0.63     & 0.46     \\
XY       &          &          &          &          &          &          &          &           & 0.28     & 0.22     & 0.27     & 0.30     & 0.41     & 0.23     & 0.27     & 0.34     \\
BJ       &          &          &          &          &          &          &          &           &          & 0.61     & 0.67     & 0.71     & 0.63     & 0.69     & 0.64     & 0.52     \\
NY       &          &          &          &          &          &          &          &           &          &          & 0.89     & 0.92     & 0.61     & 0.89     & 0.88     & 0.40     \\
JT       &          &          &          &          &          &          &          &           &          &          &          & 0.87     & 0.66     & 0.88     & 0.89     & 0.51     \\
GS       &          &          &          &          &          &          &          &           &          &          &          &          & 0.64     & 0.88     & 0.87     & 0.47     \\
GD       &          &          &          &          &          &          &          &           &          &          &          &          &          & 0.60     & 0.59     & 0.57     \\
JS       &          &          &          &          &          &          &          &           &          &          &          &          &          &          & 0.91     & 0.45     \\
ZG       &          &          &          &          &          &          &          &           &          &          &          &          &          &          &          & 0.43    \\ \bottomrule
\end{tabular}
{\footnotesize Note: The full names of the abbreviations PA, NB, \ldots, and ZX are listed in Table \ref{tab9}.}
\end{threeparttable}
\end{table}

\begin{figure}[!htbp]
\centering
\includegraphics[width=0.8\textwidth]{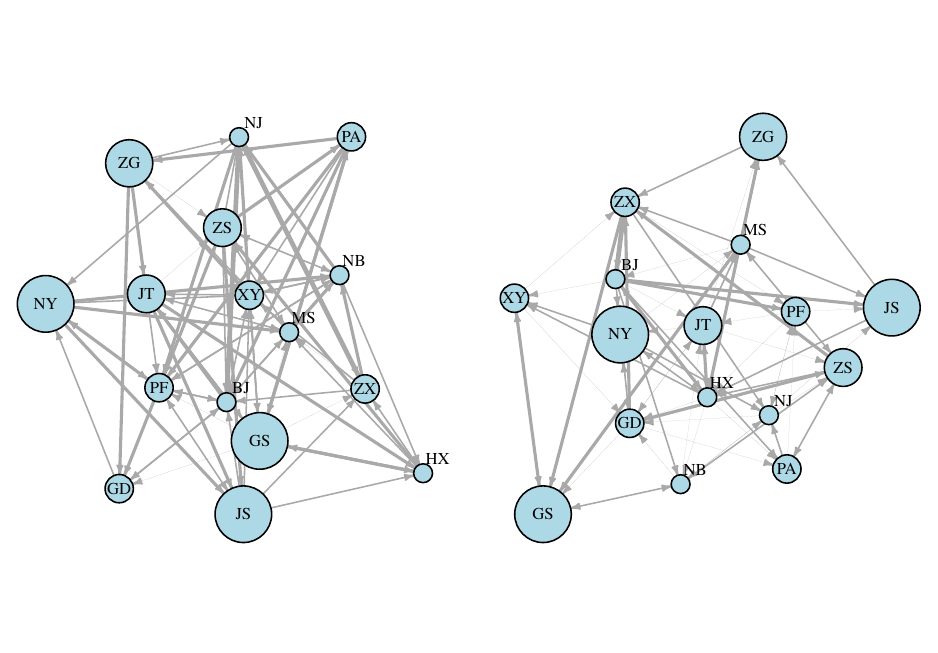}
\caption{ The network structures of estimated matrices $\widehat\bD$ (left) and $\widehat\bB$ (right). The thickness of an edge represents the relative magnitude of a coefficient, and the relative size of a vertex embodies the market value of a bank}\label{fig4}
\end{figure}

Recall that in the simulation in Section \ref{subsec:sim_MARCH}
a disturbed true value was used as an initial value for PEL.
Here for real data, we follow \cite{yao20241} to obtain an initial value. 
Specifically, we set the initial values for diagonal elements of $\bC,\bD$ and $\bB$ as square roots of estimated parameters of fitting each component series into a univariate GARCH(1,1) model, and the off-diagonal elements as random values sampled from $N(0, 0.5^2)$.
Figure \ref{fig4} plots the estimated structure of $\widehat\bD$ and $\widehat\bB$. Each arrow corresponds to a nonzero estimated off-diagonal element of $\widehat\bD$ or $\widehat\bB$. Specifically, if $\widehat\bD$'s $(i,j)$-th element $\hat{d}_{i,j}\neq 0$, the directional line shoots from $i$ to $j$. 

To check the estimated patterns of connection, we further compute the numbers of outgoing links, incoming links, and net links of the estimated $\widehat\bD$ graph, as suggested by \cite{dhaene2022volatility}. 
Table \ref{tab9} shows that the major national banks are the sources of shocks. They include the four biggest ones (Industrial and Commercial Bank of China (GS), Agricultural Bank of China (NY), Bank of China (ZG), China Construction Bank  (JS)), as well as Ping An Bank (PA), China Merchants Bank (ZS), and China CITIC Bank (ZX).
These banks are central to China's financial system and play a significant role in sustaining the country's economic growth.
On the other hand, several joint-stock commercial banks rank as top  destinations of shocks. They mostly feature a higher level of market orientation and substantial exposure to market forces. These characteristics contribute to heightened sensitivity to market shocks and greater interconnectedness, resulting in more significant volatility spillover.
The large national banks are the main focus in hedging the risk of China's banking industry. Our empirical exercise quantifies the volatility spillover effects, which can be of interest for both practitioners of risk management and policymakers.



\begin{table}[]
\setlength\tabcolsep{1.5pt}
\centering
\caption{Rank of 16 stocks in terms of net spillover links}
	\label{tab9}
	\smallskip\small
	\begin{threeparttable}
\begin{tabular}{cccccccccc}\toprule
Ticker & Company & Out & In & Out-In & Ticker & Company & Out & In & Out-In \\ \midrule
GS     & Industrial and Commercial Bank of China  & 9   & 4  & 5      & JS  &  China Construction Bank  & 5   & 4  & 1      \\
ZG &  Bank of China   & 6   & 2  & 4      & HX &  Huaxia Bank    & 4   & 5  & -1     \\
PA & Ping An Bank    & 5   & 2  & 3      & NJ &  Bank of Nanjing    & 5   & 6  & -1     \\
NY &  Agricultural Bank of China   & 5   & 3  & 2      & GD  & China Everbright Bank   & 3   & 4  & -1     \\
ZX &  China CITIC Bank   & 6   & 4  & 2      & BJ &  Bank of Beijing   & 6   & 9  & -3     \\
NB &  Ningbo Bank    & 7   & 6  & 1      & XY  & Industrial Bank   & 5   & 9  & -4     \\
ZS  & China Merchants Bank   & 7   & 6  & 1      & PF &  Shanghai Pudong Development Bank   & 5   & 10 & -5     \\
JT &  Bank of Communications   & 6   & 5  & 1      & MS &  China Minsheng Bank   & 3   & 8  & -5   
\\ \bottomrule
\end{tabular}
\end{threeparttable}
\end{table}

\section{Conclusion}\label{conclusion}

This study investigates the PEL method in the analysis of multivariate time series models characterized by high-dimensional moments and parameters. These models are prevalent in the fields of economics and finance, as seen in VAR, local projection, and volatility models. We develop the marginal EL and demonstrate the consistency of PEL. Additionally, we introduce PPEL for the inference of low-dimensional parameters, which effectively removes the impact of nuisance parameters and maintains asymptotic normality centered at zero. 
Comprehensive Monte Carlo simulations provide evidence for the validity of our approach. We illustrate its practical application in three empirical examples using real data: the dynamics of the USA's inflation, the IRF of news shocks to government spending and GDP, and the stock price volatility spillovers among Chinese banks.

The PEL/PPEL method provides a flexible and adaptable strategy. It serves as a promising technique for regulating the high dimensionality in parameters and moments. 
This approach can be extended in various directions. For example, though this paper does not address time series models with endogeneity due to space constraints, instruments can be easily integrated into our framework using moment conditions.

\bigskip
\bigskip


\begin{appendices}
\setcounter{equation}{0}
\renewcommand{\thesection}{Appendix \Alph{section}} 
\renewcommand{\thesubsection}{\Alph{section}.\arabic{subsection}} 
\renewcommand{\theequation}{\Alph{section}.\arabic{equation}}

\section{The main proofs}\label{sec:proofs}
In the sequel, we use $C$ to denote a generic {finite positive}
constant that may be different in different uses. 
For any given index set $\mathcal{L}\subset[r]$ and $\btheta\in\bTheta$, we write   $\widehat{\bGamma}_{\calL}(\btheta) = \nabla_{\btheta_{\calS}} \bar{\bfg}_{\calL}(\btheta)$  and   $\widehat{\bGamma}(\btheta) =\widehat{\bGamma}_{[r]}(\btheta)$   for simplicity.
 For any $\btheta\in\bTheta$ and $\blambda \in \hat{\Lambda}_n(\btheta)$, we define
\begin{align*}
f(\blambda;\btheta)&= \frac{1}{n}\sum_{t=1}^{n}\log\{1+\blambda^{\T} {\bfg_t(\btheta)}\} -\sum_{j=1}^{r}P_{2,\nu}(|\lambda_j|)\,,\\ S_n(\btheta)&=\max_{\blambda\in\hat{\Lambda}_n(\btheta)}f(\blambda;\btheta)+\sum_{k=1}^{p}P_{1,\pi}(|\theta_k|)\,,
\end{align*}
and $\hat{\blambda}(\btheta)=\arg \max_{\blambda\in\hat{\Lambda}_n(\btheta)}f(\blambda;\btheta)$. For some constant $c\in(C_*,1)$ with $C_*$ specified in \eqref{C*}, define $\mathcal{M}_{\btheta}(c)=\{j\in[r]:|\bar{g}_j(\btheta)|\geq c\nu\rho_2'(0^+)\}$ for $\btheta\in\bTheta$. 

\subsection{Proof of Theorem \ref{The.1}}\label{sec:prothe1}

Recall $\aleph_n=n^{-3\varphi/(6\varphi+2)}(\log r)^{1/2}$. To prove Theorem \ref{The.1}, we need Lemmas \ref{tailprabog0}--\ref{lambda0}, whose proofs are given in Sections \ref{sec:la:tailprabog0}--\ref{sec:la:lambda0} of the supplementary material, respectively.


\begin{lemma}\label{tailprabog0}
	Under Conditions {\rm \ref{con:mixingdecay}}, {\rm \ref{con:moments1}(a) and \ref{con:moments1}(b)},  it holds that $|\bar{\bg}(\btheta_{0})|_{\infty}=O_{\p}(\aleph_n)$ and 
	\begin{align*}
	\max_{j_1,j_2\in[r]}\bigg|\frac{1}{n}\sum_{t=1}^{n}\big[g_{t,j_1}(\btheta_0)g_{t,j_2}(\btheta_0)-\mathbb{E}\{g_{t,j_1}(\btheta_0)g_{t,j_2}(\btheta_0)\}\big]\bigg|=O_{\p}(\aleph_n)
	\end{align*}
provided that $L_n^{\varphi}\log (rn)\ll n^{\varphi/(3\varphi+1)}$.
\end{lemma}

\begin{lemma}\label{hatV-barV}
	Let $\mathscr{T}=\{\mathcal{T} \subset [r]:|\mathcal{T}|\leq \ell_n\}$ and $\bTheta_{n}=\{\btheta=(\btheta_\calS^{\T},\btheta_{\calS^{\rm c}}^{\T})^{\T} \in \bTheta: |\btheta_{\calS}-\btheta_{0,\calS}|_{\infty} \leq  O_{\p}(b_n^{1/2}),|\btheta_{\calS^{\rm c}}|_1\leq \aleph_n\}$. Under Conditions {\rm \ref{con:mixingdecay}, \ref{con:moments1}} and  {\rm \ref{con:moments2}(a)}, if  $L_n^{\varphi}\log (rn)\ll n^{\varphi/(3\varphi+1)}$, $s^2\ell_nb_n=o(1)$ and $\ell_n\aleph_n=o(1)$, then
	\begin{align*}
	\sup_{\btheta \in \bTheta_{n}} \sup_{\mathcal{T} \in \mathscr{T}} \| \widehat{\bfV}_{\mT} (\btheta) - \mathbb{E}\{\widehat{\bfV}_{\mT}(\btheta_0)\}\|_{2} 
	= O_{\p}(s\ell_n^{1/2}b_n^{1/2})+O_{\p}(\ell_n\aleph_n)\,.
	\end{align*}
\end{lemma}

Recall $\mathcal{M}_{\btheta}^*=\{j\in[r] : |\bar {g}_{j}(\btheta)|\geq   C_*\nu \rho_2'(0^{+})\}$ for some constant $C_*\in(0,1)$. Lemma \ref{lambdahat} provides a general result for the property of the Lagrange multiplier $\hat{\blambda}(\btheta)$, and Lemma \ref{lambda0} specifies the property of $\hat\blambda(\btheta_0)$.

\begin{lemma}\label{lambdahat}
	Let $\{\btheta_n\}$ be a sequence in $\bTheta$ and $P_{2,\nu}(\cdot)\in \mathcal{P}$ be a convex function 
 with bounded second-order derivatives around $0$, where $\mathcal{P}$ is defined in \eqref{penalty}. For some constant  $c\in(C_*,1)$, assume that all the eigenvalues of $\widehat{\bfV}_{\calM_{\btheta_n}(c)}(\btheta_n)$ are uniformly bounded away from zero and infinity w.p.a.1. Let $|\bar{\bfg}_{\calM_{{\btheta}_n}(c)}({\btheta}_n)-\nu\rho_2'(0^+)${\rm{sgn}}$\{\bar{\bfg}_{\calM_{{\btheta}_n}(c)}({\btheta}_n)\}|_2=O_{\p}(u_n)$ for some $u_n\to 0$. Assume there exists some non-random sequence $\{m_n\}$ such that $\mathbb{P}(|\mathcal{M}_{\btheta_n}^*|\leq m_n)\to 1$ as $n\to \infty$.
	If $\max_{j\in[r]}\mathbb{E}_n\{|g_{t,j}(\btheta_n)|^{\gamma}\}=O_{\p}(1)$ and  $m_n^{1/2}u_n\ll\min\{\nu,n^{-1/\gamma}\}$, then w.p.a.1 there is a sparse global maxmizer $\hat{\blambda}(\btheta_n)$ for $f(\blambda;\btheta_n)$ satisfying the three results: $\rm{(i)}$ $|\hat{\blambda}(\btheta_n)|_2=O_{\p}(u_n)$, $\rm{(ii)}$ ${\rm{supp}}\{\hat{\blambda}({\btheta}_n)\}\subset\mathcal{M}_{\btheta_n}(c)$, and $\rm{(iii)}$ ${\rm{sgn}}(\hat{\lambda}_{n,j})=$ ${\rm{sgn}}\{\bar{g}_j(\btheta_n)\}$ for any $j\in \mathcal{M}_{\btheta_n}(c)$ with $\hat{\lambda}_{n,j}\neq 0$, where $\hat{\blambda}(\btheta_n)=(\hat{\lambda}_{n,1},\dots,\hat{\lambda}_{n,r})^{\T}$.
\end{lemma}

\begin{lemma}\label{lambda0}
	Let $P_{2,\nu}(\cdot)$ be a convex function for $\mathcal{P}$ defined in $\eqref{penalty}$.    Under Conditions {\rm \ref{con:mixingdecay}}  and {\rm \ref{con:moments1}}, if $L_n^{\varphi}\log (rn)\ll n^{\varphi/(3\varphi+1)}$ and $\ell_n\aleph_n\ll \min\{n^{-1/\gamma},\nu\}$, then w.p.a.1 there exists a sparse global maximizer $\hat{\blambda}(\btheta_{0})=(\hat{\lambda}_{0,1},\dots,\hat{\lambda}_{0,r})^\T$ for $f(\blambda;\btheta_{0})$ satisfying ${\rm{supp}}\{\hat{\blambda}(\btheta_{0})\}\subset\mathcal{M}_{\btheta_{0}}(c)$ for some constant $c\in(C_*,1)$.
\end{lemma}
Now we begin to prove Theorem \ref{The.1}.
Let $\mathcal{G}_0=\text{supp}\{\hat{\blambda}(\btheta_0)\}$. It holds that 
\begin{equation*}
\begin{split}
\max_{\blambda\in \hat{\Lambda}_n(\btheta_0)}f(\blambda;\btheta_0) &=\max_{\bfeta\in\hat{\Lambda}_n^{\dagger}(\btheta_0)}\bigg[\frac{1}{n}\sum_{t=1}^{n}\log\{1+\bfeta^{\T}\bfg_{t,\calG_0}(\btheta_0)\}-\sum_{j=1}^{|\mathcal{G}_0|}P_{2,\nu}(|\eta_j|)   \bigg]\\
&\leq \max_{\bfeta\in\hat{\Lambda}_n^{\dagger}(\btheta_0)}\frac{1}{n}\sum_{t=1}^{n}\log\{1+\bfeta^{\T}\bfg_{t,\calG_0}(\btheta_0)\}\,,
\end{split}
\end{equation*}
where $\hat{\Lambda}_n^{\dagger}(\btheta_0)=\{\bfeta\in\mathbb{R}^{|\mathcal{G}_0|}:\bfeta^{\T}\bfg_{t,\calG_0}(\btheta_0)\in \mathcal{V},t\in[n]\}$ for some open interval $\mathcal{V}$ containing zero.
Define $A_n(\btheta_0;\bfeta)=n^{-1}\sum_{t=1}^{n}\log\{1+\bfeta^{\T}\bfg_{t,\calG_0}(\btheta_0)\}$ for any $\bfeta\in\hat{\Lambda}_n^{\dagger}(\btheta_0)$.
Let $\tilde{\bfeta}= \arg\max_{\bfeta\in\hat{\Lambda}_n^{\dagger}(\btheta_0)}A_n(\btheta_0;\bfeta)$. Due to $\ell_n\aleph_n\ll n^{-1/\gamma}$,
we can select 
$\delta_n$ satisfying  
$\ell_n^{1/2}\aleph_n\ll\delta_n\ll\ell_n^{-1/2}n^{-1/\gamma}$. Let $\bar{\bfeta} = \arg\max_{\bfeta\in\Lambda_n}A_n(\btheta_0;\bfeta)$ with $\Lambda_n=\{\bfeta\in \mathbb{R}^{|\mathcal{G}_0|}:|\bfeta|_2\leq\delta_n\}$. 
Lemma \ref{lambda0} ensures $|\mathcal{G}_0|\leq |\mathcal{M}_{\btheta_0}(c)|\leq \ell_n$ w.p.a.1. By Condition \ref{con:moments1}(a),  $\max_{t\in[n]}|\bfg_{t,\calG_0}(\btheta_0)|_2=O_{\p}(\ell_n^{1/2}n^{1/\gamma})$, which implies $\sup_{t\in[n],\bfeta\in\Lambda_n }|\bfeta^{\T}\bfg_{t,\calG_0}(\btheta_0)|=o_{\p}(1)$. 
By Condition \ref{con:moments1}(c) and \eqref{hatV-barVtheta0} in the supplementary material, if $L_n^{\varphi}\log (rn)\ll n^{\varphi/(3\varphi+1)}$ and $\ell_n\aleph_n=o(1)$,
we know $\lambda_{\min}\{\widehat{\bfV}_{\calG_0} (\btheta_0)\}$  is uniformly bounded away from zero w.p.a.1.
By Taylor expansion, it holds w.p.a.1 that 
\begin{equation}\label{bareta}
\begin{split}
0=A_n(\btheta_0;{\bf0}) \leq        A_n(\btheta_0;\bar{\bfeta})&=\bar{\bfeta}^{\T}\bar{\bfg}_{\calG_0}(\btheta_0)-   \frac{1}{2n}\sum_{t=1}^{n}\frac{\bar{\bfeta}^\T{\bfg}_{t,\calG_0}(\bthetazero)^{\otimes 2}\bar{\bfeta}}{\{1+\bar{c}\bar{\bfeta}^{\T}\bfg_{t,\calG_0}(\btheta_0)\}^2}\\
&\leq |\bar{\bfeta}|_2|\bar{\bfg}_{\calG_0}(\btheta_0)|_2-C|\bar{\bfeta}|_2^2\{1+o_{\p}(1)\}
\end{split}
\end{equation}
for some $\bar{c}\in(0,1)$. By Lemma \ref{tailprabog0}, we have $|\bar{\bfg}(\btheta_0)|_\infty=O_{\p}(\aleph_n)$ provided that $L_n^{\varphi}\log (rn)\ll n^{\varphi/(3\varphi+1)}$, which implies $|\bar{\bfg}_{\calG_0}(\btheta_0)|_2=O_{\p}(\ell_n^{1/2}\aleph_n)$. By (\ref{bareta}), we have $|\bar{\bfeta}|_2=O_{\p}(\ell_n^{1/2}\aleph_n)=o_{\p}(\delta_n)$, which implies  $\bar{\bfeta}\in \text{int}(\Lambda_n)$ w.p.a.1. Since $\Lambda_n \subset \hat{\Lambda}_n^{\dagger}(\btheta_0) $ w.p.a.1, it holds that $\tilde{\bfeta}=\bar{\bfeta}$ w.p.a.1 by the concavity of $A_n(\btheta_0;\bfeta)$ and the convexity of $\hat{\Lambda}_n^{\dagger}(\btheta_0)$. Hence, by (\ref{bareta}), we have $\max_{\bfeta\in \hat{\Lambda}_n^{\dagger}(\btheta_0)}A_n(\btheta_0;\bfeta)=O_{\p}(\ell_n\aleph_n^2)$, which implies  $\max_{\blambda\in \hat{\Lambda}_n(\btheta_0)}f(\blambda;\btheta_0)=O_{\p}(\ell_n\aleph_n^2)$.
Recall $a_n=\sum_{k=1}^{p}P_{1,\pi}(|\theta_{0,k}|)$, $b_{n}=\max\{a_n,\nu^2\}$ and $S_n(\btheta)=\max_{\blambda\in\hat{\Lambda}_n(\btheta)}f(\blambda;\btheta)+\sum_{k=1}^{p}P_{1,\pi}(|\theta_k|)$ for any $\btheta=(\theta_1,\dots,\theta_p)^\T$.
Notice that  $\hat{\btheta}_n=\arg\min_{\btheta\in\bTheta_*}S_n(\btheta)$  with $\bTheta_*=\{\btheta=(\btheta_\calS^{\T},\btheta_{\calS^{\rm c}}^{\T})^{\T} \in \bTheta: |\btheta_{\calS}-\btheta_{0,\calS}|_{\infty} \leq c_*,|\btheta_{\calS^{\rm c}}|_1\leq \aleph_n\}$ for some fixed constant $c_*>0$. As we have shown above, $S_n(\btheta_{0})=O_{\p}(\ell_n\aleph_n^2)+a_n=O_{\p}(b_n)$. Since $\btheta_{0}\in \bTheta_*$, then $S_n(\hat{\btheta}_n)\leq S_n(\btheta_0)=O_{\p}(b_n)$. We will show that $\hat{\btheta}_n \in \text{int}(\bTheta_*)$ w.p.a.1. Our proof takes two steps: (i) to show that for any $\epsilon_n\to \infty$ satisfying $b_n\epsilon_n^{2}n^{2/\gamma}=o(1) $, there exists a universal constant $K>0$ independent of $\btheta$ such that $\mathbb{P}\{S_n(\btheta)>Kb_n\epsilon_n^{2}\}\to 1$ as $n\to \infty$ for any $ \btheta=(\btheta_\calS^{\T},\btheta_{\calS^{\rm c}}^{\T})^{\T}\in\bTheta_*$ satisfying $|\btheta_{\calS}-\btheta_{0,\calS}|_\infty>\epsilon_nb_n^{1/2}$, which leads to 
$|\hat{\btheta}_{n,\calS}-\btheta_{0,\calS}|_\infty=O_{\p}(b_n^{1/2})$, and (ii) to show that $\hat{\btheta}_{n,\calS^{\rm c}}=\bf0$ w.p.a.1.

\subsubsection{Proof of {\rm(i)}} 

For any $\btheta=(\btheta_\calS^{\T},\btheta_{\calS^{\rm c}}^{\T})^{\T} \in \bTheta_*$ satisfying $|\btheta_\calS-\btheta_{0,\calS}|_\infty>\epsilon_nb_n^{1/2}$, write $\btheta^*=(\btheta_{\calS}^\T, \bf0^\T)^\T$. Define ${\mu}_{t,j_0}=\mathbb{E}\{g_{t,j_0}(\btheta)\}$, $\bar{\mu}_{j_0}=\mathbb{E}\{\bar{g}_{j_0}(\btheta)\}$ and $\bar{\mu}_{j_0}^*=\mathbb{E}\{\bar{g}_{j_0}(\btheta^*)\}$ with $j_0=\arg \max_{j\in[r]}|\mathbb{E}\{\bar{g}_{j}(\btheta^*)\}|$.  Select $\tilde{\blambda}=\delta^* b_n^{1/2}\epsilon_n\bfe_{j_0}$, where $\delta^*>0$ is a constant to be determined later, and $\bfe_{j_0}$ is an $r$-dimensional vector with the $j_0$-th component being $1$ and other components being $0$. Without loss of generality, we assume $\bar{\mu}_{j_0}^*>0$. By Condition \ref{con:moments1}(a), $\max_{t\in[n]}|g_{t,j_0}(\btheta)|=O_{\p}(n^{1/\gamma})$, which implies $\max_{t\in[n]}|\tilde{\blambda}^\T\bfg_{t}(\btheta)|=O_{\p}(b_n^{1/2}\epsilon_n n^{1/\gamma})=o_{\p}(1).$ Then $\tilde{\blambda}\in \hat{\Lambda}_n(\btheta)$ w.p.a.1. Write 
$\btheta=(\theta_1,\dots,\theta_p)^\T$ and $\tilde{\blambda}=(\tilde{\lambda}_1,\dots,\tilde{\lambda}_r)^\T.$ By Taylor expansion, it holds w.p.a.1 that 
\begin{align*}
S_n(\btheta)\geq&~ \frac{1}{n}\sum_{t=1}^{n}\log\{1+\tilde{\blambda}^{\T}\bfg_{t}(\btheta)\}-\sum_{j=1}^{r}P_{2,\nu}(|\tilde{\lambda}_j|)+\sum_{k=1}^{p}P_{1,\pi}(|\theta_k|)\\ 
\geq&~\frac{1}{n}\sum_{t=1}^{n}\tilde{\lambda}_{j_0}g_{t,j_0}(\btheta)-\frac{1}{n}\sum_{t=1}^{n}\{\tilde{\lambda}_{j_0}g_{t,j_0}(\btheta)\}^2-P_{2,\nu}(|\tilde{\lambda}_{j_0}|)\\
\geq&~\frac{1}{n}\sum_{t=1}^{n}\tilde{\lambda}_{j_0}g_{t,j_0}(\btheta)-\frac{1}{n}\sum_{t=1}^{n}\{\tilde{\lambda}_{j_0}g_{t,j_0}(\btheta)\}^2-C\nu\tilde{\lambda}_{j_0}\,,
\end{align*}
which implies 
\begin{align*}
\mathbb{P}\big\{S_n(\btheta)\leq Kb_n\epsilon_n^{2}\big\}
&\leq \bigg[\frac{1}{n}\sum_{t=1}^{n}\tilde{\lambda}_{j_0}g_{t,j_0}(\btheta)-\frac{1}{n}\sum_{t=1}^{n}\{\tilde{\lambda}_{j_0}g_{t,j_0}(\btheta)\}^2-C\nu\tilde{\lambda}_{j_0}\leq Kb_n\epsilon_n^{2}\bigg]+o(1)
\\&\leq
\mathbb{P}\bigg[\frac{1}{n}\sum_{t=1}^{n}\{g_{t,j_0}(\btheta)-\mu_{t,j_0}\}\leq b_n^{1/2}\epsilon_n\bigg\{\frac{K}{\delta^*}+\frac{\delta^*}{n}\sum_{t=1}^{n}g_{t,j_0}^2(\btheta)\bigg\}+C\nu-\bar{\mu}_{j_0}\bigg]+o(1)\,.
 \end{align*}
Write $\mathring{g}_{t,j_0}(\btheta)=g_{t,j_0}^2(\btheta)-\mathbb{E}\{g_{t,j_0}^2(\btheta)\}$. By H\"{o}lder inequality and Condition \ref{con:moments1}(a), we have $\mathbb{E}\{g_{t,j_0}^4(\btheta)\}\leq[\mathbb{E}\{|g_{t,j_0}(\btheta)|^{2\gamma}\}]^{2/\gamma}\leq K_4^{2/\gamma}$ for any $t\in[n]$. For any $t_1\neq t_2$, Davydov's inequality and Condition \ref{con:moments1}(a) yield
$
|{\rm{Cov}}\{\mathring{g}_{t_1,j_0}(\btheta),\mathring{g}_{t_2,j_0}(\btheta)\}|\lesssim  \exp(-CL_n^{-\varphi}|t_1-t_2|^{\varphi})$.
Then  there exists a universal positive constant $L>K_4^{1/\gamma}$ independent of $\btheta$ such that 
\begin{align}\label{eq:meang}
\mathbb{P}\bigg\{\frac{1}{n}\sum_{t=1}^{n}g_{t,j_0}^2(\btheta)>L\bigg\}\leq&~\mathbb{P}\bigg\{\frac{1}{n}\sum_{t=1}^{n}\mathring{g}_{t,j_0}(\btheta)>L-K_4^{1/\gamma}\bigg\}
\notag\leq (L-K_4^{1/\gamma})^{-2} \mathbb{E}\bigg[\bigg\{\frac{1}{n}\sum_{t=1}^{n}{\mathring{g}}_{t,j_0}(\btheta)\bigg\}^2\bigg]\notag
\\\lesssim&~\frac{1}{n^2}\sum_{t=1}^{n}\mathbb{E}\big[\{{\mathring{g}}_{t,j_0}(\btheta)\}^2\big]+\frac{1}{n^2}\sum_{t_2=1}^{n}\sum_{t_1=t_2+1}^{n}|{\rm Cov}\{\mathring{g}_{t_1,j_0}(\btheta),\mathring{g}_{t_2,j_0}(\btheta)\}|
\\\lesssim&~\frac{1}{n}+\frac{1}{n^2}\sum_{t_2=1}^{n}\sum_{t_1=t_2+1}^{n} \exp(-CL_n^{-\varphi}|t_1-t_2|^{\varphi})\lesssim \frac{L_n}{n}=o(1)\notag\,,
\end{align}
where the second inequality follows from the Markov inequality.  Thus, by selecting $\delta^*=(K/L)^{1/2}$ we have
$$\mathbb{P}\big\{S_n(\btheta)\leq Kb_n\epsilon_n^{2}\big\}\leq\mathbb{P}\bigg[\frac{1}{n}\sum_{t=1}^{n}\{g_{t,j_0}(\btheta)-\mu_{t,j_0}\}\leq 2b_n^{1/2}\epsilon_n(KL)^{1/2}+C\nu-\bar{\mu}_{j_0}\bigg]+o(1)\,. $$
By Condition \ref{con:identity}, we have $\bar{\mu}_{j_0}^*\geq K_3\epsilon_n b_n^{1/2}/2$. Recall $\aleph_n=n^{-3\varphi/(6\varphi+2)}(\log r)^{1/2}$.
By Condition \ref{con:moments2}(a), 
$$|\bar{\mu}_{j_0}-\bar{\mu}_{j_0}^*|\leq\max_{t\in[n]}  \max_{k\notin\mathcal{S}}\mathbb{E}\biggl\{\sup_{\btheta\in \bTheta}\bigg|\frac{\partial g_{t,j_0}(\btheta)}{\partial\theta_k}\bigg|\biggr\}|\btheta_{\calS^{\rm c}}|_1\leq K_7|\btheta_{\calS^{\rm c}}|_1=O(\aleph_n)=o(b_n^{1/2})\,.$$
Therefore, $\bar{\mu}_{j_0}\geq K_3\epsilon_nb_n^{1/2}/3$ for a sufficiently large $n$. For a sufficiently small $K$ independent of $\btheta$, we have $2b_{n}^{1/2}\epsilon_n(KL)^{1/2}+C\nu-\bar{\mu}_{j_0}\leq -c\bar{\mu}_{j_0}$, where $c\in(0,1)$ is a universal constant. 
Due to ${L}^{\varphi}_{n}\log {n}\ll {n}^{\varphi/(3\varphi+1)}$ and $\ell_n\aleph_n\ll b_n^{1/2}$, then $L_nn^{-1}\ll b_n$.
Analogous to \eqref{eq:meang}, it holds that
\begin{align*}
&\mathbb{P}\bigg[\frac{1}{n}\sum_{t=1}^{n}\{g_{t,j_0}(\btheta)-\mu_{t,j_0}\}\leq  2b_{n}^{1/2}\epsilon_n(KL)^{1/2}+C\nu-\bar{\mu}_{j_0}\bigg]\\
&~~~~~~\leq \mathbb{P}\bigg[\bigg|\frac{1}{n}\sum_{t=1}^{n}\{g_{t,j_0}(\btheta)-\mu_{t,j_0}\}\bigg|\geq  c\bar{\mu}_{j_0}\bigg]\lesssim
(\epsilon_n^2b_n)^{-1} \mathbb{E}\bigg(\bigg[\frac{1}{n}\sum_{t=1}^{n}\{g_{t,j_0}(\btheta)-\mu_{t,j_0}\}\bigg]^2\bigg)\\
&~~~~~~\lesssim(\epsilon_n^2b_nL_n^{-1}n)^{-1}=o(1)\,.
\end{align*}
Then $\mathbb{P}\{S_n(\btheta)\leq Kb_n\epsilon_n^{2}\}\to 0$ as $n\to \infty$. Hence, $|\hat{\btheta}_{n,\calS}-\btheta_{0,\calS}|_\infty=O_{\p}(b_n^{1/2})$. $\hfill\Box$

\subsubsection{Proof of {\rm(ii)}}\label{sec:the1ii}
If $\hat{\btheta}_{n,\calS^{\rm c}}\neq\bf0$, we define $\hat{\btheta}_n^*=(\hat{\btheta}_{n,\calS}^\T,\bf0^{\T})^\T$ and will show $S_n(\hat{\btheta}_n^*)<S_n(\hat{\btheta}_n)$ w.p.a.1. This will contradict the definition of $\hat{\btheta}_n$. Thus we have $\hat{\btheta}_{n,\calS^{\rm c}}=\bf0$ w.p.a.1. Write $\hat{\btheta}_n=(\hat{\theta}_{n,1},\dots,\hat{\theta}_{n,p})^\T$. Due to $b_n^{1/2}=o(\min_{k\in \mathcal{S}}|\theta_{0,k}|)$ and \eqref{eq:chi1}, it holds that
\begin{align*}
\max_{\blambda\in\hat{\Lambda}_n(\hat{\btheta}_n)}f(\blambda;\hat{\btheta}_n)\leq&~ \max_{\blambda\in \hat{\Lambda}_n(\btheta_0)}f(\blambda;\btheta_0)+\sum_{k=1}^{p}P_{1,\pi}(|\theta_{0,k}|)-\sum_{k=1}^{p}P_{1,\pi}(|\hat{\theta}_{n,k}|)\\\leq&~
O_{\p}(\ell_n\aleph_n^2)+\sum_{k=1}^{s}P_{1,\pi}^{'}\{c_k|\hat{\theta}_{n,k}|+(1-c_k)|\theta_{0,k}|\}|\hat{\theta}_{n,k}-\theta_{0,k}|
\\= &~O_{\p}(\ell_n\aleph_n^2)
 \end{align*}
 for some $c_k\in(0,1)$.
Since $\ell_n\aleph_n\ll n^{-1/\gamma}$, then we pick a 
$\delta_n$ satisfying  
$\ell_n^{1/2}\aleph_n\ll\delta_n\ll\ell_n^{-1/2}n^{-1/\gamma}$.  
Define
\begin{align*}
\bbeta_{\calM_{\hat{\btheta}_n}(\tilde{c})}(\hat{\btheta}_n):=\bar{\bfg}_{\calM_{\hat{\btheta}_n}(\tilde{c})}(\hat{\btheta}_n)-\nu\rho'_2(0^+)\sgn\{\bar{\bfg}_{\calM_{\hat{\btheta}_n}(\tilde{c})}(\hat{\btheta}_n)\}
\end{align*}
for $\tilde{c}\in (C_*,1)$ specified in Condition \ref{eq:eta}(a). 
Select $\blambda^*$ satisfying $\blambda_{\calM_{\hat{\btheta}_n}(\tilde{c})}^*=\delta_n\bbeta_{\calM_{\hat{\btheta}_n}(\tilde{c})}(\hat{\btheta}_n)/|\bbeta_{\calM_{\hat{\btheta}_n}(\tilde{c})}(\hat{\btheta}_n)|_2$ and $\blambda^*_{\calM_{\hat{\btheta}_n}^{\rm c}(\tilde{c})}=\bf0$.
Since $|\mathcal{M}_{\hat{\btheta}_n}(\tilde{c})|\leq \ell_n$ w.p.a.1, then we have
\begin{align*}
\max_{t\in[n]}|\blambda^{*,\T}\bfg_{t}(\hat{\btheta}_n)|\leq &~
|\blambda_{\calM_{\hat{\btheta}_n}(\tilde{c})}^*|_2\max_{t\in[n]}|\bfg_{t,\calM_{\hat{\btheta}_n}(\tilde{c})}(\hat{\btheta}_n)|_2=o(\ell_n^{-1/2}n^{-1/\gamma})\cdot O_{\p}(\ell_n^{1/2}n^{1/\gamma})=o_{\p}(1)\,,
\end{align*} which indicates that $\blambda^* \in \hat{\Lambda}_n(\hat{\btheta}_n)$ w.p.a.1.
Write $\blambda^*=(\lambda_1^*,\dots,\lambda_r^*)^\T$. Recall $P_{2,\nu}(t)=\nu\rho_2(t;\nu)$ for any $t\geq 0$, and $P_{2,\nu}(t)$ has bounded second-order derivatives around 0. 
By  Taylor expansion, it holds w.p.a.1 that 
\begin{align*}
o_{\p}(\delta_n^2)=\max_{\blambda\in\hat{\Lambda}_n(\hat{\btheta}_n)}f(\blambda;\hat{\btheta}_n)
\geq&~
\frac{1}{n}\sum_{t=1}^{n}\log\{1+\blambda^{*,\T}\bfg_{t}(\hat{\btheta}_n)\}-\sum_{j=1}^{r}P_{2,\nu}(|\lambda_j^*|)\\
=&~\blambda_{\calM_{\hat{\btheta}_n}(\tilde{c})}^{*,\T}\bar{\bfg}_{\calM_{\hat{\btheta}_n}(\tilde{c})}(\hat{\btheta}_n)-\frac{1}{2n}\sum_{t=1}^{n}\frac{\blambda_{\calM_{\hat{\btheta}_n}(\tilde{c})}^{*,\T}{\bfg}_{\calM_{\hat{\btheta}_n}(\tilde{c})}(\hat{\btheta}_n)^{\otimes2}\blambda_{\calM_{\hat{\btheta}_n}(\tilde{c})}^*}{\{1+c^*\blambda_{\calM_{\hat{\btheta}_n}(\tilde{c})}^{*,\T}{\bfg}_{\calM_{\hat{\btheta}_n}(\tilde{c})}(\hat{\btheta}_n)\}^2}\\&~~~~~~~-\sum_{j\in\mm_{\hat{\btheta}_n}(\tilde{c})}\nu\rho'_{2}(0^+)|\lambda_j^*|-\frac{1}{2}\sum_{j\in\mm_{\hat{\btheta}_n}(\tilde{c})}\nu\rho_2''(c_j|\lambda_j^*|;\nu)|\lambda_j^*|^2
\\\geq&~
\blambda_{\calM_{\hat{\btheta}_n}(\tilde{c})}^{*,\T}\big[\bar{\bfg}_{\calM_{\hat{\btheta}_n}(\tilde{c})}(\hat{\btheta}_n)-\nu\rho'_2(0^+)\sgn\{\blambda_{\calM_{\hat{\btheta}_n}(\tilde{c})}^*\}\big]-C\delta_n^2\{1+o_{\p}(1)\}
\end{align*}
for some $c^*,c_j\in(0,1)$. 
For any $j\in\mathcal{M}_{\hat{\btheta}_n}(\tilde{c})$, we have $\sgn(\lambda_j^*)=\sgn\{\bar{g}_j(\hat{\btheta}_n)\}$ if $|\bar{g}_j(\hat{\btheta}_n)|>\nu\rho_2'(0^+)$,  and $\bar{g}_j(\hat{\btheta}_n)-\nu\rho_2'(0^+)\sgn\{\bar{g}_j(\hat{\btheta}_n)\}=0=\lambda_j^*$ if 
$|\bar{g}_j(\hat{\btheta}_n)|=\nu\rho_2'(0^+)$. Then 
\begin{align*}
\lambda_j^*\{\bar{g}_j(\hat{\btheta}_n)-\nu\rho_2'(0^+)\sgn(\lambda_j^*)\}=\lambda_j^*\big[\bar{g}_j(\hat{\btheta}_n)-\nu\rho_2'(0^+)\sgn\{\bar{g}_j(\hat{\btheta}_n)\}\big]
\end{align*}
for any $j\in\mathcal{M}_{\hat{\btheta}_n}(\tilde{c})$ with 
$|\bar{g}_j(\hat{\btheta}_n)|\geq \nu\rho_2'(0^+)$. By Condition \ref{eq:eta}(a),  $\{j\in[r]:\tilde{c}\nu\rho_2'(0^+)\leq|\bar{g}_{j}(\hat{\btheta}_n)|< \nu\rho_2'(0^+) \}=\emptyset$ w.p.a.1.
We then have w.p.a.1 that 
\begin{align*}
o_{\p}(\delta_n^2)\geq&~
\blambda_{\calM_{\hat{\btheta}_n}(\tilde{c})}^{*,\T}\big[\bar{\bfg}_{\calM_{\hat{\btheta}_n}(\tilde{c})}(\hat{\btheta}_n)-\nu\rho'_2(0^+)\sgn\{\blambda_{\calM_{\hat{\btheta}_n}(\tilde{c})}^*\}\big]-C\delta_n^2\{1+o_{\p}(1)\}
\\=&~\delta_n|\bbeta_{\calM_{\hat{\btheta}_n}(\tilde{c})}(\hat{\btheta}_n)|_2-C\delta_n^2\{1+o_{\p}(1)\}\,.
\end{align*}
Thus $| \bbeta_{\calM_{\hat{\btheta}_n}(\tilde{c})}(\hat{\btheta}_n)|_2=O_{\p}(\delta_n)$. For any $\epsilon_n\to 0$, choose $\blambda^{**}$ satisfying $\blambda^{**}_{\calM_{\hat{\btheta}_n}(\tilde{c})}=\epsilon_n\bbeta_{\calM_{\hat{\btheta}_n}(\tilde{c})}(\hat{\btheta}_n)$ and $ \blambda^{**, \rm c}_{\calM_{\hat{\btheta}_n}(\tilde{c})}=\bf0$. Then $|\blambda^{**}|_2=o_{\p}(\delta_n)$. 
Due to $f(\blambda^{**};\hat{\btheta}_n)\leq \max_{\blambda\in\hat{\Lambda}_n(\hat{\btheta}_n)}f(\blambda;\hat{\btheta}_n)\leq \max_{\blambda\in\hat{\Lambda}_n(\btheta_0)}f(\blambda;\btheta_0)=O_{\p}(\ell_n\aleph_n^2)$,
using the same arguments given above, we can obtain $\epsilon_n|\bbeta_{\calM_{\hat{\btheta}_n}(\tilde{c})}(\hat{\btheta}_n)|_2^2-C\epsilon_n^2|\bbeta_{\calM_{\hat{\btheta}_n}(\tilde{c})}(\hat{\btheta}_n)|_2^2\{1+o_{\p}(1)\}=O_{\p}(\ell_n\aleph_n^2)$, which implies that $\epsilon_n|\bbeta_{\calM_{\hat{\btheta}_n}(\tilde{c})}(\hat{\btheta}_n)|_2^2=O_{\p}(\ell_n\aleph_n^2)$. Notice that we can select an arbitrary slow $\epsilon_n\to 0$. Following a standard result from probability theory, we have $|\bbeta_{\calM_{\hat{\btheta}_n}(\tilde{c})}(\hat{\btheta}_n)|_2^2=O_{\p}(\ell_n\aleph_n^2)$. Therefore, by Lemmas $\ref{hatV-barV}$ and $\ref{lambdahat}$, we obtain $|\hat{\blambda}(\hat{\btheta}_n)|_2=O_{\p}(\ell_n^{1/2}\aleph_n)$.

Recall $\hat{\blambda}({\btheta})=\arg\max_{\blambda\in\hat{\Lambda}_n(\btheta)}f(\blambda;\btheta)$ for any $\btheta \in \bTheta$. Write $\hat{\blambda}=\hat{\blambda}(\hat{\btheta}_n)$ and $\hat{\blambda}^*=\hat{\blambda}(\hat{\btheta}_n^*)$. Notice that
$S_n(\btheta)=\max_{\blambda\in\hat{\Lambda}_n(\btheta)}f(\blambda;\btheta)+\sum_{k=1}^{p}P_{1,\pi}(|\theta_k|)$. Then
\begin{align*}
S_n(\hat{\btheta}_n^*)-S_n(\hat{\btheta}_n)=&~f(\hat{\blambda}^*;\hat{\btheta}_n^*)-f(\hat{\blambda};\hat{\btheta}_n)-\sum_{k\in\mathcal{S}^{\rm c}}P_{1,\pi}(|\hat{\theta}_{n,k}|)\\\leq&~
f(\hat{\blambda}^*;\hat{\btheta}_n^*)-f(\hat{\blambda}^*;\hat{\btheta}_n)-\sum_{k\in\mathcal{S}^{\rm c}}P_{1,\pi}(|\hat{\theta}_{n,k}|)
\\=&~
\frac{1}{n}\sum_{t=1}^{n}\log\{1+\hat{\blambda}^{*,\T}\bfg_{t}(\hat{\btheta}_n^*)\}-\frac{1}{n}\sum_{t=1}^{n}\log\{1+\hat{\blambda}^{*,\T}\bfg_{t}(\hat{\btheta}_n)\}-\sum_{k\in\mathcal{S}^{\rm c}}P_{1,\pi}(|\hat{\theta}_{n,k}|)\,.
\end{align*}
It follows from Taylor expansion that
\begin{align*}
S_n(\hat{\btheta}_n^*)\leq S_n(\hat{\btheta}_n)-\underbrace{\frac{1}{n}\sum_{t=1}^{n}\frac{\hat{\blambda}^{*,\T}\nabla_{\btheta_{\calS^{\rm c}}}\bfg_{t}(\check{\btheta})}{1+\hat{\blambda}^{*,\T}\bfg_{t}(\check{\btheta})}\hat{\btheta}_{n,\calS^{\rm c}}}_{\textrm{I}}-\underbrace{\sum_{k\in\mathcal{S}^{\rm c}}P_{1,\pi}(|\hat{\theta}_{n,k}|)}_{\textrm{II}}\,,
\end{align*}
where $\check{\btheta}$ lies on the jointing line between $\hat{\btheta}_n$ and $\hat{\btheta}_n^*$. 

In the sequel, we will show I + II $>0$ w.p.a.1. To do this, we first use Lemma $\ref{lambdahat}$ to bound $|\hat{\blambda}^*|_2$. Given some $\tilde{c}_1\in(\tilde{c},1)$, we define 
\begin{align*}
\bbeta_{\calM_{\hat{\btheta}_n^*}(\tilde{c}_1)}(\hat{\btheta}_n^*):=\bar{\bfg}_{\calM_{\hat{\btheta}_n^*}(\tilde{c}_1)}(\hat{\btheta}_n^*)-\nu\rho_2'(0^+)\sgn\{\bar{\bfg}_{\calM_{\hat{\btheta}_n^*}(\tilde{c}_1)}(\hat{\btheta}_n^*)\}\,.
\end{align*}
It holds that
\begin{align*}
|\bbeta_{\calM_{\hat{\btheta}_n^*}(\tilde{c}_1)}(\hat{\btheta}_n^*)|_2\leq
|&\underbrace{\bar{\bfg}_{\calM_{\hat{\btheta}_n^*}(\tilde{c}_1)\cap\calM_{\hat{\btheta}_n}(\tilde{c})}(\hat{\btheta}_n^*)-\nu\rho_2'(0^+)\sgn\{\bar{\bfg}_{\calM_{\hat{\btheta}_n^*}(\tilde{c}_1)\cap\calM_{\hat{\btheta}_n}(\tilde{c})}(\hat{\btheta}_n^*)\}|_2}_{T_1}\notag\\
&~~~+\underbrace{|\bar{\bfg}_{\calM_{\hat{\btheta}_n^*}(\tilde{c}_1)\cap\calM_{\hat{\btheta}_n}^{\rm c}(\tilde{c})}(\hat{\btheta}_n^*)-\nu\rho_2'(0^+)\sgn\{\bar{\bfg}_{\calM_{\hat{\btheta}_n^*}(\tilde{c}_1)\cap\calM_{\hat{\btheta}_n}^{\rm c}(\tilde{c})}(\hat{\btheta}_n^*)\}|_2}_{T_2}\,.
\end{align*}
For the term $T_1$, we have
\begin{align*}
T_1&\leq |\bar{\bfg}_{\calM_{\hat{\btheta}_n}(\tilde{c})}(\hat{\btheta}_n^*)-\nu\rho_2'(0^+)\sgn\{\bar{\bfg}_{\calM_{\hat{\btheta}_n}(\tilde{c})}(\hat{\btheta}_n^*)\}|_2\\&\leq 
|\bar{\bfg}_{\calM_{\hat{\btheta}_n}(\tilde{c})}(\hat{\btheta}_n)-\nu\rho_2'(0^+)\sgn\{\bar{\bfg}_{\calM_{\hat{\btheta}_n}(\tilde{c})}(\hat{\btheta}_n)\}|_2+
|\bar{\bfg}_{\calM_{\hat{\btheta}_n}(\tilde{c})}(\hat{\btheta}_n^*)-\bar{\bfg}_{\calM_{\hat{\btheta}_n}(\tilde{c})}(\hat{\btheta}_n)|_2
\\&~~~~~~~+\nu\rho_2'(0^+)|\sgn\{\bar{\bfg}_{\calM_{\hat{\btheta}_n}(\tilde{c})}(\hat{\btheta}_n)\}-\sgn\{\bar{\bfg}_{\calM_{\hat{\btheta}_n}(\tilde{c})}(\hat{\btheta}_n^*)\}|_2
\\&\leq 
O_{\p}(\ell_n^{1/2}\aleph_n)+|\bar{\bfg}_{\calM_{\hat{\btheta}_n}(\tilde{c})}(\hat{\btheta}_n^*)-\bar{\bfg}_{\calM_{\hat{\btheta}_n}(\tilde{c})}(\hat{\btheta}_n)|_2+\nu\rho_2'(0^+)|\sgn\{\bar{\bfg}_{\calM_{\hat{\btheta}_n}(\tilde{c})}(\hat{\btheta}_n)\}-\sgn\{\bar{\bfg}_{\calM_{\hat{\btheta}_n}(\tilde{c})}(\hat{\btheta}_n^*)\}|_2\,.
\end{align*}
By Condition \ref{con:moments2}(a) and Taylor expansion,
\begin{align*}
\max_{j\in[r]}|\bar{g}_{j}(\hat{\btheta}_n^*)-\bar{g}_{j}(\hat{\btheta}_n)|\leq |\hat{\btheta}_{n,\calS^{\rm c}}|_1\cdot O_{\p}(1)\,,
\end{align*}
which implies $|\bar{\bfg}_{\calM_{\hat{\btheta}_n}(\tilde{c})}(\hat{\btheta}_n^*)-\bar{\bfg}_{\calM_{\hat{\btheta}_n}(\tilde{c})}(\hat{\btheta}_n)|_2\leq\ell_n^{1/2}|\hat{\btheta}_{n,\calS^{\rm c}}|_1\cdot O_{\p}(1)$. Moreover, for any $j\in \mathcal{M}_{\hat{\btheta}_n}(\tilde{c})$, we have $|\bar{g}_j(\hat{\btheta}_n)|\geq \tilde{c}\nu\rho_2'(0^+)$. Due to $|\hat{\btheta}_{n,\calS^{\rm c}}|_1\leq \aleph_n=o(\nu)$, it holds w.p.a.1 that $\sgn\{\bar{g}_{j}(\hat{\btheta}_n^*)\}=\sgn\{\bar{g}_{j}(\hat{\btheta}_n)\}$ for any $j\in \mathcal{M}_{\hat{\btheta}_n}(\tilde{c})$. Hence, $T_1=O_{\p}(\ell_n\aleph_n^{1/2})$. For the term $T_2$, notice that for any $j\in\mm_{\hat{\btheta}_n^*}(\tilde{c}_1)\cap\mathcal{M}_{\hat{\btheta}_n}^{\rm c}(\tilde{c})$, we have $|\bar{g}_{j}(\hat{\btheta}_n^*)|\geq \tilde{c}_1\nu\rho_2'(0^+)$ and $|\bar{g}_{j}(\hat{\btheta}_n)|< \tilde{c}\nu\rho_2'(0^+)$ for some $\tilde{c}_1>\tilde{c}$. Since $\max_{j\in[r]}|\bar{g}_{j}(\hat{\btheta}_n)-\bar{g}_{j}(\hat{\btheta}_n^*)|\leq |\hat{\btheta}_{n,\calS^{\rm c}}|_1\cdot O_{\p}(1)=o_{\p}(\nu)$, it holds that $\mathcal{M}_{\hat{\btheta}_n^*}(\tilde{c}_1)\cap\mathcal{M}_{\hat{\btheta}_n}^{\rm c}(\tilde{c})=\emptyset$ w.p.a.1, which implies that $T_2=0$ w.p.a.1.
Thus $|\bbeta_{\calM_{\hat{\btheta}_n^*}(\tilde{c}_1)}(\hat{\btheta}_n^*)|_2=O_{\p}(\ell_n^{1/2}\aleph_n)$. Together with  Lemma \ref{lambdahat}, we have $|\hat{\blambda}^*|_2=O_{\p}(\ell_n^{1/2}\aleph_n)$, which implies 
$
\max_{t\in[n]}|\hat{\blambda}^{*,\T}\bfg_{t}(\check{\btheta})|=\max_{t\in[n]}|\hat{\blambda}^{*,\T}_{\calM_{\hat{\btheta}_n^*}(\tilde{c}_1)}\bfg_{t,\calM_{\hat{\btheta}_n^*}(\tilde{c}_1)}(\check{\btheta})|=o_{\p}(1)$.
For I, by Condition \ref{con:moments2}(a), 
we have
\begin{align*}
|{\rm I}|=\bigg|\frac{1}{n}\sum_{t=1}^{n}\frac{\hat{\blambda}^{*,\T}\nabla_{\btheta_{\calS^{\rm c}}}\bfg_{t}(\check{\btheta})}{1+\hat{\blambda}^{*,\T}\bfg_{t}(\check{\btheta})}\hat{\btheta}_{n,\calS^{\rm c}}\bigg|\leq \ell_n^{1/2}|\hat{\blambda}^{*}|_2|\hat{\btheta}_{n,\calS^{\rm c}}|_1\{1+o_{\p}(1)\}=|\hat{\btheta}_{n,	\calS		 ^{\rm c}}|_1\cdot O_{\p}(\ell_n\aleph_n) \,.
\end{align*}
For II, we have 
$${\rm II}=\sum_{k\in \mathcal{S}^{\rm c}}\pi\rho_1'(c_k|\hat{\theta}_k|;\pi)|\hat{\theta}_{k}|\geq C\pi|\hat{\btheta}_{n,\calS^{\rm c}}|_1$$ for some
$c_k\in (0,1)$. Due to $\ell_n\aleph_n\ll\pi$, we can obtain I $+$ II $>0$ w.p.a.1, which implies that 
$S_n(\hat{\btheta}_n^*)<S_n(\hat{\btheta}_n)$ w.p.a.1. We complete the proof of part (ii).
$\hfill\Box$

\subsection{Proof of Theorem \ref{The.2} }\label{sec:profthe2}
Recall \begin{align*}
\hat{\blambda}(\btheta)=\arg\max_{\blambda\in\hat{\Lambda}(\btheta)}\bigg[\frac{1}{n}\sum_{t=1}^{n}\log \{1+\blambda^{\T}\bg_{t}(\btheta)\}-\sum_{j=1}^{r}P_{2,\nu}(|\lambda_j|)\bigg] \,.
\end{align*}
Write $\hat{\blambda}=\hat{\blambda}(\hat{\btheta}_n)$ and $\mathcal{R}_{n}={\rm supp}\{\hat{\blambda}(\hat{\btheta}_n)\}$.
To prove Theorem 2, we need Lemmas \ref{hatVL2norm}--\ref{cltZ}. As shown in the proof of  Theorem $\ref{The.1}$,   $|\hat{\blambda}|_2=O_{\p}(\ell_n^{1/2}\aleph_n)$. 
The proof of Lemma \ref{hatVL2norm} follows that of Lemma 2 in \cite{CTW_2018} by replacing  $\omega_n$ there by $1$
and using the convergence rate of $|\hat{\blambda}|_2$ derived in Theorem \ref{The.1}, and  
the proof of Lemma  \ref{eq:etaj0} resembles that of Lemma 4 in \cite{bel2024} by replacing  $\alpha_n$ there by $\aleph_n$. The proofs of Lemmas \ref{gammahatL2norm} and \ref{cltZ}  are given in Sections \ref{sec:la:gammahatL2norm} and \ref{sec:la:cltZ} of the supplementary material, respectively, where the proof of Lemma \ref{cltZ} invokes Theorem 1 of \cite{Sunklodas(1984)}.

\begin{lemma}\label{hatVL2norm}
	Assume the conditions of Theorem {\rm \ref{The.1}} hold. 
 Then it holds that 
\begin{align*}
&\sup_{c\in(0,1)}\bigg\|\frac{1}{n}\sum_{t=1}^{n}\frac{\bfg_{t,\calR_n}(\hat{\btheta}_n)^{\otimes2}}{\{1+c\hat{\blambda}_{\calR_n}^\T\bfg_{t,\calR_n}(\hat{\btheta}_n)\}^2}-\widehat{\bfV}_{\calR_n}(\hat{\btheta}_n)\bigg\|_2=O_{\p}(\ell_n\aleph_nn^{1/\gamma})\,, \\
& \mbox{and}~~~\bigg|\biggl\{\frac{1}{n}\sum_{t=1}^{n}\frac{\nabla_{\btheta_{\calS}}\bfg_{t,\calR_n}(\hat{\btheta}_n)}{1+\hat{\blambda}_{\calR_n}^\T\bfg_{t,\calR_n}(\hat{\btheta}_n)}-\widehat{\bGamma}_{\calR_n}(\hat{\btheta}_n)\biggr\}\bfz\bigg|_2=|\bfz|_2\cdot O_{\p}(\ell_ns^{1/2}\aleph_n)
\end{align*}
uniformly over $\bfz\in\mathbb{R}^s$.
\end{lemma}

\begin{lemma}\label{eq:etaj0}
Assume Condition {\rm \ref{eq:eta}(b)} and the conditions of Theorem {\rm \ref{The.1}} hold. It then holds w.p.a.1 that $\hat{\blambda}(\btheta)$ is continuously differentiable at $\hat{\btheta}_n$ and $\nabla_{\btheta}\hat{\blambda}_{\calR^{\rm c}_n}(\hat{\btheta}_n)={\bf0}$.
\end{lemma}

\begin{lemma}\label{gammahatL2norm}
	Let $\mathscr{T}=\{\mathcal{T} \subset [r]:|\mathcal{T}|\leq \ell_n\}$. Assume the conditions of Theorem {\rm \ref{The.1}}, Conditions {\rm \ref{con:moments2}(b)} and {\rm \ref{con:moments2}(c)} hold. Then 
\begin{align*}
\sup_{\mathcal{T}\in\mathscr{T}}\big|\big[\widehat{\bGamma}_{\mT}(\hat{\btheta}_n)-\mathbb{E}\{\widehat{\bGamma}_{\mT}(\btheta_0)\}\big]\bfz\big|_2=|\bfz|_2\cdot\{O_{\p}(\ell_n^{1/2}s^{3/2}b_n^{1/2})+O_{\p}(\ell_n^{1/2}s^{1/2}\aleph_n)\}
\end{align*}
	holds uniformly over $\bfz\in\mathbb{R}^s$. 
\end{lemma}

\begin{lemma}\label{cltZ}
	Let $\mathscr{T}=\{\mathcal{T} \subset [r]:s\le|\mathcal{T}|\leq \ell_n\}$ and $\ell_n^{3}L_{n}^{4\varphi}(\log n)^{4}\ll n$. Write  $\bXi_{\mT}(\btheta_0)={\rm Var}\{n^{1/2}\bar{\bg}_{\mT}(\btheta_0)\}$	and   ${\bJ}_{\mT}=\{[\mathbb{E}\{\widehat{\bGamma}_{\mT}(\btheta_0)\}]^{\T}[\mathbb{E}\{\widehat{\bV}_{\mT}(\btheta_0)\}]^{-1}\bXi_{\mT}^{1/2}(\btheta_{0})\}^{\otimes2}$ for any $\mathcal{T}\in \mathscr{T}$.  Under Conditions {\rm \ref{con:mixingdecay}, \ref{con:moments1}(a), \ref{con:moments1}(c)  and \ref{con:Gamma}}, for  any $\bfz\in\mathbb{R}^s$ with $|\bfz|_2=1$,  it holds that
	\begin{equation*}
	\sup_{\mathcal{T}\in\mathscr{T}}\sup_{u\in\mathbb{R}}\big|\mathbb{P}\{n^{1/2}\bfz^\T\bJ_{\mT}^{-1/2}[\mathbb{E}\{\widehat{\bGamma}_{\mT}(\btheta_0)\}]^{\T}[\mathbb{E}\{\widehat{\bV}_{\mT}(\btheta_0)\}]^{-1}\bar{\bfg}_{\mT}(\btheta_{0})\leq u\}-{\Phi}(u)\big|\to 0
	\end{equation*}
    as $n\to\infty$, 
 where $\Phi(\cdot)$ is the cumulative distribution function of the standard normal distribution.
\end{lemma}

Now we begin to prove Theorem \ref{The.2}.
 Write $\blambda=({\lambda}_1,\dots,{\lambda}_r)^\T$.  Define 
\begin{align*}
J_n(\btheta;\blambda)=\frac{1}{n}\sum_{t=1}^{n}\log\{1+\blambda^{\T}\bfg_t(\btheta)\}-\sum_{j=1}^{r}P_{2,\nu}(|\lambda_j|)+\sum_{k=1}^{p}P_{1,\pi}(|\theta_k|)\,.
\end{align*}
 Notice that  $\nabla_{\blambda}J_n(\hat{\btheta}_n;\hat{\blambda})={\bf0}$, i.e.,
\begin{align*}
{\bf0}=\frac{1}{n}\sum_{t=1}^{n}\frac{\bfg_{t}(\hat{\btheta}_n)}{1+\hat{\blambda}^\T\bfg_{t}(\hat{\btheta}_n)}-\hat{\bfeta}\,,
\end{align*}
where $\hat{\bfeta}=(\hat{\eta}_1,\dots,\hat{\eta}_{r})^\T$ with $\hat{\eta}_j=\nu\rho_2'(|\hat{\lambda}_j|;\nu)\sgn(\hat{\lambda}_j)$ for $\hat{\lambda}_j\neq 0$ and $\hat{\eta}_j\in[-\nu\rho_2'(0^+),\nu\rho_2'(0^+)]$ for $\hat{\lambda}_j=0$. By Taylor expansion, we know
\begin{align}\label{eq:Atheta}
{\bf0}&=\frac{1}{n}\sum_{t=1}^{n}\bfg_{t,\calR_n}(\hat{\btheta}_n)-\frac{1}{n}\sum_{t=1}^{n}\frac{\bfg_{t,\calR_n}(\hat{\btheta}_n)^{\otimes2}}{\{1+c\hat{\blambda}_{\calR_n}^\T\bfg_{t,\calR_n}(\hat{\btheta}_n)\}^2}\hat{\blambda}_{\calR_n}-\hat{\bfeta}_{\calR_n}
\notag\\
&=:
\bar{\bfg}_{\calR_n}(\hat{\btheta}_n)-\bfA(\hat{\btheta}_n)\hat{\blambda}_{\calR_n}-\hat{\bfeta}_{\calR_n}
\end{align}
for some $c\in(0,1)$, which implies $\hat{\blambda}_{\calR_n}=\bfA^{-1}(\hat{\btheta}_n)\{\bar{\bfg}_{\calR_n}(\hat{\btheta}_n)-\hat{\bfeta}_{\calR_n}\}$. By the definition of $\hat{\btheta}_n$, we have ${\bf0}=\nabla_{\btheta}J_n\{\btheta;\hat{\blambda}(\btheta)\}|_{\btheta=\hat{\btheta}_n}$. Notice that 
\begin{align*}
\nabla_{\btheta}J_n\{\btheta;\hat{\blambda}(\btheta)\}|_{\btheta=\hat{\btheta}_n}=\frac{\partial J_n(\hat{\btheta}_n;\hat{\blambda})}{\partial \btheta}+\biggl\{\underbrace{
\frac{\partial J_n(\hat{\btheta}_n;\hat{\blambda})}{\partial \blambda_{\calR_n}^{\T}}}_{\rm I}\frac{\partial\hat{\blambda}_{\calR_n}(\hat{\btheta}_n)}{\partial\btheta}+\frac{\partial J_n(\hat{\btheta}_n;\hat{\blambda})}{\partial \blambda_{\calR_n^{\rm c}}^{\T}}\underbrace{\frac{\partial\hat{\blambda}_{\calR_n^{\rm c}}(\hat{\btheta}_n)}{\partial\btheta}}_{\rm II}\biggr\}^{\T}\,.
\end{align*}
Due to $\hat{\blambda}(\hat{\btheta}_n)=\arg \max_{\blambda\in \hat{\Lambda}_n(\hat{\btheta}_n)}f(\blambda;\hat{\btheta}_n)=\arg \max_{\blambda\in \hat{\Lambda}_n(\hat{\btheta}_n)}J_n(\hat{\btheta}_n;\blambda)$, then ${\rm I}={\bf0}$ w.p.a.1. On the other hand, Lemma \ref{eq:etaj0} implies that ${\rm II}={\bf0}$ w.p.a.1. Thus, ${\bf0}={\partial J_n(\hat{\btheta}_n;\hat{\blambda})}/{\partial \btheta}$. Together with \eqref{eq:Atheta}, we have
\begin{align}\label{T2eq:1}
{\bf0}&=\bigg\{\frac{1}{n}\sum_{t=1}^{n}\frac{\nabla_{\btheta_{\calS}}\bfg_{t,\calR_n}(\hat{\btheta}_n)}{1+\hat{\blambda}_{\calR_n}^\T\bfg_{t,\calR_n}(\hat{\btheta}_n)}\bigg\}^\T\bfA^{-1}(\hat{\btheta}_n)\{\bar{\bfg}_{\calR_n}(\hat{\btheta}_n)-\hat{\bfeta}_{\calR_n}\}+\hat{\mathbf{h}}_{}
\notag\\&=:
\bfB^{\T}(\hat{\btheta}_n)\bfA^{-1}(\hat{\btheta}_n)\{\bar{\bfg}_{\calR_n}(\hat{\btheta}_n)-\hat{\bfeta}_{\calR_n}\}+\hat{\mathbf{h}}\,,
\end{align}
where $\hat{\mathbf{h}}=\{\sum_{k=1}^{p}\nabla_{\btheta_{\calS}}P_{1,\pi}(|\theta_k|)\}|_{\btheta=\hat{\btheta}_n}$.  Theorem $\ref{The.1}$ and \eqref{eq:chi1} imply that  $\hat{\mathbf{h}}=\bzero$ w.p.a.1.
Recall ${\bJ}_{\calR_n}=\{\mathbb{E}\{\widehat{\bGamma}^{\T}_{\calR_n}(\btheta_0)\}[\mathbb{E}\{\widehat{\bV}_{\calR_n}(\btheta_0)\}]^{-1}\bXi_{\calR_n}^{1/2}(\btheta_{0})\}^{\otimes2}$  with $\bXi_{\calR_n}(\btheta_0)={\rm Var}\{n^{1/2}\bar{\bg}_{\calR_n}(\btheta_0)\}$.
For any $\bfz\in \mathbb{R}^s$ with $|\bfz|_2=1$, let $\bdelta={\bJ}_{\calR_n}^{-1/2}\bfz$. Since  $\|{\bJ}_{\calR_n}^{-1}\|_2$ is bounded, then  $|\bdelta|_2=O(1)$. By the Cauchy-Schwarz inequality and  Condition \ref{con:Gamma}(a), it holds that
\begin{align*}
\big|\mathbb{E}\{\widehat{\bGamma}_{\calR_n}(\btheta_0)\}\bdelta\big|_2
\leq\lambda_{\max}^{1/2}([\mathbb{E}\{\widehat{\bGamma}^{\T}_{\calR_n}(\btheta_0)\}]^{\otimes2})\cdot|\bdelta|_2=O(1)\,.
\end{align*}
 Therefore, by the triangle inequality, Lemmas $\ref{hatVL2norm}$ and $\ref{gammahatL2norm}$ yield that  $|\bfB(\hat{\btheta}_n)\bdelta|_2=O_{\p}(1)$ provided that $\ell_ns^{1/2}\aleph_n=o(1)$ and $\ell_ns^{3}b_n=o(1)$. Recall \begin{align*}
\bbeta_{\calM_{\hat{\btheta}_n}(\tilde{c})}(\hat{\btheta}_n)=\bar{\bfg}_{\calM_{\hat{\btheta}_n}(\tilde{c})}(\hat{\btheta}_n)-\nu\rho'_2(0^+)\sgn\{\bar{\bfg}_{\calM_{\hat{\btheta}_n}(\tilde{c})}(\hat{\btheta}_n)\}
\end{align*}
for some constant $\tilde{c}\in (C_*,1)$ specified in Condition \ref{eq:eta}(a). 
As shown in Appendix \ref{sec:the1ii}, $|\bbeta_{\calM_{\hat{\btheta}_n}(\tilde{c})}(\hat{\btheta}_n)|_2=O_{\p}(\ell_n^{1/2}\aleph_n)$. By  Lemma \ref{lambdahat}, it holds w.p.a.1 that  (i) $|\hat{\blambda}|_2=O_{\p}(\ell_n^{1/2}\aleph_n)$,
(ii) $\sgn\{\bar{g}_{j}(\hat{\btheta}_n)\}=\sgn(\hat{\lambda}_{j})$ for any $j\in\mathcal{R}_n$, and 
(iii) $\mathcal{R}_n\subset \mathcal{M}_{\hat{\btheta}_n}(\tilde{c})$.
 Since $P_{2,\nu}(\cdot)$ has bounded second-order derivatives around $0$,
 it holds w.p.a.1 that 
\begin{align*}
\big|\nu\rho_2'(0^+)\sgn\{\bar{\bfg}_{\calR_n}(\hat{\btheta}_n)\}-\hat{\bfeta}_{\calR_n}\big|_2^2&=
\sum_{j\in\mathcal{R}_n}\{\nu\rho_2'(0^+)\sgn(\hat{\lambda}_j)-\nu\rho_2'(|\hat{\lambda}_j|;\nu)\sgn(\hat{\lambda}_j)\}^2
\\&=
\sum_{j\in\mathcal{R}_n}\{\nu\rho_2''(c_j|\hat{\lambda}_j|;\nu)|\hat{\lambda}_j|\}^2=C\cdot o_{\p}(|\hat{\blambda}_{\calR_n}|_2^2)
\end{align*}
for some $c_j\in(0,1)$, which implies
\begin{align*}
|\bar{\bfg}_{\calR_n}(\hat{\btheta}_n)-\hat{\bfeta}_{\calR_n}|_2&\leq
\big|\bar{\bfg}_{\calR_n}(\hat{\btheta}_n)-\nu\rho_2'(0^+)\sgn\{\bar{\bfg}_{\calR_n}(\hat{\btheta}_n)\}\big|_2+\big|\nu\rho_2'(0^+)\sgn\{\bar{\bfg}_{\calR_n}(\hat{\btheta}_n)\}-\hat{\bfeta}_{\calR_n}\big|_2 \\&=
O_{\p}(\ell_n^{1/2}\aleph_n)\,.
\end{align*}
Due to $\hat{\mathbf{h}}=\bzero$ w.p.a.1, by \eqref{T2eq:1} it holds w.p.a.1 that 
\begin{align*}
\big|\bdelta^\T\widehat{\bGamma}_{\calR_n}^\T(\hat{\btheta}_n)\widehat{\bfV}_{\calR_n}^{-1}(\hat{\btheta}_n)\{\bar{\bfg}_{\calR_n}(\hat{\btheta}_n)-\hat{\bfeta}_{\calR_n}\}\big|
&=
\big|\bdelta^\T\{\widehat{\bGamma}_{\calR_n}^\T(\hat{\btheta}_n)\widehat{\bfV}_{\calR_n}^{-1}(\hat{\btheta}_n)-\bfB^\T(\hat{\btheta}_n)\bfA^{-1}(\hat{\btheta}_n)\}\{\bar{\bfg}_{\calR_n}(\hat{\btheta}_n)-\hat{\bfeta}_{\calR_n}\}\big|
\\&\leq 
\underbrace{\big|\bdelta^\T\widehat{\bGamma}_{\calR_n}^\T(\hat{\btheta}_n)\{\widehat{\bfV}_{\calR_n}^{-1}(\hat{\btheta}_n)-\bfA^{-1}(\hat{\btheta}_n)\}\{\bar{\bfg}_{\calR_n}(\hat{\btheta}_n)-\hat{\bfeta}_{\calR_n}\}\big|}_{H_1}
\\&~~~+\underbrace{\big|\bdelta^\T\{\widehat{\bGamma}_{\calR_n}^\T(\hat{\btheta}_n)-\bfB^\T(\hat{\btheta}_n)\}\bfA^{-1}(\hat{\btheta}_n)\{\bar{\bfg}_{\calR_n}(\hat{\btheta}_n)-\hat{\bfeta}_{\calR_n}\}\big|}_{H_2}\,.
\end{align*}
By Lemma \ref{hatV-barV} and Condition \ref{con:moments1}(c), we know that
the eigenvalues of $\widehat{\bfV}_{\calR_n}(\hat{\btheta}_n)$
are uniformly bounded away from zero and infinity w.p.a.1. By Lemma \ref{hatVL2norm}, if $\ell_n\aleph_n\ll n^{-1/\gamma}$, we have $\|\bfA^{-1}(\hat{\btheta}_n)\|_2=O_{\p}(1)$.
Thus
\begin{align*}
\|\widehat{\bfV}_{\calR_n}^{-1}(\hat{\btheta}_n)-\bfA^{-1}(\hat{\btheta}_n)\|_2\leq\|\widehat{\bfV}_{\calR_n}^{-1}(\hat{\btheta}_n)\|_2\|\bfA(\hat{\btheta}_n)-\widehat{\bfV}_{\calR_n}(\hat{\btheta}_n)\|_2\|\bfA^{-1}(\hat{\btheta}_n)\|_2=
O_{\p}(\ell_n\aleph_nn^{1/\gamma})\,.
\end{align*}
By Condition \ref{con:Gamma}(a) and Lemma \ref{gammahatL2norm}, we have $|\widehat{\bGamma}_{\calR_n}(\hat{\btheta}_n)\bdelta|_2=O_{\p}(1)$ provided that $\ell_ns^{3}b_n=o(1)$ and $\ell_ns\aleph_n^2=o(1)$.
Then 
\begin{align*}
H_1\leq 
|\widehat{\bGamma}_{\calR_n}(\hat{\btheta}_n)\bdelta|_2\|\widehat{\bfV}_{\calR_n}^{-1}(\hat{\btheta}_n)-\bfA^{-1}(\hat{\btheta}_n)\|_2|\bar{\bfg}_{\calR_n}(\hat{\btheta}_n)-\hat{\bfeta}_{\calR_n}|_2
=
O_{\p}(\ell_n^{3/2}\aleph_n^2n^{1/\gamma})\,.
\end{align*}
Moreover, by Lemma \ref{hatVL2norm},
\begin{align*}
H_2\leq 
|\bdelta|_2\|\bfA^{-1}(\hat{\btheta}_n)\|_2|\bar{\bfg}_{\calR_n}(\hat{\btheta}_n)-\hat{\bfeta}_{\calR_n}|_2\cdot O_{\p}(\ell_ns^{1/2}\aleph_n)
=
O_{\p}(\ell_n^{3/2}s^{1/2}\aleph_n^2)\,.
 \end{align*}
Therefore, it holds that $$ \bdelta^\T\widehat{\bGamma}_{\calR_n}^\T(\hat{\btheta}_n)\widehat{\bfV}_{\calR_n}^{-1}(\hat{\btheta}_n)\{\bar{\bfg}_{\calR_n}(\hat{\btheta}_n)-\hat{\bfeta}_{\calR_n}\}=O_{\p}\big[\ell_n^{3/2}\aleph_n^2\max\{s^{1/2},n^{1/\gamma}\}\big]$$
provided that $L_n^{\varphi}\log (rn)\ll n^{\varphi/(3\varphi+1)}$, $b_n\ll \min\{n^{-2/\gamma},s^{-3}\ell_n^{-1}\}$ and $\ell_n\aleph_n\ll \min\{\nu,\pi,s^{-1/2}\}$.

By Lemma \ref{hatV-barV} and Condition \ref{con:moments1}(c),
\begin{align*}
\|\widehat{\bfV}^{-1}_{\calR_n} (\hat{\btheta}_n)-[\mathbb{E}\{\widehat{\bfV}_{\calR_n}(\btheta_0)\}]^{-1}\|_2 
&\leq \|\widehat{\bfV}^{-1}_{\calR_n} (\hbthetan)\|_2 \cdot \|\mathbb{E}\{\widehat{\bfV}_{\calR_n}(\btheta_0)\}-\widehat{\bfV}_{\calR_n} (\hbthetan)\|_2 \cdot \|[\mathbb{E}\{\widehat{\bfV}_{\calR_n}(\btheta_0)\}]^{-1}\|_2 \\
&=O_{\p}(s\ell_n^{1/2}b_n^{1/2})+O_{\p}(\ell_n\aleph_n)\,.
\end{align*}
Thus, by the  triangle inequality and  Lemma \ref{gammahatL2norm},
\begin{align*}
&\big|\bdelta^{\T}\mathbb{E}\{\widehat{\bGamma}^{\T}_{\calR_n}(\btheta_0)\}
[\mathbb{E}\{\widehat{\bfV}_{\calR_n}(\btheta_0)\}]^{-1}\{\bar{\bfg}_{\calR_n}(\hat{\btheta}_n)-\hat{\bfeta}_{\calR_n}\}\big|
\\ &~~~~~~~~\leq\big|\bdelta^\T\widehat{\bGamma}_{\calR_n}^\T(\hat{\btheta}_n)\widehat{\bfV}_{\calR_n}^{-1}(\hat{\btheta}_n)\{\bar{\bfg}_{\calR_n}(\hat{\btheta}_n)-\hat{\bfeta}_{\calR_n}\}\big|\\&~~~~~~~~~~~~~+
\big|\bdelta^\T\mathbb{E}\{\widehat{\bGamma}^{\T}_{\calR_n}(\btheta_0)\}\{[\mathbb{E}\{\widehat{\bfV}_{\calR_n}(\btheta_0)\}]^{-1}-\widehat{\bfV}_{\calR_n}^{-1}(\hat{\btheta}_n)\}\{\bar{\bfg}_{\calR_n}(\hat{\btheta}_n)-\hat{\bfeta}_{\calR_n}\}\big|
\\&~~~~~~~~~~~~~+\big|\bdelta^\T\{\mathbb{E}\{\widehat{\bGamma}^{\T}_{\calR_n}(\btheta_0)\}-\widehat{\bGamma}_{\calR_n}^\T(\hat{\btheta}_n)\}\widehat{\bfV}_{\calR_n}^{-1}(\hat{\btheta}_n)\{\bar{\bfg}_{\calR_n}(\hat{\btheta}_n)-\hat{\bfeta}_{\calR_n}\}\big|
\\&~~~~~~~~=O_{\p}\big[\ell_n^{3/2}\aleph_n^2\max\{s^{1/2},n^{1/\gamma}\}\big]+
O_{\p}(\ell_ns^{3/2}\aleph_nb_n^{1/2})+O_{\p}(\ell_ns^{1/2}\aleph_n^2)
\\&~~~~~~~~=
O_{\p}\big[\ell_n^{3/2}\aleph_n^2\max\{s^{1/2},n^{1/\gamma}\}\big]+O_{\p}(\ell_ns^{3/2}\aleph_nb_n^{1/2})\,.
\end{align*}
By Taylor expansion, it holds w.p.a.1 that
\begin{align*}
&\bdelta^\T\mathbb{E}\{\widehat{\bGamma}^{\T}_{\calR_n}(\btheta_0)\}[\mathbb{E}\{\widehat{\bfV}_{\calR_n}(\btheta_0)\}]^{-1}\{\widehat{\bGamma}_{\calR_n}(\tilde{\btheta})(\hat{\btheta}_{n,\calS}-\btheta_{0,\calS})-\hat{\bfeta}_{\calR_n}\}\notag
\\&~~~~=
-\bdelta^\T\mathbb{E}\{\widehat{\bGamma}^{\T}_{\calR_n}(\btheta_0)\}[\mathbb{E}\{\widehat{\bfV}_{\calR_n}(\btheta_0)\}]^{-1}\bar{\bfg}_{\calR_n}(\btheta_{0})+O_{\p}\big[\ell_n^{3/2}\aleph_n^2\max\{s^{1/2},n^{1/\gamma}\}\big]+O_{\p}(\ell_ns^{3/2}\aleph_nb_n^{1/2})\,,\notag
\end{align*}
where $\tilde{\btheta}$ is on the line joining $\btheta_{0}$ and $\hat{\btheta}_n$. Since $|\bar{\bfg}_{\calR_n}(\hat{\btheta}_n)|_2\leq |\bar{\bfg}_{\calR_n}(\hat{\btheta}_n)-\hat{\bfeta}_{\calR_n}|_2+|\hat{\bfeta}_{\calR_n}|_2$, then $|\bar{\bfg}_{\calR_n}(\hat{\btheta}_n)|_2=O_{\p}(\ell_n^{1/2}\nu)$. By Lemma \ref{tailprabog0}, 
$|\bar{\bfg}_{\calR_n}({\btheta}_0)|_2=O_{\p}(\ell_n^{1/2}\aleph_n)$ provided that $L_n^{\varphi}\log (rn)\ll n^{\varphi/(3\varphi+1)}$. Thus $|\bar{\bfg}_{\calR_n}(\hat{\btheta}_n)-\bar{\bfg}_{\calR_n}({\btheta}_0)|_2=O_{\p}(\ell_n^{1/2}\nu)$. By Taylor expansion, we have $|\bar{\bfg}_{\calR_n}(\hat{\btheta}_n)-\bar{\bfg}_{\calR_n}({\btheta}_0)|_2=|\widehat{\bGamma}_{\calR_n}(\check{\btheta})(\hat{\btheta}_{n,\calS}-\btheta_{0,\calS})|_2\geq \lambda_{\min}^{1/2}[\{\widehat{\bGamma}_{\calR_n}^\T(\check{\btheta})\}^{\otimes2}]|\hat{\btheta}_{n,\calS}-\btheta_{0,\calS}|_2$ for some $\check{\btheta}$ lying on the line joining $\btheta_{0}$ and $\hat{\btheta}_n$.  By similar arguments of Lemma \ref{gammahatL2norm}, 
we have 
\begin{align}\label{eq:chechtheta}
\big|\big[\widehat{\bGamma}_{\calR_n}(\check{\btheta})-\mathbb{E}\{\widehat{\bGamma}_{\calR_n}(\btheta_0)\}\big]\bfz\big|_2=|\bfz|_2\cdot\{O_{\p}(\ell_n^{1/2}s^{3/2}b_n^{1/2})+O_{\p}(\ell_n^{1/2}s^{1/2}\aleph_n)\}
\end{align} 
uniformly over $\bfz\in\mathbb{R}^s$. Together with 
Condition \ref{con:Gamma}(a), we have  $\lambda_{\min}[\{\widehat{\bGamma}_{\calR_n}^\T(\check{\btheta})\}^{\otimes2}]\geq C$ w.p.a.1 provided that $\ell_ns^{3}b_n=o(1)$ and $\ell_ns\aleph_n^2=o(1)$. 
Hence, under Conditions \ref{con:moments2}(b), \ref{con:moments2}(c) and  \ref{con:Gamma}(a),
\begin{align}\label{eq:thetas}
    |\hat{\btheta}_{n,\calS}-\btheta_{0,\calS}|_2=O_{\p}(\ell_n^{1/2}\nu)\,.
\end{align}
provided that $\ell_ns^{3}b_n=o(1)$ and $\ell_ns\aleph_n^2=o(1)$.
Recall $b_n=\max\{a_n,\nu^2\}$ and $\nu\gg\aleph_n$.  Analogous to \eqref{eq:chechtheta}, we have 
\begin{align*}
\big|[\widehat{\bGamma}_{\calR_n}(\tilde{\btheta})-\mathbb{E}\{\widehat{\bGamma}_{\calR_n}(\btheta_0)\}](\hat{\btheta}_{n,\calS}-\btheta_{0,\calS}) \big|_2=O_{\p}(\ell_ns^{3/2}\nu b_n^{1/2})+O_{\p}(\ell_ns^{1/2}\nu\aleph_n)
=O_{\p}(\ell_ns^{3/2}\nu b_n^{1/2})\,.
\end{align*}
Let 
$
\hat{\bpsi}_{\calR_n}={\bM}_{\calR_n}^{-1}\mathbb{E}\{\widehat{\bGamma}^{\T}_{\calR_n}(\btheta_0)\}[\mathbb{E}\{\widehat{\bfV}_{\calR_n}(\btheta_0)\}]^{-1}\hat{\bfeta}_{\calR_n}$ with 
\begin{align*}
     &{\bJ}_{\calR_n}=\{\mathbb{E}\{\widehat{\bGamma}^{\T}_{\calR_n}(\btheta_0)\}[\mathbb{E}\{\widehat{\bV}_{\calR_n}(\btheta_0)\}]^{-1}\bXi_{\calR_n}^{1/2}(\btheta_{0})\}^{\otimes2}\,,
\\&\bM_{\calR_n}=\mathbb{E}\{\widehat{\bGamma}^{\T}_{\calR_n}(\btheta_0)\}[\mathbb{E}\{\widehat{\bfV}_{\calR_n}(\btheta_0)\}]^{-1}\mathbb{E}\{\widehat{\bGamma}_{\calR_n}(\btheta_0)\}\,.
 \end{align*} 
Hence, if  $L_n^{\varphi}\log (rn)\ll n^{\varphi/(3\varphi+1)}$, $b_n\ll \min\{n^{-2/\gamma},s^{-3}\ell_n^{-1}\}$ and $\ell_n\aleph_n\ll \min\{\nu,\pi,s^{-1/2}\}$, we have 
\begin{align*}
n^{1/2}\bdelta^{\T}\bM_{\calR_n}(\hat{\btheta}_{n,\calS}-\btheta_{0,\calS}-\hat{\bpsi}_{\calR_n})
=&-n^{1/2}\bfz^\T{\bJ}_{\calR_n}^{-1/2}\mathbb{E}\{\widehat{\bGamma}^{\T}_{\calR_n}(\btheta_0)\}[\mathbb{E}\{\widehat{\bfV}_{\calR_n}(\btheta_0)\}]^{-1}\bar{\bfg}_{\calR_n}(\btheta_{0})\\&+O_{\p}(n^{1/2}\ell_ns^{3/2}\nu b_n^{1/2})
+O_{\p}\big[n^{1/2}\ell_n^{3/2}\aleph_n^2\max\{s^{1/2},n^{1/\gamma}\}\big]\,.
\end{align*}
By Lemma \ref{cltZ}, it holds that
\begin{align}\label{eq:the2lct}
n^{1/2}\bfz^\T{\bJ}_{\calR_n}^{-1/2}\bM_{\calR_n}(\hat{\btheta}_{n,\calS}-\btheta_{0,\calS}-\hat{\bpsi}_{\calR_n}) \rightarrow N(0,1)
\end{align}
in distribution
as $n\rightarrow\infty$, provided that
$L_n^{\varphi}\log (rn)\ll n^{\varphi/(3\varphi+1)}$, $b_n\ll \min\{n^{-2/\gamma},s^{-3}\ell_n^{-1}, n^{-1}\ell_n^{-2}s^{-3}\nu^{-2}\}$,  $\ell_n\aleph_n\ll \min\{\nu,\pi,s^{-1/2}\}$, $\ell_n^{3}L_{n}^{4\varphi}(\log n)^{4}\ll n$,  and  $n\ell_n^{3}\aleph_n^4\max\{s,n^{2/\gamma}\}=o(1)$.

Due to $\aleph_n=n^{-3\varphi/(6\varphi+2)}(\log r)^{1/2}$ and $\nu\gg \ell_n\aleph_n$, then $n^{-1}\ell_n^{-2}s^{-3}\nu^{-2}\ll s^{-3}\ell_n^{-1}$, which implies $b_n\ll \min\{n^{-2/\gamma}, n^{-1}\ell_n^{-2}s^{-3}\nu^{-2}\}$. Recall $b_n=\max\{a_n,\nu^2\}$. Due to $\ell_n\aleph_n\ll \nu$, then $s^2\nu^2b_n\geq s^2\nu^4\gg \ell_n\aleph_n^4$, which implies 
$n\ell_n^{3}\aleph_n^4s\ll n\ell_n^2s^3\nu^2b_n$. Hence, the restriction $n\ell_n^3\aleph_n^4s=o(1)$ holds automatically if $b_n\ll \min\{n^{-2/\gamma}, n^{-1}\ell_n^{-2}s^{-3}\nu^{-2}\}$. Since  $\varphi\geq 1$
and 
$\ell_n^{3}L_{n}^{4\varphi}(\log n)^{4}\ll n$, we know  
$L_n^{\varphi}\log n \ll n^{\varphi/(3\varphi+1)}$ holds automatically if $\ell_n^{3}L_{n}^{4\varphi}(\log n)^{4}\ll n$.
Hence, \eqref{eq:the2lct} holds provided that 
$L_n^{\varphi}\log r\ll n^{\varphi/(3\varphi+1)}$, $b_n\ll \min\{n^{-2/\gamma}, n^{-1}\ell_n^{-2}s^{-3}\nu^{-2}\}$, 
$\ell_n\aleph_n\ll \min\{\nu,\pi,s^{-1/2}\}$, $\ell_n^{3}L_{n}^{4\varphi}(\log n)^{4}\ll n$ and 
$\ell_n^3(\log r)^2\ll n^{(3\varphi-1)/(3\varphi+1)-2/\gamma}$.
Thus, we complete the proof of Theorem \ref{The.2}.
$\hfill\Box$

\subsection{Proof of Proposition \ref{p.1} }\label{sec:p.1}
Recall $\bfA=(\ba_1^0,\dots, \ba_m^0)^{\T}$ and let $\bff_t^{\bA}(\btheta)=\bA\bg_t(\btheta)$. Write $\widehat{\bGamma}_{\bff^{\bfA}}(\btheta) = \nabla_{\btheta_{\calM}} \bar{\bff}^{\bfA}(\btheta)$,  $\widehat{\bfV}_{\bff^{\bfA}}(\btheta)=\mathbb{E}_n\{\bff^{\bfA}_t( \btheta)^{\otimes2}\}$, $\widehat{\bGamma}_{\bff^{\bfA_n}}(\btheta) = \nabla_{\btheta_{\calM}} \bar{\bff}^{\bfA_n}(\btheta)$ and $\widehat{\bfV}_{\bff^{\bfA_n}}(\btheta)=\mathbb{E}_n\{\bff^{\bfA_n}_t( \btheta)^{\otimes2}\}$.
Define 
 $$H_n({\btheta_{\calM}},\blambda)=\frac{1}{n}\sum_{t=1}^{n}\log\{1+\blambda^\T\bff_t^{\bfA_n}(\btheta_{\calM},\hat{\btheta}_{n,\calM^{\rm c}})\}$$ and  $\tilde{\blambda}^*=\arg\max_{\blambda\in
 	\tilde{\Lambda}_n(\tilde{\btheta}_{\calM})}H_n(\tilde{\btheta}_{\calM},\blambda)$ with 
$\tilde{\btheta}_{\calM}=\arg\min_{\btheta_{\calM}\in\widehat{\bTheta}_{\calM}}\max_{\blambda\in{\tilde{\Lambda}}_n(\btheta_{\calM})}H_n({\btheta_{\calM}},\blambda)$.
 To prove Proposition \ref{p.1}, we need Lemmas \ref{hVhatL2norm}--\ref{gammahatstarL2norm}, whose proofs are  
 given  in Sections \ref{sec:la:hVhatL2norm}--\ref{sec:la:gammahatstarL2norm} of the supplementary material, respectively.

\begin{lemma}\label{hVhatL2norm}
	Assume the conditions of Theorem $\ref{The.1}$ hold. Under Conditions {\rm \ref{con:moments2}(b)}, {\rm \ref{con:moments2}(c)}, {\rm \ref{con:Gamma}(a)} and  {\rm \ref{con:alpha}}, if $\ell_ns^{3}b_n=o(1)$, then  
	\begin{align*}
&\|\widehat{\bfV}_{\bff^{\bfA_n}}(\btheta_{0,\calM},\hat{\btheta}_{n,\calM^{\rm c}})-\mathbb{E}\{\widehat{\bfV}_{\bff^{\bfA}}({\btheta}_0)\}\|_2=O_{\p}(L_n^{1/2}n^{-1/2})+O_{\p}(\omega_n)+O_{\p}(s^{1/2}\ell_n^{1/2}\nu)\,,
 \\&~~~~~~~~~~~~~~~
\|\widehat{\bfV}_{\bff^{\bfA_n}}(\tilde{\btheta}_{\calM},\hat{\btheta}_{n,\calM^{\rm c}})-\widehat{\bfV}_{\bff^{\bfA_n}}({\btheta}_{0,\calM},\hat{\btheta}_{n,\calM^{\rm c}})\|_2=O_{\p}(\ell_n^{1/2}\nu)\,.
 \end{align*}
\end{lemma}

\begin{lemma}\label{lambdahatstarL2norm}
Assume the conditions of Theorem {\rm \ref{The.1}} hold. Under Conditions {\rm \ref{con:moments2}(b)}, {\rm \ref{con:moments2}(c)}, {\rm \ref{con:Gamma}(a)} and  {\rm \ref{con:alpha}},  if $\ell_ns^{3}b_n=o(1)$,
$n\aleph_n^2\omega_n^2=O(1)$ and 
$ns\ell_n\nu^2\max\{s\ell_n\nu^2, \varsigma^2\}=O(1)$, then
\begin{align*}
|\bar{\bff}^{\bfA_n}(\tilde{\btheta}_{\calM},\hat{\btheta}_{n,\calM^{\rm c}})|_2=O_{\p}(L_n^{1/2}n^{-1/2})=|\bar{\bff}^{\bfA_n}(\btheta_{0,\calM},\hat{\btheta}_{n,\calM^{\rm c}})|_2
\end{align*} and $|\tilde{\blambda}^*|_2=O_{\p}(L_n^{1/2}n^{-1/2})$.
\end{lemma}

\begin{lemma}\label{gammahatstarL2norm}
	 Assume the conditions of Theorem $\ref{The.1}$ hold. Under Conditions {\rm \ref{con:moments2}(b)}, {\rm \ref{con:moments2}(c)}, {\rm \ref{con:Gamma}(a)} and  {\rm \ref{con:alpha}},
 if $\ell_ns^{3}b_n=o(1)$, then 
	\begin{align*}
	\big|\big[\widehat{\bGamma}_{\bff^{\bfA_n}}(\tilde{\btheta}_{\calM},\hat{\btheta}_{n,\calM^{\rm c}})-\mathbb{E}\{\widehat{\bGamma}_{\bff^{\bfA}}(\btheta_{0})\}\big ]\bfz\big|_2=|\bfz|_2\cdot\{O_{\p}(\omega_n)+O_{\p}(s^{1/2}\ell_n^{1/2}\nu)+O_{\p}(L_n^{1/2}n^{-1/2})\}\,
	\end{align*}
   holds uniformly over $\bz\in\mathbb{R}^m$.
\end{lemma}

Now we begin to prove Proposition \ref{p.1}.
By  Lemma \ref{lambdahatstarL2norm}, it holds that 
\begin{equation}\label{eq:fgamma}
\begin{split}
O_{\p}(L_n^{1/2}n^{-1/2})&=|\bar{\bff}^{\bfA_n}(\tilde{\btheta}_{\calM},\hat{\btheta}_{n,\calM^{\rm c}})-\bar{\bff}^{\bfA_n}({\btheta_{0,\calM}},\hat{\btheta}_{n,\calM^{\rm c}})|_2\\&\geq \lambda_{\min}^{1/2}\{\widehat{\bGamma}^{\T,\otimes2}_{\bff^{\bfA_n}}(\check{\btheta}_{\calM},\hat{\btheta}_{n,\calM^{\rm c}})\}|\tilde{\btheta}_{\calM}-\btheta_{0,\calM}|_2\,,
\end{split}
\end{equation}
where $\check{\btheta}_{\calM}$ is  on the jointing line between $\tilde{\btheta}_{\calM}$ and $\btheta_{0,\calM}$. Let  $\bSigma=(\bxi_{1},\ldots,\bxi_{m})^{\T}$. Then $\bfA \mathbb{E}\{\nabla_{\btheta}\bar{\bfg}(\btheta_{0})\}=\bSigma$. Denote by $\bSigma_{\cdot,[m]}$ the columns of $\bSigma$ that are indexed in $[m]$. Recall $\bxi_{k}$ is a $p$-dimensional vector with its $k$-th component being $1$ and all other components being $0$. Then $\bSigma_{\cdot,[m]}={\rm \bI}_{m}$. Hence, the eigenvalues of $[\mathbb{E}\{\widehat{\bGamma}^{\T}_{\bff^{\bfA}}(\btheta_0)\}]^{\otimes2}=\bSigma^{\T,\otimes2}_{\cdot,[m]}$ are uniformly bounded away from zero. Following the proof of Lemma  \ref{gammahatstarL2norm}, 
\begin{align*}
    \big|\big[\widehat{\bGamma}_{\bff^{\bfA_n}}(\check{\btheta}_{\calM},\hat{\btheta}_{n,\calM^{\rm c}})-\mathbb{E}\{\widehat{\bGamma}_{\bff^{\bfA}}(\btheta_{0})\}\big]\bfz\big|_2=|\bfz|_2\cdot\{O_{\p}(\omega_n)+O_{\p}(s^{1/2}\ell_n^{1/2}\nu)+O_{\p}(L_n^{1/2}n^{-1/2})\}
\end{align*}holds uniformly over $\bfz\in \mathbb{R}^{m}$. 
Due to $\omega_n=o(1)$, $s\ell_n\nu^2=o(1)$ and  $L_n\ll n$, it then holds that $\lambda_{\min}\{\widehat{\bGamma}^{\T,\otimes 2}_{\bff^{\bfA_n}}(\check{\btheta}_{\calM},\hat{\btheta}_{n,\calM^{\rm c}})\} $ is uniformly bounded away from zero
 w.p.a.1. By \eqref{eq:fgamma}, we have $|\tilde{\btheta}_{\calM}-\btheta_{0,\calM}|_2=O_{\p}(L_n^{1/2}n^{-1/2})$. We complete the proof of  Proposition \ref{p.1}.
$\hfill\Box$

\subsection{Proof of Theorem \ref{The.3}}
Let $\bXi_{\bff^{\bA}}(\btheta)={\rm Var}\{n^{1/2}\bar{\bff}^{\bA}(\btheta)\}$ for any $\btheta\in\bTheta$. To prove Theorem 3, we need Lemmas \ref{hhat-hatVstarL2norm}--\ref{cltZhat}, whose proofs are given in Sections  \ref{sec:la:hhat-hatVstarL2norm}--\ref{sec:la:cltZhat} of the supplementary material, respectively. Recall 
 $H_n({\btheta_{\calM}},\blambda)=n^{-1}\sum_{t=1}^{n}\log\{1+\blambda^\T\bff_t^{\bfA_n}(\btheta_{\calM},\hat{\btheta}_{n,\calM^{\rm c}})\}$ and  $\tilde{\blambda}^*=\arg\max_{\blambda\in
 	\tilde{\Lambda}_n(\tilde{\btheta}_{\calM})}H_n(\tilde{\btheta}_{\calM},\blambda)$.

\begin{lemma}\label{hhat-hatVstarL2norm}
	Assume the conditions of Proposition $\ref{p.1}$ hold. 
 It then holds that
	\begin{align*}
	&\sup_{c\in(0,1)}\bigg\|\frac{1}{n}\sum_{t=1}^{n}\frac{\bff_t^{\bfA_n}(\tilde{\btheta}_{\calM},\hat{\btheta}_{n,\calM^{\rm c}})^{\otimes2}}{\{1+c\tilde{\blambda}^{*,\T}\bff_t^{\bfA_n}(\tilde{\btheta}_{\calM},\hat{\btheta}_{n,\calM^{\rm c}})\}^2}-\widehat{\bfV}_{\bff^{\bfA_n}}(\tilde{\btheta}_{\calM},\hat{\btheta}_{n,\calM^{\rm c}})\bigg\|_2 =O_{\p}(L_n^{1/2}n^{-1/2+1/\gamma})\,,
\\&	\mbox{and}~~~~~
	\bigg|\biggl\{\frac{1}{n}\sum_{t=1}^{n}\frac{\nabla_{\btheta_{\calM}}\bff_t^{\bfA_n}(\tilde{\btheta}_{\calM},\hat{\btheta}_{n,\calM^{\rm c}})}{1+\tilde{\blambda}^{*,\T}\bff_t^{\bfA_n}(\tilde{\btheta}_{\calM},\hat{\btheta}_{n,\calM^{\rm c}})}-\widehat{\bGamma}_{\bff^{\bfA_n}}  (\tilde{\btheta}_{\calM},\hat{\btheta}_{n,\calM^{\rm c}})\biggr\}\bfz\bigg|_2=|\bfz|_2\cdot O_{\p}(L_n^{1/2}n^{-1/2}) 
	\end{align*} 
	 uniformly over  $\bfz\in\mathbb{R}^m$.
\end{lemma}



{\begin{lemma}\label{longruncov}
	Assume the conditions of Proposition $\ref{p.1}$ hold.	Under Condition {\rm \ref{con:kernel}}, 
 it holds that
\begin{align*}
	\|\widehat{\bXi}_{\bff^{\bA_n}}(\tilde{\btheta}_{\calM},\hat{\btheta}_{n,\calM^{\rm c}})-\bXi_{\bff^{\bA}}(\btheta_0)\|_2=O_{\p}(s^{1/2}\ell_n^{1/2}\nu h_n)+O_{\p}(h_n\omega_n)+O_{\p}(L_n^{1/2}h_nn^{-1/2})+O(L_n^{2}h_n^{-1})\,.
\end{align*}
\end{lemma}}

\begin{lemma}\label{cltZhat}
	Assume the conditions of Proposition $\ref{p.1}$ hold. Under Conditions {\rm \ref{con:Gamma}(b)} and  {\rm \ref{con:kernel}}, if  $L_{n}^{4\varphi}(\log n)^{4}\ll n$, $n\aleph_n^2\omega_n^2=o(1)$,
    $L_n h_n^2\max\{s\ell_n\nu^2,\omega_n^2,L_nn^{-1}\}=o(1)$ and $L_n^5\ll h_n^2$  	for any $\bfz\in\mathbb{R}^{m}$ with $|\bfz|_2=1$, it then holds that
\begin{align*}
n^{1/2}\bfz^{\T}\widehat{\bJ}_{\bff^{\bfA_{n}}}^{-1/2}\widehat{\bGamma}_{{\bff}^{\bfA_n}}^{\T}(\tilde{\btheta}_{\calM},\hat{\btheta}_{n,\calM^{\rm c}})\widehat{\bfV}^{-1}_{\bff^{\bfA_n}}(\tilde{\btheta}_{\calM},\hat{\btheta}_{n,\calM^{\rm c}})\bar{\bff}^{\bfA_n}(\btheta_{0})\rightarrow N(0,1)
\end{align*}
in distribution as $n\to\infty$.
\end{lemma}

Now we begin to prove Theorem \ref{The.3}.
By the definition of $\tilde{\btheta}_{\calM}$ and $\tilde{\blambda}^*$,
 we have $\nabla_{\blambda}H_n(\tilde{\btheta}_{\calM},\tilde{\blambda}^*)={\bf0}$, i.e.,
\begin{align*}
{\bf0}=\frac{1}{n}\sum_{t=1}^{n}\frac{\bff_t^{\bfA_n}(\tilde{\btheta}_{\calM},\hat{\btheta}_{n,\calM^{\rm c}})}{1+\tilde{\blambda}^{*,\T}\bff_t^{\bfA_n}(\tilde{\btheta}_{\calM},\hat{\btheta}_{n,\calM^{\rm c}})}\,.
\end{align*}
By Taylor expansion, it holds that
\begin{align*}
{\bf0}=\frac{1}{n}\sum_{t=1}^{n}\bff_t^{\bfA_n}(\tilde{\btheta}_{\calM},\hat{\btheta}_{n,\calM^{\rm c}})-\frac{1}{n}\sum_{t=1}^{n}\frac{\bff_t^{\bfA_n}(\tilde{\btheta}_{\calM},\hat{\btheta}_{n,\calM^{\rm c}})^{\otimes2}\tilde{\blambda}^*}{\{1+c\tilde{\blambda}^{*,\T}\bff_t^{\bfA_n}(\tilde{\btheta}_{\calM},\hat{\btheta}_{n,\calM^{\rm c}})\}^2}
\end{align*}
for some $c\in(0,1)$, which implies 
\begin{align*}
\tilde{\blambda}^*=\bigg[\frac{1}{n}\sum_{t=1}^{n}\frac{\bff_t^{\bfA_n}(\tilde{\btheta}_{\calM},\hat{\btheta}_{n,\calM^{\rm c}})^{\otimes2}}{\{1+c\tilde{\blambda}^{*,\T}\bff_t^{\bfA_n}(\tilde{\btheta}_{\calM},\hat{\btheta}_{n,\calM^{\rm c}})\}^2}\bigg]^{-1}\bar{\bff}^{\bfA_n}(\tilde{\btheta}_{\calM},\hat{\btheta}_{n,\calM^{\rm c}})\,.
\end{align*}
By the implicit function theorem [Theorem 9.28 of \cite{Rudin1976}], for all ${\btheta_{\calM}}$ in a $|\cdot|_2$-neighborhood of $\tilde{\btheta}_{\calM}$, there is a $\tilde{\blambda}^*({\btheta}_{\calM})$ such that $\nabla_{\blambda}H_n\{\btheta_{\calM},\tilde{\blambda}^*(\btheta_{\calM})\}={\bf0}$ and $\tilde{\blambda}^*({\btheta}_{\calM})$ is continuously differentiable in ${\btheta}_{\calM}$. By the concavity of $H_n({\btheta_{\calM}},\blambda)$ w.r.t $\blambda$, $H_n\{{\btheta}_{\calM},\tilde{\blambda}^*(\btheta_{\calM})\}=\max_{\blambda\in\tilde{\Lambda}_n(\btheta_{\calM})}H_n(\btheta_{\calM},\blambda)$. It follows from the envelope theorem that 
\begin{align*}
{\bf0}=\nabla_{\btheta_{\calM}}H_n\{\btheta_{\calM},\tilde{\blambda}^*(\btheta_{\calM})\}\bigg|_{{\btheta_{\calM}}=\tilde{\btheta}_{\calM}}=\bigg\{\frac{1}{n}\sum_{t=1}^{n}\frac{\nabla_{\btheta_{\calM}}\bff_t^{\bfA_n}(\tilde{\btheta}_{\calM},\hat{\btheta}_{n,\calM^{\rm c}})}{1+\tilde{\blambda}^{*,\T}\bff_t^{\bfA_n}(\tilde{\btheta}_{\calM},\hat{\btheta}_{n,\calM^{\rm c}})}\bigg\}^\T\tilde{\blambda}^*.
\end{align*}
Therefore, we have 
\begin{align}\label{eq:CD}
{\bf0}&=\bigg\{\frac{1}{n}\sum_{t=1}^{n}\frac{\nabla_{\btheta_{\calM}}\bff_t^{\bfA_n}(\tilde{\btheta}_{\calM},\hat{\btheta}_{n,\calM^{\rm c}})}{1+\tilde{\blambda}^{*,\T}\bff_t^{\bfA_n}(\tilde{\btheta}_{\calM},\hat{\btheta}_{n,\calM^{\rm c}})}\bigg\}^\T\notag\bigg[\frac{1}{n}\sum_{t=1}^{n}\frac{\bff_t^{\bfA_n}(\tilde{\btheta}_{\calM},\hat{\btheta}_{n,\calM^{\rm c}})^{\otimes2}}{\{1+c\tilde{\blambda}^{*,\T}\bff_t^{\bfA_n}(\tilde{\btheta}_{\calM},\hat{\btheta}_{n,\calM^{\rm c}})\}^2}\bigg]^{-1}\bar{\bff}^{\bfA_n}(\tilde{\btheta}_{\calM},\hat{\btheta}_{n,\calM^{\rm c}})\notag\\&=:\bfC^{\T}(\tilde{\btheta}_{\calM})\bfD^{-1}(\tilde{\btheta}_{\calM})\bar{\bff}^{\bfA_n}(\tilde{\btheta}_{\calM},\hat{\btheta}_{n,\calM^{\rm c}})\,.
\end{align}
Recall \begin{align*}
\widehat{\bJ}_{\bff^{\bfA_{n}}}=\{\widehat{\bGamma}_{{\bff}^{\bfA_n}}^{\T}(\tilde{\btheta}_{\calM},\hat{\btheta}_{n,\calM^{\rm c}})\widehat{\bfV}^{-1}_{\bff^{\bfA_n}}(\tilde{\btheta}_{\calM},\hat{\btheta}_{n,\calM^{\rm c}})\widehat{\bXi}^{1/2}_{\bff^{\bA_n}}(\tilde{\btheta}_{\calM},\hat{\btheta}_{n,\calM^{\rm c}})\}^{\otimes2}\,.
\end{align*} 
As shown in the proof of Lemma \ref{cltZhat}, we know the eigenvalues of $\widehat{\bGamma}^{\T}_{{\bff}^{\bfA_n}}(\tilde{\btheta}_{\calM},\hat{\btheta}_{n,\calM^{\rm c}})^{\otimes2}$, $\widehat{\bfV}^{-1}_{\bff^{\bfA_n}}(\tilde{\btheta}_{\calM},\hat{\btheta}_{n,\calM^{\rm c}})$, $\widehat{\bXi}_{\bff^{\bA_n}}(\tilde{\btheta}_{\calM},\hat{\btheta}_{n,\calM^{\rm c}})$ and $\widehat{\bJ}_{\bff^{\bfA_{n}}}$
are uniformly bounded away from zero and infinity
 w.p.a.1.
For any $\bfz\in \mathbb{R}^m$ with $|\bfz|_2=1$, let $\bdelta=\widehat{\bJ}_{\bff^{\bfA_{n}}}^{-1/2}\bfz$. Then
\begin{align*}
|\widehat{\bGamma}_{{\bff}^{\bfA_n}}(\tilde{\btheta}_{\calM},\hat{\btheta}_{n,\calM^{\rm c}})\bdelta|_2^2\leq &~
\lambda_{\max}(\widehat{\bJ}^{-1}_{\bff^{\bfA_{n}}})\lambda_{\max}\{\widehat{\bGamma}^{\T}_{{\bff}^{\bfA_n}}(\tilde{\btheta}_{\calM},\hat{\btheta}_{n,\calM^{\rm c}})^{\otimes 2}\}\cdot|\bfz|_2^2=O_{\p}(1)\,.
\end{align*}
By \eqref{eq:CD}, it holds that 
\begin{align*}
\big|&\bdelta^{\T}\widehat{\bGamma}^{\T}_{{\bff}^{\bfA_n}}(\tilde{\btheta}_{\calM},\hat{\btheta}_{n,\calM^{\rm c}})\widehat{\bfV}^{-1}_{\bff^{\bfA_n}}(\tilde{\btheta}_{\calM},\hat{\btheta}_{n,\calM^{\rm c}})\bar{\bff}^{\bfA_n}(\tilde{\btheta}_{\calM},\hat{\btheta}_{n,\calM^{\rm c}})\big|\\&~~~~~~~~=
\big|\bdelta^{\T}\{\widehat{\bGamma}^{\T}_{{\bff}^{\bfA_n}}(\tilde{\btheta}_{\calM},\hat{\btheta}_{n,\calM^{\rm c}})\widehat{\bfV}^{-1}_{\bff^{\bfA_n}}(\tilde{\btheta}_{\calM},\hat{\btheta}_{n,\calM^{\rm c}})-\bfC^{\T}(\tilde{\btheta}_{\calM})\bfD^{-1}(\tilde{\btheta}_{\calM})\}\bar{\bff}^{\bfA_n}(\tilde{\btheta}_{\calM},\hat{\btheta}_{n,\calM^{\rm c}})\big|\\&~~~~~~~~\leq 
\underbrace{\big|\bdelta^{\T}\widehat{\bGamma}^{\T}_{{\bff}^{\bfA_n}}(\tilde{\btheta}_{\calM},\hat{\btheta}_{n,\calM^{\rm c}})\{\widehat{\bfV}^{-1}_{\bff^{\bfA_n}}(\tilde{\btheta}_{\calM},\hat{\btheta}_{n,\calM^{\rm c}})-\bfD^{-1}(\tilde{\btheta}_{\calM})\}\bar{\bff}^{\bfA_n}(\tilde{\btheta}_{\calM},\hat{\btheta}_{n,\calM^{\rm c}})\big|}_{T_1}\\&~~~~~~~~~~~~~~+
\underbrace{\big|\bdelta^{\T}\{\widehat{\bGamma}^{\T}_{{\bff}^{\bfA_n}}(\tilde{\btheta}_{\calM},\hat{\btheta}_{n,\calM^{\rm c}})-\bfC^{\T}(\tilde{\btheta}_{\calM})\}\bfD^{-1}(\tilde{\btheta}_{\calM})\bar{\bff}^{\bfA_n}(\tilde{\btheta}_{\calM},\hat{\btheta}_{n,\calM^{\rm c}})\big|}_{T_2}\,.
\end{align*}
Due to $L_n\ll n^{1-2/\gamma}$,
by Lemma \ref{hhat-hatVstarL2norm},  $\|\bfD^{-1}(\tilde{\btheta}_{\calM})\|_2=O_{\p}(1)$, which implies 
\begin{align*}
&\|\widehat{\bfV}^{-1}_{\bff^{\bfA_n}}(\tilde{\btheta}_{\calM},\hat{\btheta}_{n,\calM^{\rm c}})-\bfD^{-1}(\tilde{\btheta}_{\calM})\|_2\\&~~~~~~~\leq \|\widehat{\bfV}^{-1}_{\bff^{\bfA_n}}(\tilde{\btheta}_{\calM},\hat{\btheta}_{n,\calM^{\rm c}})\|_2\cdot \|\widehat{\bfV}_{\bff^{\bfA_n}}(\tilde{\btheta}_{\calM},\hat{\btheta}_{n,\calM^{\rm c}})-\bfD(\tilde{\btheta}_{\calM})\|_2\cdot \|\bfD^{-1}(\tilde{\btheta}_{\calM})\|_2\\&~~~~~~~=
O_{\p}(L_n^{1/2}n^{-1/2+1/\gamma})\,.
\end{align*}
By Lemma \ref{lambdahatstarL2norm}, we know $|\bar{\bff}^{\bfA_n}(\tilde{\btheta}_{\calM},\hat{\btheta}_{n,\calM^{\rm c}})|_2=O_{\p}(L_n^{1/2}n^{-1/2})$. Then  
\begin{align*}
T_1\leq&~
|\widehat{\bGamma}_{{\bff}^{\bfA_n}}(\tilde{\btheta}_{\calM},\hat{\btheta}_{n,\calM^{\rm c}})\bdelta|_2\cdot \|\widehat{\bfV}^{-1}_{\bff^{\bfA_n}}(\tilde{\btheta}_{\calM},\hat{\btheta}_{n,\calM^{\rm c}})-\bfD^{-1}(\tilde{\btheta}_{\calM})\|_2\cdot |\bar{\bff}^{\bfA_n}(\tilde{\btheta}_{\calM},\hat{\btheta}_{n,\calM^{\rm c}})|_2\\=&~
O_{\p}(L_nn^{-1+1/\gamma})\,.
\end{align*}
Moreover, by Lemma \ref{hhat-hatVstarL2norm}, we  also  have
\begin{align*}
T_2	\leq &~
|\bdelta|_2\cdot \|\bfD^{-1}(\tilde{\btheta}_{\calM})\|_2\cdot|\bar{\bff}^{\bfA_n}(\tilde{\btheta}_{\calM},\hat{\btheta}_{n,\calM^{\rm c}})|_2 \cdot O_{\p}(L_n^{1/2}n^{-1/2}) =
O_{\p}(L_nn^{-1})\,.
\end{align*} 
Hence,   
\begin{align*}
    |\bdelta^{\T}\widehat{\bGamma}^{\T}_{{\bff}^{\bfA_n}}(\tilde{\btheta}_{\calM},\hat{\btheta}_{n,\calM^{\rm c}})\widehat{\bfV}^{-1}_{\bff^{\bfA_n}}(\tilde{\btheta}_{\calM},\hat{\btheta}_{n,\calM^{\rm c}})\bar{\bff}^{\bfA_n}(\tilde{\btheta}_{\calM},\hat{\btheta}_{n,\calM^{\rm c}})|=O_{\p}(L_nn^{-1+1/\gamma})\,.
\end{align*}
By Taylor expansion,
\begin{equation}\label{eq:bGammaex}
\begin{split}
&\bdelta^{\T}\widehat{\bGamma}^{\T}_{{\bff}^{\bfA_n}}(\tilde{\btheta}_{\calM},\hat{\btheta}_{n,\calM^{\rm c}})\widehat{\bfV}^{-1}_{\bff^{\bfA_n}}(\tilde{\btheta}_{\calM},\hat{\btheta}_{n,\calM^{\rm c}})\{\widehat{\bGamma}_{{\bff}^{\bfA_n}}(\check{\btheta}_{\calM},\hat{\btheta}_{n,\calM^{\rm c}})(\tilde{\btheta}_{\calM}-\btheta_{0,\calM})\}\\&~~~~~~=
-\bdelta^{\T}\widehat{\bGamma}^{\T}_{{\bff}^{\bfA_n}}(\tilde{\btheta}_{\calM},\hat{\btheta}_{n,\calM^{\rm c}})\widehat{\bfV}^{-1}_{\bff^{\bfA_n}}(\tilde{\btheta}_{\calM},\hat{\btheta}_{n,\calM^{\rm c}})\bar{\bff}^{\bfA_n}(\btheta_{0,\calM},\hat{\btheta}_{n,\calM^{\rm c}})+O_{\p}(L_nn^{-1+1/\gamma})\,,
\end{split}
\end{equation}
where $\check{\btheta}_{\calM}$ is on the jointing line  between $\btheta_{0,\calM}$ and $\tilde{\btheta}_{\calM}$. As shown in \eqref{eq:fantheta} of supplementary material,
 we have 
\begin{align*}
|\bar{\bff}^{\bfA_n}(\btheta_{0,\calM},\hat{\btheta}_{n,\calM^{\rm c}})-\bar{\bff}^{\bfA_n}(\btheta_{0})|_2=&~	O_{\p}(s\ell_n\nu^2)+O_{\p}\{s^{1/2}\ell_n^{1/2}\nu(\varsigma+\ell_n^{1/2}\nu)\}=
o_{\p}(n^{-1/2})
\end{align*}
provided that $ns\ell_n\nu^2\max\{s\ell_n\nu^2, \varsigma^2\}=o(1)$.
By Taylor expansion, Jensen's inequality and the Cauchy-Schwarz inequality, it holds that 
\begin{align*}
&\big|\{\widehat{\bGamma}_{{\bff}^{\bfA_n}}(\tilde{\btheta}_{\calM},\hat{\btheta}_{n,\calM^{\rm c}})-\widehat{\bGamma}_{{\bff}^{\bfA_n}}(\check{\btheta}_{\calM},\hat{\btheta}_{n,\calM^{\rm c}})\}(\tilde{\btheta}_{\calM}-\btheta_{0,\calM})\big|_2^2\\&~~~~~~\leq |\tilde{\btheta}_{\calM}-\btheta_{0,\calM}|_2^4\cdot \frac{1}{n}\sum_{j=1}^{m}\sum_{t=1}^{n}\sum_{k\in \mm}
\sum_{l\in\mm}\bigg|\frac{\partial ^2f_{t,j}^{\bfA_{n}}(\dot{\btheta})}{\partial \theta_{k}\partial \theta_{l}}\bigg|^2\,,
\end{align*}
where $\dot{\btheta}$ lies on the jointing line between $\bar{\btheta}_1=(\tilde{\btheta}_{\calM}^{\T},\hat{\btheta}_{n,\calM^{\rm c}}^{\T})^{\T}$ and $\bar{\btheta}_2=(\check{\btheta}_{\calM}^{\T},\hat{\btheta}_{n,\calM^{\rm c}}^{\T})^{\T}$. By Lemma \ref{hL2norm} in the supplementary material and Proposition \ref{p.1}, we have $|\{\widehat{\bGamma}_{{\bff}^{\bfA_n}}(\tilde{\btheta}_{\calM},\hat{\btheta}_{n,\calM^{\rm c}})-\widehat{\bGamma}_{{\bff}^{\bfA_n}}(\check{\btheta}_{\calM},\hat{\btheta}_{n,\calM^{\rm c}})\}(\tilde{\btheta}_{\calM}-\btheta_{0,\calM})|_2=O_{\p}(L_nn^{-1})$. Recall $\bdelta=\widehat{\bJ}_{\bff^{\bfA_{n}}}^{-1/2}\bfz$. 
Therefore, by \eqref{eq:bGammaex}, it holds that 
\begin{align*}
&\sqrt{n}\bdelta^{\T}\widehat{\bGamma}^{\T}_{{\bff}^{\bfA_n}}(\tilde{\btheta}_{\calM},\hat{\btheta}_{n,\calM^{\rm c}})\widehat{\bfV}^{-1}_{\bff^{\bfA_n}}(\tilde{\btheta}_{\calM},\hat{\btheta}_{n,\calM^{\rm c}})\widehat{\bGamma}_{{\bff}^{\bfA_n}}(\tilde{\btheta}_{\calM},\hat{\btheta}_{n,\calM^{\rm c}})(\tilde{\btheta}_{\calM}-\btheta_{0,\calM})\\&~~~=-\sqrt{n}\bdelta^{\T}\widehat{\bGamma}^{\T}_{{\bff}^{\bfA_n}}(\tilde{\btheta}_{\calM},\hat{\btheta}_{n,\calM^{\rm c}})\widehat{\bfV}^{-1}_{\bff^{\bfA_n}}(\tilde{\btheta}_{\calM},\hat{\btheta}_{n,\calM^{\rm c}})\bar{\bff}^{\bfA_n}(\btheta_{0})+O_{\p}(L_nn^{-1/2+1/\gamma})+o_{\p}(1)\\&~~~=
-\sqrt{n}\bfz^{\T}\widehat{\bJ}_{\bff^{\bfA_{n}}}^{-1/2}\widehat{\bGamma}^{\T}_{{\bff}^{\bfA_n}}(\tilde{\btheta}_{\calM},\hat{\btheta}_{n,\calM^{\rm c}})\widehat{\bfV}^{-1}_{\bff^{\bfA_n}}(\tilde{\btheta}_{\calM},\hat{\btheta}_{n,\calM^{\rm c}})\bar{\bff}^{\bfA_n}(\btheta_{0})+O_{\p}(L_nn^{-1/2+1/\gamma})+o_{\p}(1)\,.
\end{align*}
Recall $\widehat{\bM}_{\bff^{\bA_n}}=\widehat{\bGamma}^{\T}_{{\bff}^{\bfA_n}}(\tilde{\btheta}_{\calM},\hat{\btheta}_{n,\calM^{\rm c}})\widehat{\bfV}^{-1}_{\bff^{\bfA_n}}(\tilde{\btheta}_{\calM},\hat{\btheta}_{n,\calM^{\rm c}})\widehat{\bGamma}_{{\bff}^{\bfA_n}}(\tilde{\btheta}_{\calM},\hat{\btheta}_{n,\calM^{\rm c}})$.
 Due to $L_n\ll n^{1/2-1/\gamma}$, by Lemma \ref{cltZhat},  we have  Theorem \ref{The.3}.
$\hfill\Box$

\section{Supplementary material}
The proofs of the auxiliary lemmas are deferred to the supplementary material of the paper.

\end{appendices}

\begingroup
\singlespacing
\small
\bibliography{ref}
\endgroup

\newpage

\setcounter{page}{1}
 
\rhead{\bfseries\thepage}
\lhead{\bfseries SUPPLEMENTARY MATERIAL}

\setcounter{page}{1}
\begin{center}
	{\bf\Large Supplementary material for ``Empirical likelihood approach for high-dimensional moment restrictions with dependent data'' }
\end{center}


\begin{center}
	Jinyuan Chang$^{1,2}$, Qiao Hu$^{1}$, Zhentao Shi$^{3}$, and Jia Zhang$^{1}$ \\
 \medskip   
{\it \small $^{1}$Joint Laboratory of Data Science and Business Intelligence, Southwestern University of Finance and Economics, Chengdu, Sichuan, China} \\
{\it \small $^2$Academy of Mathematics and Systems Science, Chinese Academy of Sciences, Beijing, China} \\
{\it \small $^3$Department of Economics, Chinese University of
Hong Kong, Hong Kong SAR, China}
\end{center}

\bigskip

\setcounter{equation}{0} 

\setcounter{section}0

\renewcommand{\thesection}{S\arabic{section}}
\renewcommand{\thelemma}{S\arabic{lemma}} 
\numberwithin{equation}{section}
\setcounter{lemma}{0} 

In the sequel, we use $C$ and $\tilde{C}$ to denote generic finite positive universal constants that may be different in different uses. For any real number $x$, define
$
\lceil x \rceil = \min \{ q \in \mathbb{Z} : q \geq  x \}
$,
where \( \mathbb{Z} \) denotes the set of all integers.
\section{Proofs of  Lemmas  \ref{tailprabog0}--\ref{cltZhat}}\label{proofs of  lemmas}

\subsection{Proof of Lemma \ref{tailprabog0}}\label{sec:la:tailprabog0}

To prove Lemma \ref{tailprabog0}, we need Lemmas \ref{partial sum ex21} and \ref{self-normalized alpha-mixing}, whose proofs are given in Sections \ref{secla:partial sum ex2} and \ref{sec:la:self-normalized alpha-mixing}, respectively.

\begin{lemma}\label{partial sum ex21}
Let $\{\xi_t\}_{t=1}^{\tilde{n}} $ be an $\alpha$-mixing sequence of  centered random variables with $\alpha$-mixing coefficients $\{\tilde{\alpha}_{\tilde{n}}(k)\}_{k\geq1}$.
Assume there exist some universal constants $\tilde{\gamma}>4$, $b_1>1$, $b_2>0$, $c_1>0$ and $\varphi\geq 1$ such that {\rm (i)} $  \mathbb{E}(|\xi_t|^{\tilde{\gamma}})\leq c_1^{\tilde{\gamma}}$ for any $t\in[\tilde{n}]$, and {\rm(ii)} $\tilde{\alpha}_{\tilde{n}}(k)\leq b_1\exp\{-b_2(\tilde{L}_{\tilde{n}}^{-1}k)^{\varphi}\}$ for any integer $k\geq 1$, where $0<\tilde{L}_{\tilde{n}}\ll \tilde{n}$ may diverge with $\tilde{n}$. For any  integers $l\geq 1$ and $h\geq 1$ satisfying $ l+h\leq \tilde{n}$,
let $S_{l,h}=\sum_{t=l}^{l+h}\xi_t$. Then 
$$\max_{l\in[\tilde{n}-h]}\mathbb{E}(S_{l,h}^2)\lesssim h\tilde{L}_{\tilde{n}}\,.$$
Furthermore,
if $\tilde{L}_{\tilde{n}}^{\varphi}\log h\ll h^{\varphi}$, it also  holds that 
\begin{align*}
\max_{l\in[\tilde{n}-h]}\mathbb{E}(S_{l,h}^{4})\lesssim h^{2}\tilde{L}_{\tilde{n}}^{2}\,.
\end{align*}
\end{lemma}

\begin{lemma}\label{self-normalized alpha-mixing}
	Let $\{\xi_t\}_{t=1}^{\tilde{n}} $ be an $\alpha$-mixing sequence of  centered random variables with $\alpha$-mixing coefficients $\{\tilde{\alpha}_{\tilde{n}}(k)\}_{k\geq1}$.
	Assume there exist some universal constants $\tilde{\gamma}>4$, $b_1>1$, $b_2>0$, $c_1>0$, $c_2>0$ and $\varphi\geq 1$ such that {\rm (i)} $\mathbb{E}(|\xi_t|^{\tilde{\gamma}})\leq c_1^{\tilde{\gamma}}$ for any $t\in[\tilde{n}]$, {\rm (ii)} $\tilde{\alpha}_{\tilde{n}}(k)\leq b_1\exp\{-b_2(\tilde{L}_{\tilde{n}}^{-1}k)^{\varphi}\}$ for any integer $k\geq 1$, where $0<\tilde{L}_{\tilde{n}}\ll \tilde{n}$ may diverge with $\tilde{n}$, and {\rm(iii)} $
	\mathbb{E}(S_{l,h}^2)\geq c_2^2h$ for any integers $l\geq 1$ and $h\geq 1$, where $S_{l,h}=\sum_{t=l}^{l+h}\xi_t$. Select   $\tilde{m}=\lfloor {\tilde{n}}^{1/(3\varphi+1)}\rfloor$, $\tilde{k}_{\tilde{n}}= \lceil \tilde{n}/\tilde{m}\rceil$ and  
	$ \tilde{H}_j=\{t\in[\tilde{n}]:\tilde{m}(j-1)+1\leq t \leq \{\tilde{m}(j-1)+\tilde{m}\}\wedge\tilde{n}\}$ for any $j\in[\tilde{k}_{\tilde{n}}]$. 
    Write $S_{\tilde{n}}=S_{1,\tilde{n}-1}$ and $V_{\tilde{k}_{\tilde{n}}}^2= \sum_{j=1}^{\tilde{k}_{\tilde{n}}}Y_j^2$ with $Y_j=\sum_{t\in \tilde{H}_j}\xi_{t}$.
	If $\tilde{L}^{\varphi}_{\tilde{n}}\log\tilde{n}\ll \tilde{n}^{\varphi/(3\varphi+1)}$, then
	\begin{align*}
	\mathbb{P}\big(S_{\tilde{n}}\geq 4xV_{\tilde{k}_{\tilde{n}}}\big)\lesssim \exp(-x^2/2)
	\end{align*}
	holds uniformly over $c_2^{1/2}\leq x\leq c_*{\tilde{n}}^{\varphi/(6\varphi+2)}\tilde{L}_{\tilde{n}}^{-\varphi/2}$, where $c_*>0$ is a sufficiently small constant.
\end{lemma}

Write $\mathring{g}_{t,j_1,j_2}(\btheta_{0})=g_{t,j_1}(\btheta_0)g_{t,j_2}(\btheta_0)-\mathbb{E}\{g_{t,j_1}(\btheta_0)g_{t,j_2}(\btheta_0)\}$ and   
$ k_n= \lceil n/m\rceil$ with  $m=\lfloor n^{1/(3\varphi+1)}\rfloor$. Let $Y_{i,j_1,j_2}=\sum_{t\in H_i}\mathring{g}_{t,j_1,j_2}(\btheta_{0})$ for $ H_i=\{t\in[n]:m(i-1)+1\leq t \leq \{m(i-1)+m\}\wedge n\}$.  
By Condition \ref{con:moments1}(a) and the Cauchy-Schwarz inequality, it holds that 
\begin{align*}
\max_{t\in[n]}\max_{j_1,j_2\in[r]}\mathbb{E}\{|\mathring{g}_{t,j_1,j_2}(\btheta_{0})|^{{\gamma}}\}\lesssim \max_{t\in[n]}\max_{j\in[r]}\mathbb{E}\{|g_{t,j}(\btheta_0)|^{{2\gamma}}\}\lesssim 1\,.
\end{align*}
By Condition \ref{con:moments1}(b), we have 
\begin{align*}
\min_{j_1,j_2\in[r]}\mathbb{E}\bigg[\bigg\{\sum_{t=l}^{l+h}\mathring{g}_{t,j_1,j_2}(\btheta_{0})\bigg\}^2\bigg]\geq K_5 h
\end{align*}
for any integer $h\geq 1$ and $l\geq 1$. Write $V_{k_n,j_1,j_2}^2=\sum_{i=1}^{k_n}Y_{i,j_1,j_2}^2$. Applying  Lemma \ref{self-normalized alpha-mixing} with $\tilde{n}=n$, $\tilde{L}_{\tilde{n}}=L_n$, $\xi_t=\mathring{g}_{t,j_1,j_2}(\btheta_{0})$ and $\tilde{\gamma}=\gamma$,  then 
\begin{align*}
\mathbb{P}\bigg\{\frac{1}{V_{k_n,j_1,j_2}}\sum_{t=1}^{n}\mathring{g}_{t,j_1,j_2}(\btheta_{0})\geq C(\log r)^{1/2}\bigg\}\lesssim \exp(-\tilde{C}\log r)
\end{align*}
provided that $L_n^{\varphi}\log (rn)\ll n^{\varphi/(3\varphi+1)}$. Analogously, we also show 
\begin{align*}
	\mathbb{P}\bigg\{\frac{1}{V_{k_n,j_1,j_2}}\sum_{t=1}^{n}\mathring{g}_{t,j_1,j_2}(\btheta_{0})\leq - C(\log r)^{1/2}\bigg\}\lesssim \exp(-\tilde{C}\log r)
\end{align*}
provided that $L_n^{\varphi}\log (rn)\ll n^{\varphi/(3\varphi+1)}$.
Hence, if $L_n^{\varphi}\log (rn)\ll n^{\varphi/(3\varphi+1)}$, by the Bonferroni inequality, 
\begin{align*}
&\mathbb{P}\bigg\{\max_{j_1,j_2\in[r]}\bigg|\frac{1}{V_{k_n,j_1,j_2}}\sum_{t=1}^{n}\mathring{g}_{t,j_1,j_2}(\btheta_{0})\bigg|\geq C(\log r)^{1/2} \bigg\}
\\&~~~~~~\leq r^2\max_{j_1,j_2\in[r]}\mathbb{P}\bigg\{\bigg|\frac{1}{V_{k_n,j_1,j_2}}\sum_{t=1}^{n}\mathring{g}_{t,j_1,j_2}(\btheta_{0})\bigg|\geq C(\log r)^{1/2}\bigg\} \lesssim \exp(-\tilde{C}\log r)\,,
\end{align*}
which implies 
\begin{align}\label{eq:mathringg}
\max_{j_1,j_2\in[r]}\bigg|\frac{1}{V_{k_n,j_1,j_2}}\sum_{t=1}^{n}\mathring {g}_{t,j_1,j_2}(\btheta_{0})\bigg|=O_{\p}\{(\log r)^{1/2}\}\,.
\end{align}
Recall $V_{k_n,j_1,j_2}^2=\sum_{i=1}^{k_n}Y_{i,j_1,j_2}^2$ with $Y_{i,j_1,j_2}=\sum_{t\in H_i}\mathring{g}_{t,j_1,j_2}(\btheta_{0})$.
By Condition \ref{con:moments1}(a),  Jensen's inequality, the Cauchy-Schwarz inequality, and the inequality of geometric and quadratic means, we have
\begin{align*}
\max_{j_1,j_2\in[r]}V_{k_n,j_1,j_2}^2&=\max_{j_1,j_2\in[r]} \sum_{i=1}^{k_n
}\bigg\{\sum_{t\in H_i}\mathring{g}_{t,j_1,j_2}(\btheta_{0})\bigg\}^2\leq m\max_{j_1,j_2\in[r]}\sum_{i=1}^{k_n}\sum_{t\in H_i}\mathring{g}_{t,j_1,j_2}^2(\btheta_{0})
\\&\lesssim m\max_{j\in[r]}\sum_{t=1}^{n}\big[\mathbb{E}\{{g}_{t,j}^2(\btheta_{0})\}\big]^2+ m\max_{j\in[r]}\sum_{t=1}^n{g}_{t,j}^4(\btheta_{0})
\\&=  
O_{\p}(mn)\,.
\end{align*}
Recall $\aleph_n=n^{-3\varphi/(6\varphi+2)}(\log r)^{1/2}$.
Due to $m\asymp n^{1/(3\varphi+1)}$, by \eqref{eq:mathringg}, 
\begin{align*}
\max_{j_1,j_2\in[r]}\bigg|\frac{1}{n}\sum_{t=1}^{n}\mathring{g}_{t,j_1,j_2}(\btheta_{0})\bigg|\leq&~ \frac{1}{n}\max_{j_1,j_2\in[r]}\bigg|\frac{1}{V_{k_n,j_1,j_2}}\sum_{t=1}^{n}\mathring{g}_{t,j_1,j_2}(\btheta_{0})\bigg| \cdot  \max_{j_1,j_2\in[r]}V_{k_n,j_1,j_2}
\\\leq &~O_{\p}\{(n^{-1}m\log r)^{1/2}\}=O_{\p}\{n^{-3\varphi/(6\varphi+2)}(\log r)^{1/2}\}=O_{\p}(\aleph_n)
\end{align*}
provided that
$L_n^{\varphi}\log (rn)\ll n^{\varphi/(3\varphi+1)}$.
Following the same arguments,  we can also show  
\begin{align*}
	\max_{j\in[r]}\bigg|\frac{1}{n}\sum_{t=1}^{n}g_{t,j}(\btheta_0)\bigg|=O_{\p}(\aleph_n)\,
\end{align*}
provided that $L_n^{\varphi}\log (rn)\ll n^{\varphi/(3\varphi+1)}$.
We complete the proof of Lemma \ref{tailprabog0}.
$\hfill\Box$

\subsection{Proof of Lemma \ref{hatV-barV}}
Recall $\bTheta_{n}=\{\btheta=(\btheta_\calS^{\T},\btheta_{\calS^{\rm c}}^{\T})^{\T} \in \bTheta: |\btheta_{\calS}-\btheta_{0,\calS}|_{\infty} = O_{\p}(b_n^{1/2}),|\btheta_{\calS^{\rm c}}|_1\leq \aleph_n\}$ and $\mathscr{T}=\{\mathcal{T} \subset [r]:|\mathcal{T}|\leq \ell_n\}$. By the  triangle inequality, 
\begin{align*}
\| \widehat{\bfV}_{\mT} (\btheta) - \mathbb{E}\{\widehat{\bfV}_{\mT}(\btheta_0)\}\|_{2} \leq \| \widehat{\bfV}_{\mT} (\btheta) - \widehat{\bfV}_{\mT} (\bthetazero)\|_{2} + \| \widehat{\bfV}_{\mT} (\bthetazero) - \mathbb{E}\{\widehat{\bfV}_{\mT}(\btheta_0)\|_{2}
\end{align*} for any $\mathcal{T} \in \mathscr{T}$ and $\btheta \in \bTheta_{n}$. Recall $\widehat{\bfV}_{\mT}(\btheta)=\mathbb{E}_n\{ \bfg_{t,\mT} (\btheta)^{\otimes2}\}$. By Lemma \ref{tailprabog0}, 
\begin{align*}
\max_{j_1,j_2\in[r]}\bigg|\frac{1}{n}\sum_{t=1}^{n}\big[g_{t,j_1}(\btheta_0)g_{t,j_2}(\btheta_0)-\mathbb{E}\{g_{t,j_1}(\btheta_0)g_{t,j_2}(\btheta_0)\}\big]\bigg|=O_{\p}(\aleph_n)
\end{align*}
provided that $L_n^{\varphi}\log (rn)\ll n^{\varphi/(3\varphi+1)}$,
which implies
\begin{equation}\label{hatV-barVtheta0}
\sup_{\mathcal{T} \in \mathscr{T}} \| \widehat{\bfV}_{\mT} (\bthetazero) - \mathbb{E}\{\widehat{\bfV}_{\mT}(\btheta_0)\}\|_{2}=O_{\p}(\ell_n\aleph_n)\,. 
\end{equation}
 For any $\bfz \in \mathbb{R}^{|\mathcal{T}|}$ with unit $L_{2}$-norm, by the triangle inequality and the Cauchy-Schwarz inequality, 
\begin{align*}
&|\bfz ^\T \{\widehat{\bfV}_{\mT} (\btheta) - \widehat{\bfV}_{\mT} (\bthetazero)\} \bfz | \\
&~~~~~= \bigg | \bfz^\T \bigg [  \frac{1}{n} \sum_{t=1}^{n} \{\bfg_{t,\mT}(\btheta)-\bfg_{t,\mT}(\bthetazero)\}^{\otimes2} +\frac{1}{n} \sum_{t=1}^{n} \{\bfg_{t,\mT}(\btheta)-\bfg_{t,\mT}(\bthetazero)\} {\bfg_{t,\mT}(\bthetazero)}^\T \\
&~~~~~~~~~~~~+ \frac{1}{n} \sum_{t=1}^{n} \bfg_{t,\mT}(\bthetazero) \{\bfg_{t,\mT}(\btheta)-\bfg_{t,\mT}(\bthetazero)\} ^\T  \bigg ] \bfz \bigg| \\
 &~~~~~\leq\frac{1}{n} \sum_{t=1}^{n} \big[\bfz^\T \{\bfg_{t,\mT}(\btheta)-\bfg_{t,\mT}(\bthetazero)\}\big]^{2} +  2\bigg | \frac{1}{n} \sum_{t=1}^{n} \bfz^\T  \bfg_{t,\mT}(\bthetazero) \{\bfg_{t,\mT}(\btheta)-\bfg_{t,\mT}(\bthetazero)\} ^\T \bfz \bigg| \\
 &~~~~~\leq \frac{1}{n} \sum_{t=1}^{n} |\bfg_{t,\mT}(\btheta)-\bfg_{t,\mT}(\bthetazero)|_{2}^{2}+2 \lambda_{\max} ^{1/2} \{ \widehat{\bfV}_{\mT} (\bthetazero)\} \bigg \{ \frac{1}{n} \sum_{t=1}^{n} |\bfg_{t,\mT}(\btheta)-\bfg_{t,\mT}(\bthetazero)|_{2}^{2} \bigg\}^{1/2} \,.
\end{align*}
Notice that  $\widehat{\bfV}_{\mT} (\btheta) - \widehat{\bfV}_{\mT} (\bthetazero)$ is a symmetric matrix. Then 
\begin{align}\label{eq:hatVt}
\sup_{\mathcal{T} \in \mathscr{T}}\|\widehat{\bfV}_{\mT} (\btheta) - \widehat{\bfV}_{\mT} (\bthetazero)\|_{2} 
\leq & \sup_{\mathcal{T} \in \mathscr{T}} \bigg \{\frac{1}{n} \sum_{t=1}^{n} |\bfg_{t,\mT}(\btheta)-\bfg_{t,\mT}(\bthetazero)|_{2}^{2}\bigg\} \\
&+ 2 \sup_{\mathcal{T} \in \mathscr{T}} \lambda_{\max} ^{1/2} \{ \widehat{\bfV}_{\mT} (\bthetazero)\} \cdot \sup_{\mathcal{T} \in \mathscr{T}} \bigg \{ \frac{1}{n} \sum_{t=1}^{n} |\bfg_{t,\mT}(\btheta)-\bfg_{t,\mT}(\bthetazero)|_{2}^{2} \bigg\}^{1/2}\,.\notag
\end{align}
Write $ \btheta=(\btheta_\calS^{\T},\btheta_{\calS^{\rm c}}^{\T})^{\T}$ with $\btheta_\calS \in \mathbb{R}^{s}$. By Taylor expansion and  the Cauchy-Schwarz inequality, we have
\begin{align*}
\frac{1}{n}\sum_{t=1}^{n}|\bfg_{t,\mT}(\btheta)-\bfg_{t,\mT}(\bthetazero)|_2^2
&\leq \frac{2}{n}\sum_{t=1}^{n}\bigg|\frac{\partial\bfg_{t,\mT}(\tilde{\btheta})}{\partial\btheta_{\calS}}(\btheta_{\calS}-\btheta_{0,\calS})\bigg|_2^2+
\frac{2}{n}\sum_{t=1}^{n}\bigg|\frac{\partial\bfg_{t,\mT}(\tilde{\btheta})}{\partial\btheta_{\calS^{\rm c}}}\btheta_{\calS^{\rm c}}\bigg|_2^2 \\
&\leq 
2|\btheta_{\calS}-\btheta_{0,\calS}|_1^2\max_{k_1,k_2\in[s]}\bigg|\frac{1}{n}\sum_{t=1}^{n}\bigg\{\frac{\partial\bfg_{t,\mT}(\tilde{\btheta})}{\partial\theta_{k_1}}\bigg\}^\T\frac{\partial\bfg_{t,\mT}(\tilde{\btheta})}{\partial\theta_{k_2}}\bigg|\\&~~~+
2|\btheta_{\calS^{\rm c}}|_1^2\max_{s+1\leq k_1,k_2\leq p}\bigg|\frac{1}{n}\sum_{t=1}^{n}\bigg\{\frac{\partial\bfg_{t,\mT}(\tilde{\btheta})}{\partial\theta_{k_1}}\bigg\}^\T\frac{\partial\bfg_{t,\mT}(\tilde{\btheta})}{\partial\theta_{k_2}}\bigg|
\end{align*} 
for some $\tilde{\btheta}$ lying on the jointing line between $\bthetazero$ and $\btheta$. By Condition \ref{con:moments2}(a), 
\begin{align*}
\sup_{\btheta\in\bTheta}\max_{k_1,k_2\in[p]}\bigg|\frac{1}{n}\sum_{t=1}^{n}\bigg\{\frac{\partial\bfg_{t,\mT}(\btheta)}{\partial\theta_{k_1}}\bigg\}^\T\frac{\partial\bfg_{t,\mT}(\btheta)}{\partial\theta_{k_2}}\bigg| &\leq \sum_{j\in \mathcal{T}}\sup_{\btheta\in\bTheta}\max_{k\in [p]}\bigg\{\frac{1}{n}\sum_{t=1}^{n}\bigg|\frac{\partial g_{t,j}(\btheta)}{\partial\theta_k}\bigg|^2\bigg\} \\ &\leq 
|\mathcal{T}|\sup_{\btheta\in\bTheta}\max_{j\in[r]} \max_{k\in[p] }\bigg\{\frac{1}{n}\sum_{t=1}^{n}\bigg|\frac{\partial g_{t,j}(\btheta)}{\partial\theta_k}\bigg|^2\bigg\}= 
O_{\p}(\ell_n)\,.
\end{align*}
Therefore,
\begin{align*}
\sup_{\btheta\in\bTheta_n}\bigg\{\frac{1}{n}\sum_{t=1}^{n}|\bfg_{t,\mT}(\btheta)-\bfg_{t,\mT}(\bthetazero)|_2^2\bigg\}=O_{\p}(s^2\ell_nb_n)+O_{\p}(\ell_n\aleph_n^2)\,.
\end{align*}
Since $\ell_n\aleph_n=o(1)$, by  Condition \ref{con:moments1}(c) and \eqref{hatV-barVtheta0}, we know  $\sup_{\mathcal{T}\in\mathscr{T}}\lambda_{\max}\{\widehat{\bfV}_{\mT}(\btheta_0)\}\leq C$ w.p.a.1.
Therefore, due to $s^2\ell_nb_n=o(1)$, \eqref{eq:hatVt} implies that 
\begin{align*}
\sup_{\btheta\in\bTheta_n}\sup_{\mathcal{T} \in \mathscr{T}}\|\widehat{\bfV}_{\mT} (\btheta) - \widehat{\bfV}_{\mT} (\bthetazero)\|_{2} =O_{\p}\{s(\ell_nb_n)^{1/2}\}+O_{\p}(\ell_n^{1/2}\aleph_n)\,.
\end{align*}
Together with \eqref{hatV-barVtheta0}, we have
\begin{align*}
\sup_{\btheta\in\bTheta_n}\sup_{\mathcal{T} \in \mathscr{T}}	\| \widehat{\bfV}_{\mT} (\btheta) - \mathbb{E}\{\widehat{\bfV}_{\mT}(\btheta_0)\}\|_{2} = O_{\p}\{s(\ell_nb_n)^{1/2}\}+O_{\p}(\ell_n\aleph_n)\,.
\end{align*}
  We complete the proof of Lemma \ref{hatV-barV}.
$\hfill\Box$

\subsection{Proof of Lemma \ref{lambdahat}}
Recall that
\begin{align*}
f(\blambda;\btheta)= \frac{1}{n}\sum_{t=1}^{n}\log\{1+\blambda^{\T} \bfg_t(\btheta)\} -\sum_{j=1}^{r}P_{2,\nu}(|\lambda_j|)\,.
\end{align*}
To simplify the notation, we write $\mathcal{M}_{\btheta_n}(c)$ as $\mathcal{M}_{\btheta_n}$.
Due to the convexity of $P_{2,\nu}(\cdot)$, we know that $f(\blambda;\btheta_n)$ is a concave 
function w.r.t $\blambda$. We only need to show there exists a sparse local maximizer $\hat{\blambda}(\btheta_n)$ satisfying the three results stated in Lemma \ref{lambdahat}. By the definition of $\mathcal{M}_{\btheta_n}$ and $\mathcal{M}_{\btheta_n}^*$, we have $\mathcal{M}_{\btheta_n}\subset \mathcal{M}_{\btheta_n}^*$, which implies $|\mathcal{M}_{\btheta_n}|\leq m_n$ w.p.a.1. Notice that $m_n^{1/2}u_nn^{1/\gamma}=o(1)$. Given $\mathcal{M}_{\btheta_n}$, we select $\delta_n$ satisfying 
$u_n\ll\delta_n\ll m_n^{-1/2}n^{-1/\gamma}$. Let $\bar{\blambda}_n=\arg\max_{\blambda\in\Lambda_n}f(\blambda;\btheta_n)$ where $\Lambda_n=\{\blambda=(\blambda_{\calM_{\btheta_n}}^\T,\blambda_{\calM_{\btheta_n}^{\rm c}}^\T)^\T\in\mathbb{R}^r:|\blambda_{\calM_{\btheta_n}}|_2\leq\delta_n,\blambda_{\calM_{\btheta_n}^{\rm c}}=\bf0\}$. Due to $\max_{j\in[r]}n^{-1}\sum_{t=1}^{n}|g_{t,j}(\btheta_n)|^{\gamma}=O_{\p}(1)$, we have $\max_{t\in[n],j\in[r]}|g_{t,j}(\btheta_n)|=O_{\p}(n^{1/\gamma})$, which implies $\max_{t\in[n]}|\bfg_{t,\calM_{\btheta_n}}(\btheta_n)|_2=O_{\p}(m_n^{1/2}n^{1/\gamma})$. Therefore, $\max_{t\in[n]}|\bar{\blambda}_n^\T\bfg_t(\btheta_n)|=o_{\p}(1)$. By Taylor expansion, it holds that 
\begin{align}\label{eq:tyexp1}
0=f({\bf0};\btheta_{n})\leq&~ f(\bar{\blambda}_n;\btheta_n)\notag\\\leq &~
\frac{1}{n}\sum_{t=1}^{n}\bar{\blambda}_n^\T\bfg_t(\btheta_n)-\frac{1}{2n}\sum_{t=1}^{n}\frac{\bar{\blambda}_n^\T\bfg_t(\btheta_n)^{\otimes2}\bar{\blambda}_n}{\{1+\bar{c}\bar{\blambda}_n^\T\bfg_t(\btheta_n)\}^2}-\sum_{j=1}^{r}P_{2,\nu}(|\bar{\lambda}_{n,j}|)\,,
\end{align}
where $\bar{\blambda}_n=(\bar{\lambda}_{n,1},\dots,\bar{\lambda}_{n,r})^\T$ and $\bar{c}\in (0,1)$. Recall $P_{2,\nu}(t)=\nu\rho_2(t;\nu)$. By the convexity of $P_{2,\nu}(\cdot)$, we have $\rho_2'(t;\nu)\geq \rho_2'(0^+)$ for any $t>0$. Notice that $\lambda_{\min}\{\widehat{\bfV}_{\calM_{{\btheta}_n}}(\btheta_n)\}$ is uniformly bounded away from zero w.p.a.1, and $|\bar{\lambda}_{n,j}|\geq \bar{\lambda}_{n,j}$ \rm{\sgn}$\{\bar{g}_j(\btheta_n)\}$. Thus, $\eqref{eq:tyexp1}$ leads to 
\begin{align*}
0\leq \bar{\blambda}_{n,\calM_{\btheta_n}}^\T[\bar{\bfg}_{\calM_{{\btheta}_n}}({\btheta}_n)-\nu\rho_2'(0^+)\sgn\{\bar{\bfg}_{\calM_{{\btheta}_n}}({\btheta}_n)\}]-C|{\bar{\blambda}}_{n,\calM_{\btheta_n}}|_2^2\{1+o_{\p}(1)\}\,.
\end{align*}
Due to $|\bar{\bfg}_{\calM_{{\btheta}_n}}({\btheta}_n)-\nu\rho_2'(0^+)\sgn\{\bar{\bfg}_{\calM_{{\btheta}_n}}({\btheta}_n)\}|_2=O_{\p}(u_n)$, then $|\bar{\blambda}_{n,\calM_{\btheta_n}}|_2=O_{\p}(u_n)=o_{\p}(\delta_n)$.
Write $\bar{\blambda}_{n,\calM_{\btheta_n}}=(\bar{\lambda}_1,\dots,\bar{\lambda}_{|\calM_{\btheta_n}|})^\T$. It holds  w.p.a.1 that
\begin{align}\label{hateta}
{\bf0}=\frac{1}{n}\sum_{t=1}^{n}\frac{\bfg_{t,\calM_{\btheta_n}}(\btheta_n)}{1+\bar{\blambda}_{n,\calM_{\btheta_n}}^\T\bfg_{t,\calM_{\btheta_n}}(\btheta_n)}-\hat{\bfeta}\,,
\end{align}
where $\hat{\bfeta}=(\hat{\eta}_1,\dots,\hat{\eta}_{|\calM_{\btheta_n}|})^\T$ with $\hat{\eta}_j=\nu\rho_2'(|\bar{\lambda}_j|;\nu)\sgn(\bar{\lambda}_j)$ for $\bar{\lambda}_j\neq 0$ and $\hat{\eta}_j\in[-\nu\rho_2'(0^+),\nu\rho_2'(0^+)]$ for $\bar{\lambda}_j=0$. It follows from \eqref{hateta} that $\hat{\bfeta}=\bar{\bfg}_{\calM_{\btheta_n}}(\btheta_n)+\bfR$ with 
\begin{align*}
|\bfR|_{\infty}^2=&~\bigg|\frac{1}{n}\sum_{t=1}^{n}\frac{\bar{\blambda}_{n,\calM_{\btheta_n}}^\T\bfg_{t,\calM_{\btheta_n}}(\btheta_n)\bfg_{t,\calM_{\btheta_n}}(\btheta_n)}{1+\bar{\blambda}_{n,\calM_{\btheta_n}}^\T\bfg_{t,\calM_{\btheta_n}}(\btheta_n)}\bigg|_\infty^2\\
\leq &~
\max_{j\in\mm_{\btheta_n}}\bigg\{\frac{1}{n}\sum_{t=1}^{n}|\bar{\blambda}_{n,\calM_{\btheta_n}}^\T\bfg_{t,\calM_{\btheta_n}}(\btheta_n)||g_{t,j}(\btheta_n)|\bigg\}^2\cdot \{1+o_{\p}(1)\} \\\leq &~
\big\{\bar{\blambda}_{n,\calM_{\btheta_n}}^\T\widehat{\bfV}_{\calM_{\btheta_n}}(\btheta_n)\bar{\blambda}_{n,\calM_{\btheta_n}}\big\}\max_{j\in\mm_{\btheta_n}}\bigg\{\frac{1}{n}\sum_{t=1}^{n}|g_{t,j}(\btheta_n)|^2\bigg\}\cdot \{1+o_{\p}(1)\}
\\= &~
O_{\p}(|\bar{\blambda}_{n,\calM_{\btheta_n}}|_2^2)\,,
\end{align*}
which indicates that $|\bfR|_\infty=O_{\p}(u_n)=o_{\p}(\nu)$. Hence, we  have w.p.a.1  that $\sgn(\bar{\lambda}_{n,j})=\sgn\{\bar{g}_{j}(\btheta_n)\}$ for any $j\in \mathcal{M}_{\btheta_n}$ with $\bar{\lambda}_{n,j}\neq 0$. In the sequel, we will show $\bar{\blambda}_n$ is a local maximizer of $f(\blambda;\btheta_n)$ w.p.a.1, which includes two steps.

\underline{\it Step 1.} Define $\Lambda_n^*=\{
\blambda=(\blambda_{\calM_{\btheta_n}^*}^\T,\blambda_{\calM_{\btheta_n}^{*,{\rm c}}}^\T)^\T\in\mathbb{R}^r:|\blambda_{\calM_{\btheta_n}^*}|_2\leq \epsilon,\blambda_{\calM_{\btheta_n}^{*,{\rm c}}}=\bzero\}$ for some sufficiently small $\epsilon>0$. For $\bar{\blambda}_n$ defined before, we will show that $\bar{\blambda}_n=\arg\max_{\blambda\in \Lambda_n^*}f(\blambda;\btheta_n)$ w.p.a.1. Due to $\bar{\blambda}_n\in\Lambda_n$ and $\mathcal{M}_{\btheta_n}\subset \mathcal{M}_{\btheta_n}^*$, we know $\bar{\blambda}_n\in\Lambda_n^*$ for a sufficiently large $n$.
Restricted on $\blambda\in\Lambda_n^*$, by the concavity of $f(\blambda;\btheta_n)$ w.r.t $\blambda_{\calM_{\btheta_n}^*}$, it suffices
to show that w.p.a.1 for any $j\in\mathcal{M}_{\btheta_n}^*$ it holds that 
\begin{align}\label{partial f}
\frac{\partial f(\bar{\blambda}_n;\btheta_n)}{\partial \lambda_j}=0\,.
\end{align}
Due to $\bar{\blambda}_n\in\Lambda_n$ and $|\bar{\blambda}_n|_2=o_{\p}(\delta_n)$, then $\bar{\blambda}_{n,\calM_{\btheta_n}}$ is an interior point of the set 
$\{\blambda_{\calM_{\btheta_n}}\in 
\mathbb{R}^{|\mathcal{M}_{\btheta_n}|}:
|\blambda_{\calM_{\btheta_n}}|_2\leq \delta_n\}$. Restricted on $\blambda\in \Lambda_n$, we know $f(\blambda;\btheta_n)$ is  concave w.r.t $\blambda_{\calM_{\btheta_n}}$. Notice that $\bar{\blambda}_n=\arg \max_{\blambda\in\Lambda_n}f(\blambda;
\btheta_n)$. Therefore, $\eqref{partial f}$ holds for any $ j\in\mathcal{M}_{\btheta_n}$. Recall $\bar{\blambda}_n=(\bar{\lambda}_{n,1},\dots,\bar{\lambda}_{n,r})^\T$. For any $j\in \mathcal{M}_{\btheta_n}^*\backslash\mathcal{M}_{\btheta_n}$, we have $\bar{\lambda}_{n,j}=0$ and 
$$\frac{1}{n}\sum_{t=1}^{n}\frac{g_{t,j}(\btheta_n)}{1+\bar{\blambda}_{n,\calM_{\btheta_n}^*}^{\T}\bfg_{t,\calM_{\btheta_n}^*}(\btheta_n)}=\bar{g}_j(\btheta_n)+O_{\p}(u_n)\,,$$
where the term $O_{\p}(u_n)=o_{\p}(\nu)$ is uniform over $j\in \mathcal{M}_{\btheta_n}^*\backslash\mathcal{M}_{\btheta_n}$. Such a conclusion can be obtained by the same arguments for deriving the convergence rate of $|\bfR|_\infty$ stated above. Notice that for any $j\in \mathcal{M}_{\btheta_n}^*\backslash \mathcal{M}_{\btheta_n}$, we have $C_*\nu\rho_2'(0^+)\leq |\bar{g}_{j}(\btheta_n)|<c\nu\rho_2'(0^+)$ with some constant $c<1$. Hence, we have w.p.a.1 that
\begin{align*}
\max_{j\in \mathcal{M}_{\btheta_n}^*\backslash\mathcal{M}_{\btheta_n}}\bigg|\frac{1}{n}\sum_{t=1}^{n}\frac{g_{t,j}(\btheta_n)}{1+\bar{\blambda}_{n,\calM_{\btheta_n}^*}^{\T}\bfg_{t,\calM_{\btheta_n}^*}(\btheta_n)}\bigg|<
\nu\rho_2'(0^+)\,,
\end{align*}
which implies that there exists some $\hat{\eta}_j^*\in[-\nu\rho_2'(0^+),\nu\rho_2'(0^+)]$ such that 
\begin{align*}
0=\frac{1}{n}\sum_{t=1}^{n}\frac{g_{t,j}(\btheta_n)}{1+\bar{\blambda}_{n,\calM_{\btheta_n}^*}^{\T}\bfg_{t,\calM_{\btheta_n}^*}(\btheta_n)}-\hat{\eta}_j^*\,.
\end{align*}
Hence, \eqref{partial f} holds for any $j\in \mathcal{M}_{\btheta_n}^*\backslash \mathcal{M}_{\btheta_n}$. Then we have $\bar{\blambda}_n=\arg\max_{\blambda\in\Lambda_n^*}f(\blambda;\btheta_n)$ w.p.a.1.

\underline{\it Step 2.} Define $\tilde{\Lambda}_n=\{\blambda=(\blambda_{\calM_{\btheta_n}^*}^\T,\blambda_{\calM_{\btheta_n}^{*,{\rm c}}}^\T)^\T\in\mathbb{R}^r:|\blambda_{\calM_{\btheta_n}^*}-\bar{\blambda}_{n,\calM_{\btheta_n}^*}|_2\leq o(u_n),|\blambda_{\calM_{\btheta_n}^{*,{\rm c}}}|_1\leq O(m_n^{1/2}u_n)\}$. We will prove $\bar{\blambda}_n=\arg\max_{\blambda\in\tilde{\Lambda}_n}f(\blambda;\btheta_n)$ w.p.a.1. Due to $m_n^{1/2}u_nn^{1/\gamma}=o(1)$ and $\max_{t\in[n],j\in[r]}|g_{t,j}(\btheta_n)|=O_{\p}(n^{1/\gamma})$, then  $\max_{t\in[n],\blambda\in\tilde{\Lambda}_n(\btheta_n)}|\blambda^\T\bfg_{t}(\btheta_n)|=o_{\p}(1)$. For any $\blambda=(\lambda_1,\dots,\lambda_r)^\T\in \tilde{\Lambda}_n$, denote by $\tilde{\blambda}=(\blambda_{\calM_{\btheta_n}^*}^\T,{\bf0}^\T)^\T$ the projection of $\blambda=(\blambda_{\calM_{\btheta_n}^*}^\T,\blambda_{\calM_{\btheta_n}^{*,{\rm c}}}^\T)^\T$ onto $\Lambda_n^*$. Then it holds that
\begin{align*}
\sup_{\blambda\in \tilde{\Lambda}_n}\{f(\blambda;\btheta_n)-f(\tilde{\blambda};\btheta_n)\}=\sup_{\blambda\in \tilde{\Lambda}_n}\bigg\{\frac{1}{n}\sum_{t=1}^{n}\frac{\bfg_{t}(\btheta_n)^\T(\blambda-\tilde{\blambda})}{1+\blambda_*^\T\bfg_{t}(\btheta_n)}-\sum_{j\in\mathcal{M}_{\btheta_n}^{*,{\rm c}}}P_{2,\nu}(|\lambda_j|)\bigg\}\,,
\end{align*}
where $\blambda_*$ lies between $\blambda$ and $\tilde{\blambda}$. By  Taylor expansion, due to $\sum_{j\in\mathcal{M}_{\btheta_n}^{*,{\rm c}}}P_{2,\nu}(|\lambda_j|)\geq \nu\rho_2'(0^+)|\blambda_{\calM_{\btheta_n}^{*,{\rm c}}}|_1$ and $\sup_{\blambda\in\tilde{\Lambda}_n}|\blambda|_1=O_{\p}(m_n^{1/2}u_n)$,
we have 
\begin{align*}
&~\sup_{\blambda\in \tilde{\Lambda}_n}\bigg\{\frac{1}{n}\sum_{t=1}^{n}\frac{\bfg_{t}(\btheta_n)^\T(\blambda-\tilde{\blambda})}{1+\blambda_*^\T\bfg_{t}(\btheta_n)}-\sum_{j\in\mathcal{M}_{\btheta_n}^{*,{\rm c}}}P_{2,\nu}(|\lambda_j|)\bigg\}\\\leq&{~
\sup_{\blambda\in \tilde{\Lambda}_n}\bigg[\blambda_{\calM_{\btheta_n}^{*,{\rm c}}}^\T\bar{\bfg}_{\calM_{\btheta_n}^{*,{\rm c}}}(\btheta_n)+\bigg\{\frac{1}{n}\sum_{t=1}^{n}|\blambda_*^\T\bfg_{t}(\btheta_n){\bfg}_{t,\calM_{\btheta_n}^{*,{\rm c}}}(\btheta_n)^\T\blambda_{\calM_{\btheta_n}^{*,{\rm c}}}|\bigg\}\{1+o_{\p}(1)\}-\sum_{j\in\mathcal{M}_{\btheta_n}^{*,{\rm c}}}P_{2,\nu}(|\lambda_j|)\bigg]}
\\\leq&~
\sup_{\blambda\in \tilde{\Lambda}_n}\bigg[|\bar{\bfg}_{\calM_{\btheta_n}^{*,{\rm c}}}(\btheta_n)|_\infty|\blambda_{\calM_{\btheta_n}^{*,{\rm c}}}|_1+\bigg\{\frac{1}{n}\sum_{t=1}^{n}\sum_{j=1}^{r}\sum_{k\in\mathcal{M}_{\btheta_n}^{*,{\rm c}}}|\lambda_{*,j}g_{t,j}(\btheta_n)\lambda_kg_{t,k}(\btheta_n)|\bigg\}\{1+o_{\p}(1)\}\\&~~~~~~~~~~~~~~~~~~~~~~~~~~~~~~~~~~-\nu\rho_2'(0^+)|\blambda_{\calM_{\btheta_n}^{*,{\rm c}}}|_1\bigg]
\\\leq&~
\sup_{\blambda\in \tilde{\Lambda}_n}\bigg[C_*\nu\rho_2'(0^+)|\blambda_{\calM_{\btheta_n}^{*,{\rm c}}}|_1+\max_{j\in[r]}\bigg\{\frac{1}{n}\sum_{t=1}^{n}|g_{t,j}(\btheta_n)|^2\bigg\}|\blambda_{\calM_{\btheta_n}^{*,{\rm c}}}|_1|
\blambda_*|_1\{1+o_{\p}(1)\}
-
\nu\rho_2'(0^+)|\blambda_{\calM_{\btheta_n}^{*,{\rm c}}}|_1\bigg]\\\leq&~
\{-(1-C_*)\nu\rho_2'(0^+)+O_{\p}(m_n^{1/2}u_n)\}|\blambda_{\calM_{\btheta_n}^{*,{\rm c}}}|_1\,,
\end{align*}
where $\blambda_*=(\lambda_{*,1},\ldots,\lambda_{*,r})^{\T}$, and 
the term $O_{\p}(m_n^{1/2}u_n)$ holds uniformly over $\blambda\in\tilde{\Lambda}_n$. Since   $m_n^{1/2}u_n=o(\nu)$ and $C_*<1$, then 
$\sup_{\blambda\in \tilde{\Lambda}_n}\{f(\blambda;\btheta_n)-f(\tilde{\blambda};\btheta_n)\}<0$ w.p.a.1, which implies $\bar{\blambda}_n$ is a local maximizer of $f(\blambda;\btheta_n)$ w.p.a.1. We complete the proof of Lemma \ref{lambdahat}.
$\hfill\Box$

\subsection{Proof of Lemma \ref{lambda0}}\label{sec:la:lambda0}
Due to $f(\blambda;\bthetazero)$ is concave w.r.t $\blambda$, we only need to show that there exists a local maximizer $\hat{\blambda}(\bthetazero)$ satisfying the results in the lemma.
To simplify notation, we write $\mathcal{M}_{\bthetazero}(c)$ as $\mathcal{M}_{\bthetazero}$. 

\subsubsection{Case 1: $\mathcal{M}_{\bthetazero}\neq \emptyset$ }
\medskip

 Recall $\mathcal{M}_{\btheta_0}^*=\{j\in[r]: |\bar {g}_{j}(\btheta_0)|\geq   C_*\nu \rho_2 '(0^{+})\}$. By  Lemma \ref{tailprabog0}, if $L_n^{\varphi}\log (rn)\ll n^{\varphi/(3\varphi+1)}$, then   $|\bar{\bfg}(\btheta_{0})|_{\infty}=O_{\p}(\aleph_n)$.
Since $\nu\gg\aleph_n$, we know $\mathbb{P}(\mathcal{M}_{\btheta_0}^*=\emptyset)\to 1$, which implies  $|\mathcal{M}_{\btheta_0}^*|\leq \ell_n$ w.p.a.1. Due to $\ell_n\aleph_n\ll n^{-1/\gamma}$,
we can select 
$\delta_n$ satisfying  
$\ell_n^{1/2}\aleph_n\ll\delta_n\ll\ell_n^{-1/2}n^{-1/\gamma}$. Recall $\mathcal{M}_{\btheta_0}=\{j\in[r] : |\bar {g}_{j}(\btheta_0)|\geq   c\nu \rho_2 '(0^{+})\}$ for some constant $c\in(C_*,1)$. Then $|\mathcal{M}_{\btheta_0}|\leq |\mathcal{M}_{\btheta_0}^*|\leq \ell_n$ w.p.a.1. Let $\Lambda_0=\{\blambda=(\blambda^{\T}_{\mm_{\btheta_{0}}},\blambda^{\T}_{\mm^{\rm c}_{\btheta_{0}}})^{\T}\in \mathbb{R}^r:|\blambda_{\calM_{\btheta_0}}|_2\leq \delta_n,\blambda_{\calM^{\rm c}_{\btheta_0}}={\bf0}\}$ and $\bar{\blambda}_0=\arg\max_{\blambda\in\Lambda_0}f(\blambda;\btheta_{0})$. Due to $\max_{j\in[r],t\in[n]}|g_{t,j}(\btheta_{0})|=O_{\p}(n^{1/\gamma})$, then  $\max_{t\in[n]}|\bar{\blambda}_0^{\T}\bfg_t(\btheta_{0})|=o_{\p}(1)$. Similar to (\ref{eq:tyexp1}), we have
\begin{equation}\label{eq:lamd_0}
\begin{split}
0=f({\bf0};\btheta_{0})\leq&~ f(\bar{\blambda}_0;\btheta_0)\\\leq &~
\frac{1}{n}\sum_{t=1}^{n}\bar{\blambda}_0^\T\bfg_t(\btheta_0)-\frac{1}{2n}\sum_{t=1}^{n}\frac{\bar{\blambda}_0^\T\bfg_t(\btheta_0)^{\otimes2}\bar{\blambda}_0}{\{1+\check{c}\bar{\blambda}_0^\T\bfg_t(\btheta_{0})\}^2}-\sum_{j=1}^{r}P_{2,\nu}(|\bar{\lambda}_{0,j}|)
\end{split}
\end{equation}
for some $\check{c}\in (0,1)$. By \eqref{hatV-barVtheta0} and Condition \ref{con:moments1}(c), if  $L_n^{\varphi}\log (rn)\ll n^{\varphi/(3\varphi+1)}$ and $\ell_n\aleph_n=o(1)$, we have    $\lambda_{\min}\{\widehat{\bfV}_{\mm_{\btheta_{0}}}(\btheta_{0})\}$ is uniformly bounded away from zero w.p.a.1. Therefore, (\ref{eq:lamd_0}) leads to
\begin{align*}
0&\leq  \bar{\blambda}_{0,\calM_{\btheta_{0}}}^\T\bar{\bfg}_{\calM_{\btheta_{0}}}(\btheta_{0})-\frac{1}{2}\lambda_{\min}\{\widehat{\bfV}_{\calM_{\btheta_{0}}}(\btheta_{0})\}|\bar{\blambda}_{0,\calM_{\btheta_{0}}}|_2^2\{1+o_{\p}(1)\}\\
&\leq 
\bar{\blambda}_{0,\calM_{\btheta_{0}}}^\T\bar{\bfg}_{\calM_{\btheta_{0}}}(\btheta_{0})-C|\bar{\blambda}_{0,\calM_{\btheta_{0}}}|_2^2\{1+o_{\p}(1)\}\,.
\end{align*}
Due to $|\bar{\bfg}(\btheta_{0})|_\infty=O_{\p}(\aleph_n)$, then 
$
|\bar{\bfg}_{\calM_{\btheta_{0}}}(\btheta_{0})|_2=O_{\p}(\ell_n^{1/2}\aleph_n)$,
which implies $|\bar{\blambda}_{0,\calM_{\btheta_{0}}}|_2=O_{\p}(\ell_n^{1/2}\aleph_n)=o_{\p}(\delta_n)$. Due to the same arguments as the proof of $\bar{\blambda}_n$ being a local maximizer of $f(\blambda;\btheta_n)$ w.p.a.1 in Lemma \ref{lambdahat}, we can also show  $\bar{\blambda}_0$ defined this way is a local maximizer of $f(\blambda;\btheta_{0})$ w.p.a.1. 
The concavity of $f(\blambda;\bthetazero)$  w.r.t $\blambda$ yields $\hat{\blambda}(\bthetazero)=\bar{\blambda}_0$ and $\supp\{\hat{\blambda}(\bthetazero)\}\subset \mathcal{M}_{\btheta_0}$ w.p.a.1. $\hfill\Box$

\subsubsection{Case 2: $\mathcal{M}_{\bthetazero}=\emptyset$ }
\medskip

In this case, we will show $\bzero \in \mathbb{R}^r$ is a local maximizer for $f(\blambda;\btheta_0)$ w.p.a.1. Due to the concavity of $f_n(\blambda;\btheta_0)$ w.r.t $\blambda$, we then have $\hat{\blambda}(\bthetazero) = \bzero$ w.p.a.1, which implies ${\rm supp}\{\hat{\blambda}(\bthetazero)\} \subset \mathcal{M}_{\bthetazero}$ w.p.a.1. Let $j_0 = \arg\max_{j\in[r]}\mathbb{E}[\mathbb{E}_n\{g^2_{t,j}(\bthetazero)\}]$. By Conditions \ref{con:mixingdecay}, \ref{con:moments1}(a) and 
Davydov's inequality, we have
\begin{align*}
    \mathbb{E}\{|\bar{g}_{j_0}(\btheta_0)|^2\}\leq&~ \frac{1}{n^2}\sum_{t_1=1}^{n}\sum_{t_2=1}^{n}|{\rm Cov}\{g_{t_1,j_0}(\btheta_0),g_{t_2,j_0}(\btheta_0)\}|
    \\ \lesssim &~\frac{1}{n^2}\sum_{t=1}^{n}\mathbb{E}\{g^2_{t,j_0}(\btheta_0)\}+\frac{1}{n^2}\sum_{t_1=1}^{n-1}\sum_{t_2=t_1+1}^{n}\exp\{-CL_{n}^{-\varphi}|t_1-t_2|^{\varphi}\}
    \\ \lesssim &~L_nn^{-1}\,,
\end{align*}
which implies $|\bar{g}_{j_0}(\btheta_0)|=O_{\p}(L_n^{1/2}n^{-1/2})$. Analogously, we can also show $|\mathbb{E}_n[\{g_{t,j_0}^2(\btheta_0)\} - \mathbb{E}\{g_{t,j_0}^2(\btheta_0)\}]|= O_{\p}(L_n^{1/2}n^{-1/2})$. By Lemma \ref{tailprabog0}, we have $|\bar{\bg}(\btheta_0)|_{\infty}=O_{\p}(\aleph_n)$ provided that $L_n^{\varphi}\log (rn)\ll n^{\varphi/(3\varphi+1)}$. Let $\breve{\blambda}_0 = \arg\max_{\blambda \in \breve{\Lambda}_0} f_n(\blambda;\btheta_0)$ with $\breve{\Lambda}_0 =\{ \blambda\in\mathbb{R}^r: |\lambda_{j_0}|\leq (\log n)^{-1}n^{-1/\gamma}, \blambda_{[r]\setminus \{j_0\}}=\bzero \}$, and write $\breve{\blambda}_0 =(\breve{\lambda}_{0,1},\ldots,\breve{\lambda}_{0,r})^{\T}$. 

The rest of the proof closely resembles Section J.2.2 in the supplementary material of \cite{bel2024} but with replacing $\alpha_n$ there by $\aleph_n$ and using the convergence rates of $|\mathbb{E}_n[\{g_{t,j_0}^2(\btheta_0)\} - \mathbb{E}\{g_{t,j_0}^2(\btheta_0)\}]|$ and $|\bar{g}_{j_0}(\btheta_0)|$ derived above. Hence, we can obtain $|\breve{\lambda}_{0,j_0}|=O_{\p}(L_n^{1/2}n^{-1/2})$, which causes an additional restriction 
$L_n(\log n)^2\ll n^{1-2/\gamma}$ from the requirement $|\breve{\lambda}_{0,j_0}|=o_{\p}\{(\log n)^{-1}n^{-1/\gamma}\}$. Due to $L_n^{\varphi}\log (rn)\ll n^{\varphi/(3\varphi+1)}$, we know $L_n\ll n^{1/(3\varphi+1)}$. Since $\varphi\geq 1$ and $\gamma>4$, then $n^{1/(3\varphi+1)}\ll n^{1-2/\gamma}(\log n)^{-2}$, which implies $L_n(\log n)^2\ll n^{1-2/\gamma}$ holds
automatically. For $\breve{\eta}_{j_0}$ specified in Section J.2.2 in the supplementary material of \cite{bel2024}, we have $|\breve{\eta}_{j_0}|=O_{\p}(L_n^{1/2}n^{-1/2})$ under current setting. Based on the requirement $|\breve{\eta}_{j_0}|=o_{\p}(\nu)$, we will have another additional restriction $L_nn^{-1}\ll \nu^2$. Recall  $\aleph_n=n^{-3\varphi/(6\varphi+2)}(\log r)^{1/2}$. Due to $L_n^{\varphi}\log (rn)\ll n^{\varphi/(3\varphi+1)}$ and $\ell_n\aleph_n\ll \nu$, we know $L_nn^{-1}\ll\nu^2$ holds automatically. 
$\hfill\Box$

\subsection{Proof of Lemma \ref{gammahatL2norm}}\label{sec:la:gammahatL2norm}
Recall $\widehat{\bGamma}_{\mT}(\btheta)=\nabla_{\btheta_{\calS}}\bar{\bfg}_{\mT}(\btheta)$ for any $\btheta\in\bTheta$.
By the triangle inequality, we have
\begin{align*}
\big|[\widehat{\bGamma}_{\mT}(\hat{\btheta}_n)-\mathbb{E}\{\widehat{\bGamma}_{\mT}(\btheta_0)\}]\bfz\big|_2\leq \big|\{\widehat{\bGamma}_{\mT}(\hat{\btheta}_n)-\widehat{\bGamma}_{\mT}(\btheta_0)\}\bfz\big|_2+\big|[\widehat{\bGamma}_{\mT}({\btheta}_0)-\mathbb{E}\{\widehat{\bGamma}_{\mT}(\btheta_0)\}]\bfz\big|_2
\end{align*}
for any $\bfz\in\mathbb{R}^s$. 
Write $\hat{\btheta}_n=(\hat{\theta}_{n,1},\ldots,\hat{\theta}_{n,p})^{\T}$ and ${\btheta}_0=({\theta}_{0,1},\ldots,{\theta}_{0,p})^{\T}$.
By  Taylor expansion, Jensen's inequality and the Cauchy-Schwarz inequality, it holds  w.p.a.1 that
\begin{align*}
&\big|\{\widehat{\bGamma}_{\mT}(\hat{\btheta}_n)-\widehat{\bGamma}_{\mT}(\btheta_0)\}\bfz\big|_2^2
=\sum_{j\in\mathcal{T}}\biggl\{\frac{1}{n}\sum_{t=1}^{n}\sum_{k=1}^{s}z_k\sum_{l=1}^{s}\frac{\partial^2g_{t,j}(\tilde{\btheta})}{\partial\theta_k\partial\theta_l}(\hat{\theta}_{n,l}-\theta_{0,l})\biggr\}^2
\\
&~~~~~~~~~~\leq 
\sum_{j\in\mathcal{T}}\frac{1}{n}\sum_{t=1}^{n}\biggl\{\sum_{k=1}^{s}z_k\sum_{l=1}^{s}\frac{\partial^2g_{t,j}(\tilde{\btheta})}{\partial\theta_k\partial\theta_l}(\hat{\theta}_{n,l}-\theta_{0,l})\biggr\}^2\leq 
\frac{|\bfz|_2^2}{n}\sum_{j\in\mathcal{T}}\sum_{t=1}^{n}\sum_{k=1}^{s}\biggl\{\sum_{l=1}^{s}\frac{\partial^2g_{t,j}(\tilde{\btheta})}{\partial\theta_k\partial\theta_l}(\hat{\theta}_{n,l}-\theta_{0,l})\biggr\}^2
\\
&~~~~~~~~~~\leq 
\frac{|\bfz|_2^2}{n}\sum_{j\in\mathcal{T}}\sum_{t=1}^{n}\sum_{k=1}^{s}\sum_{l=1}^{s}\bigg|\frac{\partial^2g_{t,j}(\tilde{\btheta})}{\partial\theta_k\partial\theta_l}\bigg|^2|\hat{\btheta}_{n,\calS}-\btheta_{0,\calS}|_2^2\,,
\end{align*}
where $\tilde{\btheta}$ lies on the jointing line between $\btheta_{0}$ and $\hat{\btheta}_n$. 
By Theorem $\ref{The.1}$, we have $|\hat{\btheta}_{n,\calS}-\btheta_{0,\calS}|_2=O_{\p}(s^{1/2}b_n^{1/2})$.
Due to $|\mathcal{T}|\leq \ell_n$,
by Condition \ref{con:moments2}(c), we have 
\begin{align*}
\sup_{\mathcal{T}\in\mathscr{T}}\big|\{\widehat{\bGamma}_{\mT}(\hat{\btheta}_n)-\widehat{\bGamma}_{\mT}(\btheta_0)\}\bfz\big|_2=|\bfz|_2\cdot O_{\p}(\ell_n^{1/2}s^{3/2}b_n^{1/2})\,.
\end{align*}
By the Cauchy-Schwarz inequality, it holds that
\begin{align*}
\big|[\widehat{\bGamma}_{\mT}({\btheta}_0)-\mathbb{E}\{\widehat{\bGamma}_{\mT}(\btheta_0)\}]\bfz\big|_2^2
&=
\sum_{j\in\mathcal{T}}\bigg(\frac{1}{n}\sum_{t=1}^{n}\sum_{k=1}^{s}z_k\bigg[\frac{\partial g_{t,j}(\btheta_{0})}{\partial\theta_k}-\mathbb{E}\bigg\{\frac{\partial g_{t,j}(\btheta_{0})}{\partial\theta_k}\bigg\}\bigg]\bigg)^2
\\ &\leq
|\bfz|_2^2\sum_{j\in\mathcal{T}}\sum_{k=1}^{s}\bigg(\frac{1}{n}\sum_{t=1}^{n}\bigg[\frac{\partial g_{t,j}(\btheta_{0})}{\partial\theta_k}-\mathbb{E}\bigg\{\frac{\partial g_{t,j}(\btheta_{0})}{\partial\theta_k}\bigg\}\bigg]\bigg)^2.
\end{align*}
Following the proof of  Lemma \ref{tailprabog0}, by
 Conditions \ref{con:mixingdecay}, \ref{con:moments2}(a) and \ref{con:moments2}(b), we have 
\begin{align*}
\max_{j\in[r]}\max_{k\in\mathcal{S}}\bigg|\frac{1}{n}\sum_{t=1}^{n}\bigg[\frac{\partial g_{t,j}(\btheta_{0})}{\partial\theta_k}-\mathbb{E}\bigg\{\frac{\partial g_{t,j}(\btheta_{0})}{\partial\theta_k}\bigg\}\bigg]\bigg|=O_{\p}(\aleph_n)
\end{align*}
provided that
$L_n^{\varphi}\log (rn)\ll n^{\varphi/(3\varphi+1)}$.
Then
\begin{align*}
\sup_{\mathcal{T}\in\mathscr{T}}\big|[\widehat{\bGamma}_{\mT}({\btheta}_0)-\mathbb{E}\{\widehat{\bGamma}_{\mT}(\btheta_0)\}]\bfz\big|_2=|\bfz|_2\cdot O_{\p}(\ell_n^{1/2}s^{1/2}\aleph_n)\,.
\end{align*}
Hence,
\begin{align*}
\sup_{\mathcal{T}\in\mathscr{T}}\big|[\widehat{\bGamma}_{\mT}(\hat{\btheta}_n)-\mathbb{E}\{\widehat{\bGamma}_{\mT}(\btheta_0)\}]\bfz\big|_2=|\bfz|_2\cdot\{O_{\p}(\ell_n^{1/2}s^{3/2}b_n^{1/2})+O_{\p}(\ell_n^{1/2}s^{1/2}\aleph_n)\}\,.
\end{align*}
We complete the proof of  Lemma \ref{gammahatL2norm}.
$\hfill\Box$

\subsection{Proof of Lemma \ref{cltZ}}\label{sec:la:cltZ}
To prove Lemma \ref{cltZ}, we need Lemma \ref{berry-essen mixng}.
\begin{lemma}[Theorem 1 of \cite{Sunklodas(1984)}]\label{berry-essen mixng}
	Let $\{\xi_t\}_{t=1}^{\tilde{n}} $ be an $\alpha$-mixing sequence of  centered random variables with $\alpha$-mixing coefficients $\{\tilde{\alpha}_{\tilde{n}}(k)\}_{k\geq1}$. Write $S_{\tilde{n}}=\sum_{t=1}^{\tilde{n}}\xi_{t}$ and $\sigma_{\tilde{n}}^2=\mathbb{E}(S_{\tilde{n}}^2)$.  Assume there exist some universal constants $b_0>0$, $b_1>0$ and $b_2>0$
	such that $ \sigma^2_{\tilde{n}}\geq b_0\tilde{n}$, and 
	$\tilde{\alpha}_{\tilde{n}}(k)\leq b_1\exp(-b_2\tilde{L}_{\tilde{n}}^{-\varphi}k)$ for any integer $k\geq 1$,  where $\tilde{L}_{\tilde{n}}$ may diverge with $\tilde{n}$. Let $F_{\tilde{n}}(x)=\mathbb{P}(S_{\tilde{n}}/\sigma_{\tilde{n}}< x)$ and $d_{\tilde{n}}=\max_{t\in[\tilde{n}]}\mathbb{E}(|\xi_t|^3)$.
	For all $\tilde{n}>1$, there exist some universal constants $c_1>0$, $c_2>0$ and $c_3>4$ such that
	\begin{align*}
		\sup_{x\in\mathbb{R}}|F_{\tilde{n}}(x)-\Phi(x)|\leq \frac{c_1}{b_0b_2^2}\frac{d_{\tilde{n}}\tilde{L}_{\tilde{n}}^{2\varphi}\log^2(b_0^{-1/2}\sigma_{\tilde{n}})}{\sigma_{\tilde{n}}}
	\end{align*}
provided that $l_{1,\tilde{n}}\leq b_2\tilde{L}_{\tilde{n}}^{-\varphi}\leq l_{2,\tilde{n}}$ with
$
		l_{1,\tilde{n}}=c_2\log^{c_3}(b_0^{-1/2}\sigma_{\tilde{n}})\tilde{n}^{-1}$ and $ l_{2,\tilde{n}}=12\log(b_0^{-1/2}\sigma_{\tilde{n}})$, where $\Phi(\cdot)$ is the cumulative distribution function of standard normal distribution.
\end{lemma}
For any given  $\mathcal{T}\in\mathscr{T}$, let $Z_{t,\mT}=\bfz^\T\bJ_{\mT}^{-1/2}[\mathbb{E}\{\widehat{\bGamma}_{\mT}(\btheta_0)\}]^{\T}
[\mathbb{E}\{\widehat{\bV}_{\mT}(\btheta_0)\}]^{-1}\bfg_{t,\mT}(\btheta_{0})$
and  $\bar{Z}_\mT=\mathbb{E}_n(Z_{t,\mT})$. Recall $\bXi_{\mT}(\btheta_0)={\rm Var}\{n^{1/2}\bar{\bg}_{\mT}(\btheta_0)\}$	and   ${\bJ}_{\mT}=\{[\mathbb{E}\{\widehat{\bGamma}_{\mT}(\btheta_0)\}]^{\T}[\mathbb{E}\{\widehat{\bV}_{\mT}(\btheta_0)\}]^{-1}\bXi_{\mT}^{1/2}(\btheta_{0})\}^{\otimes2}$. Due to $|\bz|_2=1$, 
then  
\begin{align*}
    \mathbb{E}\{(\sqrt{n}\bar{Z}_{\mT})^2\}&=\bz^{\T}\bJ_{\mT}^{-1/2}[\mathbb{E}\{\widehat{\bGamma}_{\mT}(\btheta_0)\}]^{\T}
[\mathbb{E}\{\widehat{\bV}_{\mT}(\btheta_0)\}]^{-1}\bXi_{\mT}(\btheta_0)[\mathbb{E}\{\widehat{\bV}_{\mT}(\btheta_0)\}]^{-1}\mathbb{E}\{\widehat{\bGamma}_{\mT}(\btheta_0)\}\bJ_{\mT}^{-1/2}\bz\\&=\bz^{\T}\bJ_{\mT}^{-1/2}\bJ_{\mT}\bJ_{\mT}^{-1/2}\bz=1\,.
\end{align*}
By the Cauchy-Schwarz inequality,   Conditions \ref{con:moments1}(c)  and \ref{con:Gamma}, 
\begin{align*}
|Z_{t,\mT}|^2=&~|\bfz^\T\bJ_{\mT}^{-1/2}[\mathbb{E}\{\widehat{\bGamma}_{\mT}(\btheta_0)\}]^{\T}[\mathbb{E}\{\widehat{\bV}_{\mT}(\btheta_0)\}]^{-1}\bfg_{t,\mT}(\btheta_{0})|^2
\\\leq &~
|\bfz|_2^2\cdot\lambda_{\min}^{-1}(\bJ_{\mT})\lambda_{\max}([\mathbb{E}\{\widehat{\bGamma}_{\mT}(\btheta_0)\}]^{\otimes2})\lambda_{\min}^{-2}[\mathbb{E}\{\widehat{\bV}_{\mT}(\btheta_0)\}]\cdot |\bfg_{t,\mT}(\btheta_{0})|_2^2\leq C|\bfg_{t,\mT}(\btheta_{0})|_2^2\,.
\end{align*}
By  Jensen's inequality and Condition \ref{con:moments1}(a), we obtain that 
\begin{align*}
\mathbb{E}\{|\bfg_{t,\mT}(\btheta_{0})|_2^3\}=\mathbb{E}\bigg[\biggl\{\frac{1}{|\mathcal{T}|}\sum_{j\in\mathcal{T}}|g_{t,j}(\btheta_{0})|^{2}\biggr\}^{3/2}\bigg]\cdot |\mathcal{T}|^{3/2}\leq \mathbb{E}\biggl\{\frac{1}{|\mathcal{T}|}\sum_{j\in\mathcal{T}}|g_{t,j}(\btheta_{0})|^{3}\biggr\}\cdot \ell_n^{3/2}
\leq C\ell_n^{3/2}\,,
\end{align*}
which implies 
$
    \max_{t\in[n]}
\mathbb{E}(|Z_{t,\mT}|^3) \leq C\ell_n^{3/2}
$
for a universal positive constant $C$ independent of $\mathcal{T}$.
Applying  Lemma \ref{berry-essen mixng} with $n=\tilde{n}, \tilde{L}_{\tilde{n}}=L_n$, $\xi_t=Z_{t,\mT}$ and $\sigma^2_{\tilde{n}}=n$, it holds that
\begin{align*}
\sup_{x\in\mathbb{R}}\big|\mathbb{P}(\sqrt{n}\bar{Z}_{\mT}\leq x)-{\Phi}(x)\big|\leq Cn^{-1/2}L_{n}^{2\varphi}\ell_n^{3/2}(\log n)^2=o(1)
\end{align*}
provided that $\ell_n^{3}L_{n}^{4\varphi}(\log n)^{4}\ll n$, where $C$ is a universal positive constant independent of $\mathcal{T}$. 
We complete the proof of Lemma \ref{cltZ}.
$\hfill\Box$

\subsection{Proof of Lemma \ref{hVhatL2norm}}\label{sec:la:hVhatL2norm}

To prove Lemma \ref{hVhatL2norm}, we need Lemma \ref{hL2norm}, whose proof is identical to that of Lemma 1  in \cite{Chang2020}. Hence, we omit the proof of Lemma \ref{hL2norm}. 

\begin{lemma}\label{hL2norm}
	Under Conditions {\rm \ref{con:moments2}(a) and \ref{con:alpha}}, it holds that
	\begin{align*}
		\sup_{\btheta\in\bTheta}\max_{k\in[m]}\max_{l\in[p]}\frac{1}{n}\sum_{t=1}^{n}\bigg|\frac{\partial f_{t,k}^{\bfA_n} (\btheta)}{\partial\theta_l}\bigg|^2=O_{\p}(1)\,.
	\end{align*} 
	Meanwhile, under Conditions {\rm \ref{con:moments2}(c)  and \ref{con:alpha}}, it holds that
	\begin{align*}
		\sup_{\btheta\in\bTheta}\max_{k\in[m]}\max_{l_1,l_2\in[p]}\frac{1}{n}\sum_{t=1}^{n}\bigg|\frac{\partial^2f_{t,k}^{\bfA_n}(\btheta)}{\partial\theta_{l_1}\partial\theta_{l_2}}\bigg|=O_{\p}(1)\,.
	\end{align*}
\end{lemma}
By the triangle inequality, 
\begin{align}\label{eq:Vf}
\|\widehat{\bfV}_{\bff^{\bfA_n}}(\btheta_{0,\calM},\hat{\btheta}_{n,\calM^{\rm c}})-\mathbb{E}\{\widehat{\bfV}_{\bff^{\bfA}}(\btheta_0)\}\|_2&\leq \underbrace{\|\widehat{\bfV}_{\bff^{\bfA}}(\btheta_0)-\mathbb{E}\{\widehat{\bfV}_{\bff^{\bfA}}(\btheta_0)\}\|_2}_{\rm I}+\underbrace{\|\widehat{\bfV}_{\bff^{\bfA_n}}(\btheta_0)-\widehat{\bfV}_{\bff^{\bfA}}(\btheta_0)\|_2}_{ \rm II}\notag\\&~~~~+\underbrace{\|\widehat{\bfV}_{\bff^{\bfA_n}}(\btheta_{0,\calM},\hat{\btheta}_{n,\calM^{\rm c}})-\widehat{\bfV}_{\bff^{\bfA_n}}(\btheta_0)\|_2}_{ \rm III}
\,. 
\end{align}
As we will show in Sections \ref{sec:I}--\ref{sec:III}, 
\begin{align*}
  {\rm I}=  O_{\p}(L_n^{1/2}n^{-1/2})\,, ~{\rm II}=O_{\p}(\omega_n)~~\textrm{and}~ ~{\rm III}=O_{\p}(s^{1/2}\ell_n^{1/2}\nu)\,.
\end{align*}
By \eqref{eq:Vf}, 
we have the first result of Lemma \ref{hVhatL2norm}.
Recall $\tilde{\btheta}_{\calM}=\arg\min_{\btheta_{\calM}\in\widehat{\bTheta}_{\calM}}\max_{\blambda\in{\tilde{\Lambda}}_n(\btheta_{\calM})}H_n({\btheta_{\calM}},\blambda)$ with $\widehat{\bTheta}_{\calM}=\{\btheta_{\calM}\in\mathbb{R}^m:|\btheta_{\calM}-\hat{\btheta}_{n,\calM}|_{\infty}\leq O_{\p}(\nu)\}$.
 By the same arguments for  deriving \eqref{eq:hatVt}, we have 
\begin{align}\label{eq:hatVAn02}
&\|\widehat{\bfV}_{\bff^{\bfA_n}}(\tilde{\btheta}_{\calM},\hat{\btheta}_{n,\calM^{\rm c}})-\widehat{\bfV}_{\bff^{\bfA_n}}({\btheta}_{0,\calM},\hat{\btheta}_{n,\calM^{\rm c}})\|_2\notag\\&~~~~~~~\leq \frac{1}{n}\sum_{t=1}^{n}|\bff_t^{\bfA_n}(\tilde{\btheta}_{\calM},\hat{\btheta}_{n,\calM^{\rm c}})-\bff_t^{\bfA_n}({\btheta}_{0,\calM},\hat{\btheta}_{n,\calM^{\rm c}})|_2^2\\&~~~~~~~~~~+
\notag2\lambda^{1/2}_{\max}\{\widehat{\bfV}_{\bff^{\bfA_n}}({\btheta}_{0,\calM},\hat{\btheta}_{n,\calM^{\rm c}})\}\bigg\{\frac{1}{n}\sum_{t=1}^{n}|\bff_t^{\bfA_n}(\tilde{\btheta}_{\calM},\hat{\btheta}_{n,\calM^{\rm c}})-\bff_t^{\bfA_n}({\btheta}_{0,\calM},\hat{\btheta}_{n,\calM^{\rm c}})|_2^2\bigg\}^{1/2}\,.
\end{align}	
By Jensen's inequality, we have
\begin{align*}
&\frac{1}{n}\sum_{t=1}^{n}|\bff_t^{\bfA_n}(\tilde{\btheta}_{\calM},\hat{\btheta}_{n,\calM^{\rm c}})-\bff_t^{\bfA_n}(\btheta_{0,\calM},\hat{\btheta}_{n,\calM^{\rm c}})|_2^2=\sum_{k=1}^{m}\frac{1}{n}\sum_{t=1}^{n}\bigg|\sum_{l\in\mathcal{M}}({\theta}_{l}-\theta_{0,l})\frac{\partial f_{t,k}^{\bfA_n}(\dot{\btheta})}{\partial \theta_l}\bigg|^2\\&~~~~~~~\leq 
\sum_{k=1}^{m}|\tilde{\btheta}_{\calM}-{\btheta}_{0,\calM}|_1^2\max_{l\in\mathcal{M}}\frac{1}{n}\sum_{t=1}^{n}\bigg|\frac{\partial f_{t,k}^{\bfA_n}(\dot{\btheta})}{\partial \theta_l}\bigg|^2
\leq 
m|\tilde{\btheta}_{\calM}-{\btheta}_{0,\calM}|_1^2\max_{k\in[m],l\in\mathcal{M}}\frac{1}{n}\sum_{t=1}^{n}\bigg|\frac{\partial f_{t,k}^{\bfA_n}(\dot{\btheta})}{\partial \theta_l}\bigg|^2\,,
\end{align*}	
where $\dot{\btheta}$ is on the line joining $\btheta$ and ${\btheta}_{0}$.  Recall $b_n=\max\{a_n,\nu^2\}$. Due to $\nu\gg\aleph_n$,
under Conditions \ref{con:moments2}(b), \ref{con:moments2}(c) and  \ref{con:Gamma}(a),
by \eqref{eq:thetas}, we know $|\hat{\btheta}_{n,\calS}-\btheta_{0,\calS}|_2=O_{\p}(\ell_n^{1/2}\nu)$ provided that $\ell_ns^{3}b_n=o(1)$.
Since
$\hat{\btheta}_{n,\calS^{\rm c}}=\bzero$ w.p.a.1, 
$|\tilde{\btheta}_{\calM}-\hat{\btheta}_{n,\calM}|_{\infty}\leq O_{\p}(\nu)$ and $m=|\mathcal{M}|$ is a fixed constant,  then by the triangle inequality, 
\begin{align*}
    |\tilde{\btheta}_{\calM}-\btheta_{0,\calM}|_1\leq |\tilde{\btheta}_{\calM}-\hat{\btheta}_{n,\calM}|_1+|\hat{\btheta}_{n,\calM}-\btheta_{0,\calM}|_1\leq O_{\p}(\nu)+m^{1/2}|\hat{\btheta}_{n,\calS}-\btheta_{0,\calS}|_2=O_{\p}(\ell_n^{1/2}\nu)\,.
\end{align*}
By Lemma \ref{hL2norm}, we  know 
\begin{align*}
\frac{1}{n}\sum_{t=1}^{n}|\bff_t^{\bfA_n}(\tilde{\btheta}_{\calM},\hat{\btheta}_{n,\calM^{\rm c}})-\bff_t^{\bfA_n}(\btheta_{0,\calM},\hat{\btheta}_{n,\calM^{\rm c}})|_2^2=O_{\p}(\ell_n\nu^2)\,.
\end{align*}
Therefore, from \eqref{eq:hatVAn02} we have $\|\widehat{\bfV}_{\bff^{\bfA_n}}(\tilde{\btheta}_{\calM},\hat{\btheta}_{n,\calM^{\rm c}})-\widehat{\bfV}_{\bff^{\bfA_n}}({\btheta}_{0,\calM},\hat{\btheta}_{n,\calM^{\rm c}})\|_2=O_{\p}(\ell_n^{1/2}\nu)$ provided that $\ell_n\nu^2=o(1)$.
 We complete the proof of Lemma \ref{hVhatL2norm}. $\hfill\Box$

\subsubsection{ Convergence rate of  ${\rm I}$}\label{sec:I}
Write $\ba_k^0=(a_{k,1}^0,\ldots, a_{k,r}^0)^{\T}$ for any $k\in[m]$. By Jensen's inequality, for any $k\in[m]$, we have 
\begin{align*}
	|f_{t,k}^{\bfA}(\btheta_{0})|^{\gamma}=\bigg|\sum_{j=1}^{r}a_{k,j}^0g_{t,j}(\btheta_0)\bigg|^{\gamma}\leq |\ba_k^0|_1^{\gamma}\sum_{j=1}^{r}\frac{|a_{k,j}^0|}{|\ba_{k}^0|_1}|g_{t,j}(\btheta_{0})|^{\gamma}=|\ba_k^0|_1^{\gamma-1}\sum_{j=1}^{r}{|a_{k,j}^0|}|g_{t,j}(\btheta_{0})|^{\gamma}\,.
\end{align*}
By Conditions \ref{con:moments1}(a) and \ref{con:alpha}, it holds that 
\begin{align}\label{eq:eftka}	\max_{t\in[n]}\max_{k\in[m]}\mathbb{E}\{|f_{t,k}^{\bfA}(\btheta_{0})|^{\gamma}\}\leq&~ \max_{k\in[m]}|\ba_k^0|_1^{\gamma-1}\sum_{j=1}^{r}{|a_{k,j}^0|}\cdot\max_{t\in[n]}\max_{j\in[r]}\mathbb{E}\{|g_{t,j}(\btheta_{0})|^{\gamma}\} \lesssim 1\,.
\end{align}
Let  $\mathring{f}_{t,k_1,k_2}^{\bA}(\btheta_0)=f^{\bA}_{t,k_1}(\btheta_0)f^{\bA}_{t,k_2}(\btheta_0)-\mathbb{E}\{f^{\bA}_{t,k_1}(\btheta_0)f^{\bA}_{t,k_2}(\btheta_0)\}$. Then, by the Cauchy-Schwarz inequality,
\begin{align*}
\max_{t\in[n]}\max_{k_1,k_2\in[m]}\mathbb{E}\{|\mathring{f}_{t,k_1,k_2}^{\bA}(\btheta_0)|^{\gamma/2}\}\lesssim \max_{t\in[n]}\max_{k\in[m]}\mathbb{E}\{|f_{t,k}^{\bfA}(\btheta_{0})|^{\gamma}\}\lesssim 1\,.
\end{align*}
For any $u>0$, by the
 Markov inequality, Davydov's inequality, and Condition \ref{con:mixingdecay}, 
\begin{align*}
	&\mathbb{P}\bigg\{\sum_{k_1,k_2\in[m]}\bigg|\frac{1}{n}\sum_{t=1}^{n}\mathring{f}_{t,k_1,k_2}^{\bA}(\btheta_0)\bigg|^2\geq u\bigg\}\leq  u^{-1}\sum_{k_1,k_2\in[m]}\mathbb{E}\bigg\{\bigg|\frac{1}{n}\sum_{t=1}^{n}\mathring{f}_{t,k_1,k_2}^{\bA}(\btheta_0)\bigg|^2\bigg\}
	\\&~~~~~~\lesssim u^{-1}\sum_{k_1,k_2\in[m]}\bigg[\frac{1}{n^2}\sum_{t=1}^{n}\mathbb{E}\big\{\mathring{f}_{t,k_1,k_2}^{\bA,2}(\btheta_0)\big\}+\frac{1}{n^2}\sum_{t_2=1}^{n-1}\sum_{t_1=t_2+1}^{n}\big|\mathbb{E}\big\{\mathring{f}_{t_1,k_1,k_2}^{\bA}(\btheta_0)\mathring{f}_{t_2,k_1,k_2}^{\bA}(\btheta_0)\big\}\big|\bigg]
		\\&~~~~~~\lesssim  u^{-1}m^2\bigg\{\frac{1}{n}+\frac{1}{n^2}\sum_{t_2=1}^{n-1}\sum_{t_1=t_2+1}^{n}\exp(-CL_n^{-\varphi}|t_1-t_2|^{\varphi})\bigg\}
			\\&~~~~~~\lesssim u^{-1}m^2(n^{-1}+n^{-1}L_n)\lesssim u^{-1}m^2n^{-1}L_n\,,
	\end{align*}
which implies  
\begin{align*}
\sum_{k_1,k_2\in[m]}\bigg|\frac{1}{n}\sum_{t=1}^{n}\mathring{f}_{t,k_1,k_2}^{\bA}(\btheta_0)\bigg|^2=O_{\p}(m^2L_n n^{-1})\,.
\end{align*}
Since $m$ is a fixed constant, then  
\begin{align*}
{\rm I}=\|\widehat{\bfV}_{\bff^{\bfA}}(\btheta_0)-\mathbb{E}\{\widehat{\bfV}_{\bff^{\bfA}}(\btheta_0)\}\|_2\leq&~ \bigg\{\sum_{k_1,k_2\in[m]}\bigg|\frac{1}{n}\sum_{t=1}^{n}\mathring{f}_{t,k_1,k_2}^{\bA}(\btheta_0)\bigg|^2\bigg\}^{1/2}=O_{\p}(L_n^{1/2}n^{-1/2})\,.
\end{align*}
We have constructed the convergence rate of ${\rm I}$. $\hfill\Box$
\subsubsection{ Convergence rate of   ${\rm II}$}\label{sec:II}
 By the same arguments for  deriving \eqref{eq:hatVt}, we have
\begin{align}\label{eq:hvhat}
\|\widehat{\bfV}_{\bff^{\bfA_n}}(\btheta_0)-\widehat{\bfV}_{\bff^{\bfA}}(\btheta_0)\|_2&\leq \frac{1}{n}\sum_{t=1}^{n}|\bff_t^{\bfA_n}(\btheta_0)-\bff_t^{\bfA}(\btheta_0)|_2^2\notag\\&~~~~+
2\lambda^{1/2}_{\max}\{\widehat{\bfV}_{\bff^{\bfA}}(\btheta_0)\}\bigg\{\frac{1}{n}\sum_{t=1}^{n}|\bff_t^{\bfA_n}(\btheta_0)-\bff_t^{\bfA}(\btheta_0)|_2^2\bigg\}^{1/2}\,.
\end{align}
Due to  $\bff_t^{\bfA_n}(\btheta_0)-\bff_t^{\bfA}(\btheta_0)=(\bfA_n-\bfA)\bfg_t(\btheta_0)$, by Conditions \ref{con:moments1}(a) and \ref{con:alpha} we have 
\begin{align*}
\frac{1}{n}\sum_{t=1}^{n}|\bff_t^{\bfA_n}(\btheta_0)-\bff_t^{\bfA}(\btheta_0)|_2^2=&~\sum_{k=1}^{m}(\ba_k^n-\ba_k^0)^\T\bigg\{\frac{1}{n}\sum_{t=1}^{n}\bfg_t(\btheta_0)^{\otimes 2}\bigg\}(\ba_k^n-\ba_k^0)\\\leq &~
\bigg|\frac{1}{n}\sum_{t=1}^{n}\bfg_t(\btheta_0)^{\otimes 2}\bigg|_\infty\sum_{k=1}^{m}|\ba_k^n-\ba_k^0|_1^2=O_{\p}(\omega_n^2)\,.
\end{align*}
We have shown in  Section \ref{sec:I} that $\|\widehat{\bfV}_{\bff^{\bfA}}(\btheta_0)-\mathbb{E}\{\widehat{\bfV}_{\bff^{\bfA}}(\btheta_0)\}\|_2=O_{\p}(L_n^{1/2}n^{-1/2})$. Due to $L_nn^{-1}=o(1)$, by Conditions \ref{con:moments1}(c) and \ref{con:alpha} we know $\lambda_{\max}\{{\widehat{\bfV}_{\bff^{\bfA}}}(\btheta_0)\}$ is uniformly bounded away from infinity w.p.a.1. Together with $\omega_n=o(1)$, by \eqref{eq:hvhat},  \begin{align*}
 {\rm II}=\|\widehat{\bfV}_{\bff^{\bfA_n}}(\btheta_0)-\widehat{\bfV}_{\bff^{\bfA}}(\btheta_0)\|_2=O_{\p}(\omega_n)\,.
\end{align*}
We have constructed the convergence rate of ${\rm II}$. $\hfill\Box$

\subsubsection{Convergence rate of   ${\rm III}$}\label{sec:III}
Analogous to \eqref{eq:hvhat},  it holds that
\begin{align}\label{eq:hatVAn0}
\|\widehat{\bfV}_{\bff^{\bfA_n}}(\btheta_{0,\calM},\hat{\btheta}_{n,\calM^{\rm c}})-\widehat{\bfV}_{\bff^{\bfA_n}}({\btheta}_0)\|_2&\leq \frac{1}{n}\sum_{t=1}^{n}|\bff_t^{\bfA_n}(\btheta_{0,\calM},\hat{\btheta}_{n,\calM^{\rm c}})-\bff_t^{\bfA_n}(\btheta_0)|_2^2\\&~~~+
\notag2\lambda^{1/2}_{\max}\{\widehat{\bfV}_{\bff^{\bfA_n}}(\btheta_0)\}\bigg\{\frac{1}{n}\sum_{t=1}^{n}|\bff_t^{\bfA_n}(\btheta_{0,\calM},\hat{\btheta}_{n,\calM^{\rm c}})-\bff_t^{\bfA_n}(\btheta_0)|_2^2\bigg\}^{1/2}\,.
\end{align}	
By Jensen's inequality, we have
\begin{align*}
&\frac{1}{n}\sum_{t=1}^{n}|\bff_t^{\bfA_n}(\btheta_{0,\calM},\hat{\btheta}_{n,\calM^{\rm c}})-\bff_t^{\bfA_n}(\btheta_0)|_2^2=\sum_{k=1}^{m}\frac{1}{n}\sum_{t=1}^{n}\bigg|\sum_{l\in\mathcal{M}^{\rm c}}(\hat{\theta}_{n,l}-\theta_{0,l})\frac{\partial f_{t,k}^{\bfA_n}(\check{\btheta})}{\partial \theta_l}\bigg|^2\\&~~~~\leq 
\sum_{k=1}^{m}|\hat{\btheta}_{n,\calM^{\rm c}}-{\btheta}_{0,\calM^{\rm c}}|_1^2\max_{ l\in\mathcal{M}^{\rm c}}\frac{1}{n}\sum_{t=1}^{n}\bigg|\frac{\partial f_{t,k}^{\bfA_n}(\check{\btheta})}{\partial \theta_l}\bigg|^2
\leq 
m|\hat{\btheta}_{n,\calM^{\rm c}}-{\btheta}_{0,\calM^{\rm c}}|_1^2\max_{k\in[m]}\max_{l\in\mathcal{M}^{\rm c}}\frac{1}{n}\sum_{t=1}^{n}\bigg|\frac{\partial f_{t,k}^{\bfA_n}(\check{\btheta})}{\partial \theta_l}\bigg|^2\,,
\end{align*}	
where $\check{\btheta}$ is on the line joining $\btheta_0$ and $\bar{\btheta}:=(\btheta_{0,\calM}^{\T},\hat{\btheta}_{n,\calM^{\rm c}}^{\T})^{\T}$. Recall $b_n=\max\{a_n,\nu^2\}$. Due to $\nu\gg\aleph_n$,
under Conditions \ref{con:moments2}(b), \ref{con:moments2}(c) and  \ref{con:Gamma}(a),
by \eqref{eq:thetas}, we have $|\hat{\btheta}_{n,\calS}-\btheta_{0,\calS}|_2=O_{\p}(\ell_n^{1/2}\nu)$ provided that $\ell_ns^{3}b_n=o(1)$.
Since   
$\hat{\btheta}_{n,\calS^{\rm c}}=\bzero$ w.p.a.1 and $|\mathcal{S}\,\cap\,\mathcal{M}^{\rm c}|\asymp s$, then   
$
|\hat{\btheta}_{n,\calM^{\rm c}}-{\btheta}_{0,\calM^{\rm c}}|_1 \leq O_{\p}(s^{1/2}\ell_n^{1/2}\nu)$. 
By Lemma \ref{hL2norm}, we  know 
\begin{align*}
\frac{1}{n}\sum_{t=1}^{n}|\bff_t^{\bfA_n}(\btheta_{0,\calM},\hat{\btheta}_{n,\calM^{\rm c}})-\bff_t^{\bfA_n}(\btheta_0)|_2^2=O_{\p}(s\ell_n\nu^2)\,.
\end{align*}
Therefore, from \eqref{eq:hatVAn0} we have $ {\rm III}=O_{\p}(s^{1/2}\ell_n^{1/2}\nu)$ provided that $s\ell_n\nu^2=o(1)$. 
 $\hfill\Box$

\subsection{Proof of Lemma \ref{lambdahatstarL2norm}}\label{sec:la:lambdahatstarL2norm}
To prove Lemma \ref{lambdahatstarL2norm}, we need  Lemmas \ref{bargm} and  \ref{AnR1Linfitynorm}. As  shown in Lemma \ref{hVhatL2norm}, under Conditions \ref{con:moments2}(b), \ref{con:moments2}(c) and  \ref{con:Gamma}(a), $|\hat{\btheta}_{n,\calM}-\btheta_{0,\calM}|_1=O_{\p}(\ell_n^{1/2}\nu)$  and $
|\hat{\btheta}_{n,\calM^{\rm c}}-{\btheta}_{0,\calM^{\rm c}}|_1 = O_{\p}(s^{1/2}\ell_n^{1/2}\nu)$ provided that $\ell_ns^{3}b_n=o(1)$.
The proofs of Lemmas \ref{bargm} and \ref{AnR1Linfitynorm} are, respectively, almost identical to those for Lemmas 2 and 3 in \cite{Chang2020} but with replacing $(\btheta^*, \xi_{1,n},\xi_{2,n},\tau)$ there by $(\hat{\btheta}_{n},\ell_n^{1/2}\nu,s^{1/2}\ell_n^{1/2}\nu,\varsigma)$. Hence, we omit the proofs of Lemmas \ref{bargm} and \ref{AnR1Linfitynorm}.

\begin{lemma}\label{bargm}
	Assume the conditions of Theorem $\ref{The.1}$ hold. Under Conditions {\rm \ref{con:moments2}(b)}, {\rm \ref{con:moments2}(c)}, {\rm \ref{con:Gamma}(a)} and {\rm \ref{con:alpha}}, if $\ell_ns^{3}b_n=o(1)$, it holds that
	\begin{align*}
		\max_{k\in[m]}|\{\nabla_{\btheta_{\calM^{\rm c}}}\bar{\bfg}(\btheta_{0,\calM},\hat{\btheta}_{n,\calM^{\rm c}})\}^\T\ba_k^n|_\infty\leq \varsigma+O_{\p}(\ell_n^{1/2}\nu)\,.
	\end{align*}
\end{lemma}

\begin{lemma}\label{AnR1Linfitynorm}
	Assume the conditions of Theorem $\ref{The.1}$ hold. Under  Conditions {\rm \ref{con:moments2}(b)}, {\rm \ref{con:moments2}(c)}, {\rm \ref{con:Gamma}(a)} and {\rm \ref{con:alpha}}, if $\ell_ns^{3}b_n=o(1)$, it holds that
	$
		|\bfA_n\bfR_1|_\infty=O_{\p}(s\ell_n\nu^2)$,
	where \begin{align*}
	    \bfR_1=\bar{\bfg}(\btheta_{0,\calM},\hat{\btheta}_{n,\calM^{\rm c}})-\bar{\bfg}(\btheta_{0})-\{\nabla_{\btheta_{\calM^{\rm c}}}\bar{\bfg}(\btheta_{0,\calM},\hat{\btheta}_{n,\calM^{\rm c}})\}(\hat{\btheta}_{n,\calM^{\rm c}}-\btheta_{0,\calM^{\rm c}})\,.
	\end{align*}
\end{lemma}

Let $\tilde{\blambda}=\arg\max_{\blambda\in\tilde{\Lambda}_n(\btheta_{0,\calM})}H_n({\btheta}_{0,\calM},\blambda)$. Due to $L_n\ll n^{1-2/\gamma}$, we can pick 
$\delta_n$ satisfying 
$L_n^{1/2}n^{-1/2}\ll\delta_n\ll n^{-1/\gamma}$.  Let $\bar{\blambda}=\arg\max_{\blambda\in\Lambda_n}H_n({\btheta}_{0,\calM},\blambda)$ where $\Lambda_n=\{\blambda\in \mathbb{R}^m:|\blambda|_2\leq \delta_n\}$. Recall  $m=|\mathcal{M}|$ is a fixed constant and $\bff^{\bfA_n}(\cdot;\cdot)=\bfA_n\bfg(\cdot;\cdot)$ with $\bfA_n=(\ba_1^n,\dots, \ba_m^n)^{\T}$. By Condition \ref{con:moments1}(a), we have $\max_{j\in[r],t\in[n]}|g_{t,j}(\btheta_{0,\calM},\hat{\btheta}_{n,\calM^{\rm c}})|=O_{\p}(n^{1/\gamma})$, which implies 
\begin{align}\label{fttheta}
\max_{t\in[n]}|\bff_t^{\bfA_n}(\btheta_{0,\calM},\hat{\btheta}_{n,\calM^{\rm c}})|_2^2&=\max_{t\in[n]}\sum_{k=1}^{m}\big|\ba_k^{n,\T}\bfg_t(\btheta_{0,\calM},\hat{\btheta}_{n,\calM^{\rm c}})\big|^2
\notag\\&\leq 
\max_{t\in[n]}\max_{j\in[r] }g_{t,j}^2(\btheta_{0,\calM},\hat{\btheta}_{n,\calM^{\rm c}})\sum_{k=1}^{m}|\ba_k^n|_1^2=O_{\p}(n^{2/\gamma})\,.
\end{align}
Hence, $\max_{t\in[n]}|\bff_t^{\bfA_n}(\btheta_{0,\calM},\hat{\btheta}_{n,\calM^{\rm c}})|_2=O_{\p}(n^{1/\gamma})$, which implies $\sup_{t\in[n],\blambda\in \Lambda_n}|\blambda^\T\bff_t^{\bfA_n}(\btheta_{0,\calM},\hat{\btheta}_{n,\calM^{\rm c}})|=o_{\p}(1)$. Under  Conditions \ref{con:moments1}(c)
and  {\rm \ref{con:alpha}}, by  Lemma \ref{hVhatL2norm},  we know  $\lambda_{\min}\{\widehat{\bfV}_{\bff^{\bfA_n}}(\btheta_{0,\calM},\hat{\btheta}_{n,\calM^{\rm c}})\}$ is uniformly bounded away from zero w.p.a.1 provided that  $\ell_ns^{3}b_n=o(1)$. By   Taylor expansion, it holds w.p.a.1 that
\begin{align}\label{eq:H0}
0=H_n(\btheta_{0,\calM},{\bf0})\leq&~ H_n(\btheta_{0,\calM},\bar{\blambda})\notag\\=&~\bar{\blambda}^\T\bar{\bff}^{\bfA_n}(\btheta_{0,\calM},\hat{\btheta}_{n,\calM^{\rm c}})-\frac{1}{2n}\sum_{t=1}^{n}\frac{\bar{\blambda}^\T\bff_t^{\bfA_n}(\btheta_{0,\calM},\hat{\btheta}_{n,\calM^{\rm c}})^{\otimes 2}\bar{\blambda}}{\{1+c\bar{\blambda}^\T\bff_t^{\bfA_n}(\btheta_{0,\calM},\hat{\btheta}_{n,\calM^{\rm c}})\}^2}
\notag\\\leq &~
|\bar{\blambda}|_2|\bar{\bff}^{\bfA_n}(\btheta_{0,\calM},\hat{\btheta}_{n,\calM^{\rm c}})|_2-C|\bar{\blambda}|_2^2\{1+o_{\p}(1)\}
\end{align}
for some $|c|\leq 1$.
By \eqref{eq:eftka},  $\max_{t\in[n],k\in[m]}\mathbb{E}\{|f_{t,k}^{\bfA}(\btheta_{0})|^{\gamma}\}\lesssim1$.
By 
 Davydov's inequality and Condition \ref{con:mixingdecay}, we have 
\begin{align}\label{eq:hdavydov}
\mathbb{E}\{|\bar{\bff}^{\bfA}(\btheta_{0})|_2^2\}&=
\frac{1}{n^2}\sum_{k=1}^{m}\mathbb{E}\bigg[\sum_{t=1}^{n}\{f_{t,k}^{\bfA}(\btheta_{0})\}^2+\sum_{t_1\neq t_2}f_{t_1,k}^{\bfA}(\btheta_{0})f_{t_2,k}^{\bfA}(\btheta_{0})\bigg]\notag\\&\lesssim
\frac{1}{n}+\frac{1}{n^2}\sum_{t_1\neq t_2}\exp(-CL_n^{-\varphi}|t_1-t_2|^{\varphi})=
O(L_nn^{-1})\,,
\end{align}
which implies  $|\bar{\bff}^{\bfA}(\btheta_{0})|_2=O_{\p}(L_n^{1/2}n^{-1/2})$. Recall $\bA=(\ba_1^0,\dots, \ba_m^0)^{\T}$ and $\bff_t^{\bfA_n}(\btheta_{0})-\bff_t^{\bfA}(\btheta_{0})=(\bfA_n-\bfA)\bfg_t(\btheta_{0})$. 
By Lemma \ref{tailprabog0}, $|\bar{\bfg}(\btheta_{0})|_\infty=O_{\p}(\aleph_n)$.
 By the triangle inequality and Condition \ref{con:alpha},  if $n\aleph_n^2\omega_n^2=O( L_n)$, we have 
\begin{align}\label{eq:barhAn}
|\bar{\bff}^{\bfA_n}(\btheta_{0})|_2\leq |\bar{\bff}^{\bfA}(\btheta_{0})|_2+m^{1/2}|\bar{\bfg}(\btheta_{0})|_\infty\max_{k\in[m]}|\ba_k^n-\ba_k^0|_1=O_{\p}(L_n^{1/2}n^{-1/2})\,.
\end{align}
On the other hand, it holds that 
\begin{align}
|\bar{\bff}^{\bfA_n}(\btheta_{0,\calM},\hat{\btheta}_{n,\calM^{\rm c}})-\bar{\bff}^{\bfA_n}(\btheta_{0})|_2\notag
&\leq 
\big|\{\nabla_{\btheta_{\calM^{\rm c}}}\bar{\bff}^{\bfA_n}(\btheta_{0,\calM},\check{\btheta}_{\calM^{\rm c}})-\nabla_{\btheta_{\calM^{\rm c}}}\bar{\bff}^{\bfA_n}(\btheta_{0,\calM},\hat{\btheta}_{n,\calM^{\rm c}})\}(\hat{\btheta}_{n,\calM^{\rm c}}-\btheta_{0,\calM^{\rm c}})\big|_2
\\&~~~+
\big|\nabla_{\btheta_{\calM^{\rm c}}}\bar{\bff}^{\bfA_n}(\btheta_{0,\calM},\hat{\btheta}_{n,\calM^{\rm c}})(\hat{\btheta}_{n,\calM^{\rm c}}-\btheta_{0,\calM^{\rm c}})\big|_2\,,\notag
\end{align}
where  $\check{\btheta}_{\calM^{\rm c}}$ is on the line between $\hat{\btheta}_{n,\calM^{\rm c}}$ and $\btheta_{0,\calM^{\rm c}}$.
By Lemma  \ref{AnR1Linfitynorm}, 
\begin{align*}
\big|\{\nabla_{\btheta_{\calM^{\rm c}}}\bar{\bff}^{\bfA_n}(\btheta_{0,\calM},\check{\btheta}_{\calM^{\rm c}})-\nabla_{\btheta_{\calM^{\rm c}}}\bar{\bff}^{\bfA_n}(\btheta_{0,\calM},\hat{\btheta}_{n,\calM^{\rm c}})\}(\hat{\btheta}_{n,\calM^{\rm c}}-\btheta_{0,\calM^{\rm c}})\big|_2=O_{\p}(s\ell_n\nu^2)\,.
\end{align*}
Due to $|\hat{\btheta}_{n,\calM^{\rm c}}-\btheta_{0,\calM^{\rm c}}|_1=O_{\p}(s^{1/2}\ell_n^{1/2}\nu)$,
by Lemma \ref{bargm}, 
\begin{align*}
&\big|\nabla_{\btheta_{\calM^{\rm c}}}\bar{\bff}^{\bfA_n}(\btheta_{0,\calM},\hat{\btheta}_{n,\calM^{\rm c}})(\hat{\btheta}_{n,\calM^{\rm c}}-\btheta_{0,\calM^{\rm c}})\big|_2 \leq m^{1/2}	\big|\nabla_{\btheta_{\calM^{\rm c}}}\bar{\bff}^{\bfA_n}(\btheta_{0,\calM},\hat{\btheta}_{n,\calM^{\rm c}})(\hat{\btheta}_{n,\calM^{\rm c}}-\btheta_{0,\calM^{\rm c}})\big|_\infty
\\&~~~~~~~~= m^{1/2}\max_{k
\in[m]}\big|\ba_k^{n,\T}\nabla_{\btheta_{\calM^{\rm c}}}\bar{\bg}(\btheta_{0,\calM},\hat{\btheta}_{n,\calM^{\rm c}})(\hat{\btheta}_{n,\calM^{\rm c}}-\btheta_{0,\calM^{\rm c}})\big|
\\&~~~~~~~~\leq  m^{1/2}\max_{k
	\in[m]}\big|\{\nabla_{\btheta_{\calM^{\rm c}}}\bar{\bfg}(\btheta_{0,\calM},\hat{\btheta}_{n,\calM^{\rm c}})\}^\T\ba_k^n\big|_\infty|\hat{\btheta}_{n,\calM^{\rm c}}-\btheta_{0,\calM^{\rm c}}|_1	=O_{\p}\{s^{1/2}\ell_n^{1/2}\nu(\varsigma+\ell_n^{1/2}\nu)\}\,.
\end{align*}
Hence, we have 
\begin{align}\label{eq:fantheta}
    |\bar{\bff}^{\bfA_n}(\btheta_{0,\calM},\hat{\btheta}_{n,\calM^{\rm c}})-\bar{\bff}^{\bfA_n}(\btheta_{0})|_2=&~	O_{\p}(s\ell_n\nu^2)+O_{\p}\{s^{1/2}\ell_n^{1/2}\nu(\varsigma+\ell_n^{1/2}\nu)\}\\=&~
O_{\p}(L_n^{1/2}n^{-1/2})\notag
\end{align}
provided that 
$ns\ell_n\nu^2\max\{s\ell_n\nu^2, \varsigma^2\}=O(L_n)$.
Together with $|\bar{\bff}^{\bfA_n}(\btheta_{0})|_2=O_{\p}(L_n^{1/2}n^{-1/2})$, it holds that $|\bar{\bff}^{\bfA_n}(\btheta_{0,\calM},\hat{\btheta}_{n,\calM^{\rm c}})|_2=O_{\p}(L_n^{1/2}n^{-1/2})$. Then (\ref{eq:H0}) implies  $|\bar{\blambda}|_2=O_{\p}(L_n^{1/2}n^{-1/2})=o_{\p}(\delta_n)$. Therefore, $\bar{\blambda}\in \text{int}(\Lambda_n)$ w.p.a.1. Since $\Lambda_n \subset \tilde{\Lambda}_n(\btheta_{0,\calM})$ w.p.a.1, $\bar{\blambda}=\tilde{\blambda}$ w.p.a.1 by the concavity of $H_n(\btheta_{0,\calM},\blambda)$ and the convexity of $\tilde{\Lambda}_n(\btheta_{0,\calM})$. Hence, by (\ref{eq:H0}), \begin{align*}
\max_{\blambda\in\tilde{\Lambda}_n(\btheta_{0,\calM})}H_n(\btheta_{0,\calM},\blambda)=O_{\p}(L_nn^{-1})\,.
\end{align*}

Recall $\tilde{\blambda}^*=\arg\max_{\blambda\in
	\tilde{\Lambda}_n(\tilde{\btheta}_{\calM})}H_n(\tilde{\btheta}_{\calM},\blambda)$. In the sequel, we will show $|\tilde{\blambda}^*|_2=O_{\p}(L_n^{1/2}n^{-1/2})$. For $\delta_n$ specified above, let $\blambda^*=\delta_n\bar{\bff}^{\bfA_n}(\tilde{\btheta}_{\calM},\hat{\btheta}_{n,\calM^{\rm c}})/|\bar{\bff}^{\bfA_n}(\tilde{\btheta}_{\calM},\hat{\btheta}_{n,\calM^{\rm c}})|_2$. Then $\blambda^*\in \Lambda_n=\{\blambda\in \mathbb{R}^m:|\blambda|_2\leq \delta_n\}$. Under Conditions \ref{con:moments1}(c), {\rm \ref{con:moments2}(b)}, {\rm \ref{con:moments2}(c)}, {\rm \ref{con:Gamma}(a)} and  {\rm \ref{con:alpha}}, by  Lemma \ref{hVhatL2norm}, 
 $\lambda_{\max}\{\widehat{\bfV}_{\bff^{\bfA_n}}(\tilde{\btheta}_{\calM},\hat{\btheta}_{n,\calM^{\rm c}})\} $ is uniformly bounded away from infinity w.p.a.1  provided that $\ell_ns^{3}b_n=o(1)$. By Taylor expansion, it holds w.p.a.1 that
\begin{align*}
H_n(\tilde{\btheta}_{\calM},\blambda^*)=&~\blambda^{*,\T}\bar{\bff}^{\bfA_n}(\tilde{\btheta}_{\calM},\hat{\btheta}_{n,\calM^{\rm c}})-\frac{1}{2n}\sum_{t=1}^{n}\frac{\blambda^{*,\T}\bff_t^{\bfA_n}(\tilde{\btheta}_{\calM},\hat{\btheta}_{n,\calM^{\rm c}})^{\otimes 2}\blambda^*}{\{1+c\blambda^{*,\T}\bff_t^{\bfA_n}(\tilde{\btheta}_{\calM},\hat{\btheta}_{n,\calM^{\rm c}})\}^2}\\\geq &~
\delta_n|\bar{\bff}^{\bfA_n}(\tilde{\btheta}_{\calM},\hat{\btheta}_{n,\calM^{\rm c}})|_2-C\delta_n^2\{1+o_{\p}(1)\}
\end{align*}
for some $|c|<1$. Due to \begin{align*}
H_n(\tilde{\btheta}_{\calM},\blambda^*)\leq \max_{\blambda\in\tilde{\Lambda}_n(\tilde{\btheta}_{\calM})}H_n(\tilde{\btheta}_{\calM},\blambda)\leq \max_{\blambda\in\tilde{\Lambda}_n(\btheta_{0,\calM})}H_n(\btheta_{0,\calM},\blambda)=O_{\p}(L_nn^{-1})=o_{\p}(\delta_n^2)\,, 
\end{align*} 
we know $|\bar{\bff}^{\bfA_n}(\tilde{\btheta}_{\calM},\hat{\btheta}_{n,\calM^{\rm c}})|_2=O_{\p}(\delta_n)$. Consider any $\epsilon_n\to 0$ and let $\blambda^{**}=\epsilon_n\bar{\bff}^{\bfA_n}(\tilde{\btheta}_{\calM},\hat{\btheta}_{n,\calM^{\rm c}})$. Then $|\blambda^{**}|_2=o_{\p}(\delta_n)$. Analogously, we have 
\begin{align*}
\epsilon_n|\bar{\bff}^{\bfA_n}(\tilde{\btheta}_{\calM},\hat{\btheta}_{n,\calM^{\rm c}})|_2^2 -C\epsilon_n^2|\bar{\bff}^{\bfA_n}(\tilde{\btheta}_{\calM},\hat{\btheta}_{n,\calM^{\rm c}})|_2^2\{1+o_{\p}(1)\}=O_{\p}(L_nn^{-1})\,,
\end{align*}
which implies 
$\epsilon_n|\bar{\bff}^{\bfA_n}(\tilde{\btheta}_{\calM},\hat{\btheta}_{n,\calM^{\rm c}})|_2^2=O_{\p}(L_nn^{-1})$.
Notice that we can select an arbitrary slow $\epsilon_n\to 0$. Following a standard result from probability theory, we can obtain $|\bar{\bff}^{\bfA_n}(\tilde{\btheta}_{\calM},\hat{\btheta}_{n,\calM^{\rm c}})|_2^2=O_{\p}(L_nn^{-1})$. Similar to \eqref{eq:H0}, we have $|\tilde{\blambda}^*|_2=O_{\p}(L_n^{1/2}n^{-1/2})$.  We complete the proof of Lemma \ref{lambdahatstarL2norm}.
$\hfill\Box$

\subsection{Proof of Lemma \ref{gammahatstarL2norm}}\label{sec:la:gammahatstarL2norm}
By the triangle inequality,   
\begin{align*}
\big|\big[\widehat{\bGamma}_{\bff^{\bfA_n}}(\tilde{\btheta}_{\calM},\hat{\btheta}_{n,\calM^{\rm c}})-\mathbb{E}\{\widehat{\bGamma}_{\bff^{\bfA}}(\btheta_{0})\}\big]\bfz\big|_2&\leq \underbrace{\big|\{\widehat{\bGamma}_{\bff^{\bfA_n}}(\tilde{\btheta}_{\calM},\hat{\btheta}_{n,\calM^{\rm c}})-\widehat{\bGamma}_{\bff^{\bfA}}(\tilde{\btheta}_{\calM},\hat{\btheta}_{n,\calM^{\rm c}})\}\bfz\big|_2}_{\rm I}\\&~~+\underbrace{\big|\{\widehat{\bGamma}_{\bff^{\bfA}}(\tilde{\btheta}_{\calM},\hat{\btheta}_{n,\calM^{\rm c}})-\widehat{\bGamma}_{\bff^{\bfA}}(\btheta_{0})\}\bfz\big|_2}_{\rm II}
+
\underbrace{\big|\big[\widehat{{\bGamma}}_{\bff^{\bfA}}(\btheta_{0})-\mathbb{E}\{\widehat{\bGamma}_{\bff^{\bfA}}(\btheta_{0})\}\big]\bfz\big|_2}_{\rm III}
\end{align*}
for any $\bfz\in \mathbb{R}^m$. As we will show in Sections \ref{sec:gammaI}--\ref{sec:gammaIII}, \begin{align*}
  &{\rm I}= |\bfz|_2\cdot O_{\p}(\omega_n)\,, ~{\rm II}=|\bfz|_2\cdot O_{\p}(s^{1/2}\ell_n^{1/2}\nu)~\textrm{and}~{\rm III}=|\bz|_2\cdot O_{\p}(L_n^{1/2}n^{-1/2})\,.
\end{align*}
Then Lemma \ref{gammahatstarL2norm} holds.
$\hfill\Box$

\subsubsection{Convergence rate of  ${\rm I}$}\label{sec:gammaI}
Recall $m=|\mathcal{M}|$ is a fixed constant.
By Condition \ref{con:moments2}(a), $|\nabla_{\btheta_\calM}\bar{\bfg}(\tilde{\btheta}_{\calM},\hat{\btheta}_{n,\calM^{\rm c}})\bfz|_{\infty}=|\bz|_2\cdot O_{\p}(1)$ holds uniformly over $\bz\in\mathbb{R}^m$.
Together with Condition \ref{con:alpha},
\begin{align*}
&\notag\big|\{\widehat{\bGamma}_{\bff^{\bfA_n}}(\tilde{\btheta}_{\calM},\hat{\btheta}_{n,\calM^{\rm c}})-\widehat{\bGamma}_{\bff^{\bfA}}(\tilde{\btheta}_{\calM},\hat{\btheta}_{n,\calM^{\rm c}})\}\bfz\big|_2^2=\big|(\bfA_{n}-\bfA)\nabla_{\btheta_\calM}\bar{\bfg}(\tilde{\btheta}_{\calM},\hat{\btheta}_{n,\calM^{\rm c}})\bfz\big|_2^2\notag
\\&~~~~~~~~~~~\leq\sum_{k=1}^{m}|\ba_k^n-\ba_k^0|_1^2\cdot|\nabla_{\btheta_\calM}\bar{\bfg}(\tilde{\btheta}_{\calM},\hat{\btheta}_{n,\calM^{\rm c}})\bfz|_{\infty}^2=|\bfz|_2^2\cdot O_{\p}(\omega_n^2)\,.\notag
\end{align*}
Then ${\rm I}=|\bfz|_2\cdot O_{\p}(\omega_n)$. $\hfill\Box$

\subsubsection{ Convergence rate of  ${\rm II}$}\label{sec:gammaII}
Write $\bar{\btheta}=(\tilde{\btheta}^{\T}_{\calM},\hat{\btheta}^{\T}_{n,\calM^{\rm c}})^{\T}=(\bar{\theta}_1,\ldots,\bar{\theta}_p)^{\T}$.  Under Conditions \ref{con:moments2}(b), \ref{con:moments2}(c) and  \ref{con:Gamma}(a), we know
$|\hat{\btheta}_{n,\calM}-\btheta_{0,\calM}|_1=O_{\p}(\ell_n^{1/2}\nu)$  and $
|\hat{\btheta}_{n,\calM^{\rm c}}-{\btheta}_{0,\calM^{\rm c}}|_1 = O_{\p}(s^{1/2}\ell_n^{1/2}\nu)$ provided that $\ell_ns^{3}b_n=o(1)$.
Due to $|\tilde{\btheta}_{\calM}-\hat{\btheta}_{n,\calM}|_{\infty}=O_{\p}(\nu)$, then 
$|\tilde{\btheta}_{\calM}-\btheta_{0,\calM}|_1\leq |\hat{\btheta}_{n,\calM}-\btheta_{0,\calM}|_1+ |\tilde{\btheta}_{\calM}-\hat{\btheta}_{n,\calM}|_1=O_{\p}(\ell_n^{1/2}\nu)$.
Under Conditions \ref{con:moments2}(c) and \ref{con:alpha},
by the Cauchy-Schwarz inequality and Lemma \ref{hL2norm},
\begin{align}\label{eq:hGammas}
&\big|\{\widehat{\bGamma}_{\bff^{\bfA}}(\tilde{\btheta}_{\calM},\hat{\btheta}_{n,\calM^{\rm c}})-\widehat{\bGamma}_{\bff^{\bfA}}(\btheta_{0})\}\bfz\big|_2^2=|\{\nabla_{\btheta_\calM}\bar{\bff}^{\bfA}(\tilde{\btheta}_{\calM},\hat{\btheta}_{n,\calM^{\rm c}})-\nabla_{\btheta_\calM}\bar{\bff}^{\bfA}({\btheta}_0)\}\bfz|_2^2\notag\\&~~~~~~=
\sum_{j=1}^{m}\biggl[\sum_{k\in\mathcal{M}}z_{k}\sum_{l=1}^{p}(\bar{\theta}_{l}-\theta_{0,l})\bigg\{\frac{1}{n}\sum_{t=1}^{n}\frac{\partial^2f^{\bfA}_{t,j}(\check{\btheta})}{\partial\theta_k\partial\theta_l}\bigg\}\biggr]^2
\notag\\&~~~~~~\leq |\bfz|_2^2\sum_{j=1}^{m}\sum_{k\in\mathcal{M}}\biggl[\sum_{l=1}^{p}(\bar{\theta}_{l}-\theta_{0,l})\bigg\{\frac{1}{n}\sum_{t=1}^{n}\frac{\partial^2f^{\bfA}_{t,j}(\check{\btheta})}{\partial\theta_k\partial\theta_l}\bigg\}\biggr]^2
\\&~~~~~~\leq|\bfz|_2^2\sum_{j=1}^{m}\sum_{k\in\mathcal{M}}\max_{l\in[p]}\bigg\{\frac{1}{n}\sum_{t=1}^{n}\bigg|\frac{\partial^2f^{\bfA}_{t,j}(\check{\btheta})}{\partial\theta_k\partial\theta_l}\bigg|\bigg\}^2(|\tilde{\btheta}_{\calM}-\btheta_{0,\calM}|_1+|\hat{\btheta}_{n,\calM^{\rm c}}-\btheta_{0,\calM^{\rm c}}|_1)^2 
\notag\\&~~~~~~=|\bfz|_2^2\cdot O_{\p}(s\ell_n\nu^2)\notag\,,
\end{align}
where $\check{\btheta}$ is on the line joining $\btheta_0$ and $\bar{\btheta}$.
Then ${\rm II}=|\bfz|_2\cdot O_{\p}(s^{1/2}\ell_n^{1/2}\nu) $.
$\hfill\Box$

\subsubsection{ Convergence rate of  ${\rm III}$}\label{sec:gammaIII}
By the Cauchy-Schwarz inequality,  it holds that  
\begin{align}\label{eq:hGamma}
\big|\big[\widehat{\bGamma}_{\bff^{\bfA}}(\btheta_{0})-\mathbb{E}\{\widehat{\bGamma}_{\bff^{\bfA}}(\btheta_{0})\}\big]\bfz\big|_2^2\notag&= 
\bigg|\bigg(\frac{1}{n}\sum_{t=1}^{n}\big[\nabla_{\btheta_\calM}\bff^{\bfA}_{t}(\btheta_{0})-\mathbb{E}\{\nabla_{\btheta_\calM}\bff^{\bfA}_t(\btheta_{0})\}\big]\bigg)\bfz\bigg|_2^2\\ &\leq|\bfz|_2^2\sum_{j=1}^{m}\sum_{k\in\mathcal{M}}\bigg(\frac{1}{n}\sum_{t=1}^{n}\bigg[\frac{\partial f_{t,j}^{\bfA}(\btheta_{0})}{\partial\theta_k}-\mathbb{E}\biggl\{\frac{\partial f_{t,j}^{\bfA}(\btheta_{0})}{\partial\theta_k}\biggr\}\bigg]\bigg)^2\,.
\end{align}
Write $\ba_k^0=(a_{k,1}^0,\ldots,a_{k,r}^0)^{\T}$ for any $k\in[m]$.
By Jensen's inequality, we have
\begin{align*}
\biggl|\frac{\partial f_{t,j}^{\bfA}(\btheta_{0})}{\partial\theta_k}\biggr|^{\gamma}=\bigg|\sum_{i=1}^{r}a_{j,i}^0\frac{\partial g_{t,i}(\btheta_{0})}{\partial\theta_k}\bigg|^{\gamma}\leq |\ba_j^0|_1^{\gamma}\sum_{i=1}^{r}\frac{|a_{j,i}^0|}{|\ba_j^0|_1}\bigg|\frac{\partial g_{t,i}(\btheta_{0})}{\partial\theta_k}\bigg|^{\gamma}=|\ba_j^0|_1^{\gamma-1}\sum_{i=1}^{r}{|a_{j,i}^0|}\bigg|\frac{\partial g_{t,i}(\btheta_{0})}{\partial\theta_k}\bigg|^{\gamma}\,.
\end{align*}
By Conditions \ref{con:moments2}(b) and \ref{con:alpha}, 
\begin{align*}
\max_{j\in[m]}\max_{t\in[n]}\max_{k\in[p]}\mathbb{E}\bigg\{\bigg|\frac{\partial f_{t,j}^{\bfA}(\btheta_{0})}{\partial\theta_k}\bigg|^{\gamma}\bigg\}\leq \max_{j\in[m]}|\ba_j^0|_1^{\gamma}\cdot\max_{t\in[n]}\max_{i\in[r]}\max_{k\in[p]}\mathbb{E}\biggl\{\bigg|\frac{\partial g_{t,i}(\btheta_{0})}{\partial\theta_k}\bigg|^{\gamma}\biggr\}\lesssim1\,.
\end{align*}
Analogous to \eqref{eq:hdavydov},  we can also show 
\begin{align*}
\max_{k\in\mathcal{M}}\mathbb{E}\bigg\{\sum_{j=1}^{m}\bigg(\frac{1}{n}\sum_{t=1}^{n}\bigg[\frac{\partial f_{t,j}^{\bfA}(\btheta_{0})}{\partial\theta_k}-\mathbb{E}\biggl\{\frac{\partial f_{t,j}^{\bfA}(\btheta_{0})}{\partial\theta_k}\biggr\}\bigg]\bigg)^2\bigg\}=O(L_nn^{-1})\,.
\end{align*}
Together  with  
\eqref{eq:hGamma}, we have  
  ${\rm III}=|\bz|_2\cdot O_{\p}(L_n^{1/2}n^{-1/2})$. $\hfill\Box$

\subsection{Proof of Lemma \ref{hhat-hatVstarL2norm}}\label{sec:la:hhat-hatVstarL2norm}
Analogous to \eqref{fttheta},  $\max_{t\in[n]}|                                              \bff_t^{\bfA_n}(\tilde{\btheta}_{\calM},\hat{\btheta}_{n,\calM^{\rm c}})|_2=O_{\p}(n^{1/\gamma})$. By Lemma \ref{lambdahatstarL2norm},
 $|\tilde{\blambda}^*|_2=O_{\p}(L_n^{1/2}n^{-1/2})$.
 Due to $L_n\ll n^{1-2/\gamma}$, then
$\max_{t\in[n]}|\tilde{\blambda}^{*,\T}\bff_t^{\bfA_n}(\tilde{\btheta}_{\calM},\hat{\btheta}_{n,\calM^{\rm c}})|=O_{\p}(L_n^{1/2}n^{-1/2+1/\gamma})=o_{\p}(1)$. Notice that $|(1+x)^{-2}-1|\leq C|x|$ for any $|x|<1/2$. Under Conditions \ref{con:moments1}(c) and \ref{con:alpha}, by   Lemma \ref{hVhatL2norm}, we know the eigenvalues of $\widehat{\bfV}_{\bff^{\bfA_n}}(\tilde{\btheta}_{\calM},\hat{\btheta}_{n,\calM^{\rm c}})$ are uniformly bounded away from infinity
 w.p.a.1.
Then it holds w.p.a.1 that 
\begin{align*}
\sup_{c\in(0,1)}\bigg\|\frac{1}{n}\sum_{t=1}^{n}&\frac{\bff_t^{\bfA_n}(\tilde{\btheta}_{\calM},\hat{\btheta}_{n,\calM^{\rm c}})^{\otimes2}}{\{1+c\tilde{\blambda}^{*,\T}\bff_t^{\bfA_n}(\tilde{\btheta}_{\calM},\hat{\btheta}_{n,\calM^{\rm c}})\}^2}-\widehat{\bfV}_{\bff^{\bfA_n}}(\tilde{\btheta}_{\calM},\hat{\btheta}_{n,\calM^{\rm c}})\bigg\|_2 \\=&~ \sup_{c\in(0,1)}\bigg\|\frac{1}{n}\sum_{t=1}^{n}\bff_t^{\bfA_n}(\tilde{\btheta}_{\calM},\hat{\btheta}_{n,\calM^{\rm c}})^{\otimes2}\big[\{1+c\tilde{\blambda}^{*,\T}\bff_t^{\bfA_n}(\tilde{\btheta}_{\calM},\hat{\btheta}_{n,\calM^{\rm c}})\}^{-2}-1\big] \bigg\|_2\\\leq &~
C\lambda_{\max}\{\widehat{\bfV}_{\bff^{\bfA_n}}(\tilde{\btheta}_{\calM},\hat{\btheta}_{n,\calM^{\rm c}})\}\cdot\max_{t\in[n]}|\tilde{\blambda}^{*,\T}\bff_t^{\bfA_n}(\tilde{\btheta}_{\calM},\hat{\btheta}_{n,\calM^{\rm c}})|=
O_{\p}(L_n^{1/2}n^{-1/2+1/\gamma})\,.
\end{align*}                                We have the first result    of Lemma \ref{hhat-hatVstarL2norm}.  
Recall $\widehat{\bGamma}_{\bff^{\bfA_n}}  (\tilde{\btheta}_{\calM},\hat{\btheta}_{n,\calM^{\rm c}})=\nabla_{\btheta_{\calM}}\bar{\bff} ^{\bfA_n}(\tilde{\btheta}_{\calM},\hat{\btheta}_{n,\calM^{\rm c}})$. By  Taylor expansion and the Cauchy-Schwarz inequality, it holds w.p.a.1 that
\begin{align*}
&\bigg|\biggl\{\frac{1}{n}\sum_{t=1}^{n}\frac{\nabla_{\btheta_{\calM}}\bff_t^{\bfA_n}(\tilde{\btheta}_{\calM},\hat{\btheta}_{n,\calM^{\rm c}})}{1+\tilde{\blambda}^{*,\T}\bff_t^{\bfA_n}(\tilde{\btheta}_{\calM},\hat{\btheta}_{n,\calM^{\rm c}})}-\widehat{\bGamma}_{\bff^{\bfA_n}}  (\tilde{\btheta}_{\calM},\hat{\btheta}_{n,\calM^{\rm c}})\biggr\}\bfz\bigg|_2^2\\ &~~~~~\leq
\bigg|\frac{1}{n}\sum_{t=1}^{n}\frac{\tilde{\blambda}^{*,\T}\bff_t^{\bfA_n}(\tilde{\btheta}_{\calM},\hat{\btheta}_{n,\calM^{\rm c}})^{\otimes2}\tilde{\blambda}^*}{\{1+\bar{c}\tilde{\blambda}^{*,\T}\bff_t^{\bfA_n}(\tilde{\btheta}_{\calM},\hat{\btheta}_{n,\calM^{\rm c}})\}^4}\bigg|\cdot \bigg|\frac{1}{n}\sum_{t=1}^{n}\bfz^{\T}\{\nabla_{\btheta_{\calM}}\bff_t^{\bfA_n}(\tilde{\btheta}_{\calM},\hat{\btheta}_{n,\calM^{\rm c}})\}^{\T,\otimes2}\bfz\bigg|\\ &~~~~~\leq
C|\tilde{\blambda}^{*,\T}\widehat{\bfV}_{\bff^{\bfA_n}}(\tilde{\btheta}_{\calM},\hat{\btheta}_{n,\calM^{\rm c}})\tilde{\blambda}^{*}|\cdot\bigg|\frac{1}{n}\sum_{t=1}^{n}\bfz^{\T}\{\nabla_{\btheta_{\calM}}\bff_t^{\bfA_n}(\tilde{\btheta}_{\calM},\hat{\btheta}_{n,\calM^{\rm c}})\}^{\T,\otimes2}\bfz\bigg|
\end{align*}
for some $\bar{c}\in(0,1)$. Notice that
\begin{align*}
|\tilde{\blambda}^{*,\T}\widehat{\bfV}_{\bff^{\bfA_n}}(\tilde{\btheta}_{\calM},\hat{\btheta}_{n,\calM^{\rm c}})\tilde{\blambda}^{*}|\leq \lambda_{\max}\{\widehat{\bfV}_{\bff^{\bfA_n}}(\tilde{\btheta}_{\calM},\hat{\btheta}_{n,\calM^{\rm c}})\}|\tilde{\blambda}^{*}|_2^2=  O_{\p}(L_nn^{-1})\,.
\end{align*}  
By the Cauchy-Schwarz inequality and Lemma \ref{hL2norm}, we have 
\begin{align*}
&\bigg|\frac{1}{n}\sum_{t=1}^{n}\bfz^{\T}\{\nabla_{\btheta_{\calM}}\bff_t^{\bfA_n}(\tilde{\btheta}_{\calM},\hat{\btheta}_{n,\calM^{\rm c}})\}^{\T,\otimes2}\bfz\bigg|\leq \sum_{k=1}^{m}\sum_{l\in\mathcal{M}}	 \frac{1}{n}\sum_{t=1}^{n}\bigg|\frac{\partial f_{t,k}^{\bfA_n}(\tilde{\btheta}_{\calM},\hat{\btheta}_{n,\calM^{\rm c}})}{\partial \theta_{l}}\bigg|^2|\bfz|_2^2\\&~~~~~~\leq 
m^2|\bfz|_2^2\max_{k\in [m]}	\max_{l\in\mathcal{M}}\bigg\{\frac{1}{n}\sum_{t=1}^{n}\bigg|\frac{\partial f_{t,k}^{\bfA_n}(\tilde{\btheta}_{\calM},\hat{\btheta}_{n,\calM^{\rm c}})}{\partial \theta_{l}}\bigg|^2\bigg\}=
|\bfz|_2^2\cdot O_{\p}(1)\,.
\end{align*}      
Therefore, 
\begin{align*}
\bigg|\biggl\{\frac{1}{n}\sum_{t=1}^{n}\frac{\nabla_{\btheta_{\calM}}\bff_t^{\bfA_n}(\tilde{\btheta}_{\calM},\hat{\btheta}_{n,\calM^{\rm c}})}{1+\tilde{\blambda}^{*,\T}\bff_t^{\bfA_n}(\tilde{\btheta}_{\calM},\hat{\btheta}_{n,\calM^{\rm c}})}-\widehat{\bGamma}_{\bff^{\bfA_n}}  (\tilde{\btheta}_{\calM},\hat{\btheta}_{n,\calM^{\rm c}})\biggr\}\bfz\bigg|_2=|\bfz|_2\cdot O_{\p}(L_n^{1/2}n^{-1/2})\,.
\end{align*}        
 We complete the proof of Lemma \ref{hhat-hatVstarL2norm}.                               $\hfill\Box$

            \subsection{Proof of Lemma \ref{longruncov}}
Write $\bXi^*_{\bff^{\bA}}(\btheta_0)=\sum_{j=-n+1}^{n-1}\mathcal{K}(j/h_n)\bH_j(\btheta_0)$ with 
\begin{equation*}
	\bH_j(\btheta_0)=\left\{
	\begin{aligned}
		\frac{1}{n}\sum_{t=j+1}^{n}\mathbb{E}\{\bff_{t}^{\bA}(\btheta_{0})\bff_{t-j}^{\bA}(\btheta_{0})^{\T}\}\,, ~~&\textrm{if}~j\geq 0\,, 
		\\
		\frac{1}{n}\sum_{t=1-j}^{n}\mathbb{E}\{\bff_{t+j}^{\bA}(\btheta_{0})\bff_{t}^{\bA}(\btheta_{0})^{\T}\}\,,  ~~& \textrm{if}~j<0\,.
	\end{aligned}
	\right.
\end{equation*}
 By the triangle inequality, we have 
\begin{align*}
	\|\widehat{\bXi}_{\bff^{\bA_n}}(\tilde{\btheta}_{\calM},\hat{\btheta}_{n,\calM^{\rm c}})-{\bXi}_{\bff^{\bA}}(\btheta_0)\|_2\leq \underbrace{\|\bXi^*_{\bff^{\bA}}(\btheta_{0})-{\bXi}_{\bff^{\bA}}(\btheta_0)\|_2}_{\rm I}+ \underbrace{\|\widehat{\bXi}_{\bff^{\bA_n}}(\tilde{\btheta}_{\calM},\hat{\btheta}_{n,\calM^{\rm c}})-{\bXi}^*_{\bff^{\bA}}(\btheta_0)\|_2}_{\rm II}\,.
\end{align*}
As we will show in Sections \ref{sec:xi1} and \ref{sec:xi2},
\begin{align*}
    {\rm I} = O(L_n^{2}h_n^{-1})~~\text{and}~~
     {\rm II}=O_{\p}(s^{1/2}\ell_n^{1/2}\nu h_n)+O_{\p}(h_n\omega_n)+O_{\p}(L_n^{1/2}h_nn^{-1/2})\,,
\end{align*}
which implies 
\begin{align*}
	\|\widehat{\bXi}_{\bff^{\bA_n}}(\tilde{\btheta}_{\calM},\hat{\btheta}_{n,\calM^{\rm c}})-\bXi_{\bff^{\bA}}(\btheta_0)\|_2=O_{\p}(s^{1/2}\ell_n^{1/2}\nu h_n)+O_{\p}(h_n\omega_n)+O_{\p}(L_n^{1/2}h_nn^{-1/2})+O(L_n^{2}h_n^{-1})\,.
\end{align*}
We complete the proof of Lemma \ref{longruncov}. $\hfill\Box$

\subsubsection{Convergence rate of {\rm I}}\label{sec:xi1}

 Recall $\bXi_{\bff^{\bA}}(\btheta_0)={\rm Var}\{n^{1/2}\bar{\bff}^{\bA}(\btheta_0)\}$.
Write $\bXi_{\bff^{\bA}}(\btheta_0)=(\sigma_{l_1,l_2})_{m\times m}$ and $\bH_j(\btheta_{0})=(H_{j,l_1,l_2})_{m\times m}$. For any $l_1,l_2\in[m]$, $\sigma_{l_1,l_2}=H_{0,l_1,l_2}+\sum_{j=1}^{n-1}H_{-j,l_1,l_2}+\sum_{j=1}^{n-1}H_{j,l_1,l_2}$.
Since 
$
\max_{t\in[n]}\max_{l\in[m]}\mathbb{E}\{|f_{t,l}^{\bA}(\btheta_{0})|^{\gamma}\}\lesssim 1\,,$
under Condition \ref{con:mixingdecay}  the  triangle inequality and  Davydov's inequality give
\begin{align*}
    |H_{j,l_1,l_2}|\leq \frac
    {1}{n}\sum_{t=j+1}^{n}|\mathbb{E}\{f_{t,l_1}^{\bA}(\btheta_{0})f_{t-j,l_2}^{\bA}(\btheta_{0})\}|\leq C \exp(-CL_n^{-\varphi}j^{\varphi})
\end{align*}
for any $j\geq 1$. This bound also holds for $|H_{-j,l_1,l_2}|$ with $j\geq 1$. Write $\bXi^*_{\bff^{\bA}}(\btheta_0)=(\sigma^*_{l_1,l_2})_{m\times m}$ with  $\sigma^*_{l_1,l_2}=\sum_{j=-n+1}^{n-1}\mathcal{K}(j/h_n)H_{j,l_1,l_2}$. Under Condition \ref{con:kernel}, by the triangle inequality and  Taylor expansion, 
\begin{align*}
	{\rm I}&\leq \bigg[\sum_{l_1,l_2=1}^{m}\bigg\{\sum_{j=1}^{n-1}\bigg|\mathcal{K}\bigg(\frac{j}{h_n}\bigg)-1\bigg|(|H_{j,l_1,l_2}|+|H_{-j,l_1,l_2}|)\bigg\}^2\bigg]^{1/2}
    \\&\lesssim \bigg[\sum_{l_1,l_2=1}^{m}\bigg\{\frac{1}{h_n}\sum_{j=1}^{n-1}j(|H_{j,l_1,l_2}|+|H_{-j,l_1,l_2}|)\bigg\}^2\bigg]^{1/2}
     \\&\lesssim 
     \frac{1}{h_n}\sum_{j=1}^{n-1}j\exp(-CL_n^{-\varphi}j^{\varphi})
    \lesssim L_n^{2}h_n^{-1}\,.
\end{align*}
We have constructed the convergence rate of ${\rm I}$.
 $\hfill\Box$

\subsubsection{Convergence rate of {\rm II}}\label{sec:xi2}
Define  \begin{equation*}
	\widehat{\bH}^{(1)}_j(\btheta_0)=\left\{
	\begin{aligned}
		\frac{1}{n}\sum_{t=j+1}^{n}\bff_{t}^{\bA_n}(\btheta_{0})\bff_{t-j}^{\bA_n}(\btheta_{0})^{\T}\,, ~~&\textrm{if}~j\geq 0\,, 
		\\
		\frac{1}{n}\sum_{t=1-j}^{n}\bff_{t+j}^{\bA_n}(\btheta_{0})\bff_{t}^{\bA_n}(\btheta_{0})^{\T}\,,  ~~& \textrm{if}~j<0\,,
	\end{aligned}
	\right.
\end{equation*}
and 
\begin{equation*}
	\widehat{\bH}^{(2)}_j(\btheta_0)=\left\{
	\begin{aligned}
		\frac{1}{n}\sum_{t=j+1}^{n}\bff_{t}^{\bA}(\btheta_{0})\bff_{t-j}^{\bA}(\btheta_{0})^{\T}\,, ~~&\textrm{if}~j\geq 0\,, 
		\\
		\frac{1}{n}\sum_{t=1-j}^{n}\bff_{t+j}^{\bA}(\btheta_{0})\bff_{t}^{\bA}(\btheta_{0})^{\T}\,,  ~~& \textrm{if}~j<0\,.
	\end{aligned}
	\right.
\end{equation*}
Recall $\widehat{\bXi}_{\bff^{\bA_n}}(\tilde{\btheta}_{\calM},\hat{\btheta}_{n,\calM^{\rm c}})=\sum_{j=-n+1}^{n-1}\mathcal{K}(j/h_n)\widehat{\bH}_j(\tilde{\btheta}_{\calM},\hat{\btheta}_{n,\calM^{\rm c}})$
with 
\begin{equation*}
	\widehat{\bH}_j(\tilde{\btheta}_{\calM},\hat{\btheta}_{n,\calM^{\rm c}})=\left\{
	\begin{aligned}
		\frac{1}{n}\sum_{t=j+1}^{n}\bff_t^{\bfA_n}(\tilde{\btheta}_{\calM},\hat{\btheta}_{n,\calM^{\rm c}})\bff_{t-j}^{\bfA_n}(\tilde{\btheta}_{\calM},\hat{\btheta}_{n,\calM^{\rm c}})^{\T}\,, ~~&\textrm{if}~j\geq 0\,, 
		\\
		\frac{1}{n}\sum_{t=1-j}^{n}\bff_{t+j}^{\bfA_n}(\tilde{\btheta}_{\calM},\hat{\btheta}_{n,\calM^{\rm c}})\bff_t^{\bfA_n}(\tilde{\btheta}_{\calM},\hat{\btheta}_{n,\calM^{\rm c}})^{\T}\,,  ~~& \textrm{if}~j<0\,.
	\end{aligned}
	\right.
\end{equation*}
By the  triangle inequality, we have 
\begin{align}\label{eq:longruncov}
	\bigg\|\sum_{j=0}^{n-1}\mathcal{K}\bigg(\frac{j}{h_n}\bigg)\{\widehat{\bH}_j(\tilde{\btheta}_{\calM},\hat{\btheta}_{n,\calM^{\rm c}})-\bH_j(\btheta_{0})\}\bigg\|_2
	\leq \|\bU_1\|_2+\|\bU_2\|_2+\|\bU_3\|_2\,,
\end{align}
where 
\begin{align*}
	&\bU_1=\sum_{j=0}^{n-1}\mathcal{K}\bigg(\frac{j}{h_n}\bigg)\{\widehat{\bH}_j(\tilde{\btheta}_{\calM},\hat{\btheta}_{n,\calM^{\rm c}})-\widehat{\bH}_j^{(1)}(\btheta_{0})\}\,,
	\\&\bU_2=\sum_{j=0}^{n-1}\mathcal{K}\bigg(\frac{j}{h_n}\bigg)\{\widehat{\bH}_j^{(1)}(\btheta_{0})-\widehat{\bH}_j^{(2)}(\btheta_{0})\}\,,
	\\&\bU_3=\sum_{j=0}^{n-1}\mathcal{K}\bigg(\frac{j}{h_n}\bigg)\{\widehat{\bH}_j^{(2)}(\btheta_{0})-\bH_j(\btheta_{0})\}\,.
\end{align*}
As we will show in Sections \ref{sec:u1}--\ref{sec:u3},
\begin{align*}
    \|\bU_1\|_2=O_{\p}(s^{1/2}\ell_n^{1/2}\nu h_n)\,,~\|\bU_2\|_2=O_{\p}(h_n\omega_n)~~\text{and}~~	\|\bU_3\|_2=O_{\p}(L_n^{1/2}h_nn^{-1/2})\,.
\end{align*}
By \eqref{eq:longruncov}, it holds that
\begin{align*}
	&\bigg\|\sum_{j=0}^{n-1}\mathcal{K}\bigg(\frac{j}{h_n}\bigg)\{\widehat{\bH}_j(\tilde{\btheta}_{\calM},\hat{\btheta}_{n,\calM^{\rm c}})-\bH_j(\btheta_{0})\}\bigg\|_2=O_{\p}(s^{1/2}\ell_n^{1/2}\nu h_n)+O_{\p}(h_n\omega_n)+O_{\p}(L_n^{1/2}h_nn^{-1/2})\,.
\end{align*}
Such a convergence rate also holds for $\|\sum_{j=-n+1}^{-1}\mathcal{K}({j}/{h_n})\{\widehat{\bH}_j(\tilde{\btheta}_{\calM},\hat{\btheta}_{n,\calM^{\rm c}})-\bH_j(\btheta_{0})\}\|_2$. Hence,
\begin{align}\label{eq:xiII}
   {\rm II}=O_{\p}(s^{1/2}\ell_n^{1/2}\nu h_n)+O_{\p}(h_n\omega_n)+O_{\p}(L_n^{1/2}h_nn^{-1/2})\,.
\end{align}
We have constructed the convergence rate of ${\rm II}$.
 $\hfill\Box$

\subsubsection{Convergence rate of $\|\bU_1\|_2$}\label{sec:u1}

Write $\bU_{1}=(U_{1,l_1,l_2})_{m\times m}$.
For any $l_1,l_2\in[m]$,
\begin{align*}
U_{1,l_1,l_2}&=\frac{1}{n} \sum_{j=0}^{n-1}\mathcal{K}\bigg(\frac{j}{h_n}\bigg)\sum_{t=j+1}^{n}\{f_{t,l_1}^{\bA_n}(\tilde{\btheta}_{\calM},\hat{\btheta}_{n,\calM^{\rm c}})f_{t-j,l_2}^{\bA_n}(\tilde{\btheta}_{\calM},\hat{\btheta}_{n,\calM^{\rm c}})-f_{t,l_1}^{\bA_n}(\btheta_{0})f_{t-j,l_2}^{\bA_n}(\btheta_{0})\}
	\\&~=\underbrace{\frac{1}{n} \sum_{j=0}^{n-1}\mathcal{K}\bigg(\frac{j}{h_n}\bigg)\sum_{t=j+1}^{n}f_{t,l_1}^{\bA_n}(\btheta_{0})\{f_{t-j,l_2}^{\bA_n}(\tilde{\btheta}_{\calM},\hat{\btheta}_{n,\calM^{\rm c}})-f_{t-j,l_2}^{\bA_n}(\btheta_{0})\}}_{U^{(1)}_{1,l_1,l_2}}
	\\&~~~~~~~~+\underbrace{\frac{1}{n} \sum_{j=0}^{n-1}\mathcal{K}\bigg(\frac{j}{h_n}\bigg)\sum_{t=j+1}^{n}f_{t-j,l_2}^{\bA_n}(\tilde{\btheta}_{\calM},\hat{\btheta}_{n,\calM^{\rm c}})\{f_{t,l_1}^{\bA_n}(\tilde{\btheta}_{\calM},\hat{\btheta}_{n,\calM^{\rm c}})-f_{t,l_1}^{\bA_n}(\btheta_{0})\}}_{U^{(2)}_{1,l_1,l_2}}
	\,.
\end{align*}
Write $\bar{\btheta}=(\tilde{\btheta}_{\calM}^{\T},\hat{\btheta}_{n,\calM^{\rm c}}^{\T})^{\T}$. Recall 
$|\tilde{\btheta}_{\calM}-\btheta_{0,\calM}|_1=O_{\p}(\ell_n^{1/2}\nu)$  and $
|\hat{\btheta}_{n,\calM^{\rm c}}-{\btheta}_{0,\calM^{\rm c}}|_1 = O_{\p}(s^{1/2}\ell_n^{1/2}\nu)$.
By Condition \ref{con:kernel},  $\sum_{j=0}^{n}|\mathcal{K}(j/h_n)|\lesssim h_n$. 
Under Conditions \ref{con:moments1}(a), \ref{con:moments2}(a) and \ref{con:alpha},
by Taylor expansion, the Cauchy-Schwarz inequality and    Lemma \ref{hL2norm}, we have  
\begin{align*}
	\max_{l_1,l_2\in[m]}|U^{(1)}_{1,l_1,l_2}|=&~ \max_{l_1,l_2\in[m]}\bigg|\sum_{j=0}^{n-1}\mathcal{K}\bigg(\frac{j}{h_n}\bigg)\frac{1}{n}\sum_{t=j+1}^{n}f_{t,l_1}^{\bA_n}(\btheta_{0})\frac{\partial f_{t-j,l_2}^{\bA_n}(\check{\btheta})}{\partial\btheta^{\T}}(\bar{\btheta}-\btheta_0)\bigg|
	\\ \leq &~ \max_{l_1,l_2\in[m]}\sum_{j=0}^{n-1}\bigg|\mathcal{K}\bigg(\frac{j}{h_n}\bigg)\bigg|\frac{1}{n}\sum_{t=j+1}^{n}|f_{t,l_1}^{\bA_n}(\btheta_{0})|\bigg|\frac{\partial f_{t-j,l_2}^{\bA_n}(\check{\btheta})}{\partial\btheta^{\T}}(\bar{\btheta}-\btheta_0)\bigg|
		\\ \leq &~ \max_{l_1,l_2\in[m]}\sum_{j=0}^{n-1}\bigg|\mathcal{K}\bigg(\frac{j}{h_n}\bigg)\bigg|\bigg\{\frac{1}{n}\sum_{t=j+1}^{n}|f_{t,l_1}^{\bA_n}(\btheta_{0})|^2\bigg\}^{1/2}\bigg\{\frac{1}{n}\sum_{t=j+1}^{n}\bigg|\frac{\partial f_{t-j,l_2}^{\bA_n}(\check{\btheta})}{\partial\btheta^{\T}}(\bar{\btheta}-\btheta_0)\bigg|^2\bigg\}^{1/2}
		\\ \leq &~ \sum_{j=0}^{n-1}\bigg|\mathcal{K}\bigg(\frac{j}{h_n}\bigg)\bigg|\cdot\max_{l_1\in[m]}\bigg\{\frac{1}{n}\sum_{t=1}^{n}|f_{t,l_1}^{\bA_n}(\btheta_{0})|^2\bigg\}^{1/2}\cdot\max_{l_2\in[m]}\bigg\{\frac{1}{n}\sum_{t=1}^{n}\bigg|\frac{\partial f_{t,l_2}^{\bA_n}(\check{\btheta})}{\partial\btheta^{\T}}(\bar{\btheta}-\btheta_0)\bigg|^2\bigg\}^{1/2}
	\\\leq &~O_{\p}(h_n)\cdot|\bar{\btheta}-\btheta_{0}|_1=O_{\p}(s^{1/2}\ell_n^{1/2}\nu h_n)\,,
\end{align*}
where $\check{\btheta}$ lies on the line between $\bar{\btheta}$ and $\btheta_{0}$.
Analogously, we also have 
$
\max_{l_1,l_2\in[m]}|U^{(2)}_{1,l_1,l_2}|=O_{\p}(s^{1/2}\ell_n^{1/2}\nu h_n)$.
Hence  
$
\max_{l_1,l_2\in[m]}|U_{1,l_1,l_2}|=O_{\p}(s^{1/2}\ell_n^{1/2}\nu h_n)$, which further implies
$
	\|\bU_1\|_2=O_{\p}(s^{1/2}\ell_n^{1/2}\nu h_n)$.
$\hfill\Box$

\subsubsection{Convergence rate of $\|\bU_2\|_2$}\label{sec:u2}
Write $\bU_{2}=(U_{2,l_1,l_2})_{m\times m}$.
For any $l_1,l_2\in[m]$, notice that
\begin{align*}
	U_{2,l_1,l_2}&=\frac{1}{n} \sum_{j=0}^{n-1}\mathcal{K}\bigg(\frac{j}{h_n}\bigg)\sum_{t=j+1}^{n}\{f_{t,l_1}^{\bA_n}(\btheta_{0})f_{t-j,l_2}^{\bA_n}(\btheta_{0})-f_{t,l_1}^{\bA}(\btheta_{0})f_{t-j,l_2}^{\bA}(\btheta_{0})\}
	\\&~=\underbrace{\frac{1}{n} \sum_{j=0}^{n-1}\mathcal{K}\bigg(\frac{j}{h_n}\bigg)\sum_{t=j+1}^{n}f_{t,l_1}^{\bA}(\btheta_{0})\{f_{t-j,l_2}^{\bA_n}(\btheta_{0})-f_{t-j,l_2}^{\bA}(\btheta_{0})\}}_{U^{(1)}_{2,l_1,l_2}}
	\\&~~~~~~~~+\underbrace{\frac{1}{n} \sum_{j=0}^{n-1}\mathcal{K}\bigg(\frac{j}{h_n}\bigg)\sum_{t=j+1}^{n}f_{t-j,l_2}^{\bA_n}(\btheta_{0})\{f_{t,l_1}^{\bA_n}(\btheta_{0})-f_{t,l_1}^{\bA}(\btheta_{0})\}}_{U^{(2)}_{2,l_1,l_2}}\,.
\end{align*}
Recall $f_{t,l}^{\bA_n}(\btheta_{0})-f_{t,l}^{\bA}(\btheta_{0})=(\ba^n_l-\ba_l^0)^{\T}\bg_{t}(\btheta_0)$ for any $l\in[m]$. Under Conditions \ref{con:moments1}(a) and  \ref{con:alpha}, by the Cauchy-Schwarz inequality, we have
\begin{align*}
	\max_{l_1,l_2\in[m]}|U^{(1)}_{2,l_1,l_2}|=&~\max_{l_1,l_2\in[m]} \bigg|\sum_{j=0}^{n-1}\mathcal{K}\bigg(\frac{j}{h_n}\bigg)\frac{1}{n}\sum_{t=j+1}^{n}f_{t,l_1}^{\bA}(\btheta_{0})(\ba^n_{l_2}-\ba_{l_2}^0)^{\T}\bg_{t-j}(\btheta_0)\bigg|
	\\\leq &~\max_{l_1,l_2\in[m]}\sum_{j=0}^{n-1}\bigg|\mathcal{K}\bigg(\frac{j}{h_n}\bigg)\bigg|\bigg\{\frac{1}{n}\sum_{t=1}^{n}|f_{t,l_1}^{\bA}(\btheta_{0})|^2\bigg\}^{1/2} \bigg\{\frac{1}{n}\sum_{t=1}^{n}|(\ba^n_{l_2}-\ba_{l_2}^0)^{\T}\bg_{t}(\btheta_0)|^2\bigg\}^{1/2}
	\\\leq &~O_{\p}(h_n)\cdot\max_{l_2\in[m]}|\ba^{n}_{l_2}-\ba_{l_2}^0|_1
	= O_{\p}(h_n\omega_n)\,.
\end{align*}
Analogously, we also have $\max_{l_1,l_2\in[m]}|U^{(2)}_{2,l_1,l_2}|=O_{\p}(h_n\omega_n)$. Hence, $\max_{l_1,l_2\in[m]}|U_{2,l_1,l_2}|=O_{\p}(h_n\omega_n)$, which implies $
	\|\bU_2\|_2=O_{\p}(h_n\omega_n)$.
$\hfill\Box$

\subsubsection{Convergence rate of $\|\bU_3\|_2$}\label{sec:u3}
Write $\bU_{3}=(U_{3,l_1,l_2})_{m\times m}$ and  $Q_{t,l}=f_{t,l}^{\bA}(\btheta_{0})$.
Then 
for any $l_1,l_2\in[m]$, 
\begin{align*}
U_{3,l_1,l_2}=\sum_{j=0}^{n-1}\mathcal{K}\bigg(\frac{j}{h_n}\bigg)\bigg[\frac{1}{n}\sum_{t=j+1}^{n}\{Q_{t,l_1}Q_{t-j,l_2}-\mathbb{E}(Q_{t,l_1}Q_{t-j,l_2})\}\bigg]=\sum_{j=0}^{n-1}\mathcal{K}\bigg(\frac{j}{h_n}\bigg)\bigg\{\frac{1}{n}\sum_{t=1}^{n-j}\zeta_{t,l_1,l_2}^{(j)}\bigg\}\,,
\end{align*}
where  $\zeta_{t,l_1,l_2}^{(j)}=Q_{t+j,l_1}Q_{t,l_2}-\mathbb{E}(Q_{t+j,l_1}Q_{t,l_2})$. 
By Conditions \ref{con:moments1}(a) and \ref{con:alpha},
\begin{align*}
	\max_{j\in {\{0\}}\cup[n-1]}\max_{t\in\{j+1,\ldots,n\}}\max_{l_1,l_2\in[m]}\mathbb{E}\{|\zeta_{t,l_1,l_2}^{(j)}|^{\gamma}\}\lesssim 1\,.
\end{align*}
By the triangle inequality, we have
\begin{align*}
&{\rm Var}\bigg\{\frac{1}{n}\sum_{t=1}^{n-j}\zeta_{t,l_1,l_2}^{(j)}\bigg\}\leq \frac{1}{n^2}\sum_{t_1=1}^{n-j}\sum_{t_2=1}^{n-j}\big|{\rm Cov}\{\zeta_{t_1,l_1,l_2}^{(j)},\zeta_{t_2,l_1,l_2}^{(j)}\}\big|
\leq \frac{C}{n}+\frac{C}{n^2}\sum_{t_1<t_2}\big|{\rm Cov}\{\zeta_{t_1,l_1,l_2}^{(j)},\zeta_{t_2,l_1,l_2}^{(j)}\}\big|
\\&~~~~~~~~~= \frac{C}{n}+\underbrace{\frac{C}{n^2}\sum_{\substack{1\leq t_1,t_2\leq n-j\\1\leq t_2-t_1\leq j}}\big|{\rm Cov}\{\zeta_{t_1,l_1,l_2}^{(j)},\zeta_{t_2,l_1,l_2}^{(j)}\}\big|}_{{\rm I}_{{l_1,l_2}}^{(j)}}+\underbrace{\frac{C}{n^2}\sum_{\substack{1\leq t_1,t_2\leq n-j\\j+1\leq t_2-t_1\leq n-j-1}}\big|{\rm Cov}\{\zeta_{t_1,l_1,l_2}^{(j)},\zeta_{t_2,l_1,l_2}^{(j)}\}\big|}_{{\rm II}_{{l_1,l_2}}^{(j)}}\,.
\end{align*}
Here we adopt the convention ${\rm II}_{{l_1,l_2}}^{(j)}=0$ if $j+1>n-j-1$. 

When $1\leq t_2-t_1\leq j$, by the triangle inequality, 
\begin{align*}
\big|{\rm Cov}\{\zeta_{t_1,l_1,l_2}^{(j)},\zeta_{t_2,l_1,l_2}^{(j)}\}\big|
&=\big|\mathbb{E}(Q_{t_1+j,l_1}Q_{t_1,l_2}Q_{t_2+j,l_1}Q_{t_2,l_2})-\mathbb{E}(Q_{t_1+j,l_1}Q_{t_1,l_2})\mathbb{E}(Q_{t_2+j,l_1}Q_{t_2,l_2})\big|
\\&\leq\big|\mathbb{E}(Q_{t_1+j,l_1}Q_{t_1,l_2}Q_{t_2+j,l_1}Q_{t_2,l_2})-\mathbb{E}(Q_{t_1,l_2}Q_{t_2,l_2})\mathbb{E}(Q_{t_1+j,l_1}Q_{t_2+j,l_1})\big|\\&~~~~~+\big|\mathbb{E}(Q_{t_1,l_2}Q_{t_2,l_2})\mathbb{E}(Q_{t_1+j,l_1}Q_{t_2+j,l_1})-\mathbb{E}(Q_{t_1+j,l_1}Q_{t_1,l_2})\mathbb{E}(Q_{t_2+j,l_1}Q_{t_2,l_2})\big|\,.
\end{align*}
By Davydov’s inequality and Condition \ref{con:mixingdecay}, we have 
\begin{align*}
&~~~~~~~~~~~~~~~~\max_{l\in[m]}|\mathbb{E}(Q_{k_1,l}Q_{k_2,l})|\leq C\exp(-CL_n^{-\varphi}|k_1-k_2|^{\varphi})\,,
\\
&\max_{l_1,l_2\in[m]}\big|\mathbb{E}(Q_{t_1+j,l_1}Q_{t_1,l_2}Q_{t_2+j,l_1}Q_{t_2,l_2})-\mathbb{E}(Q_{t_1,l_2}Q_{t_2,l_2})\mathbb{E}(Q_{t_1+j,l_1}Q_{t_2+j,l_1})\big|\\&~~~~~~~~~~~~~~~~\leq C \exp(-CL_n^{-\varphi}|t_1+j-t_2|^{\varphi})\,.
\end{align*}
Hence, 
\begin{align*}
{\rm I}_{{l_1,l_2}}^{(j)}\leq&~ \frac{C}{n^2}\sum_{\substack{1\leq t_1,t_2\leq n-j\\1\leq t_2-t_1\leq j}}\exp(-CL_n^{-\varphi}|t_1+j-t_2|^{\varphi})\\&~~~~
+\frac{C}{n^2}\sum_{\substack{1\leq t_1,t_2\leq n-j\\1\leq t_2-t_1\leq j}}\exp(-CL_n^{-\varphi}|t_1-t_2|^{\varphi})
+\frac{C}{n^2}\sum_{\substack{1\leq t_1,t_2\leq n-j\\1\leq t_2-t_1\leq j}}\exp(-CL_n^{-\varphi}j^{\varphi})
\\\leq &~ \frac{C}{n}\sum_{k=0}^{j}\exp(-CL_n^{-\varphi}|j-k|^{\varphi})
+\frac{C}{n}\sum_{k=1}^{j}\exp(-CL_n^{-\varphi}k^{\varphi})+\frac{C}{n}\sum_{k=1}^{j}\exp(-CL_n^{-\varphi}k^{\varphi})
\\\leq &~ \frac{C}{n}\sum_{k=0}^{j}\exp(-CL_n^{-\varphi}k^{\varphi})\leq \frac{CL_n}{n}\,.
\end{align*}
When $j+1\leq t_2-t_1\leq n-j-1$, by Davydov’s inequality and Condition \ref{con:mixingdecay},
\begin{align*}
    \max_{l_1,l_2\in[m]}\big|{\rm Cov}\{\zeta_{t_1,l_1,l_2}^{(j)},\zeta_{t_2,l_1,l_2}^{(j)}\}\big|\leq C\exp(-CL_n^{-\varphi}|t_2-t_1-j|^{\varphi})\,,
\end{align*}
 which implies 
\begin{align*}
	{\rm II}_{{l_1,l_2}}^{(j)}\leq \frac{C}{n^2}\sum_{\substack{1\leq t_1,t_2\leq n-j\\j+1\leq t_2-t_1\leq n-j-1}}\exp(-CL_n^{-\varphi}|t_2-t_1-j|^{\varphi})\leq \frac{C}{n}\sum_{k=1}^{n-2j-1}\exp(-CL_n^{-\varphi}k^{\varphi})\leq \frac{CL_n}{n}\,.
\end{align*}
Hence, we have
\begin{align*}
\max_{j\in {\{0\}}\cup[n-1]}\max_{l_1,l_2\in[m]}{\rm Var}\bigg\{\frac{1}{n}\sum_{t=1}^{n-j}\zeta_{t,l_1,l_2}^{(j)}\bigg\}\leq \frac{CL_n}{n}\,.
\end{align*}
Due to $\sum_{j=0}^{n-1}|\mathcal{K}(j/h_n)|\lesssim h_n$, by Jensen’s inequality,
\begin{align*}
\max_{l_1,l_2\in[m]}\mathbb{E}(|U_{3,l_1,l_2}|^2)\leq\sum_{j=0}^{n-1}\bigg|\mathcal{K}\bigg(\frac{j}{h_n}\bigg)\bigg|\cdot\bigg[ \sum_{j=0}^{n-1}\bigg|\mathcal{K}\bigg(\frac{j}{h_n}\bigg)\bigg|\max_{l_1,l_2\in[m]}{\rm Var}\bigg\{\frac{1}{n}\sum_{t=1}^{n-j}\zeta_{t,l_1,l_2}^{(j)}\bigg\}\bigg]\lesssim \frac{L_nh_n^2}{n}\,,
\end{align*}
which implies
$
	\|\bU_3\|_2=O_{\p}(L_n^{1/2}h_nn^{-1/2})$.
$\hfill\Box$

\subsection{Proof of Lemma \ref{cltZhat}}\label{sec:la:cltZhat}
Write
\begin{align*}
\bJ_{\bff^{\bfA}}=\{\mathbb{E}\{\widehat{\bGamma}^{\T}_{\bff^{\bfA}}(\btheta_{0})\}[\mathbb{E}\{\widehat{\bfV}_{\bff^{\bfA}}(\btheta_0)\}]^{-1}\bXi_{\bff^{\bA}}^{1/2}(\btheta_0)\}^{\otimes2}
\end{align*}
with $\bXi_{\bff^{\bA}}(\btheta_0)={\rm Var}\{n^{1/2}\bar{\bff}^{\bA}(\btheta_0)\}$.
By Conditions \ref{con:moments1}(c), \ref{con:Gamma} and \ref{con:alpha}, we know the eigenvalues of $\bJ_{\bff^{\bfA}}$ are uniformly bounded away from zero and infinity.
Let $Z_{t,\bff^{\bA}}=\bz^{\T}\bJ_{\bff^{\bfA}}^{-1/2}\mathbb{E}\{\widehat{\bGamma}^{\T}_{\bff^{\bfA}}(\btheta_{0})\}[\mathbb{E}\{\widehat{\bfV}_{\bff^{\bfA}}(\btheta_0)\}]^{-1}\bff^{\bfA}_{t}(\btheta_{0})$ and $\bar{Z}_{\bff^{\bfA}}=\mathbb{E}_n(Z_{t,\bff^{\bA}})$.
Due to $|\bz|_2=1$, 
then  
\begin{align*}
    \mathbb{E}\{(\sqrt{n}\bar{Z}_{\bff^{\bfA}})^2\}&=\bz^{\T}\bJ_{\bff^{\bfA}}^{-1/2}\mathbb{E}\{\widehat{\bGamma}^{\T}_{\bff^{\bfA}}(\btheta_{0})\}
[\mathbb{E}\{\widehat{\bV}_{\bff^{\bfA}}(\btheta_0)\}]^{-1}\bXi_{\bff^{\bfA}}(\btheta_0)[\mathbb{E}\{\widehat{\bV}_{\bff^{\bfA}}(\btheta_0)\}]^{-1}\mathbb{E}\{\widehat{\bGamma}_{\bff^{\bfA}}(\btheta_0)\}\bJ_{\bff^{\bfA}}^{-1/2}\bz\\&=\bz^{\T}\bJ_{\bff^{\bfA}}^{-1/2}\bJ_{\bff^{\bfA}}\bJ_{\bff^{\bfA}}^{-1/2}\bz=1\,.
\end{align*}
Under Conditions \ref{con:moments1}(c), \ref{con:Gamma} and \ref{con:alpha}, by the Cauchy-Schwarz inequality, 
\begin{align*}
|{Z}_{t, \bff^{\bA}}|^2&=\big|\bz^{\T}\bJ_{\bff^{\bfA}}^{-1/2}\mathbb{E}\{\widehat{\bGamma}^{\T}_{\bff^{\bfA}}(\btheta_{0})\}[\mathbb{E}\{\widehat{\bfV}_{\bff^{\bfA}}(\btheta_0)\}]^{-1}\bff^{\bfA}_{t}(\btheta_{0})\big|^2\\&\leq 
|\bfz|_2^2\cdot \lambda_{\min}^{-1}(\bJ_{\bff^{\bfA}})\lambda_{\max}([\mathbb{E}\{\widehat{\bGamma}_{\bff^{\bfA}}(\btheta_{0})\}]^{\otimes 2})\lambda_{\min}^{-2}[\mathbb{E}\{\widehat{\bfV}_{\bff^{\bfA}}(\btheta_0)\}]\cdot|\bff^{\bfA}_{t}(\btheta_{0})|_2^2\,.
\end{align*}
Due to $\gamma>4$, by \eqref{eq:eftka} we have
$\max_{t\in[n],k\in[m]}\mathbb{E}\{|f_{t,k}^{\bfA}(\btheta_{0})|^3\}\lesssim 1$.
Since $m$ is a fixed constant, under Conditions \ref{con:moments1}(a) and \ref{con:alpha}
 Jensen's inequality produces 
\begin{align*}
\max_{t\in[n]}\mathbb{E}(|{Z}_{t,\bff^{\bA}}|^3)\lesssim\max_{t\in[n]}\mathbb{E}\bigg[\bigg\{\sum_{k=1}^{m}|f_{t,k}^{\bfA}(\btheta_{0})|^2\bigg\}^{3/2}\bigg]\lesssim\max_{t\in[n]}\mathbb{E}\bigg\{\sum_{k=1}^{m}|f_{t,k}^{\bfA}(\btheta_{0})|^3\bigg\}\lesssim 1\,.
\end{align*}
Applying  Lemma \ref{berry-essen mixng} with $n=\tilde{n}, \tilde{L}_{\tilde{n}}=L_n$, $\xi_t=Z_{t,{\bff^{\bfA}}}$, $\sigma^2_{\tilde{n}}=n$ and $b_0=1$, it holds that
\begin{align}\label{eq:clt}
\sup_{x\in\mathbb{R}}\big|\mathbb{P}(n^{1/2}\bar{Z}_{\bff^{\bfA}}\leq x)-\Phi(x)\big|\leq Cn^{-1/2}L_n^{2\varphi}(\log n)^2=o(1)\,
\end{align}
provided that $L_{n}^{4\varphi}(\log n)^{4}\ll n$.
Recall \begin{align*}
	\widehat{\bJ}_{\bff^{\bfA_{n}}}=\{\widehat{\bGamma}_{{\bff}^{\bfA_n}}^{\T}(\tilde{\btheta}_{\calM},\hat{\btheta}_{n,\calM^{\rm c}})\widehat{\bfV}^{-1}_{\bff^{\bfA_n}}(\tilde{\btheta}_{\calM},\hat{\btheta}_{n,\calM^{\rm c}})\widehat{\bXi}_{\bff^{\bA_n}}^{1/2}(\tilde{\btheta}_{\calM},\hat{\btheta}_{n,\calM^{\rm c}})\}^{\otimes2}\,.
	\end{align*}
Under Conditions \ref{con:moments1}(c) and \ref{con:alpha}, by   Lemma \ref{hVhatL2norm}, we know  $\lambda_{\min}[\mathbb{E}\{\widehat{\bfV}_{\bff^{\bfA}}(\btheta_0)\}]$ and 
$\lambda_{\min}\{\widehat{\bfV}_{\bff^{\bfA_n}}(\tilde{\btheta}_{\calM},\hat{\btheta}_{n,\calM^{\rm c}})\}$
  are uniformly bounded away from zero
 w.p.a.1, which implies
\begin{equation}\label{eq:hhatVinverse}
\begin{split}
&\|\widehat{\bfV}^{-1}_{\bff^{\bfA_n}}(\tilde{\btheta}_{\calM},\hat{\btheta}_{n,\calM^{\rm c}})-[\mathbb{E}\{\widehat{\bfV}_{\bff^{\bfA}}(\btheta_0)\}]^{-1}\|_2\\&~~~\leq  \|\widehat{\bfV}^{-1}_{\bff^{\bfA_n}}(\tilde{\btheta}_{\calM},\hat{\btheta}_{n,\calM^{\rm c}})\|_2\cdot\|\widehat{\bfV}_{\bff^{\bfA_n}}(\tilde{\btheta}_{\calM},\hat{\btheta}_{n,\calM^{\rm c}})-\mathbb{E}\{\widehat{\bfV}_{\bff^{\bfA}}(\btheta_0)\}\|_2\cdot \|[\mathbb{E}\{\widehat{\bfV}_{\bff^{\bfA}}(\btheta_0)\}]^{-1}\|_2\\&~~~\leq 
\lambda_{\min}^{-1}\{\widehat{\bfV}_{\bff^{\bfA_n}}(\tilde{\btheta}_{\calM},\hat{\btheta}_{n,\calM^{\rm c}})\}\cdot \|\widehat{\bfV}_{\bff^{\bfA_n}}(\tilde{\btheta}_{\calM},\hat{\btheta}_{n,\calM^{\rm c}})-\mathbb{E}\{\widehat{\bfV}_{\bff^{\bfA}}(\btheta_0)\}\|_2\cdot \lambda_{\min}^{-1}[\mathbb{E}\{\widehat{\bfV}_{\bff^{\bfA}}(\btheta_0)\}]
\\&~~~=O_{\p}(L_n^{1/2}n^{-1/2})+O_{\p}(\omega_n)+O_{\p}(s^{1/2}\ell_n^{1/2}\nu)\,.
\end{split}
\end{equation}
By Conditions \ref{con:Gamma}(b) and \ref{con:alpha}, we know $\lambda_{\min}\{\bXi_{\bff^{\bA}}(\btheta_{0})\}$ 
 is uniformly bounded away from zero. By Lemma \ref{longruncov}, if $s\ell_n\nu^2h_n^2=o(1)$, $h_n\omega_n=o(1)$, $L_nh_n^2\ll n$ and $L_n^2\ll h_n$, we have $\lambda_{\min}\{\widehat{\bXi}_{\bff^{\bA_n}}(\tilde{\btheta}_{\calM},\hat{\btheta}_{n,\calM^{\rm c}})$  is uniformly bounded away from zero
 w.p.a.1, which implies 
\begin{equation}\label{eq:xihat}
    \begin{split}
        &\|\widehat{\bXi}^{-1}_{\bff^{\bA_n}}(\tilde{\btheta}_{\calM},\hat{\btheta}_{n,\calM^{\rm c}})-\bXi^{-1}_{\bff^{\bA}}(\btheta_{0})\|_2\\&~~~~~\leq
	\lambda_{\min}^{-1}\{\widehat{\bXi}_{\bff^{\bA_n}}(\tilde{\btheta}_{\calM},\hat{\btheta}_{n,\calM^{\rm c}})\}\cdot\|\widehat{\bXi}_{\bff^{\bA_n}}(\tilde{\btheta}_{\calM},\hat{\btheta}_{n,\calM^{\rm c}})-\bXi_{\bff^{\bA}}(\btheta_{0})\|_2\cdot\lambda_{\min}^{-1}\{\bXi_{\bff^{\bA}}(\btheta_{0})\}
\\&~~~~~=O_{\p}(s^{1/2}\ell_n^{1/2}\nu h_n)+O_{\p}(h_n\omega_n)+O_{\p}(L_n^{1/2}h_nn^{-1/2})+O_{\p}(L_n^{2}h_n^{-1})\,.
    \end{split}
\end{equation}
As shown in Appendix \ref{sec:p.1}, we know the eigenvalues of $[\mathbb{E}\{\widehat{\bGamma}^{\T}_{\bff^{\bfA}}(\btheta_{0})\}]^{\otimes2}$ are uniformly bounded away from  infinity. By Lemma \ref{gammahatstarL2norm}, $\|\widehat{\bGamma}_{{\bff}^{\bfA_n}}(\tilde{\btheta}_{\calM},\hat{\btheta}_{n,\calM^{\rm c}})\|_2=O_{\p}(1)$. Hence,
by  the triangle inequality, Lemma \ref{gammahatstarL2norm}, \eqref{eq:hhatVinverse} and \eqref{eq:xihat}, it holds that
\begin{align*}
    \|\widehat{\bJ}_{\bff^{\bfA_{n}}}-\bJ_{\bff^{\bfA}}\|_2= O_{\p}(s^{1/2}\ell_n^{1/2}\nu h_n)+O_{\p}(h_n\omega_n)+O_{\p}(L_n^{1/2}h_nn^{-1/2})+O_{\p}(L_n^{2}h_n^{-1})\,.
\end{align*}
If $s\ell_n\nu^2h_n^2=o(1)$, $h_n\omega_n=o(1)$, $L_nh_n^2\ll n$ and $L_n^2\ll h_n$, then $\|\widehat{\bJ}^{-1}_{\bff^{\bfA_{n}}}\|_2$ is bounded from infinity w.p.a.1.
 By the same arguments for deriving (\ref{eq:hhatVinverse}),  
 \begin{align*}
     \|\widehat{\bJ}^{-1}_{\bff^{\bfA_{n}}}-\bJ^{-1}_{\bff^{\bfA}}\|_2=O_{\p}(s^{1/2}\ell_n^{1/2}\nu h_n)+O_{\p}(h_n\omega_n)+O_{\p}(L_n^{1/2}h_nn^{-1/2})+O_{\p}(L_n^{2}h_n^{-1})\,.
 \end{align*}
 By the Ando-Hemmen inequality in \cite{DELMoral2018},
 it holds that \begin{align*}
\|\widehat{\bJ}^{-1/2}_{\bff^{\bfA_{n}}}-\bJ^{-1/2}_{\bff^{\bfA}}\|_2 \leq&~  \|\widehat{\bJ}^{-1}_{\bff^{\bfA_{n}}}-\bJ^{-1}_{\bff^{\bfA}}\|_2\cdot O_{\p}(1)\\=&~O_{\p}(s^{1/2}\ell_n^{1/2}\nu h_n)+O_{\p}(h_n\omega_n)+O_{\p}(L_n^{1/2}h_nn^{-1/2})+O_{\p}(L_n^{2}h_n^{-1})\,.
\end{align*}
Write
 \begin{align*}
\widehat{G}^{\bfA_n}=\bfz^{\T}\widehat{\bfJ}_{\bff^{\bfA_{n}}}^{-1/2}\widehat{\bGamma}_{{\bff}^{\bfA_n}}^{\T}(\tilde{\btheta}_{\calM},\hat{\btheta}_{n,\calM^{\rm c}})\widehat{\bfV}^{-1}_{\bff^{\bfA_n}}(\tilde{\btheta}_{\calM},\hat{\btheta}_{n,\calM^{\rm c}})\bar{\bff}^{\bfA_n}(\btheta_{0})\,.
\end{align*} 
Recall $\bar{Z}_{\bff^{\bfA}}=\mathbb{E}_n(Z_{t,\bff^{\bA}})$ with $Z_{t,\bff^{\bA}}=\bz^{\T}\bJ_{\bff^{\bfA}}^{-1/2}\mathbb{E}\{\widehat{\bGamma}^{\T}_{\bff^{\bfA}}(\btheta_{0})\}[\mathbb{E}\{\widehat{\bfV}_{\bff^{\bfA}}(\btheta_0)\}]^{-1}\bff^{\bfA}_{t}(\btheta_{0})$.
By the triangle inequality, 
\begin{align}\label{eq:G}
|\widehat{G}^{\bfA_{n}}-\bar{Z}_{\bff^{\bfA}}|\leq R_1+R_2+R_3+R_4\,,
\end{align}
where 
\begin{align*}
   &R_1=\big|\bfz^{\T}(\widehat{\bJ}_{\bff^{\bfA_{n}}}^{-1/2}-{\bJ}_{\bff^{\bfA}}^{-1/2})\widehat{\bGamma}_{{\bff}^{\bfA_n}}^{\T}(\tilde{\btheta}_{\calM},\hat{\btheta}_{n,\calM^{\rm c}})\widehat{\bfV}^{-1}_{\bff^{\bfA_n}}(\tilde{\btheta}_{\calM},\hat{\btheta}_{n,\calM^{\rm c}})\bar{\bff}^{\bfA_n}(\btheta_{0}) \big|\,,
   \\&R_2=\big|\bfz^{\T}{\bJ}_{\bff^{\bfA}}^{-1/2}\big[\widehat{\bGamma}_{{\bff}^{\bfA_n}}^{\T}(\tilde{\btheta}_{\calM},\hat{\btheta}_{n,\calM^{\rm c}})-\mathbb{E}\{\widehat{\bGamma}^{\T}_{\bff^{\bfA}}(\btheta_{0})\}\big]\widehat{\bfV}^{-1}_{\bff^{\bfA_n}}(\tilde{\btheta}_{\calM},\hat{\btheta}_{n,\calM^{\rm c}})\bar{\bff}^{\bfA_n}(\btheta_{0})\big|\,,
   \\&R_3=\big|\bfz^{\T}{\bJ}_{\bff^{\bfA}}^{-1/2}\mathbb{E}\{\widehat{\bGamma}^{\T}_{\bff^{\bfA}}(\btheta_{0})\}\{\widehat{\bfV}^{-1}_{\bff^{\bfA_n}}(\tilde{\btheta}_{\calM},\hat{\btheta}_{n,\calM^{\rm c}})-[\mathbb{E}\{\widehat{\bfV}_{\bff^{\bfA}}(\btheta_0)\}]^{-1}\}\bar{\bff}^{\bfA_n}(\btheta_{0})\big|\,,
   \\&R_4=\big|\bfz^{\T}{\bJ}_{\bff^{\bfA}}^{-1/2}\mathbb{E}\{\widehat{\bGamma}^{\T}_{\bff^{\bfA}}(\btheta_{0})\}[\mathbb{E}\{\widehat{\bfV}_{\bff^{\bfA}}(\btheta_0)\}]^{-1}\{\bar{\bff}^{\bfA_n}(\btheta_{0})-\bar{\bff}^{\bfA}(\btheta_{0})\}\big|\,.
\end{align*}
Recall $\bA=(\ba_1^0,\dots, \ba_m^0)^{\T}$ and $\bff_t^{\bfA_n}(\btheta_{0})-\bff_t^{\bfA}(\btheta_{0})=(\bfA_n-\bfA)\bfg_t(\btheta_{0})$. 
By Lemma \ref{tailprabog0}, $|\bar{\bfg}(\btheta_{0})|_\infty=O_{\p}(\aleph_n)$.
 By  Condition \ref{con:alpha},  if $n\aleph_n^2\omega_n^2=o(1)$, we have 
\begin{align*}
|\bar{\bff}^{\bfA_n}(\btheta_{0})-\bar{\bff}^{\bfA}(\btheta_{0})|_2\leq m^{1/2}|\bar{\bfg}(\btheta_{0})|_\infty\max_{k\in[m]}|\ba_k^n-\ba_k^0|_1=O_{\p}(\aleph_n\omega_n)=o_{\p}(n^{-1/2})\,.
\end{align*}
Due to $|\bar{\bff}^{\bfA}(\btheta_{0})|_2=O_{\p}(L_n^{1/2}n^{-1/2})$, then $|\bar{\bff}^{\bfA_n}(\btheta_{0})|_2=O_{\p}(L_n^{1/2}n^{-1/2})$.
By  the Cauchy-Schwarz inequality,
\begin{align*}
n^{1/2}R_1\leq &~
n^{1/2}|\bfz|_2\cdot \|\widehat{\bJ}^{-1/2}_{\bff^{\bfA_{n}}}-\bJ^{-1/2}_{\bff^{\bfA}}\|_2\cdot |\bar{\bff}^{\bfA_n}(\btheta_{0})|_2\cdot O_{\p}(1)\\=&~
O_{\p}(L_n^{1/2}s^{1/2}\ell_n^{1/2}\nu h_n)+O_{\p}(L_n^{1/2}h_n\omega_n)+O_{\p}(L_nh_nn^{-1/2})+O_{\p}(L_n^{5/2}h_n^{-1})\,.
\end{align*}
We can also obtain that
\begin{align*}
&~~~~~~~~~~~~n^{1/2}R_2=O_{\p}(L_n^{1/2}\omega_n)+O_{\p}(L_n^{1/2}s^{1/2}\ell_n^{1/2}\nu)+O_{\p}(L_nn^{-1/2})\,,
\\&n^{1/2}R_3=O_{\p}(L_n^{1/2}\omega_n)+O_{\p}(L_n^{1/2}s^{1/2}\ell_n^{1/2}\nu)+O_{\p}(L_nn^{-1/2})\,~\text{and}~
n^{1/2}R_4=o_{\p}(1)\,.
\end{align*}
 By \eqref{eq:G}, it holds that
\begin{align*}
n^{1/2}|\widehat{G}^{\bfA_{n}}-\bar{Z}_{\bff^{\bfA}}|=
O_{\p}(L_n^{1/2}s^{1/2}\ell_n^{1/2}\nu h_n)+O_{\p}(L_n^{1/2}h_n\omega_n)+O_{\p}(L_nh_nn^{-1/2})+O_{\p}(L_n^{5/2}h_n^{-1})+o_{\p}(1)\,.
\end{align*}
If $L_{n}^{4\varphi}(\log n)^{4}\ll n$, $L_ns\ell_n\nu^2 h_n^2=o(1)$, $L_nh_n^2\omega_n^2=o(1)$, $L_n^2h_n^2\ll n$, $L_n^5\ll h_n^2$ and $n\aleph_n^2\omega_n^2=o(1)$, by (\ref{eq:clt}),  
we have
$
n^{1/2}\widehat{G}^{\bfA_{n}}\rightarrow N(0,1)$ in distribution as 
$n\to \infty$.
  We complete the proof of Lemma \ref{cltZhat}.
$\hfill\Box$

	\section{Proofs of Lemmas \ref{partial sum ex21} and \ref{self-normalized alpha-mixing}}\label{proof of auxiliary lemmas}
	\renewcommand{\thelemma}{S1.\arabic{lemma}} 
\setcounter{lemma}{0} 

\subsection{Proof of Lemma \ref{partial sum ex21}}
\label{secla:partial sum ex2}
Recall $S_{l,h}=\sum_{t=l}^{l+h}\xi_t$.
We first bound $\mathbb{E}(S_{l,h}^2)$.
Due to $\mathbb{E}(S_{l,h}^2)\leq  \sum_{t_1=l}^{l+h}\sum_{t_2=l}^{l+h}|{\rm Cov}(\xi_{t_1},\xi_{t_2})|$, by Davydov's inequality, we have
\begin{align*}
		\mathbb{E}(S_{l,h}^2)\leq &~ \sum_{t=l}^{l+h}\mathbb{E}(\xi_t^2)+24b_1^{1-2/\tilde{\gamma}}\sum_{t_2=l}^{l+h-1}\sum_{t_1=t_2+1}^{l+h}\{\mathbb{E}(|\xi_{t_1}|^{\tilde{\gamma}})\}^{1/\tilde{\gamma}}\{\mathbb{E}(|\xi_{t_2}|^{\tilde{\gamma}})\}^{1/\tilde{\gamma}} \exp\{-b_2(1-2/\tilde{\gamma})\tilde{L}_{\tilde{n}}^{-\varphi}|t_1-t_2|^{\varphi}\}\notag
		\\\leq &~  C h\tilde{L}_{\tilde{n}}\,,
	\end{align*}
    where $C$ is a universal constant independent of $(l,h,\tilde{n})$. We complete the proof of the first result.

Now we begin to bound  $\mathbb{E}(S_{l,h}^{4})$.
Define $\tilde{\alpha}_{\tilde{n}}^{-1}(u)=\sum_{k=1}^{{\infty}}I\{u<\tilde{\alpha}_{\tilde{n}}(k)\}$, where $I(\cdot)$ is the indicator function. Since $\tilde{\alpha}_{\tilde{n}}(k)\leq b_1\exp\{-b_2(\tilde{L}_{\tilde{n}}^{-1}k)^{\varphi}\}$ for any integer $k\geq 1$, then $\tilde{\alpha}_{\tilde{n}}^{-1}(u)\leq b_2^{-1/\varphi}\tilde{L}_{\tilde{n}}\log^{1/\varphi}(b_1u^{-1})$ for any $0<u\leq 1$. Define $Q(u)=\sup_{t\in[\tilde{n}]}Q_t(u)$ with $Q_t(u)=\inf\{x>0:\mathbb{P}(|\xi_t|>x)\leq u\}$. Due to $  \max_{t\in[\tilde{n}]}\mathbb{E}(|\xi_t|^{\tilde{\gamma}})\leq c_1^{\tilde{\gamma}}$, by the Markov inequality, we have $\max_{t\in[\tilde{n}]}\mathbb{P}(|\xi_t|>x)\leq c_1^{\tilde{\gamma}}x^{-\tilde{\gamma}}$, which implies  $Q(u)\leq c_1u^{{-1}/\tilde{\gamma}}$. 
By Theorem 2.2 of \cite{RioE.2017}, 
\begin{align}\label{eq:tiledsn1}
   \mathbb{E}(S_{l,h}^{4})\leq C\underbrace{h^2\bigg[\int_{0}^{1}\{\tilde{\alpha}_{\tilde{n}}^{-1}(u)\wedge h\}Q^2(u)\,{\rm d}u\bigg]^2}_{T_1}+ \,C\underbrace{h\int_{0}^{1}\{\tilde{\alpha}_{\tilde{n}}^{-1}(u)\wedge h\}^3Q^4(u)\,{\rm d}u}_{T_2}\,.
\end{align}
Let $u_*=b_1\exp(-b_2h^{\varphi}\tilde{L}_{\tilde{n}}^{-\varphi})$.
Notice that $\tilde{\alpha}_{\tilde{n}}^{-1}(u)\leq b_2^{-1/\varphi}\tilde{L}_{\tilde{n}}\log^{1/\varphi}(b_1u^{-1})\leq h$ for any $u\geq u_*$.
For the term $T_1$, due to $\tilde{\gamma}>4$ and $\tilde{L}_{\tilde{n}}^{\varphi}\log h\ll h^{\varphi}$, it holds that
\begin{align*}
    |T_1^{1/2}|\leq &~h^2\int_{0}^{u_*}c_1^{2}u^{-2/\tilde{\gamma}}\,{\rm d}u+h\int_{u_*}^{1}b_2^{-1/\varphi}\tilde{L}_{\tilde{n}}\log^{1/\varphi}(b_1u^{-1})c_1^{2}u^{-2/\tilde{\gamma}}\,{\rm d}u
\\\leq &~Ch^2\int_{0}^{u_*}u^{-2/\tilde{\gamma}}\,{\rm d}u+Ch\tilde{L}_{\tilde{n}}\int_{0}^{1}\log^{1/\varphi}(b_1u^{-1})u^{-2/\tilde{\gamma}}\,{\rm d}u
\\\leq &~Ch^2\exp\{-b_2(1-2/\tilde{\gamma})h^{\varphi}\tilde{L}_{\tilde{n}}^{-\varphi}\}+Ch\tilde{L}_{\tilde{n}}\int_{0}^{\infty}u^{\varphi}\exp\{-(1-2/\tilde{\gamma})u^{\varphi}\}\,{\rm d}u\leq  Ch\tilde{L}_{\tilde{n}}\,,
\end{align*}
which implies that $T_1\leq  Ch^2\tilde{L}^2_{\tilde{n}}$.
For the term $T_2$,  due to $\tilde{\gamma}>4$ and $\tilde{L}_{\tilde{n}}^{\varphi}\log h\ll h^{\varphi}$, it holds that
\begin{align*}
   T_2\leq &~h^4\int_{0}^{u_*}c_1^{4}u^{-4/\tilde{\gamma}}\,{\rm d}u+h\int_{u_*}^{1}b_2^{-3/\varphi}\tilde{L}_{\tilde{n}}^3\log^{3/\varphi}(b_1u^{-1})c_1^{4}u^{-4/\tilde{\gamma}}\,{\rm d}u
\\\leq &~Ch^4\int_{0}^{u_*}u^{-4/\tilde{\gamma}}\,{\rm d}u+Ch\tilde{L}_{\tilde{n}}^3\int_{0}^{1}\log^{3/\varphi}(b_1u^{-1})u^{-4/\tilde{\gamma}}\,{\rm d}u
\\\leq &~Ch^4\exp\{-b_2(1-4/\tilde{\gamma})h^{\varphi}\tilde{L}_{\tilde{n}}^{-\varphi}\}+Ch\tilde{L}_{\tilde{n}}^3\int_{0}^{\infty}u^{\varphi/3}\exp\{-(1-4/\tilde{\gamma})u^{\varphi/3}\}\,{\rm d}u\\\leq &~Ch\tilde{L}_{\tilde{n}}^3\leq C h^2\tilde{L}_{\tilde{n}}^2\,.
\end{align*}
By \eqref{eq:tiledsn1}, we have the second result. $\hfill\Box$

 \subsection{Proof of Lemma \ref{self-normalized alpha-mixing}}\label{sec:la:self-normalized alpha-mixing}

	 Let  
		$\{U_j\}_{j=3}^{\tilde{k}_{\tilde{n}}}$ be a sequence of  independent random variables with uniform distribution over $[0,1]$, independent of the sequence $\{Y_j\}^{\tilde{k}_{\tilde{n}}}_{j=1}$. 
        Let $m_2=\lfloor \tilde{k}_{\tilde{n}}/2 \rfloor$ and $m_1=\tilde{k}_{\tilde{n}}-m_2$.
		By Theorem 3 of \cite{Bradley(1983)}, we can claim that there exist a sequence of  measurable functions $\{f_j\}_{j=3}^{\tilde{k}_{\tilde{n}}}$
   such that $Y_1^*=Y_1$,  $Y_2^*=Y_2$, $Y_{2j-1}^*=f_{2j-1}(Y_1,Y_3,\dots,Y_{2j-3},Y_{2j-1},U_{2j-1})$ for any $2\leq j\leq m_1$ and $Y_{2j}^*=f_{2j}(Y_2,Y_4,\dots,Y_{2j-2},Y_{2j},U_{2j})$ for any  $2\leq j\leq m_2$, satisfying three properties:
		
		$(\rm{a})$ For each $j\in[\tilde{k}_{\tilde{n}}]$, $Y_j^*$ has the same distribution as $Y_j$.
		
		$(\rm{b})$  $\{Y_{2j-1}^*\}_{j=1}^{m_1}$ are independent random variables, and 
		$\{Y_{2j}^*\}_{j=1}^{m_2}$ are also independent random variables.
		
		$(\rm{c})$ There exists a universal constant $C>0$ such that  
		\begin{align}\label{eq:Yj}
			\mathbb{P}\big(|Y_{j}^*-Y_j|\geq x\big)\leq C (\tilde{m}^{1/2}\tilde{L}_{\tilde{n}}^{1/2}x^{-1})^{2/5}\exp\{-4b_2(\tilde{L}_{\tilde{n}}^{-1}\tilde{m})^{\varphi}/5\}
            \end{align}
		for any $0<x\leq c_2$ and $j\in[\tilde{k}_{\tilde{n}}]$. 
        
 Recall $V_{\tilde{k}_{\tilde{n}}}^2=\sum_{j=1}^{\tilde{k}_{\tilde{n}}}Y_{j}^2$ and $S_{\tilde{n}}=S_{\tilde{n}}^{(1)}+S_{\tilde{n}}^{(2)}$ with
$S_{\tilde{n}}^{(1)}=\sum_{j=1}^{m_1}Y_{2j-1}$ and  $S_{\tilde{n}}^{(2)}=\sum_{j=1}^{m_2}Y_{2j}$. For any $x>0$, we have 
	\begin{align}\label{eq:tildeSn}
\mathbb{P}\big(S_{\tilde{n}}\geq 4xV_{\tilde{k}_{\tilde{n}}}\big)\leq \mathbb{P}\big\{S_{{\tilde{n}}}^{(1)}\geq 2xV_{\tilde{k}_{\tilde{n}}}\big\}+\mathbb{P}\big\{S_{\tilde{n}}^{(2)} \geq 2xV_{\tilde{k}_{\tilde{n}}}\big\}\,.
	\end{align}   Let $S_{\tilde{n}}^{(1),*}=\sum_{j=1}^{m_1}Y_{2j-1}^*$,  $S_{\tilde{n}}^{(2),*}=\sum_{j=1}^{m_2}Y_{2j}^*$, $V_{(1)}^{*,2}=\sum_{j=1}^{m_1}Y_{2j-1}^{*,2}$, $V_{(2)}^{*,2}=\sum_{j=1}^{m_2}Y_{2j}^{*,2}$, $V_{(1)}^2=\sum_{j=1}^{m_1}Y_{2j-1}^2$ and $V_{(2)}^2=\sum_{j=1}^{m_2}Y_{2j}^2$. 
	Let $\epsilon=\exp\{-\tilde{c}_1(\tilde{L}_{\tilde{n}}^{-1}\tilde{m})^{\varphi}\}$ for  some universal constant $\tilde{c}_1\in(0,b_2/5)$. Since $\tilde{m}\gg\tilde{L}_{\tilde{n}}$, then $\epsilon\leq c_2$ for sufficiently large $\tilde{n}$.  For any $x>0$,
	it holds that  \begin{align}\label{eq:Sn1}
		\mathbb{P}\big\{S_{\tilde{n}}^{(1)}\geq 2xV_{\tilde{k}_{\tilde{n}}}\big\}&\leq \mathbb{P}\bigg\{\sum_{j=1}^{m_1}Y_{2j-1}\geq 2xV_{(1)}\bigg\}\notag\\&\leq \mathbb{P}\bigg\{\sum_{j=1}^{m_1}Y_{2j-1}\geq 2xV_{(1)},|Y_{2j-1}^{*}-Y_{2j-1}|\leq  \epsilon,|Y_{2j-1}^{*,2}-Y_{2j-1}^2|\leq  \epsilon, \forall j\in [m_1]\bigg\}\notag\\&~~~~~~~+\sum_{j=1}^{m_1}\mathbb{P}\big(|Y_{2j-1}^{*}-Y_{2j-1}|> \epsilon\big)+\sum_{j=1}^{m_1}\mathbb{P}\big(|Y_{2j-1}^{*,2}-Y_{2j-1}^2|> \epsilon\big)
		\\&\leq 
	\underbrace{\mathbb{P}\big[S_{\tilde{n}}^{(1),*}\geq 2x\{V_{(1)}^{*,2}-m_1\epsilon\}^{1/2}-m_1\epsilon, V_{(1)}^{*,2}\geq m_1\epsilon\big]}_{\rm I}+\underbrace{\mathbb{P}\big\{V_{(1)}^{*,2}\leq m_1\epsilon\big\}}_{\rm II}\notag\\&~~~~~~~+\underbrace{\sum_{j=1}^{m_1}\mathbb{P}\big(|Y_{2j-1}^{*}-Y_{2j-1}|> \epsilon\big)}_{\rm III}+\underbrace{\sum_{j=1}^{m_1}\mathbb{P}\big(|Y_{2j-1}^{*,2}-Y_{2j-1}^2|> \epsilon\big)}_{\rm IV}\notag\,.
	\end{align}
	As we will show in Sections \ref{sec:proof 34} and \ref{sec:proof 12}, if $\tilde{L}_{\tilde{n}}^{\varphi}\log(m_1\tilde{m})\ll \tilde{m}^{\varphi}$,
    \begin{align}
   {\rm III}+{\rm IV}\lesssim &~m_1\tilde{m}\tilde{L}_{\tilde{n}}\exp\{-2\tilde{c}_1(\tilde{L}_{\tilde{n}}^{-1}\tilde{m})^{\varphi}\}\,,\label{eq:34}\\
    {\rm I}+{\rm II}\lesssim&~ \mathbb{P}\big\{S_{\tilde{n}}^{(1),*}\geq xV_{(1)}^*\big\}+ \exp\{-\tilde{c}_2m_1\tilde{L}_{\tilde{n}}^{-\check{\gamma}/(\check{\gamma}-2)}\}\label{eq:12}\,.
    \end{align}
Hence, by \eqref{eq:Sn1}, for any $x\geq {\epsilon}^{1/2}$, we have 
\begin{align}\label{eq:Sn11}
		\mathbb{P}\big\{S_{\tilde{n}}^{(1)}\geq 2xV_{\tilde{k}_{\tilde{n}}}\big\}\lesssim&~ \mathbb{P}\big\{S_{\tilde{n}}^{(1),*}\geq xV_{(1)}^*\big\}+m_1\tilde{m}\tilde{L}_{\tilde{n}}\exp\{-2\tilde{c}_1(\tilde{L}_{\tilde{n}}^{-1}\tilde{m})^{\varphi}\}+\exp\{-\tilde{c}_2m_1\tilde{L}_{\tilde{n}}^{-\check{\gamma}/(\check{\gamma}-2)}\}
	\end{align}
	 provided that $\tilde{L}_{\tilde{n}}^{\varphi}\log(m_1\tilde{m})\ll \tilde{m}^{\varphi}$.
	Notice that $
\sum_{j=1}^{m_1}\mathbb{E}(Y_{2j-1}^{*,2})\gtrsim m_1\tilde{m}$. By  Lemma \ref{partial sum ex21} and H\"{o}lder inequality,  if $\tilde{L}_{\tilde{n}}^{\varphi}\log \tilde{m}\ll \tilde{m}^{\varphi}$, it holds that 
	\begin{align*}
		\frac{\sum_{j=1}^{m_1}\mathbb{E}(|Y_{2j-1}^*|^{2+\tilde{\delta}})}{\{\sum_{j=1}^{m_1}\mathbb{E}(Y_{2j-1}^{*,2})\}^{(2+\tilde{\delta})/2}}\leq Cm_1^{-\tilde{\delta}/2}\tilde{L}_{\tilde{n}}^{(2+\tilde{\delta})/2}
	\end{align*}
    for any $0<\tilde{\delta} \leq 1$.
	Hence, if $\tilde{L}_{\tilde{n}}^{\varphi}\log \tilde{m}\ll \tilde{m}^{\varphi}$ and $\tilde{L}_{\tilde{n}}\ll m_1^{\tilde{\delta}/(2+\tilde{\delta})} $, by Theorem 2.3 of \cite{JingShaoWang2003},
	\begin{align}\label{self norm Sn}
		\frac{\mathbb{P}\{S_{\tilde{n}}^{(1),*}\geq xV_{(1)}^*\}}{1-\Phi(x)}=1+O(1)(1+x)^{2+\tilde{\delta}}m_1^{-\tilde{\delta}/2}\tilde{L}_{\tilde{n}}^{(2+\tilde{\delta})/2}
	\end{align}
	for any $c_2^{1/2}\leq x\leq C m_1^{\tilde{\delta}/(4+2\tilde{\delta})}\tilde{L}_{\tilde{n}}^{-1/2}$, where $\Phi(\cdot)$ is the cumulative distribution function of the standard normal distribution. Write $\phi(x)=({2\pi})^{-1/2}\exp(-x^2/2)$.
	Due to  $1-\Phi(x)\geq x\phi(x)/(1+x^2)  $ for any $x>0$, by \eqref{eq:Sn11} and \eqref{self norm Sn}, if $\tilde{L}_{\tilde{n}}^{\varphi}\log(m_1\tilde{m})\ll \tilde{m}^{\varphi}$ and $\tilde{L}_{\tilde{n}}\ll m_1^{\tilde{\delta}/(2+\tilde{\delta})}$, then 
	\begin{align*}
	\frac{\mathbb{P}\{S_{\tilde{n}}^{(1)}\geq 2xV_{\tilde{k}_{\tilde{n}}}\}}{1-\Phi(x)}\lesssim&~ 1+(1+x)^{2+\tilde{\delta}}m_1^{-\tilde{\delta}/2}\tilde{L}_{\tilde{n}}^{(2+\tilde{\delta})/2}+(x^{-1}+x)m_1\tilde{m}\tilde{L}_{\tilde{n}}\exp\{-2\tilde{c}_1(\tilde{L}_{\tilde{n}}^{-1}\tilde{m})^{\varphi}+x^2/2\}\\&~~~+(x^{-1}+x)\exp\{-\tilde{c}_2m_1\tilde{L}_{\tilde{n}}^{-\check{\gamma}/(\check{\gamma}-2)}+x^2/2\}
	\end{align*}
	holds uniformly over   $c_2^{1/2}\leq x\leq Cm_1^{\tilde{\delta}/(4+2\tilde{\delta})}\tilde{L}_{\tilde{n}}^{-1/2}$.  Let $a_*=1/(3\varphi+1)$. Due to $\tilde{m}\asymp \tilde{n}^{a_*}$, $m_1\asymp \tilde{n}^{1-a_*}$ and $\tilde{L}_{\tilde{n}}\ll \tilde{n}$,   we have 
		\begin{align*}
		\frac{\mathbb{P}\{S_{\tilde{n}}^{(1)}\geq 2xV_{\tilde{k}_{\tilde{n}}}\}}{1-\Phi(x)}\lesssim&~1
	\end{align*}
  holds uniformly over  $c_2^{1/2} \leq  x\leq c_*\min\{{\tilde{n}}^{(1-a_*)\tilde{\delta}/(4+2\tilde{\delta})}\tilde{L}_{\tilde{n}}^{-1/2},{\tilde{n}}^{a_*\varphi/2}\tilde{L}_{\tilde{n}}^{-\varphi/2}, \tilde{n}^{(1-a_*)/2}\tilde{L}_{\tilde{n}}^{-\check{\gamma}/(2\check{\gamma}-4)}\}$ with  some sufficiently small  constant $c_*>0$,
	provided that 
    $\tilde{L}_{\tilde{n}}\ll \tilde{n}^{(1-a_*)\tilde{\delta}/(2+\tilde{\delta})}$, $\tilde{L}_{\tilde{n}}^{\varphi}\log \tilde{n}\ll \tilde{n}^{a_*\varphi}$ and $\tilde{L}_{\tilde{n}}^{\check{\gamma}/(\check{\gamma}-2)}\ll \tilde{n}^{1-a_*}$.
	Due to $1-\Phi(x)\leq 2\exp(-x^2/2)/(1+x)$ for any $x\geq 0$, if $\tilde{L}_{\tilde{n}}\ll \tilde{n}^{(1-a_*)\tilde{\delta}/(2+\tilde{\delta})}$, $\tilde{L}_{\tilde{n}}^{\varphi}\log \tilde{n}\ll \tilde{n}^{a_*\varphi}$ and $\tilde{L}_{\tilde{n}}^{\check{\gamma}/(\check{\gamma}-2)}\ll \tilde{n}^{1-a_*}$, then 
	\begin{align}\label{eq:tildesn1}
		\mathbb{P}\big\{S_{\tilde{n}}^{(1)}\geq 2xV_{\tilde{k}_{\tilde{n}}}\big\}\lesssim  \exp(-x^2/2)
	\end{align}
	holds uniformly over  
    $c_2^{1/2}\leq x\leq c_*\min\{{\tilde{n}}^{(1-a_*)\tilde{\delta}/(4+2\tilde{\delta})}\tilde{L}_{\tilde{n}}^{-1/2},{\tilde{n}}^{a_*\varphi/2}\tilde{L}_{\tilde{n}}^{-\varphi/2}, \tilde{n}^{(1-a_*)/2}\tilde{L}_{\tilde{n}}^{-\check{\gamma}/(2\check{\gamma}-4)}\}$.
	Analogously, if $\tilde{L}_{\tilde{n}}\ll \tilde{n}^{(1-a_*)\tilde{\delta}/(2+\tilde{\delta})}$, $\tilde{L}_{\tilde{n}}^{\varphi}\log \tilde{n}\ll \tilde{n}^{a_*\varphi}$ and $\tilde{L}_{\tilde{n}}^{\check{\gamma}/(\check{\gamma}-2)}\ll \tilde{n}^{1-a_*}$, we also have 
	\begin{align}\label{eq:tildesn2}
		\mathbb{P}\big\{S_{\tilde{n}}^{(2)}\geq 2xV_{\tilde{k}_{\tilde{n}}}\big\}\lesssim  \exp(-x^2/2)
	\end{align}
 uniformly over   $c_2^{1/2}\leq x\leq c_*\min\{{\tilde{n}}^{(1-a_*)\tilde{\delta}/(4+2\tilde{\delta})}\tilde{L}_{\tilde{n}}^{-1/2},{\tilde{n}}^{a_*\varphi/2}\tilde{L}_{\tilde{n}}^{-\varphi/2}, \tilde{n}^{(1-a_*)/2}\tilde{L}_{\tilde{n}}^{-\check{\gamma}/(2\check{\gamma}-4)}\}$. To make $\min\{{\tilde{n}}^{(1-a_*)\tilde{\delta}/(4+2\tilde{\delta})}\tilde{L}_{\tilde{n}}^{-1/2},{\tilde{n}}^{a_*\varphi/2}\tilde{L}_{\tilde{n}}^{-\varphi/2}, \tilde{n}^{(1-a_*)/2}\tilde{L}_{\tilde{n}}^{-\check{\gamma}/(2\check{\gamma}-4)}\}$ as large as possible, we select $(\tilde{\delta},\check{\gamma})=(1,4)$. Recall $a_*=1/(3\varphi+1)$ with $\varphi\geq1$.
	Combining \eqref{eq:tildeSn}, \eqref{eq:tildesn1} and  \eqref{eq:tildesn2}, then if $\tilde{L}^{\varphi}_{\tilde{n}}\log\tilde{n}\ll \tilde{n}^{\varphi/(3\varphi+1)}$, then
	\begin{align*}
		\mathbb{P}\big(S_{\tilde{n}}\geq 4xV_{\tilde{k}_{\tilde{n}}}\big)\lesssim  \exp(-x^2/2)
	\end{align*}
	holds uniformly over $c_2^{1/2}\leq x\leq c_*{\tilde{n}}^{\varphi/(6\varphi+2)}\tilde{L}_{\tilde{n}}^{-\varphi/2}$.
	We complete the proof of Lemma \ref{self-normalized alpha-mixing}.
	$\hfill\Box$

    \subsubsection{Proof of \eqref{eq:34}} \label{sec:proof 34}
	By Lemma \ref{partial sum ex21}, $\mathbb{E}(Y_{2j-1}^{*,2})=\mathbb{E}(Y_{2j-1}^2)\lesssim \tilde{m}\tilde{L}_{\tilde{n}}$ for any $j\in[m_1]$. Recall $\epsilon=\exp\{-\tilde{c}_1(\tilde{L}_{\tilde{n}}^{-1}\tilde{m})^{\varphi}\}$ for  some universal constant $\tilde{c}_1\in(0,b_2/5)$. For a sufficiently large $\tilde{n}$, by \eqref{eq:Yj}  and the Markov inequality,
	\begin{align*}
		\mathbb{P}\big(|Y_{2j-1}^{*,2}-Y_{2j-1}^2|> \epsilon\big)\leq&~ \mathbb{P}\big(|Y_{2j-1}^{*}-Y_{2j-1}|> \epsilon^2\big)+\mathbb{P}\big(|Y_{2j-1}^{*}+Y_{2j-1}|> \epsilon^{-1}\big)\\\lesssim&~
		(\tilde{m}^{1/2}\tilde{L}_{\tilde{n}}^{1/2}\epsilon^{-2})^{2/5}\exp\{-4b_2(\tilde{L}_{\tilde{n}}^{-1}\tilde{m})^{\varphi}/5\}+\tilde{m}\tilde{L}_{\tilde{n}}\exp\{-2\tilde{c}_1(\tilde{L}_{\tilde{n}}^{-1}\tilde{m})^{\varphi}\}\\\leq&~  \tilde{m}^{1/5}\tilde{L}_{\tilde{n}}^{1/5}\exp\{-2b_2(\tilde{L}_{\tilde{n}}^{-1}\tilde{m})^{\varphi}/5\}+\tilde{m}\tilde{L}_{\tilde{n}}\exp\{-2\tilde{c}_1(\tilde{L}_{\tilde{n}}^{-1}\tilde{m})^{\varphi}\}
		\\\lesssim&~\tilde{m}\tilde{L}_{\tilde{n}}\exp\{-2\tilde{c}_1(\tilde{L}_{\tilde{n}}^{-1}\tilde{m})^{\varphi}\}\,,
	\end{align*}
	which implies   
	\begin{align*}
	{\rm IV}=\sum_{j=1}^{m_1}\mathbb{P}\big(|Y_{2j-1}^{*,2}-Y_{2j-1}^2|> \epsilon\big)\lesssim m_1\tilde{m}\tilde{L}_{\tilde{n}}\exp\{-2\tilde{c}_1(\tilde{L}_{\tilde{n}}^{-1}\tilde{m})^{\varphi}\}\,.
	\end{align*}
          	Analogously, we have ${\rm III}\lesssim m_1\tilde{m}\tilde{L}_{\tilde{n}}\exp\{-2\tilde{c}_1(\tilde{L}_{\tilde{n}}^{-1}\tilde{m})^{\varphi}\}$. $\hfill\Box$
            
     \subsubsection{Proof of \eqref{eq:12}}\label{sec:proof 12}       
            Notice that 
	\begin{align}\label{eq:I1}
		{\rm I}\leq&~
		\mathbb{P}\big[S_{\tilde{n}}^{(1),*}\geq 2x\{V_{(1)}^{*,2}-m_1\epsilon\}^{1/2}-m_1\epsilon,V_{(1)}^{*,2}>4m_1\epsilon\big]+ \mathbb{P}\big\{V_{(1)}^{*,2}\leq 4m_1\epsilon\big\}\notag
		\\\leq&~ \mathbb{P}\big\{S_{\tilde{n}}^{(1),*}\geq \sqrt{3}xV_{(1)}^*-m_1\epsilon\big\}+ \mathbb{P}\big\{V_{(1)}^{*,2}\leq 4m_1\epsilon\big\}
		\\\leq&~\mathbb{P}\big\{S_{\tilde{n}}^{(1),*}\geq \sqrt{3}xV_{(1)}^*-m_1\epsilon,xV_{(1)}^*>4m_1\epsilon\big\}\notag+\mathbb{P}\big\{xV_{(1)}^*\leq 4m_1\epsilon\big\}+\mathbb{P}\{V_{(1)}^{*,2}\leq 4m_1\epsilon\}\notag
		\\\leq&~\mathbb{P}\big\{S_{\tilde{n}}^{(1),*}\geq xV_{(1)}^*\big\}+\mathbb{P}\big\{xV_{(1)}^*\leq 4m_1\epsilon\big\}+\mathbb{P}\big\{V_{(1)}^{*,2}\leq 4m_1\epsilon\big\}\,.\notag
	\end{align} 
	For any $2<\check{\gamma}\leq 4$, if $\tilde{L}_{\tilde{n}}^{\varphi}\log \tilde{m}\ll \tilde{m}^{\varphi}$, 
    by  Lemma \ref{partial sum ex21} and H\"{o}lder inequality  $\mathbb{E}(|Y_{2j-1}^*|^{\check{\gamma}})=\mathbb{E}(|Y_{2j-1}|^{\check{\gamma}})\lesssim \tilde{m}^{\check{\gamma}/2}\tilde{L}_{\tilde{n}}^{\check{\gamma}/2}$, 
which implies  
 \begin{align*}
		\sum_{j=1}^{m_1}\mathbb{E}(|Y_{2j-1}^*|^{\check{\gamma}})=\sum_{j=1}^{m_1}\mathbb{E}(|Y_{2j-1}|^{\check{\gamma}})\lesssim	m_1\tilde{m}^{\check{\gamma}/2}\tilde{L}_{\tilde{n}}^{\check{\gamma}/2}\,.
	\end{align*}
Recall $V_{(1)}^{*,2}=\sum_{j=1}^{m_1}Y_{2j-1}^{*,2}$ and $\epsilon=\exp\{-\tilde{c}_1(\tilde{L}_{\tilde{n}}^{-1}\tilde{m})^{\varphi}\}$.	Due to  $\mathbb{E}\{V_{(1)}^{*,2}\}=\mathbb{E}\{V_{(1)}^{2}\}\gtrsim m_1\tilde{m}$, if $\tilde{L}_{\tilde{n}}^{\varphi}\log \tilde{m}\ll \tilde{m}^{\varphi}$, by Lemma 7.3 of \cite{chen2016}, 
	\begin{align}\label{Vk inq2}
		\mathbb{P}\big\{V_{(1)}^{*,2}\leq4 m_1\epsilon\big\}\notag&=\mathbb{P}\big[V_{(1)}^{*,2}-\mathbb{E}\{V_{(1)}^{*,2}\}\leq4 m_1\epsilon-\mathbb{E}\{V_{(1)}^{*,2}\}\big]\\&\leq\exp\bigg(-\frac{(\check{\gamma}-2)[\mathbb{E}\{V_{(1)}^{*,2}\}-4m_1\epsilon]^{\check{\gamma}/(\check{\gamma}-2)}}{8\{\sum_{j=1}^{m_1}\mathbb{E}(|Y_{2j-1}^*|^{\check{\gamma}})\}^{2/(\check{\gamma}-2)}}\bigg)\leq\exp\{-\tilde{c}_2m_1\tilde{L}_{\tilde{n}}^{-\check{\gamma}/(\check{\gamma}-2)}\}\,,
	\end{align}
 where $\tilde{c}_2>0$ is a universal  constant.
	Analogous to \eqref{Vk inq2}, for any $x\geq {\epsilon}^{1/2}$,  we have that
	\begin{align*}
		\mathbb{P}\big\{xV_{(1)}^*\leq 4m_1\epsilon\big\}&\leq \mathbb{P}\big\{V_{(1)}^{*,2}\leq 16m_1^2\epsilon\big\}\\&\leq\exp\bigg(-\frac{(\check{\gamma}-2)[\mathbb{E}\{V_{(1)}^{*,2}\}-16m_1^2\epsilon]^{\check{\gamma}/(\check{\gamma}-2)}}{8\{\sum_{j=1}^{m_1}\mathbb{E}(|Y_{2j-1}^*|^{\check{\gamma}})\}^{2/(\check{\gamma}-2)}}\bigg)\leq \exp\{-\tilde{c}_2m_1\tilde{L}_{\tilde{n}}^{-\check{\gamma}/(\check{\gamma}-2)}\}
	\end{align*}
	provided that $\tilde{L}_{\tilde{n}}^{\varphi}\log(m_1\tilde{m})\ll \tilde{m}^{\varphi}$.
	Together with \eqref{eq:I1} and \eqref{Vk inq2},  if $\tilde{L}_{\tilde{n}}^{\varphi}\log(m_1\tilde{m})\ll \tilde{m}^{\varphi}$, we have 
	\begin{align*}
		{\rm I}\lesssim \mathbb{P}\big\{S_{\tilde{n}}^{(1),*}\geq xV_{(1)}^*\big\}+ \exp\{-\tilde{c}_2m_1\tilde{L}_{\tilde{n}}^{-\check{\gamma}/(\check{\gamma}-2)}\}
	\end{align*}
    for any $x\geq{\epsilon}^{1/2}$.
 Parallel to \eqref{Vk inq2}, we also have ${\rm II}\lesssim \exp\{-\tilde{c}_2m_1\tilde{L}_{\tilde{n}}^{-\check{\gamma}/(\check{\gamma}-2)}\}$ provided that $\tilde{L}_{\tilde{n}}^{\varphi}\log \tilde{m}\ll \tilde{m}^{\varphi}$. $\hfill\Box$

\end{document}